\documentclass{emulateapj}
\usepackage{epsfig,graphicx,times}
\begin{document}
\submitted{Accepted to The Astrophysical Journal}
\journalinfo{Accepted to The Astrophysical Journal}

\title{Compact Quiescent Galaxies at Intermediate Redshifts\altaffilmark{1,}\altaffilmark{2}}
\shorttitle{Compact Quiescent Galaxies at Intermediate Redshifts}
\shortauthors{Hsu, Stockton \& Shih}
\author{Li-Yen Hsu, Alan Stockton, and Hsin-Yi Shih\altaffilmark{3}}
\affil{Institute of Astronomy, University of Hawaii, 2680 Woodlawn Drive, Honolulu, HI 96822, USA}
\altaffiltext{1}{Some of the data presented herein were obtained at the W.M. Keck Observatory, which is operated as a scientific partnership among the California Institute of Technology, 
the University of California and the National Aeronautics and Space Administration. The Observatory was made possible by the generous financial support of the W.M. Keck Foundation.}
\altaffiltext{2}{Based in part on data collected at Subaru Telescope, which is operated by the National Astronomical Observatory of Japan.}
\altaffiltext{3}{Now at Gemini Observatory, 670 N. Aohoku Pl., Hilo, HI 96720, USA}

\begin{abstract}

From several searches of the area common to the Sloan Digital Sky Survey and the United Kingdom Infrared Telescope Infrared Deep Sky Survey, we have selected 22 luminous galaxies between $z \sim$ 0.4 and $z \sim$ 0.9 that 
have colors and sizes similar to those of the compact quiescent galaxies at $z>2$. By exploring structural parameters and stellar populations, we found that most of these galaxies actually formed most of their stars at $z<2$ and are generally less compact than those found at $z > 2$. Several of these young objects are disk-like or possibly prolate. This lines up with several previous studies which found that massive quiescent galaxies at high redshifts often have disk-like morphologies. If these galaxies were to be confirmed to be disk-like, their formation mechanism must be able to account for both compactness and disks. On the other hand, if these galaxies were to be confirmed to be prolate, the fact that prolate galaxies do not exist in the local universe would indicate that galaxy formation mechanisms have evolved over cosmic time. We also found five galaxies forming over 80\% of their stellar masses at $z>2$. Three of these galaxies appear to have been modified to have spheroid-like morphologies, in agreement with the scenario of ``inside-out" buildup of massive galaxies. The remaining galaxies, SDSS\,J014355.21+133451.4 and SDSS\,J115836.93+021535.1, have truly old stellar populations and disk-like morphologies. These two objects would be good candidates for nearly unmodified compact quiescent galaxies from high redshifts that are worth future study.

\end{abstract}

\section{Introduction}
 
The structural evolution of massive galaxies over cosmic time provides constraints on galaxy formation models. In the local universe, galaxies with stellar masses greater than $10^{11} M_{\odot}$ 
are mostly early-type galaxies \citep{Baldry2004Quantifying-the,Buitrago2013Early-type}, which usually have old stellar populations and large sizes. However, many recent studies have found evidence that massive galaxies at $z > 2$ showing little or no recent star formation are generally very compact, with effective radii $R_e$ $<$ 2 kpc (e.g., \citealt{Stockton2004A-Disk-Galaxy-o,Daddi2005Passively-Evolv,Trujillo2006The-Size-Evolut,Trujillo2007Strong-size-evo,Toft2007Hubble-Space-Te,2008ApJ...677L...5V,Buitrago2008Size-Evolution-,2009ApJ...695..101D,Bruce2012The-morphologies,van-der-Wel20143D-HSTCANDELS:-}). Proposed mechanisms to create such compact galaxies from star-forming progenitors generally involve violent, dynamical processes such as gas-rich mergers \citep{Hopkins2006A-Unified-Merge} or dynamical instabilities fed by cold streams \citep{Dekel2009Formation-of-Ma}. 

Because these compact and quiescent galaxies are extremely rare in the present-day universe (e.g., \citealt{Trujillo2009Superdense-Mass,2010ApJ...720..723T}), significant size growth must have taken place over the past billions of years, and stochastic events such as major mergers cannot be the dominant mechanisms (e.g., \citealt{Bezanson2009The-Relation-Be,Lopez-Sanjuan2010The-Minor-Role-}). It has been proposed that minor mergers (e.g., \citealt{2006ApJ...648L..21K,Naab2009Minor-Mergers-a}) or ``puffing-up'' driven by active galactic nuclei (AGNs) feedback (e.g., \citealt{2008ApJ...689L.101F,Fan2010Cosmic-Evolutio}) have turned these high-redshift compact galaxies into present-day ellipticals. Most of the recent studies, however, favor the former mechanism, proposing the so-called “inside-out” buildup of present-day massive galaxies (e.g., \citealt{Bezanson2009The-Relation-Be,2009MNRAS.398..898H,van-Dokkum2010The-Growth-of-M,2011MNRAS.415.3903T,van-de-Sande2013Stellar-Kinemat}). In this scenario, the high-redshift compact galaxies become the cores of the most massive local galaxies after accreting envelopes over time.

It is very difficult to study the compact quiescent galaxies at $z>2$ in detail because of their faintness, resulting from cosmological dimming, and their low surface brightness, due to the turnover of the 
redshift-angular-size-distance relation at high redshifts. As a consequence, many recent studies have focused on identifying and characterizing some similar objects at lower redshifts (e.g., \citealt{Trujillo2009Superdense-Mass,2010ApJ...720..723T,2010ApJ...709L..58S,2011ApJ...733...45S,Ferre-Mateu2012Young-ages,Poggianti2013Superdense-Gala,Damjanov2013Discovery-of-Ni,Trujillo2014NGCnbsp1277:-A-,Stockton2014A-Search-for-Mo}), where images and spectra with high quality are accessible for studying morphologies and detailed properties such as stellar populations, kinematics and metallicities. Making use of the Sloan Digital Sky Survey (SDSS), \citet{Trujillo2009Superdense-Mass} found that the number of massive ($M_{*} > 8 \times 10^{10} M_{\odot}$) and compact ($r_{e} < 1.5$ kpc) galaxies is less than 0.03\% at $z<0.2$. Another search by \citet{2010ApJ...720..723T} at $0.066<z<0.12$ returned no candidates as massive and as compact as those identified at high redshifts. \cite{Trujillo2009Superdense-Mass}, \cite{Ferre-Mateu2012Young-ages} and \cite{Damjanov2013Discovery-of-Ni} all found young ages for the compact galaxies they identified, again suggesting that most of the high-redshift compact galaxies have already turned into large ellipticals in the local universe.

Nevertheless, the stochastic nature of merging events means that there should be a number of galaxies that formed at high redshifts that have remained unmodified until today \citep{Quilis2013Expected-Number}. 
Recently \cite{Trujillo2014NGCnbsp1277:-A-} and \cite{Stockton2014A-Search-for-Mo} have successfully found compact galaxies that may have survived with little or no modification from the population formed at high 
redshifts; further morphological and kinematic studies of these objects would provide us more insight into their formation mechanisms in the early universe.

Given the rareness of compact quiescent galaxies in the local universe, it would be useful to explore intermediate redshifts, where the number density of these objects is expected to 
be higher. The epoch between $z \sim$ 0.5 and $z \sim$ 1.0 covers a large period of cosmic time, but it is nearly unexplored in searches for compact quiescent galaxies. In this work, we 
present our results of 22 compact quiescent galaxies identified between $z \sim$ 0.4 and $z \sim$ 0.9 from several searches of the area common to SDSS and the United Kingdom Infrared Telescope (UKIRT) Infrared Deep Sky Survey (UKIDSS). The paper is structured as follows. Section~\ref{sec:data} describes our object selection and follow-up observations. The analyses and results are shown in detail in Section~\ref{sec:results}. In Section~\ref{sec:discussion}, we 
discuss our results and their implications. Section~\ref{sec:summary} summarizes our results. Throughout this paper, we assume a \citet{2003PASP..115..763C} initial mass functions (IMF) and the 
concordance $\Lambda$CDM cosmology with $\rm H_0=70~km~s^{-1}~Mpc^{-1}$, $\rm\Omega_M=0.3$, and $\Omega_\Lambda=0.7$. All magnitudes used are AB magnitudes.

\section{Object Selection and Observations}\label{sec:data}

\subsection{Object Selection}

Fifteen galaxies (hereafter, sample 1) we study in this work are from several searches for compact quiescent galaxies at $0.4 \lesssim z \lesssim 0.6$ since 2009; the other seven galaxies 
(hereafter, sample 2) are from a new search carried out this year for $0.4 \lesssim z \lesssim 1.0$. Our method for selecting objects at $0.4 \lesssim z \lesssim 0.6$ is given in \citet{2010ApJ...709L..58S}. 
In the following, we describe details of the new search performed this year.

We selected objects from the area common to UKIDSS DR9plus and SDSS DR8 for right ascensions between $8^{\rm h}$ and $16^{\rm h}$. In order to generate the expected colors of old stellar populations, a \citeauthor{2003MNRAS.344.1000B} (2003; hereafter BC03) instantaneous burst model with solar metallicity and an age of 5 Gyr was used as our spectral template. We then produced thirteen redshifted templates between $z$ = 0.4 and $z$ = 1.0, stepping 
in intervals of 0.05. Convolving these redshifted templates through the nine filters of SDSS/UKIDSS photometry \citep{Hewett2006The-UKIRT-Infra} then generated magnitudes and therefore colors. Using the Structured Query 
Language (SQL) for data in the Wide Field Camera Science Archive (WSA) on UKIRT, we searched the combined UKIDSS/SDSS database for objects (1) that have colors matching our template at a given redshift, (2) that are compact 
enough such that $-0.1$ $<$ UKIDSS $K$(Petrosian) -- $K$(1$''$ aperture) $<$ 0.3, and (3) that are at least one magnitude brighter than $L^{*}$\footnote{We use the term ``$L^{*}$ galaxy'' to mean a galaxy at a given redshift that will, through passive evolution alone, end up as an $L^{*}$ galaxy at the present epoch \citep{Huang2003The-HawaiiAnglo}; i.e., we do not attempt to take into account any evolution of the shape of the luminosity function. Operationally, we take a present-day early-type $L^{*}$ galaxy represented by a \cite{2003MNRAS.344.1000B} model formed at $z=9$ and follow the SED back in redshift, determining expected magnitudes at each interval in $z$.} for elliptical galaxies in $H$ band at that redshift. Separate SQL searches were made at the thirteen redshifts between 0.4 and 1.0. Based on UKIDSS $J_1$-band images of objects that passed the SQL, we selected our candidates for follow-up observations by choosing compact ones whose surface brightness profiles resemble nearby stars in the same field.

Our candidate selection procedure was not able to distinguish compact galaxies from stars and QSOs. We therefore needed to examine these candidates with deep and high-resolution images to select objects for 
spectroscopic follow-up. Based on our Keck and Subaru observations, we removed all the stars and QSOs, as well as galaxies that have effective radii $R_e$ $>$ 3 kpc. Our final sample in this paper comprises 
22 galaxies with both imaging and spectroscopic data available, allowing the determination of their redshifts and effective radii.

\subsection{Imaging}

The galaxies in sample 1 were imaged in $H$ or $K'$ band on various nights from 2009 to 2011 with the NIRC2 camera and the laser-guide-star adaptive optics system (LGSAO; \citealt{Wizinowich2006The-W.-M.-Keck-}) on the Keck II telescope. These fifteen fields were all selected to have at least one good signal-to-noise star in the field of view for determining the point-spread function (PSF) of the image. The exposure times ranged from 60 s $\times$ 5 to 180 s $\times$ 9, and the image scale is $0 \farcs04$ pixel$^{-1}$. For the galaxies in sample 2, we obtained $I$-band images with the imaging mode of the Faint Object Camera and Spectrograph (FOCAS; \citealt{Kashikawa2002FOCAS:-The-Fain}) on the Subaru telescope on March 4 and 5, 2013. The exposure times range from 60 s $\times$ 5 to 60 s $\times$ 15, and the image scale is $\sim 0 \farcs 1$ pixel$^{-1}$. We reduced our data with IRAF following standard procedures including bias subtraction and flat-fielding. Individual dithered images were then registered and combined with the drizzle algorithm \citep{Fruchter2002Drizzle:-A-Meth}.

\subsection{Spectroscopy}

We carried out ground-based spectroscopy for fourteen galaxies in sample 1 with the Low-Resolution Imaging Spectrograph (LRIS; \citealt{Oke1995The-Keck-Low-Re}) on the 
Keck I telescope and one galaxy (SDSSJ081053) with the Echellette Spectrograph and Imager (ESI; \citealt{Sheinis2002ESI-a-New-Keck-}) on the Keck II telescope. The LRIS spectra were 
obtained with the 600 line mm$^{-1}$ grating (FWHM resolution: 4.7 $\rm{\AA}$) blazed at 5000 or 7500 $\rm{\AA}$ on the red side of the spectrograph. The ESI spectrum was taken with the echellette mode using the 175 line mm$^{-1}$ grating (FWHM resolution: 1.3 $\rm{\AA}$). For the objects in sample 2, we took their spectra with the spectroscopic mode of FOCAS on the Subaru telescope using the VPH850 grism and SO58 filter (FWHM resolution: 11 $\rm{\AA}$). Standard data reduction procedures including bias subtraction, flat-fielding, sky subtraction, wavelength and flux calibrations were performed with IRAF to extract the 1D spectra. Spectroscopic redshifts of galaxies are determined by cross-correlation between reduced spectra and spectral templates in the IDL routine SPECPRO developed by \citet{2011PASP..123..638M}. The spectra of SDSSJ012942, SDSSJ081053 and SDSSJ235219 do not have 
well-calibrated continua due to calibration problems, but the detection of absorption lines and the 4000 $\rm{\AA}$ break still allows precise determination of redshifts. In Table~\ref{table1}, we summarize our imaging and spectroscopic observations.

\section{Data Analysis and Results}\label{sec:results}

\subsection{Morphologies}\label{sec:galfit}

From our Keck/NIRC2 and Subaru/FOCAS imaging, we explored the structural parameters of our objects with {\sc GALFIT} \citep{2002AJ....124..266P,Peng2010Detailed-Decomp}, a routine for determining models of the 
two-dimensional galaxy profile by minimizing the $\chi^2$ residuals. For each AO image, we determined the PSF profile from the nearest unsaturated star, usually within 25\arcsec\ of the galaxy. We have done tests in globular 
cluster fields and have found that on most nights the PSF does not change significantly for our purposes over this distance. In any case, in agreement with \cite{Carrasco2010Gemini-K-band-N}, we have found that the basic 
galaxy structural parameters are fairly robust against uncertainties in the PSF core width. For the FOCAS imaging, the PSF is essentially invariant over a large field, so a suitable PSF star can always be found. Using the PSFs 
we determined, we first fitted single-S${\rm \acute{e}}$rsic models \citep{Sersic1968Atlas-de-galaxi} to the galaxy images, as shown in the first rows of all the objects in Figures~\ref{keckimages} and \ref{focasimages}. One exception 
was for SDSSJ011004, where there is a companion very close to the galaxy, so we simply included this object in the fit using another single-S${\rm \acute{e}}$rsic model. These fits resulted in circularized effective radii of 
$R_e <$ 3 kpc for all the objects. 

We chose the best GALFIT results by both visual inspection of the residuals and the reduced $\chi^2$ of the fits. For eight galaxies, the single-component fits left significant systematic residuals. Three of these eight galaxies have unusual S${\rm \acute{e}}$rsic indices of $n > 6$, which is a clear indication that double-S${\rm \acute{e}}$rsic models are needed. We therefore performed two-component fits for these eight objects, as shown in the second rows of eight objects in Figure~\ref{keckimages}, leading to better residuals and reduced $\chi^2$. However, two of these galaxies, SDSSJ014355 and SDSSJ115836, both turned out to have best-fit S${\rm \acute{e}}$rsic indices of $n \sim 4$ and $n < 0.5$ for their two components. Since $n < 0.5$ implies an unphysical central dip in the three-dimensional stellar distribution, we forced $n = 0.5$ for their second components and ran the fits again. This resulted in $n=4.09$ and $n=3.90$ for their first components, which remain close to the $r^{1/4}$ law. We also tried adding a second component to the fits for all the other galaxies. Nevertheless, the reduced $\chi^2$ either remained roughly unchanged or became smaller 
but further away from unity (a sign of ``over-fitting"), suggesting that single-S${\rm \acute{e}}$rsic models are sufficient.

Spectroscopic redshifts determined from the reduced galaxy spectra (Table~\ref{table2}) allow us to convert the effective radii in unit of pixels to physical scales based on the assumed cosmology in this paper. Tables~\ref{table2} and \ref{table3}  summarize the best-fit parameters of single-S${\rm \acute{e}}$rsic and double-S${\rm \acute{e}}$rsic models, respectively. In the rest of this paper, we will use the circularized effective radius as a proxy for galaxy size. The practice of using the circularized effective radius has traditionally been used to account for the uncertainty of the projection of ellipticals and spheroids due to their triaxiality.

\subsection{Spectral Energy Distributions and Stellar Populations} 

To constrain the stellar populations of the galaxies, we used {\sc FAST} \citep{Kriek2009An-Ultra-Deep-N} to fit BC03 models simultaneously to SDSS/UKIDSS magnitudes and 
flux-calibrated spectra for most of our objects. For four galaxies, we only fitted models to SDSS/UKIDSS magnitudes due to low S/N or flux calibration problems of the spectra.
We used BC03 models with exponentially declining star formation rates, all with \citet{2003PASP..115..763C} IMF, Calzetti reddening law \citep{2000ApJ...533..682C}, and metallicities 
[Z/H] of $-0.4$, 0.0 and 0.4. All the magnitudes and spectra were corrected for galactic extinction according to NASA/IPAC Extragalactic Database (NED), and redshifts were fixed at the 
spectroscopic redshifts in the fits. The best-fit model parameters as well as the corresponding mean ages and stellar masses of galaxies are tabulated in Table~\ref{table4}. Notice that the ``age" 
of a BC03 model with exponentially declining star formation rate represents the age since the onset of star formation. On the other hand, the mean age of the stellar population is the age since one $\tau$ 
($e$-folding timescale for the star formation rate) after the onset of star formation.

\subsection{Full-Spectrum Fitting}\label{sec:ppxf}

\subsubsection{Stellar Populations}\label{sec:ppxf_pop}

It is well-known that an age-metallicity degeneracy exists in the determination of stellar populations from broadband photometry. Given that one old and low-metallicity spectrum may have the same shape, and therefore same broadband colors, as a young and high-metallicity spectrum, breaking this degeneracy relies on the differences in detail at absorption lines between similar spectra. To do this, the full-spectrum fitting method is recently becoming a popular alternative to using line-strength indices, thanks to the availability of high-quality spectral-synthesis models. 

We therefore used the Penalized Pixel-Fitting method (pPXF) by \citet{Cappellari2004Parametric-Reco} to constrain the stellar populations from ten spectra that have high S/N and well-calibrated continua. 
For each galaxy, a grid of BC03 instantaneous burst models were used with three metallicities and ages stepping in intervals of $\sim$0.25 Gyr from 0.005 Gyr to the maximum age younger than the age of universe at the galaxy redshift.  We shifted the spectra to the rest frame, corrected for galactic extinction, logarithmically rebinned the wavelength grid, and masked bad pixels. Model templates were broadened with a Gaussian to match the instrumental resolution of the de-redshifted spectra. The best-fit solution provided by pPXF is a distribution of the mass fraction in different ages and metallicity intervals, i.e., a linear combination of different BC03 models. This allows us to obtain probabilities of star formation at different cosmic time instead of a single age for each of the galaxies we analyzed. Running pPXF involves choosing a regularization parameter that affects the smoothness of the solutions. Here we assume that the initial starbursts of these galaxies were intense and rather brief since strong dissipation must have been involved to account for their compactness. We chose regularization parameters that led to fairly narrow distributions for the major star-formation periods, rather than the smoothest possible distribution consistent with the spectra. However, it should be noted that our choice is an assumption instead of a result.

Figure~\ref{spectra} shows the fits and the corresponding distributions of star formation over cosmic time. We calculated the mean mass-weighted as well as mean luminosity-weighted ages and metallicities of these distributions, as shown in Table~\ref{table5}. The age of a galaxy determined by FAST should be considered as a luminosity-weighted average result. Our mean luminosity-weighted ages based on the full-spectrum fits roughly agree with the mean ages given by FAST (column 4 of Table~\ref{table4}). We can also calculate the stellar mass of a galaxy from the optimized linear combination of BC03 models. In order to account for the flux loss of the slit, we scaled our best-fit model to match the broadband photometry. The derived masses are listed in column 7 of Table~\ref{table5}.

The uncertainties in the derived quantities are difficult to estimate, since it is likely that they will be dominated by systematic effects that are largely unknown, such as the accuracy of the models. In general, variations within 
an allowed range from the minimum $\chi^2$ value, governed by the regularization parameter, typically change the width (and sometimes slightly the ages) of the star-forming episodes, without changing the stellar mass by very 
much. In any case, at present, it is impractical to attempt to derive the random uncertainties for population analyses via automated Monte Carlo simulations (such as we are able to use for measuring the velocity dispersions, as 
described in Section~\ref{sec:ppxf_kinematics}) because of the required tuning of the spectral noise level and the regularization parameter. Thus, the models and the associated derived parameters can only be taken as indicative. To 
some extent (again subject to uncertainty of the models), confidence in the general accuracy of the solution can be judged by the agreement of different approaches, such as the degree of agreement of the luminosity-weighted ages 
from pPXF with the ages given by FAST, as mentioned above.

We have performed some additional tests for the case of SDSSJ014355, which, as discussed in Section~\ref{sec:old}, is the best candidate from the current samples of a mostly intact survivor from the high-redshift compact galaxy 
population. We tried running pPXF on this spectrum while constraining the stellar metallicities to single values of [Z/H] = $-$0.4, 0.0 and 0.4. Although our original fit, as well as these single-metallicity fits, depend on both the continuum 
shape and sharp absorption features, when we eliminate the effect of the continuum by subtracting a low-order fit to the residuals (observed spectrum minus model), testing only the fit to absorption lines and breaks, the residual 
noise is significantly lower in our original multi-metallicity fit than in any of the others. Because of the dominance of a low-metallicity population in the best fit, this object is useful for testing the influence of the well-known age--metallicity 
degeneracy. The [Z/H] = $-$0.4 model, as expected, is dominated by a maximally old population. The solar-metallicity model is strongly dominated by a starburst peaking at about 6.3 Gyr, or at a redshift of $\sim 2.9$. Only the 
super-solar model has a substantial fraction of the stellar mass formed at low redshifts, with about half with an age of $\sim 3$ Gyr, and most of the other half with an age of $\sim 8$ Gyr (formed at $z \sim 9.3$). However, all of 
these single-metallicity models are unrealistic in terms of chemical evolution in galaxies and they are significantly worse fits to the observed spectrum.

\subsubsection{Velocity Dispersions and Dynamical Masses}\label{sec:ppxf_kinematics}

We also used the pPXF code to calculate the velocity dispersions of these eleven galaxies. In order to estimate the errors, we followed \cite{Toft2012Deep-Absorption} and ran Monte Carlo simulations in the following way. We subtracted the best-fit model from the spectrum, and the residuals were randomly rearranged in wavelength space and added to the best-fit model to create 100 mock spectra. This led to a distribution of measured velocity dispersions, and the standard deviation is taken as the 1-$\sigma$ error. 

Combining the measurements of effective radii and velocity dispersions, we were able to estimate dynamical masses and compare the values with the stellar masses inferred from the stellar populations. 
For spheroids, the equation of the dynamical mass is 
\begin{equation}
M_{\rm dyn} = \beta R_e \sigma^2 / G ~, 
\label{eq}
\end{equation}
where $\beta$ is a parameter that depends mainly on the structure of the galaxy. \cite{Cappellari2006The-SAURON-proj} found that $\beta$ = $5.0 \pm 0.1$ accurately reproduces galaxy 
dynamical masses for local ellipticals, and this value is commonly used in the literature. In columns 8 and 9 of Table~\ref{table5}, we tabulate the velocity dispersions and the dynamical masses assuming $\beta$ = 5.
In this calculation, we adopted Equation (1) of \cite{Cappellari2006The-SAURON-proj} to correct the measured velocity dispersion $\sigma$ to $\sigma_e$, which would be the velocity 
dispersion measured within $R_e$. However, we can see clear discrepancies between the dynamical and stellar masses estimates. In agreement with what has been found in several 
recent studies for compact massive galaxies (e.g., \citealt{2010ApJ...709L..58S,Martinez-Manso2011Velocity-Disper,Ferre-Mateu2012Young-ages}), the stellar masses of most galaxies are unphysically larger than their dynamical masses. This indicates that we cannot assume homology between our galaxies and local ellipticals studied by \cite{Cappellari2006The-SAURON-proj}, and the assumption that $\beta$ = 5 may not be correct. This discrepancy caused by different galaxy structures is recently reinforced by \cite{Peralta-de-Arriba2013The-discrepancy}, who find an empirical relation between $\beta$ ($K$ in their paper) and the compactness of galaxies. In the last column of Table~\ref{table5} , we tabulate the dynamical masses with the correction in \cite{Peralta-de-Arriba2013The-discrepancy}. Most of these values are about two times the corresponding stellar masses; the median value of 
dynamical-to-stellar-mass ratios is 1.84.

\section{Discussion and Conclusion}\label{sec:discussion}

\subsection{Current Star-Formation Rates} 

Our selection procedure was designed to eliminate objects with significant current star formation, and our spectra indicate that this goal has been achieved. None of our spectra shown here have detectible [\ion{O}{2}] $\lambda3727$ emission (for SDSS J105745, we do not observe this spectral region). In order to estimate rough upper limits to the star formation rates, we consider a typical galaxy with $z=0.6$, $r=21.6$, for which the spectrum gives a S/N $= 12$ at 4200 \AA. We model a $2\sigma$ line at the position of [\ion{O}{2}] $\lambda3727$, which gives a flux of $2\times10^{-17}$ ergs s$^{-1}$ cm$^{-2}$. Using the relation between [\ion{O}{2}] flux and star-formation rate (SFR) given by \citeauthor{Kewley2004O-II-as-a-Star-} (2004; eq. 4), we obtain an upper limit to the SFR of 0.2 $M_{\sun}$ yr$^{-1}$ for this case. We have only one object at a substantially higher redshift, but for that case we also have about twice the S/N that we have assumed. Even taking into account various plausible uncertainties, we can place a conservative SFR upper limit of the galaxies in our sample of $<1 M_{\sun}$yr$^{-1}$.

\subsection{Mass-Size Relations and Ages}  

In Figure~1, we plot the mass-size relations of our galaxies and compare them with SDSS galaxies at $0.05<z<0.07$ from \citet{Franx2008Structure-and-S}, the compact quiescent 
galaxies found at $z > 2$ from \citet{2008ApJ...677L...5V}, and the five $z \sim 0.5$ galaxies from \citet{Stockton2014A-Search-for-Mo}. In this plot, most of our galaxies locate in the area between SDSS sample and the extremely compact galaxies from the other two papers; some of them are indistinguishable from the SDSS sample. This implies that most of our galaxies are not survivors of high-redshift population of compact quiescent 
galaxies. Our constraint on stellar populations with FAST suggests that half of the galaxies in our sample have mean ages $\lesssim $ 2 Gyr, which lines up with some recent studies that found young ages 
for local compact quiescent galaxies (e.g., \citealt{Trujillo2009Superdense-Mass,Ferre-Mateu2012Young-ages}).

\begin{figure*}
\begin{center}    
\includegraphics[width=13cm]{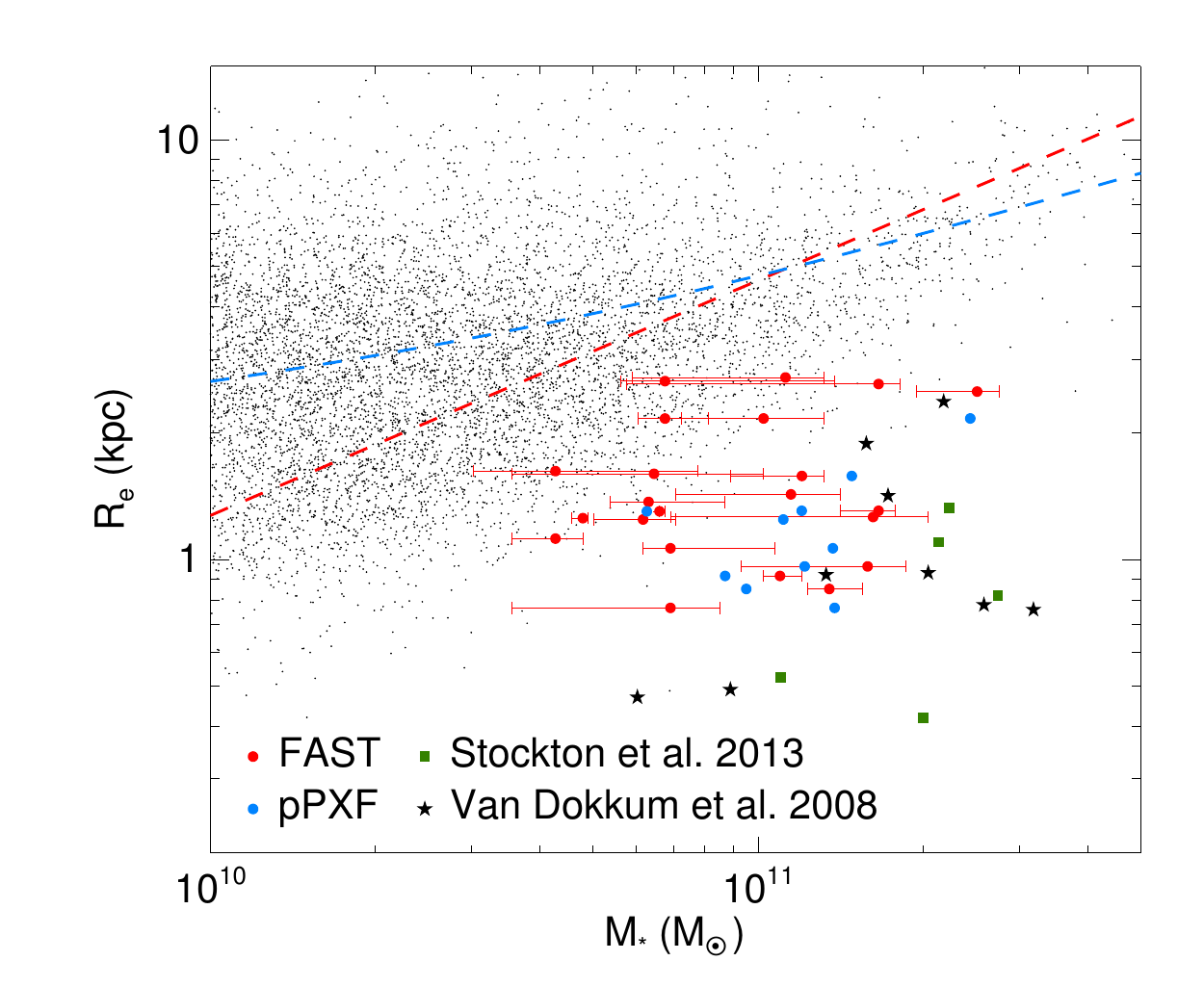}   
\caption{Relations between the stellar masses determined by FAST (red; all the 22 galaxies) or pPXF (blue; 10 galaxies) and the sizes for our sample. Notice that we omit the errors of 
the circularized effective radii since their error bars are smaller than or comparable to the sizes of red and blue data points in this plot. Small dots are SDSS galaxies at $0.05<z<0.07$ from 
\citet{Franx2008Structure-and-S}. Red and blue dashed lines indicate the mass-size relations for local early-type and late-type galaxies, respectively, from \cite{Shen2003The-size-distri}. Green squares are 
the five galaxies from \citet{Stockton2014A-Search-for-Mo}. Black stars are the $z>2$ compact quiescent galaxies from \citet{2008ApJ...677L...5V}, for 
which we omit the errors.}  
\end{center}
\label{figure1}
\end{figure*}

\subsection{Disk-Like Galaxies}\label{sec:disk}

Many of our galaxies show signs of disks based on the morphological analysis. We classify the eight galaxies fitted by two-component models (Table~\ref{table3}) as disk-like galaxies
because each of their best-fit models is either a superposition of two components with $n <$ 2.5, or a compact core with $n >$ 2.5 plus an extended component with $n <$ 2.5. 
For six out of these eight galaxies (the exception being SDSSJ011004 and SDSSJ115836), the axis ratio $b/a$ of the more extended component is much lower than the axis ratios of the other component and 
the single-S${\rm \acute{e}}$rsic model. As a result, these six objects have significant residuals along their major axes in one-component fits as shown in Figure~\ref{keckimages}. For SDSSJ115836, the 
residual in the one-component fit has a ring-like distribution, and both components in the double-S${\rm \acute{e}}$rsic model have $b/a >$ 0.5. We believe the morphology of this object is similar to the above 
six galaxies except that it is relatively face-on. In addition, there are four galaxies (SDSSJ115027, SDSSJ124257, SDSSJ135342 and SDSSJ164916) described by single-S${\rm \acute{e}}$rsic models 
with $n < 2.5$. We therefore classify these four objects as another four disk-like galaxies.

In total, we have 12 out of 22 (55\%) galaxies that are disk-like. We can compare this fraction with the result of \cite{Buitrago2013Early-type}, who calculated the fractions of massive 
galaxies showing disk-like surface brightness profiles ($n<$ 2.5) and spheroid-like ones ($n>$ 2.5) as a function of redshift between $z$ = 0 and $z$ = 3. The median redshift of our sample is $\sim$ 0.6, which 
corresponds to a disk-like fraction of $\sim$ 25\% according to Figure 7 in \cite{Buitrago2013Early-type}. Our sample of compact quiescent galaxies has a higher disk-like fraction than do massive 
galaxies generally at the same redshift range. This high disk-like fraction could be a result of young ages we found in our sample. Besides SDSSJ014355, SDSSJ115836, SDSSJ135342 and SDSSJ235219, 
two-thirds of our disk-like galaxies have fairly young ages, or at least have significant star formation in the past 2 Gyr as indicated by our full-spectrum fits (SDSSJ084223 and SDSSJ155037). Our result lines up with 
previous studies \citep{Stockton2004A-Disk-Galaxy-o,Stockton2008Morphologies-of,2008ApJ...677L...5V,2008ApJ...682..303M,van-der-Wel2011The-Majority-of,Chang2013Shape-Evolution,Chang2013Structural-Evol,2013MNRAS.428.1088M} which found that massive quiescent galaxies at high redshifts often have disk-like morphologies.

Some simulations have shown that gas-rich mergers can produce disky remnants (e.g., \citealt{Hopkins2009How-do-Disks-Su,Puech2012Galaxy-Disks-Do}). However, what these simulations produced are large thin disks, 
which do not resemble the compact galaxies discovered at high redshifts or in this work. This is because a significant amount of gas needs to be driven to large radii, where it does not feel strong torques from the merger and retains its angular momentum to form disks. If these galaxies indeed have rotating and disky structures, their formation mechanism should involve extreme dissipation to have gas settle onto a disk before converting into stars, but it also has to be rapid enough to account for their compactness.

Another possibility is that these galaxies are actually prolate, with radial orbits mostly aligned with the long axis. As suggested by \citet{Stockton2014A-Search-for-Mo}, such a morphology 
could probably explain the discrepancy between the calculated stellar mass and the dynamical mass estimated from the relation that works well for local elliptical galaxies. Also, the fact 
that prolate galaxies do not exist locally is an indication that galaxy formation mechanisms may be different at different cosmic times. Further work, including high-spatial-resolution 
spectroscopy, is needed to examine whether or not these compact galaxies are rotationally supported. Based on integral field spectroscopic measurement of H$\alpha$ emission 
lines, \citet{2014MNRAS.439.1494B} have found evidence of rotational support for 10 massive galaxies at $z \sim$ 1.4 and confirmed that half of them are rotating disks. Although their 
sample is not selected to be compact and quiescent, as ours is (only having absorption-line spectra), this work does support the idea that disks are common among massive galaxies at 
high redshifts.

\subsection{Old Galaxies}\label{sec:old}

Our pPXF fits show that SDSSJ014355, SDSSJ101009, SDSSJ115836 and SDSSJ105745 have more than 90\% (by mass) of their stellar populations formed at $z > 2$. The best-fit BC03 model with 
exponentially declining star formation rate from FAST for SDSSJ134412 also suggests formation of $\sim$ 80\% of the stellar mass at $z>2$. This indicates that these galaxies are possibly slightly modified survivors of the high-redshift population of massive compact quiescent galaxies. On the other hand, the old ages of SDSSJ235219, SDSSJ091515 and SDSSJ135342 determined by FAST have large uncertainties, so deeper spectroscopic data for full-spectrum fits are needed to better constrain their stellar population properties.

\citet{van-Dokkum2010The-Growth-of-M} suggest that the mass growth of massive galaxies is due to a gradual buildup of outer envelopes over time around the compact cores observed at high redshifts. 
Parameterizing the surface brightness profiles of massive galaxies, they found that both the effective radius and S${\rm \acute{e}}$rsic index increase towards low redshifts. Based on this scenario of “inside-out” 
growth of massive galaxies, old quiescent galaxies in the local universe are expected to have high S${\rm \acute{e}}$rsic indices and not be disk-like. The S${\rm \acute{e}}$rsic indices of SDSSJ101009, SDSSJ134412 and SDSSJ105745 are consistent with this scenario. Therefore, these three galaxies had likely been modified and do not show their original structural properties. The absence of stellar disks of these galaxies may be a result of major mergers or a sequence of many minor mergers \citep{Bournaud2007Multiple-minor-}. 

SDSSJ014355 and SDSSJ115836, in contrast, have different morphologies; they are classified as disk-like galaxies that are fitted by two-component models (Section~\ref{sec:disk}). The pPXF model for SDSSJ014355 
has $\sim$ 3\% of $\sim$ 500-Myr-old star formation being added to a dominant population formed at $z>6$, and it has a much smaller size than the other four old galaxies we discuss here ($R_e = 0.77$ kpc). 
The pPXF model for SDSSJ115836 also indicates $\sim$ 3\% of $\sim$ 1-Gyr-old star formation and a dominant population formed at $z>3$. It is possible that these two galaxies are nearly unaltered relics from the high-redshift 
massive compact galaxy population. Especially, the extreme compactness of SDSSJ014355 might be a result of formation in dense environment at the very early universe. If these galaxies were indeed disk-like survivors 
from high redshift, it would again support the idea that massive quiescent galaxies at high redshifts are often disk-like. However, the GALFIT modes of these two objects both consist of cores with $n \sim 4$ and fainter, extended 
disks. Therefore, it is also possible that the extended components are caused by recent gas accretion and star formation added onto the old compact cores. In any case, it would be very useful to perform spatially resolved 
spectroscopy or multi-wavelength imaging to compare the ages or colors of the cores and the outer envelopes, determining whether such structures are intrinsic to the formation process or modified by more recent star formation.

\section{Summary}\label{sec:summary}

Our search for luminous compact quiescent galaxies at $0.4 < z <1.0$ has returned a sample of objects that allows detailed studies with imaging and spectroscopic follow-up. Although this search 
is far from being complete and unbiased, it does offer a way for us to infer the formation of massive galaxies at $z > 2$ and to study their structural change at lower redshifts. Our result again suggests 
that unmodified relics from the population of high-redshift compact quiescent galaxies are indeed rare; most of the massive galaxies we found actually formed at $z<2$, where the density of 
universe was lower than high redshifts. These young galaxies are less compact than those found at $z > 2$ in the literature. Interestingly, several of these young objects appear to be disk-like or 
possibly prolate. If these galaxies were to be confirmed to be disks, their formation mechanism should involve extreme dissipation to have gas settle onto a disk before converting into stars. The 
models for forming massive compact galaxies through major mergers would need more tests in order to account for both compact and disk-like structures. On the other hand, if these galaxies were to be confirmed to be 
prolate, it would indicate that galaxy formation mechanisms have evolved over cosmic time. More studies need to be done to determine the actual morphologies of these galaxies.

Three out of five of the galaxies that formed more than 80\% of their masses at $z>2$ based on our analysis have morphologies similar to the local ellipticals. Building up of these galaxies by mergers or accretion might have already happened and therefore altered their structures. In contrast,  SDSSJ014355 and SDSSJ115836 are likely nearly unmodified disk-like galaxies that are worth more future work. Recently, \cite{Trujillo2014NGCnbsp1277:-A-} has discovered that NGC1277, a lenticular galaxy at a distance of only $\sim$ 73 Mpc, is an unmodified relic from the high-redshift massive compact galaxy population. One interesting characteristic of  NGC1277 is its low S${\rm \acute{e}}$rsic index of $n=2.2$, again indicating a disk-like structure. It would be of great interest to obtain detailed IFU observations to examine the stellar dynamics of this galaxy. This discovery also raises some hope of finding more such type of galaxies in the local universe where we can perform very detailed analysis.

\acknowledgments

We thank the anonymous referee for a careful and thoughtful reading of the original version of this paper and for offering numerous suggestions to improve both its substance and presentation. We also thank Allison Man for 
helpful discussions. This research has been partially supported by NSF grant AST-0807900. The UKIDSS project is defined in \cite{Lawrence2007The-UKIRT-Infra}. UKIDSS uses the UKIRT Wide 
Field Camera (WFCAM; \citealt{Casali2007The-UKIRT-wide-}) and a photometric system described in \cite{Hewett2006The-UKIRT-Infra}. The pipeline processing and science archive are described 
in \cite{Hambly2008The-WFCAM-Scien}. Funding for the SDSS has been provided by the Alfred P. Sloan Foundation, the Participating Institutions, the National Science Foundation, the U.S. Department 
of Energy, the National Aeronautics and Space Administration, the Japanese Monbukagakusho, the Max Planck Society, and the Higher Education Funding Council for England. The SDSS website 
is found at http://www.sdss.org/. 

\emph{Facilities}: Keck:I (LRIS), Keck:II (LGSAO/NIRC2/ESI), Sloan (SDSS), Subaru (FOCAS), UKIRT (UKIDSS)

% ====================================================================
%                       TABLES
% ====================================================================

\begin{table*}
\caption{Summary of Observations} 
\begin{center}
\begin{tabular}{cccccccccccccc}
\hline \hline
                      & SDSS & \multicolumn{5}{c}{Imaging}  &   \multicolumn{3}{c}{Spectroscopy}  \\
Object Name & $r$\footnote{SDSS $r$-band magnitude} & Date & Exposure & Filter & mag\footnote{Magnitude of the object in the corresponding filter} & 1$\sigma$\footnote{1$\sigma$ limiting magnitude of each image. A 2$''$ diameter aperture and a 5$''$ diameter aperture are used to calculate the magnitudes for NIRC2 and FOCAS images, respectively.}     & Date & Exposure & S/N\footnote{Median value of the signal-to-noise ratio per pixel at the red side of 4000 $\rm{\AA}$ break.} \\
                      &  & (UT)  &     (s)      &     &   &    &   (UT)  &   (s) & \\
\hline
SDSS\,J011004.73+140933.0 & 21.40 &20090915 & 1080 &$H$   & 18.81 &23.66  &20090822  & 3600 & 23.6 \\   
SDSS\,J012942.10+132420.8 & 21.35 &20110826 &   300  & $H$ & 19.00 &23.48 &20100816   & 1200 & 11.6 \\    
SDSS\,J014355.21+133451.4 & 20.93 &20110826 &  300 &$H$    & 18.83 &23.48  &20100906   & 3600 & 24.9 \\     
SDSS\,J081053.07+230443.7 & 21.36 &20120328 &  1080 &$K'$ & 18.86 &24.29 & 20120314(ESI)  & 900  & 10.8 \\   
SDSS\,J084223.93+050223.4 & 20.72 &20110426 &  1080 &$H$  & 18.72 &24.30 & 20110426   & 2400 & 19.0 \\    
SDSS\,J084616.69+052833.6 & 20.45 &20110426 &  1080 &$H$  & 18.61 &24.09 & 20110426   & 2400 & 19.5 \\    
SDSS\,J101009.25+062324.8 & 21.22 &20110426 &  1620 &$H$  & 18.83 &25.28 & 20110426   & 3600 & 13.5\\     
SDSS\,J115836.93+021535.1 & 21.46 &20100406 &  1080 &$H$  & 19.03 &24.31  & 20100408   & 3420 & 11.3 \\  
SDSS\,J121954.26+003025.2 & 20.21 &20100406 &  960 &$H$    & 18.54 &24.24  &20100320   & 600 & 6.0  \\  
SDSS\,J123106.94+053347.6 & 20.93 &20110426 &  1440 &$H$  & 18.73 &24.19 &20110426   & 3600 & 19.9\\   
SDSS\,J124257.04+102400.7 & 20.73 &20110426 &  1080 &$H$  & 18.61 &23.78 & 20110426  & 4800 & 23.6 \\   
SDSS\,J134412.30+010906.6 & 20.56 &20110426 &  1620 &$H$  & 18.48 &23.86 &20110426  & 3600 & 9.3 \\   
SDSS\,J155037.84+024746.5 & 21.04 &20100406 &  1080 &$H$  & 18.74 &24.67  &20100320   & 3600 & 15.5 \\    
SDSS\,J164916.12+294309.0 & 20.54 &20100406 &  840 &$H$    & 18.79 &23.73  &20100408   & 600 &   10.5 \\     
SDSS\,J235219.98$-$004855.7 & 21.08 & 20090915  & 840 &$H$& 18.66 &23.98  & 20090822   & 1200 & 5.2\\     
\hline
SDSS\,J091515.64+055256.9 & 23.76 &20130304  & 300 & $I$&  21.62 &26.32 & 20130304  &600 & 3.0 \\
SDSS\,J104224.46+022225.0 & 21.85 &20130305 &  300 & $I$&  20.74 &25.99 & 20130305   &900 & 10.2 \\
SDSS\,J104630.88$-$010759.0 & 21.60 & 20130304  & 300 & $I$& 20.36 &26.08 &20130304   &600 & 6.0 \\
SDSS\,J105745.85$-$005818.6 & 21.53 & 20130305  & 600 & $I$& 20.41 &26.15 &20130304\&0305   &4200 & 24.9 \\   
SDSS\,J115027.99+025118.1 & 21.54 &20130304  & 300 & $I$& 20.58  &26.29  &20130304   &600 & 5.2 \\
SDSS\,J132953.78+295140.0 & 22.89 &20130304  & 300 & $I$& 21.44 &26.33  &20130304   &900 & 4.8 \\
SDSS\,J135342.06+262157.1 & 21.56 &20130304  & 300 & $I$& 20.31 &26.22  &20130304   &900 & 5.3 \\
\hline\hline
\end{tabular} 
\end{center}
\tablecomments{For the upper block (sample 1), all the objects were imaged with Keck II/NIRC2, and spectra were taken with Keck I/LRIS except for SDSS\,J081053.07+230443.7, which 
was taken with Keck II/ESI. All the objects in the lower block (sample 2) were observed with Subaru/FOCAS for both imaging and spectroscopic modes.}
\label{table1}
\end{table*}

\begin{table*}
\caption{Spectroscopic redshifts and parameters of single-S${\rm \acute{e}}$rsic models}
\begin{center}
\begin{tabular}{ccccc}
\hline \hline
Object Name & $z_{spec}$ & $R_e$ & $n$ & $b/a$  \\
                         &                      &  (kpc) &            &     \\
\hline
SDSS\,J011004.73+140933.0 & 0.801 & 0.91$\pm$0.00 & 2.92$\pm$0.01 & 0.37   \\   
SDSS\,J012942.10+132420.8 & 0.582 & 1.12$\pm$0.01 & 3.05$\pm$0.03 & 0.56  \\   
SDSS\,J014355.21+133451.4 & 0.487 & 0.77$\pm$0.01 & 3.77$\pm$0.04 & 0.29   \\ 
SDSS\,J081053.07+230443.7 & 0.636 & 1.37$\pm$0.01 & 6.51$\pm$0.05 & 0.27   \\  
SDSS\,J084223.93+050223.4 & 0.557 & 1.25$\pm$0.00 & 1.92$\pm$0.01 & 0.26   \\ 
SDSS\,J084616.69+052833.6 & 0.589 & 1.06$\pm$0.01 & 3.81$\pm$0.05 & 0.91   \\ 
SDSS\,J101009.25+062324.8 & 0.542 & 1.31$\pm$0.02 & 5.12$\pm$0.05 & 0.45   \\
SDSS\,J115836.93+021535.1 & 0.587 & 2.17$\pm$0.04 & 7.63$\pm$0.08 & 0.56   \\ 
SDSS\,J121954.26+003025.2 & 0.412 & 1.60$\pm$0.02 & 5.87$\pm$0.05 & 0.34   \\ 
SDSS\,J123106.94+053347.6 & 0.642 & 0.96$\pm$0.00 & 2.98$\pm$0.02 & 0.45   \\ 
SDSS\,J124257.04+102400.7 & 0.680 & 0.85$\pm$0.00 & 2.11$\pm$0.01 & 0.26   \\ 
SDSS\,J134412.30+010906.6 & 0.532 & 2.51$\pm$0.02 & 4.07$\pm$0.02 & 0.76   \\ 
SDSS\,J155037.84+024746.5 & 0.553 & 1.30$\pm$0.01 & 3.31$\pm$0.01 & 0.45   \\ 
SDSS\,J164916.12+294309.0 & 0.531 & 1.26$\pm$0.01 & 2.21$\pm$0.01 & 0.53   \\  
SDSS\,J235219.98$-$004855.7 & 0.437 & 2.72$\pm$0.09 & 9.70$\pm$0.15 & 0.43  \\   
\hline
SDSS\,J091515.64+055256.9 & 0.790 & 2.62$\pm$0.04 & 2.55$\pm$0.10 & 0.43   \\  
SDSS\,J104224.46+022225.0 & 0.668 & 2.17$\pm$0.02 & 4.28$\pm$0.10 & 0.84   \\  
SDSS\,J104630.88$-$010759.0 & 0.630 & 2.66$\pm$0.05 & 3.20$\pm$0.12 & 0.84  \\
SDSS\,J105745.85$-$005818.6 & 0.655 & 1.58$\pm$0.01 & 3.56$\pm$0.06 & 0.53  \\ 
SDSS\,J115027.99+025118.1 & 0.483 & 1.62$\pm$0.01 & 2.10$\pm$0.06 & 0.51   \\   
SDSS\,J132953.78+295140.0 & 0.862 & 1.43$\pm$0.02 & 4.45$\pm$0.18 & 0.35   \\   
SDSS\,J135342.06+262157.1 & 0.538 & 1.26$\pm$0.01 & 1.90$\pm$0.04 & 0.44  \\    
\hline\hline
\end{tabular} 
\end{center}
\tablecomments{Column 2: spectroscopic redshift. Column 3: circularized effective radius. Column 4: S${\rm \acute{e}}$rsic index. Column 5:  axis ratio $b/a$. Upper and lower blocks represent 
sample 1 and sample 2, respectively.}  
\label{table2}
\end{table*}

\begin{table*}
\caption{Parameters of double-S${\rm \acute{e}}$rsic models}
\begin{center}
\begin{tabular}{ccccc}
\hline \hline
Object Name &  $R_e$ & $n$ & $b/a$  & light \\
                         &   (kpc)   &         &              &  (\%)  \\
\hline
SDSS\,J011004.73+140933.0 & 0.43$\pm$0.00 & 1.52$\pm$0.02 & 0.38   & 56\\
                                                       & 2.06$\pm$0.01 & 0.57$\pm$0.01 & 0.37   & 44\\
SDSS\,J014355.21+133451.4  & 0.58$\pm$0.01 & 4.09$\pm$0.06 & 0.42   & 76 \\
                                                        & 1.00$\pm$0.01 & [0.50] & 0.12   & 24 \\
SDSS\,J081053.07+230443.7  & 0.25$\pm$0.00 & 1.66$\pm$0.03 & 0.53   & 40 \\
                                                        & 1.71$\pm$0.01 & 0.82$\pm$0.01 & 0.21   &  60 \\
SDSS\,J084223.93+050223.4 &   0.57$\pm$0.00 & 0.64$\pm$0.02 & 0.36  & 35 \\
                                                       &   1.65$\pm$0.01 & 0.86$\pm$0.01 & 0.18  & 65 \\
                                                       
SDSS\,J115836.93+021535.1  & 0.48$\pm$0.01 & 3.90$\pm$0.10 & 0.54   & 61 \\
                                                        & 3.03$\pm$0.02 & [0.50] & 0.62   & 39 \\
                                                                                                               
SDSS\,J121954.26+003025.2  & 0.23$\pm$0.00 & 1.25$\pm$0.03 & 1.00   & 27\\
                                                        & 1.63$\pm$0.01 & 1.66$\pm$0.02 & 0.25   & 73\\
SDSS\,J155037.84+024746.5  & 0.32$\pm$0.00 & 1.00$\pm$0.02 & 0.75   & 27\\
                                                        & 1.67$\pm$0.01 & 1.20$\pm$0.01 & 0.37  & 73\\
SDSS\,J235219.98$-$004855.7  & 0.27$\pm$0.00 & 2.84$\pm$0.06 & 0.75   & 47\\
                                                        & 2.32$\pm$0.01 & 0.58$\pm$0.01& 0.24   &  53\\
\hline\hline
\end{tabular} 
\end{center}
\tablecomments{Column 2: circularized effective radius. Column 3: S${\rm \acute{e}}$rsic index. Column 4: axis ratio $b/a$. 
Column 5: fraction of light from each component. As described in the text, for SDSS\,J014355.21+133451.4 and SDSS\,J115836.93+021535.1 we force $n = 0.5$ for their second components. Notice that we allowed the centroids of the two components to float in the fits. Only the best-fit centroids of SDSS\,J011004.73+140933.0 have a large offset of $\sim 0 \farcs 1$. For the other seven galaxies, the two components essentially share the same centroid because 
the offset is less than the pixel scale ($0 \farcs 02$).}
\label{table3}
\end{table*}

\begin{table*}
\caption{Stellar population properties from {\sc FAST}}
\begin{center}
\begin{tabular}{ccccccc}
\hline \hline
Object Name &  Age$_{0}$    &  $\tau$  & $\left \langle \mathrm{Age} \right \rangle$ & [Z/H] & $A_V$    & M$_{*}$  \\ 
                         & (Gyr)   &  (Gyr)   &  (Gyr)  &      & (mag)     & ($10^{11} M_{\odot}$) \\
\hline
SDSS\,J011004.73+140933.0 & 1.05$_{-0.02}^{+0.05}$  & 0.16$_{-0.01}^{+0.02}$ & 0.89$_{-0.03}^{+0.05}$ & 0.4 & 0.31$_{-0.08}^{+0.06}$   & 1.10$_{-0.07}^{+0.11}$ \\ 

SDSS\,J012942.10+132420.8\tablenotemark{a} & 1.45$_{-0.75}^{+0.25}$ &  0.16$_{-0.16}^{+0.08}$ & 1.29$_{-0.77}^{+0.26}$  & 0.4 & 0.00$_{-0.00}^{+0.53}$    & 0.43$_{-0.08}^{+0.05}$ \\ 

SDSS\,J014355.21+133451.4 & 4.57$_{-2.87}^{+1.74}$  & 0.79$_{-0.57}^{+0.52}$  & 3.78$_{-2.93}^{+1.82}$ &0.0 &0.12$_{-0.12}^{+0.13}$   &  0.69$_{-0.34}^{+0.16}$ \\ 

SDSS\,J081053.07+230443.7\tablenotemark{a} & 1.05$_{-0.65}^{+2.75}$  & 0.13$_{-0.12}^{+0.40}$  &  0.92$_{-0.66}^{+2.78}$ & 0.4 & 0.33$_{-0.33}^{+1.12}$  &  0.63$_{-0.11}^{+0.49}$\\ 

SDSS\,J084223.93+050223.4 & 1.58$_{-0.35}^{+0.28}$ & 0.16$_{-0.16}^{+0.08}$ &  1.43$_{-0.39}^{+0.29}$ & 0.4  & 0.04$_{-0.04}^{+0.13}$   & 0.62$_{-0.12}^{+0.09}$ \\ 

SDSS\,J084616.69+052833.6 & 1.14$_{-0.21}^{+0.55}$  & 0.13$_{-0.12}^{+0.06}$ & 1.02$_{-0.25}^{+0.55}$ & 0.4 & 0.19$_{-0.13}^{+0.29}$    & 0.69$_{-0.08}^{+0.38}$ \\ 

SDSS\,J101009.25+062324.8 & 7.59$_{-1.42}^{+0.36}$  & 2.00$_{-0.55}^{+0.14}$ & 5.59$_{-1.52}^{+0.38}$ & 0.0 & 1.06$_{-0.15}^{+0.12}$  & 1.66$_{-0.25}^{+0.12}$ \\ 

SDSS\,J115836.93+021535.1 & 5.01$_{-2.61}^{+2.57}$  & 0.79$_{-0.79}^{+0.49}$ & 4.21$_{-2.73}^{+2.62}$ & 0.0 & 0.08$_{-0.08}^{+0.18}$  & 1.02$_{-0.30}^{+0.29}$ \\ 

SDSS\,J121954.26+003025.2 & 3.47$_{-2.27}^{+5.24}$  & 0.13$_{-0.12}^{+0.95}$ & 3.34$_{-2.27}^{+5.33}$ & 0.0 & 0.32$_{-0.29}^{+0.37}$  & 0.65$_{-0.29}^{+0.38}$ \\ 

SDSS\,J123106.94+053347.6 & 3.47$_{-1.69}^{+0.80}$  & 0.63$_{-0.33}^{+0.30}$ & 2.84$_{-1.72}^{+0.85}$ &0.0 & 0.61$_{-0.22}^{+0.16}$  & 1.58$_{-0.65}^{+0.28}$ \\ 

SDSS\,J124257.04+102400.7 & 1.20$_{-0.08}^{+0.12}$  & 0.20$_{-0.03}^{+0.05}$ & 1.00$_{-0.09}^{+0.13}$ &0.0 & 1.13$_{-0.09}^{+0.16}$  & 1.35$_{-0.12}^{+0.20}$  \\ 

SDSS\,J134412.30+010906.6 & 7.59$_{-2.80}^{+0.36}$  & 1.58$_{-0.79}^{+0.46}$ & 6.00$_{-2.91}^{+0.58}$ &0.0 & 1.28$_{-0.14}^{+0.19}$    & 2.51$_{-0.56}^{+0.24}$ \\ 

SDSS\,J155037.84+024746.5 & 1.00$_{-0.00}^{+0.00}$  & 0.01$_{-0.00}^{+0.01}$ & 0.99$_{-0.00}^{+0.02}$ &0.4 & 0.00$_{-0.00}^{+0.00}$   & 0.66$_{-0.00}^{+0.02}$ \\ 

SDSS\,J164916.12+294309.0 & 0.95$_{-0.08}^{+0.07}$  & 0.06$_{-0.06}^{+0.04}$ & 0.89$_{-0.10}^{+0.08}$ &0.4 & 0.00$_{-0.00}^{+0.01}$   & 0.48$_{-0.02}^{+0.01}$ \\ 

SDSS\,J235219.98$-$004855.7\tablenotemark{a} & 8.71$_{-7.84}^{+0.00}$  & 1.00$_{-1.00}^{+0.41}$ & 7.71$_{-7.71}^{+0.41}$ & 0.0 & 0.35$_{-0.35}^{+1.21}$  & 1.12$_{-0.74}^{+0.20}$ \\  
\hline
SDSS\,J091515.64+055256.9\tablenotemark{a} & 6.61$_{-5.58}^{+0.00}$  & 0.00$_{-0.00}^{+0.91}$ & 6.60$_{-5.58}^{+0.91}$ & 0.0 & 0.08$_{-0.08}^{+0.55}$     & 1.66$_{-1.10}^{+0.25}$\\ 

SDSS\,J104224.46+022225.0 & 1.45$_{-0.22}^{+0.46}$  & 0.00$_{-0.00}^{+0.17}$ & 1.44$_{-0.22}^{+0.49}$ & 0.4  & 0.01$_{-0.01}^{+0.14}$   & 0.68$_{-0.07}^{+0.14}$ \\ 

SDSS\,J104630.88$-$010759.0 & 2.00$_{-0.79}^{+4.17}$  & 0.25$_{-0.25}^{+0.68}$ & 1.74$_{-0.83}^{+4.23}$ & 0.4 & 0.29$_{-0.29}^{+0.49}$  & 0.68$_{-0.11}^{+0.70}$ \\  

SDSS\,J105745.85$-$005818.6 & 4.57$_{-2.17}^{+0.80}$  & 0.79$_{-0.75}^{+0.28}$ & 3.78$_{-2.30}^{+0.85}$ & 0.0 & 0.00$_{-0.00}^{+0.04}$   & 1.20$_{-0.31}^{+0.12}$ \\ 

SDSS\,J115027.99+025118.1 & 3.80$_{-2.10}^{+4.52}$  & 0.79$_{-0.45}^{+2.23}$  & 3.01$_{-2.15}^{+5.03}$ & 0.4 & 0.48$_{-0.42}^{+0.69}$  & 0.43$_{-0.12}^{+0.35}$  \\ 

SDSS\,J132953.78+295140.0 & 2.19$_{-1.23}^{+1.36}$  & 0.32$_{-0.31}^{+0.21}$ & 1.87$_{-1.27}^{+1.38}$ & 0.0 & 0.73$_{-0.43}^{+0.45}$   & 1.15$_{-0.44}^{+0.26}$ \\ 

SDSS\,J135342.06+262157.1 & 5.75$_{-4.55}^{+2.19}$   & 0.79$_{-0.79}^{+0.49}$ & 4.96$_{-4.62}^{+2.24}$ & 0.0 & 0.76$_{-0.26}^{+0.58}$  &  1.62$_{-0.93}^{+0.42}$ \\ 
\hline\hline
\end{tabular} 
\end{center}
\tablecomments{Column 2: the age since the onset (peak) of star formation. Column 3: $e$-folding timescale for the star formation rate. Column 4: the mean age of the stellar population, which is the age since one $\tau$ after the onset of star formation ($=$ Age$_{0} - \tau$). Column 5: metallicity. Column 5: magnitude of rest-frame visual extinction. Column 6: stellar mass. Upper and lower blocks represent sample 1 and sample 2, respectively.}  
\footnotetext{The stellar populations of these four galaxies are only constrained by the photometry.}
\label{table4}
\end{table*}

\begin{table*}
\caption{Stellar population properties, velocity dispersions and dynamical masses based on {\sc pPXF} full-spectrum fits} 
\begin{center}
\begin{tabular}{cccccccccc}
\hline \hline
Object Name & $\left \langle \mathrm{Age} \right \rangle_{\mathrm M}$ & $\left \langle \mathrm{Age} \right \rangle_{\mathrm L}$ & [$\left \langle \mathrm{Z/H} \right \rangle_{\rm M}$] & [$\left \langle \mathrm{Z/H} \right \rangle_{\rm L}$] &   $A_V$ & M$_{*}$ &  $\sigma$   & M$_{\mathrm{dyn}}$\footnote{Calculated from the virial relation M$_{\rm dyn}$ = $\beta \sigma_{e}^{2} R_{e} / G$, with $\beta$ = 5.} &  M$_{\rm{dyn}}$\footnote{Calculated from the virial relation, but with $\beta = 6(R_e / 3.185)^{-0.81}(M_{*}/10^{11})^{0.45}$, following \cite{Peralta-de-Arriba2013The-discrepancy}, where $R_e$ is in kpc and $M_{*}$ is the stellar mass from column 3, corrected for the difference between our assumed \citet{2003PASP..115..763C} IMF and the Salpeter IMF assumed by \cite{Peralta-de-Arriba2013The-discrepancy}, following the Equation (12) in \cite{Longhetti2009Stellar-mass-es}.}\\  
                         & (Gyr) & (Gyr) &  &   & (mag) & ($10^{11}$$M_{\odot}$) &(km s$^{-1}$)   & ($10^{11}$$M_{\odot}$) &   ($10^{11}$$M_{\odot}$)    \\
\hline
SDSS\,J011004.73+140933.0 & 1.03 & 0.95 & 0.34 & 0.34 & 0.00 & 0.87 & 271$\pm$8  &0.94$\pm$0.05 & 2.08$\pm$0.12\\ 
SDSS\,J014355.21+133451.4 & 8.15 & 6.71 & -0.10 & 0.00 & 0.18 &1.38 & 247$\pm$9 & 0.65$\pm$0.05 & 2.05$\pm$0.15\\   
SDSS\,J084223.93+050223.4 & 3.18 & 2.59 & 0.14 & 0.22 & 0.06 & 1.11 & 268$\pm$5 & 1.18$\pm$0.04 & 2.29$\pm$0.08\\     
SDSS\,J084616.69+052833.6 & 2.06 & 1.48 & 0.03  & 0.10 & 0.36 & 1.37 & 249$\pm$11 & 0.89$\pm$0.08& 2.15$\pm$0.18\\    
SDSS\,J101009.25+062324.8 & 5.56 & 4.20 & 0.05 & 0.14 & 0.28 & 1.20 & 142$\pm$14 & 0.34$\pm$0.07 & 0.66$\pm$0.13\\    
SDSS\,J115836.93+021535.1 & 6.04 & 4.28 & -0.01 & -0.03 & 0.77 & 2.44 &  172$\pm$9 & 0.79$\pm$0.08 & 1.39$\pm$0.14\\     
SDSS\,J123106.94+053347.6 & 2.55 & 1.85 & 0.13 & 0.22 & 0.29 &1.22 & 261$\pm$7 &  0.90$\pm$0.05 & 2.24$\pm$0.11\\    
SDSS\,J124257.04+102400.7 & 0.80 & 0.74 & 0.30 & 0.29 & 0.57 & 0.95 & 237$\pm$12 &  0.67$\pm$0.07 & 1.65$\pm$0.17\\ 
SDSS\,J155037.84+024746.5 & 3.34 & 1.93 & 0.22 & 0.31 & 0.00 & 0.63 & 227$\pm$9 &  0.88$\pm$0.07 & 1.27$\pm$0.10\\  
\hline
SDSS\,J105745.85$-$005818.6 & 5.00 & 4.09 & -0.03 & -0.08 & 0.00 &1.48 & 278$\pm$11   & 1.57$\pm$0.12 &  2.86$\pm$0.22\\ 
\hline\hline
\end{tabular} 
\end{center}
\tablecomments{Columns 2 and 3: mean mass-weighted and luminosity-weighted ages. Columns 4 and 5: mean mass-weighted and luminosity-weighted metallicities. Column 6: magnitude of rest-frame visual extinction. Column 7: stellar mass. Column 8: velocity dispersion. Columns 9 and 10: dynamical masses. Upper and lower blocks include objects from sample 1 and sample 2, respectively.}
\label{table5}
\end{table*}

% =================== B I B L I O G R A P H Y ========================= %
\bibliographystyle{apj}
\bibliography{paper-compact}  

\appendix

We present all the 22 galaxy images and their corresponding GALFIT fits in Figures~\ref{keckimages} and~\ref{focasimages}. The ten spectra we used for pPXF full-spectrum fits and their results 
are shown in Figure~\ref{spectra}.

\renewcommand{\thefigure}{A1}

\begin{figure*}
\centering
\includegraphics[width=4cm]{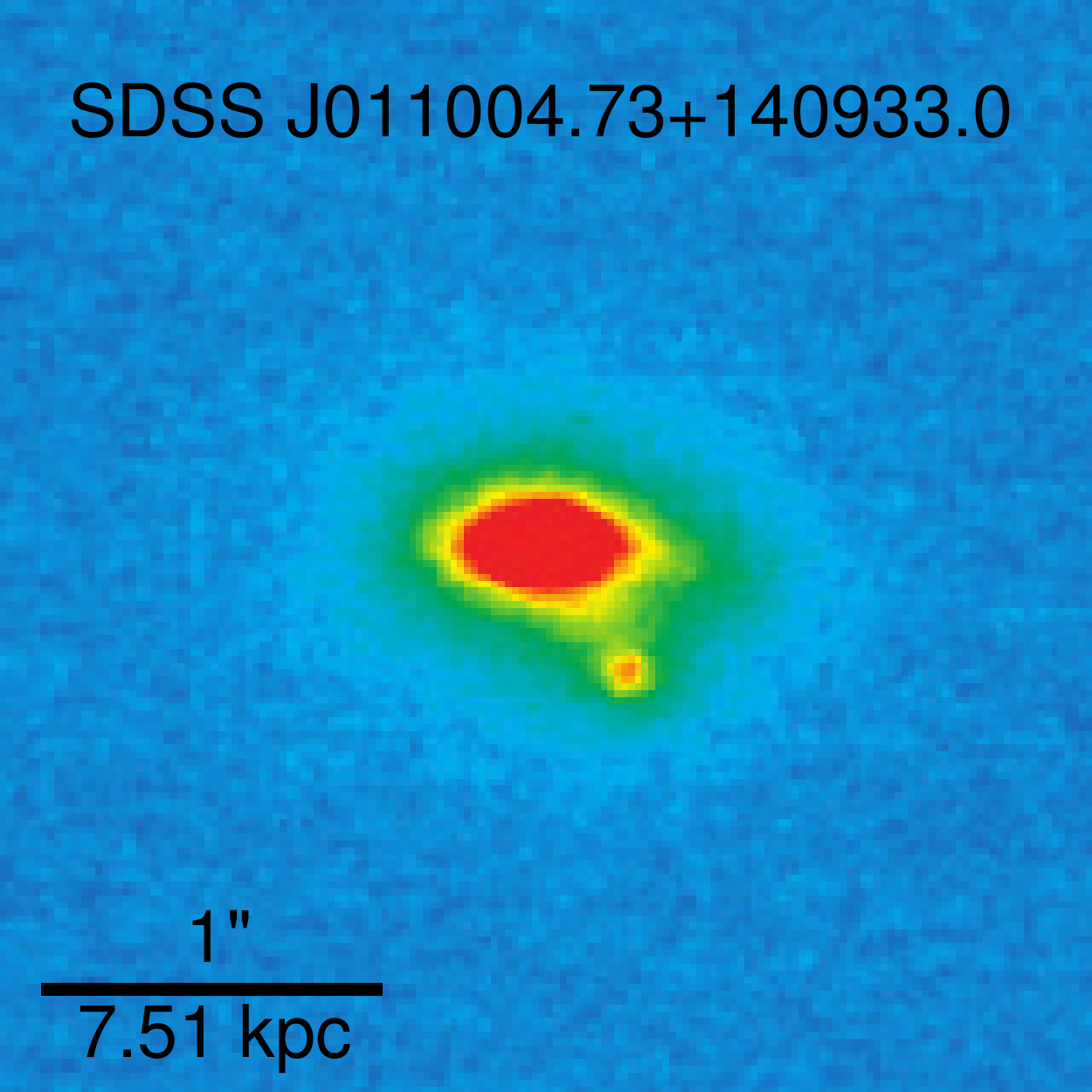}
\includegraphics[width=4cm]{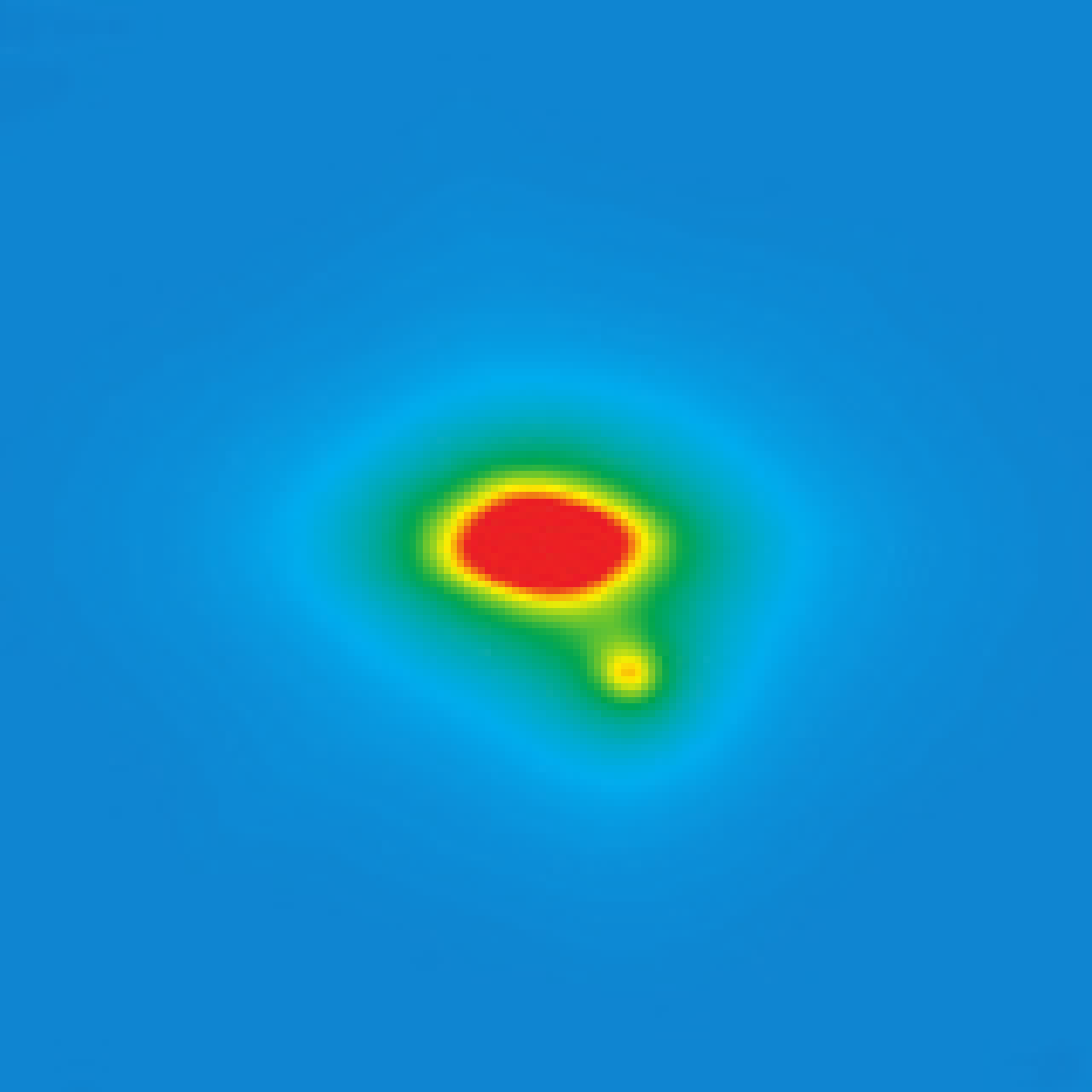}    
\includegraphics[width=4cm]{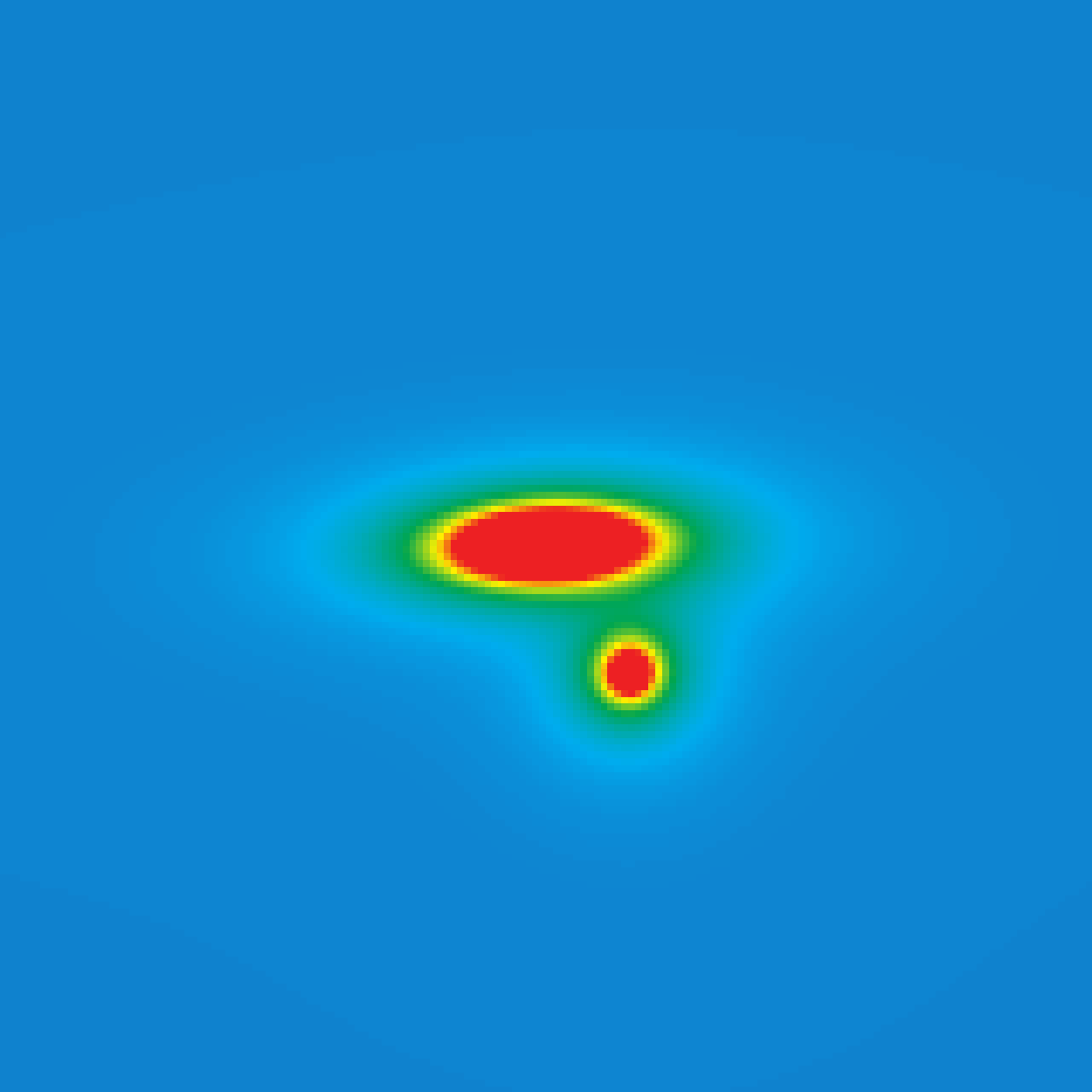}
\includegraphics[width=4cm]{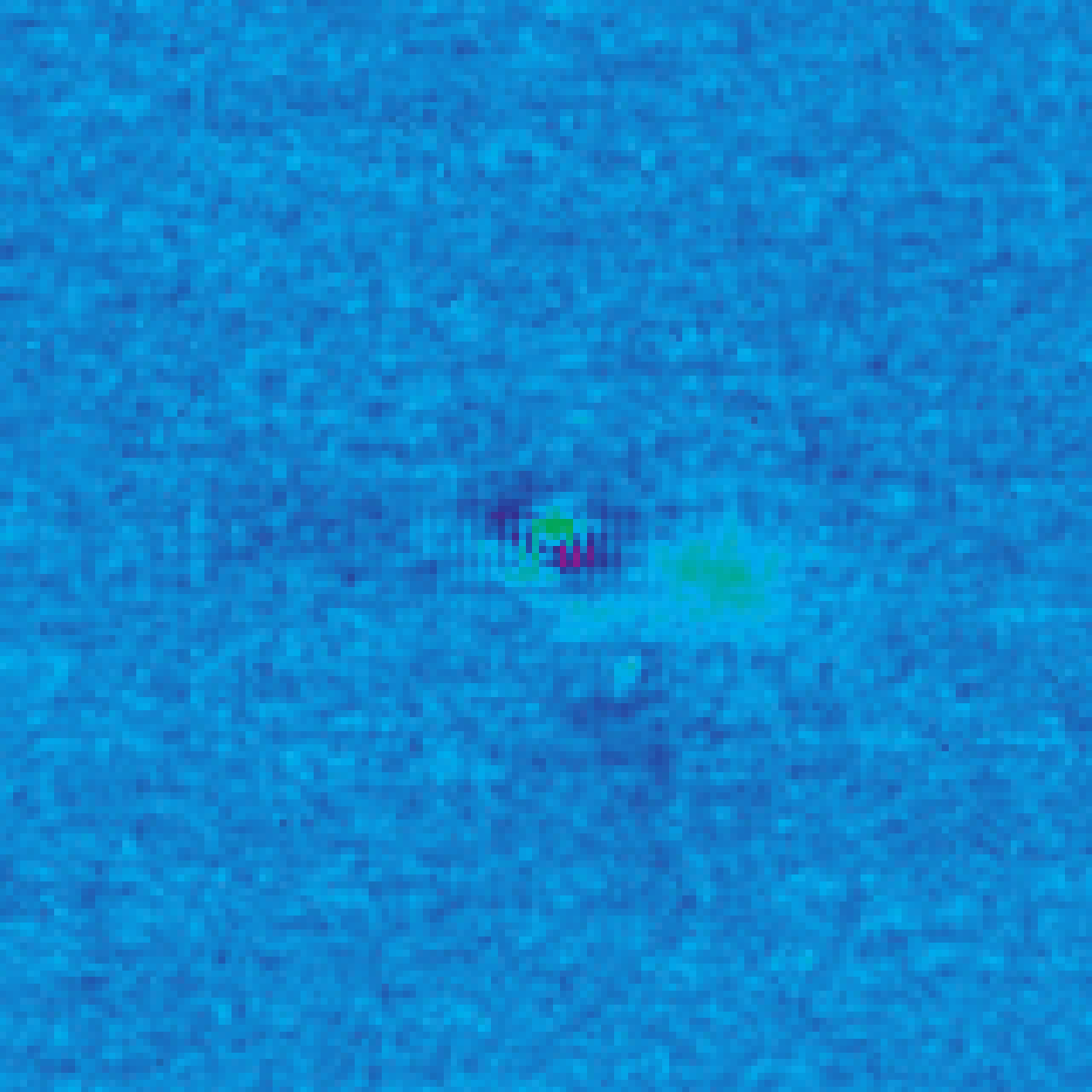}\\[1mm]
\hspace{40.7mm}\includegraphics[width=4cm]{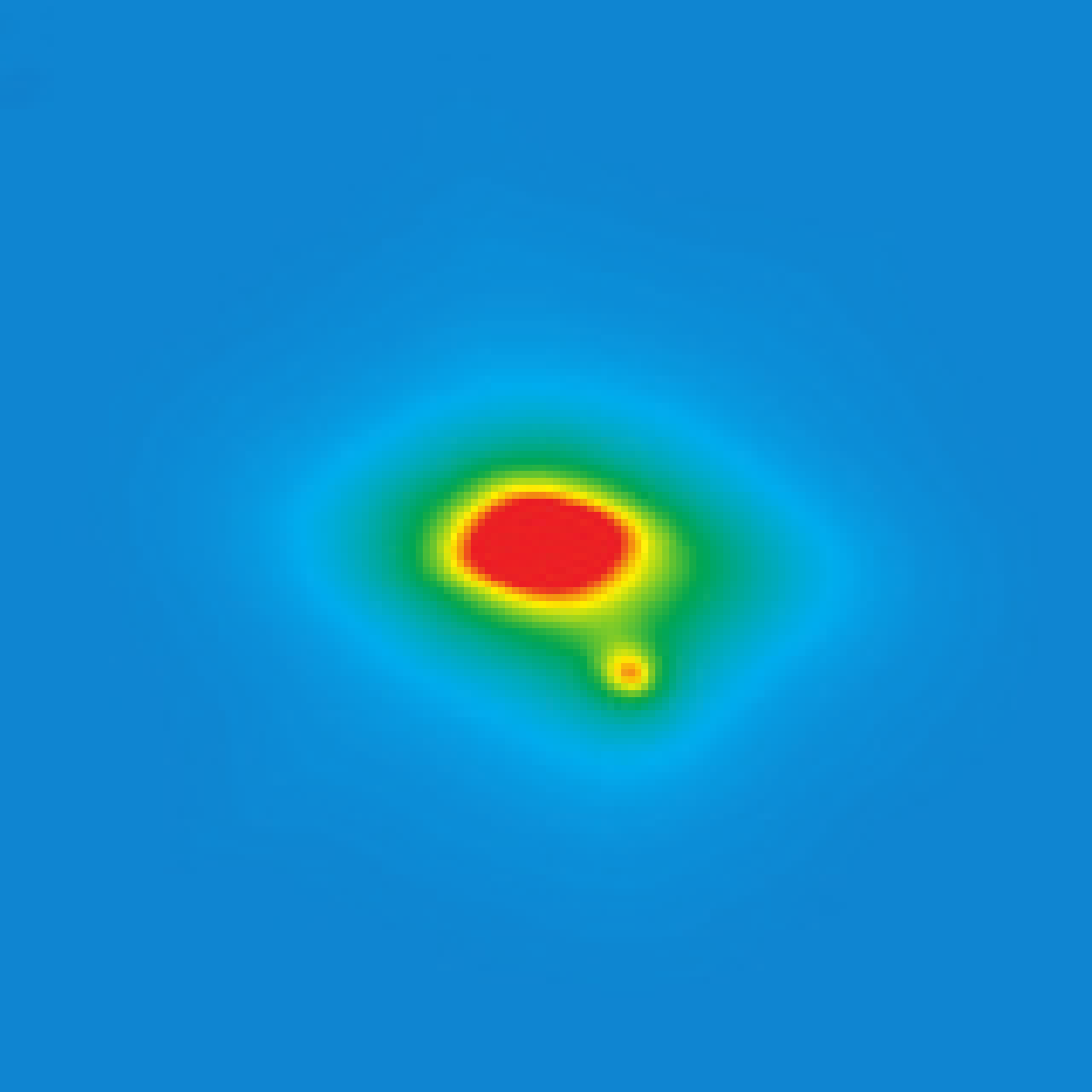} 
\includegraphics[width=4cm]{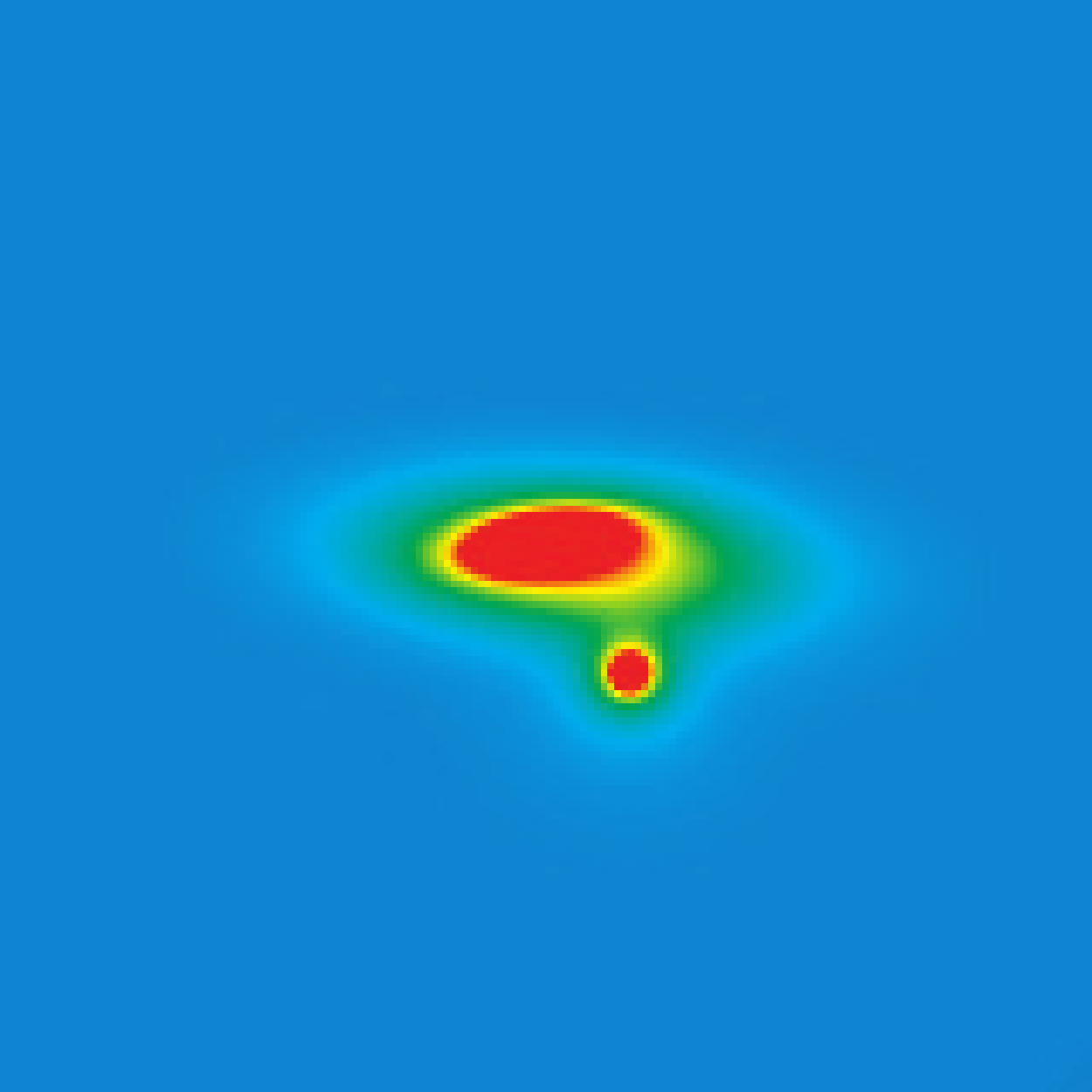}
\includegraphics[width=4cm]{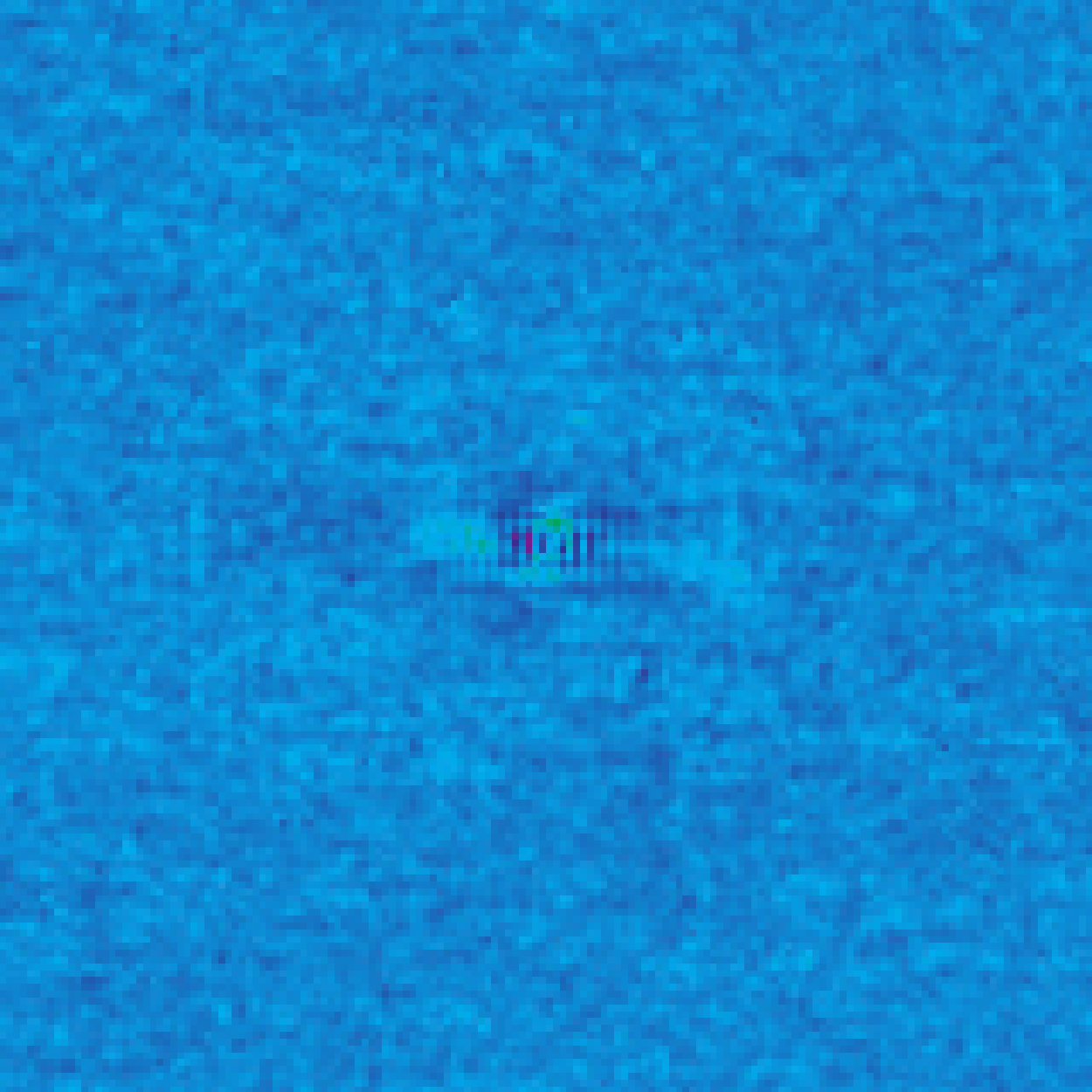}\\[2mm]
\includegraphics[width=4cm]{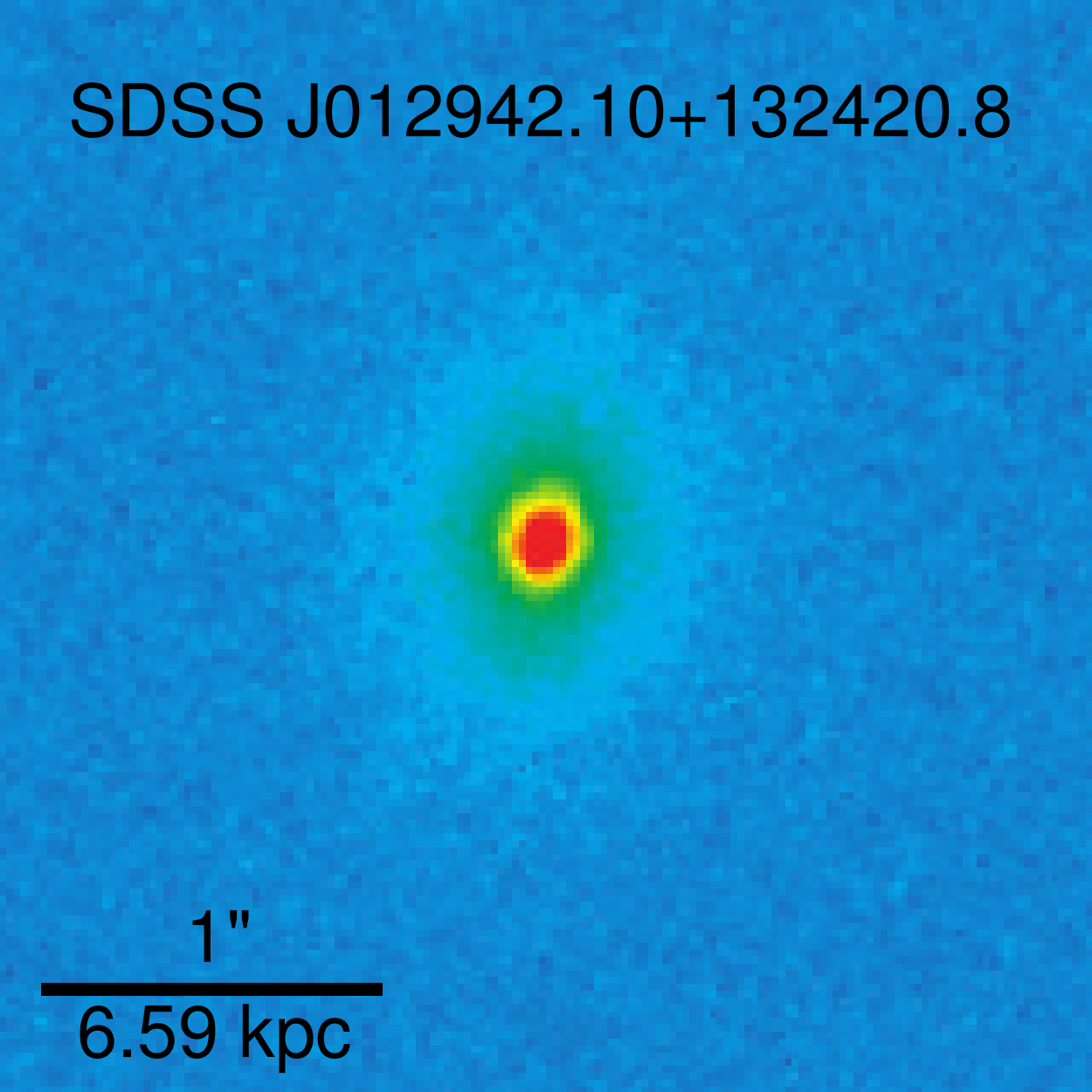}
\includegraphics[width=4cm]{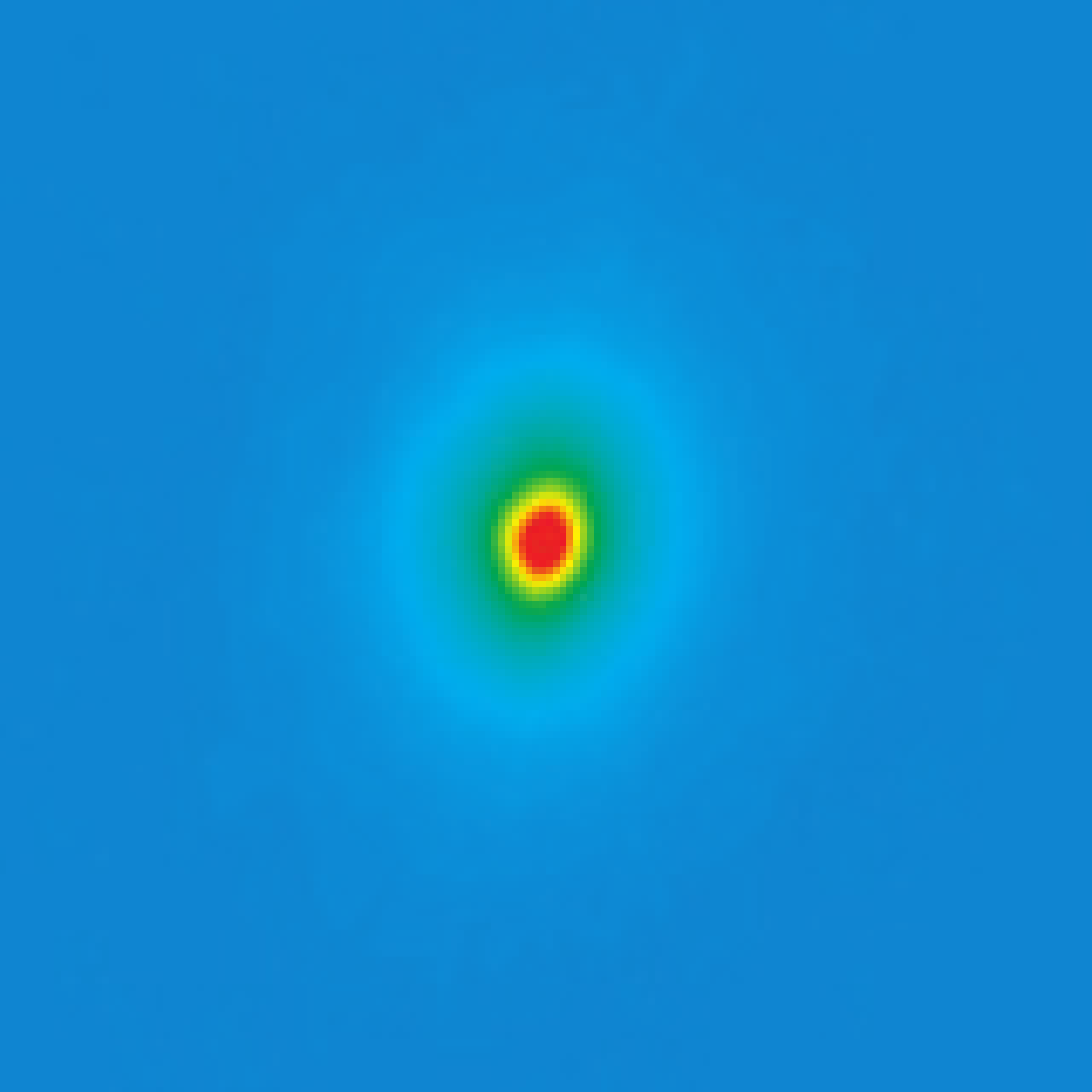}    
\includegraphics[width=4cm]{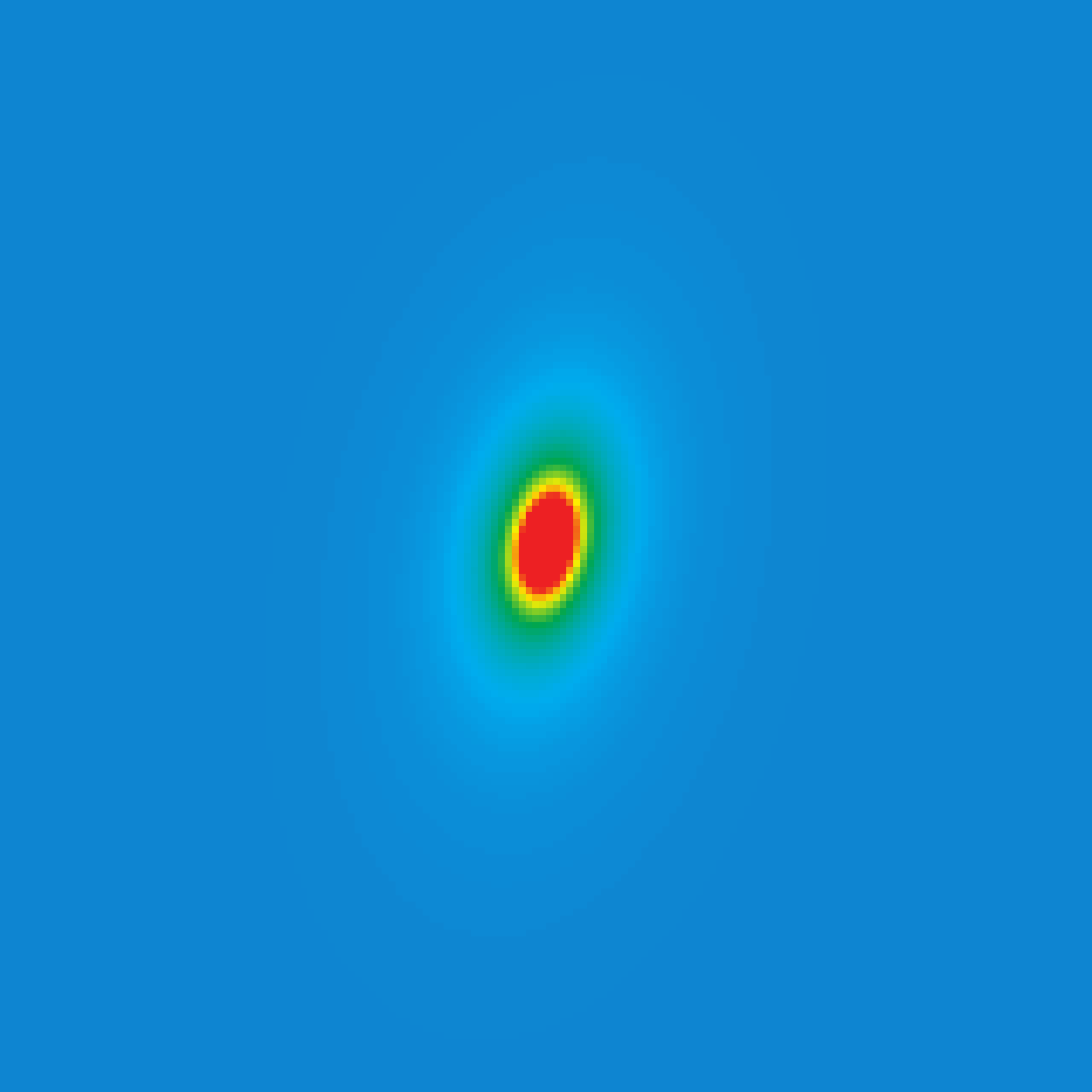}
\includegraphics[width=4cm]{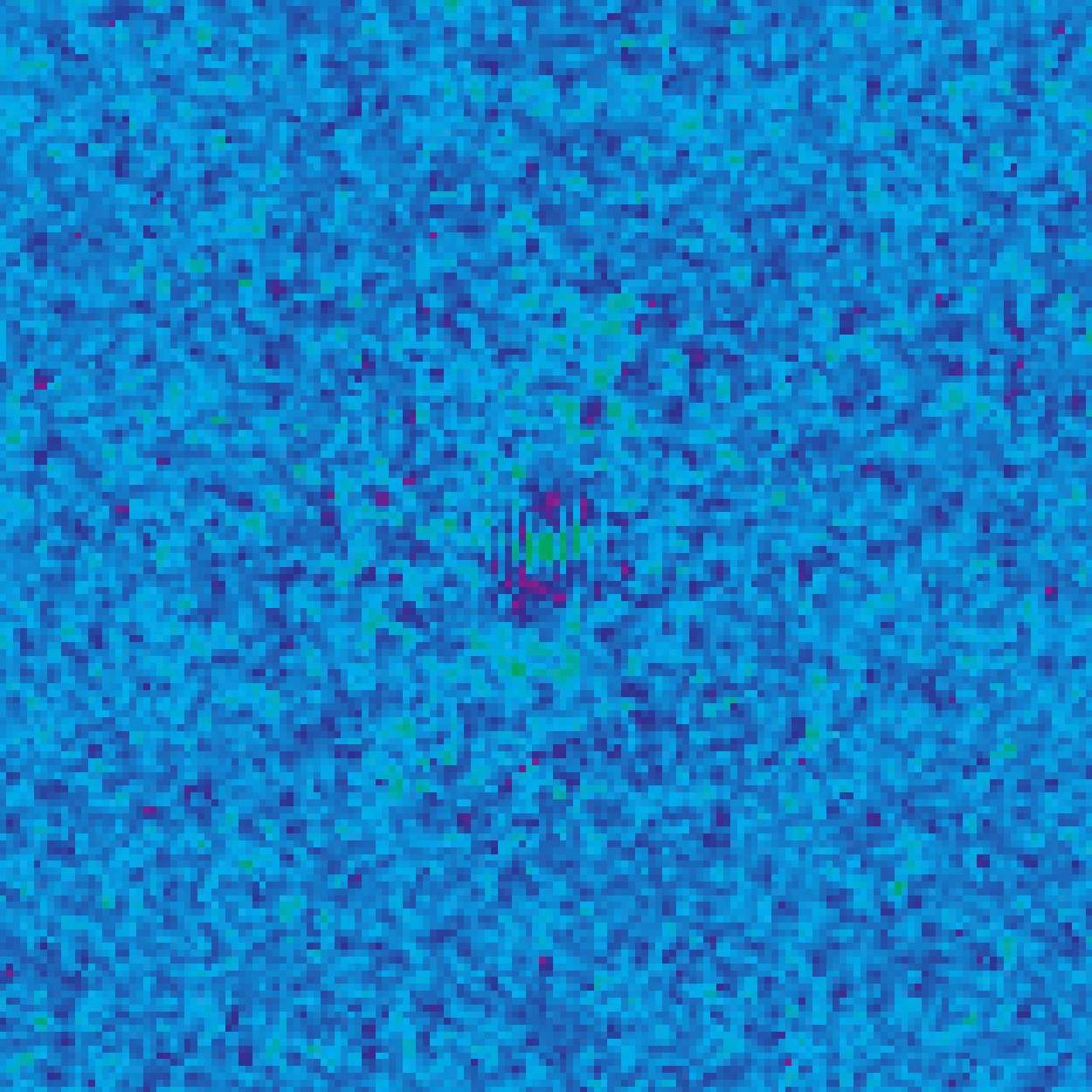}\\[2mm]
\includegraphics[width=4cm]{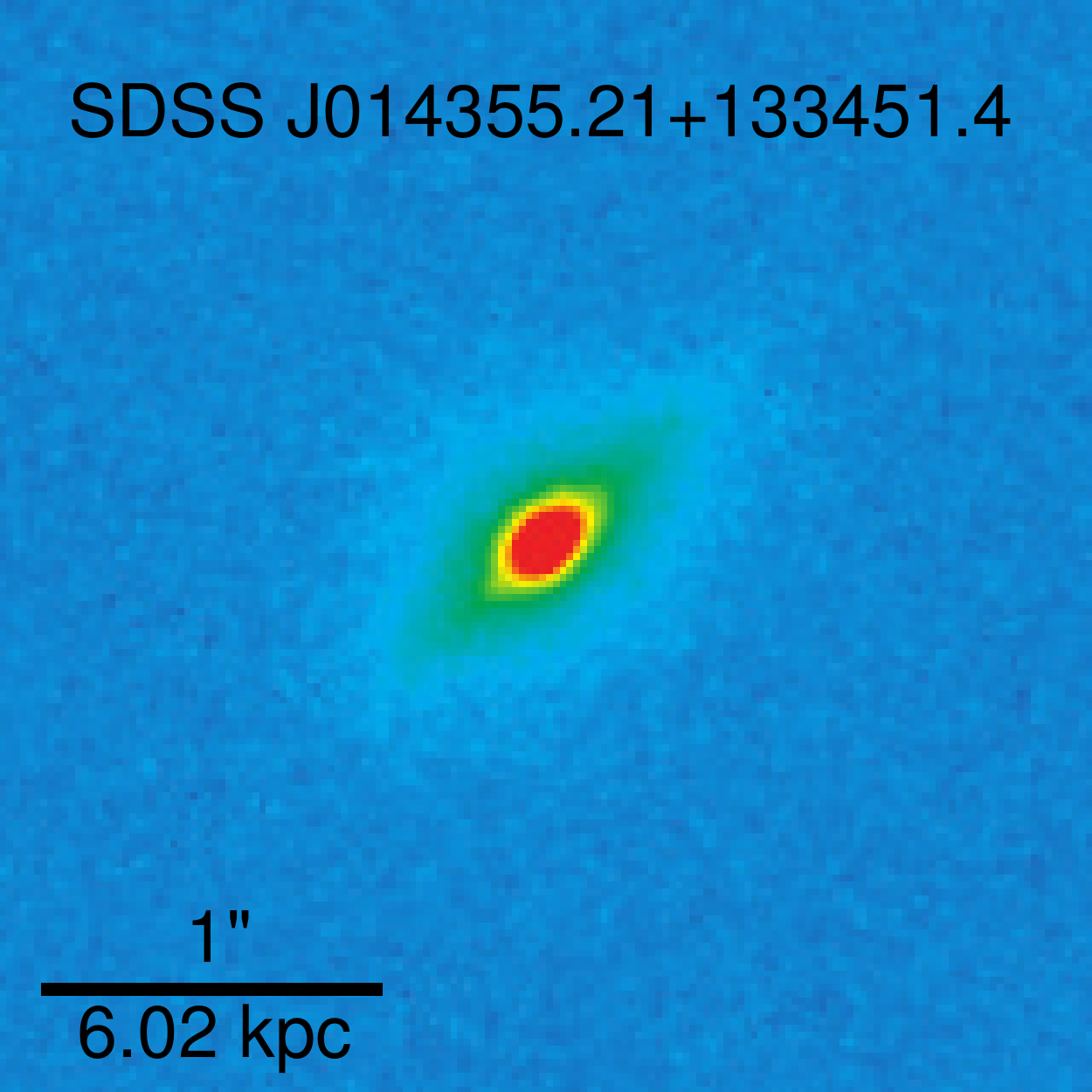}
\includegraphics[width=4cm]{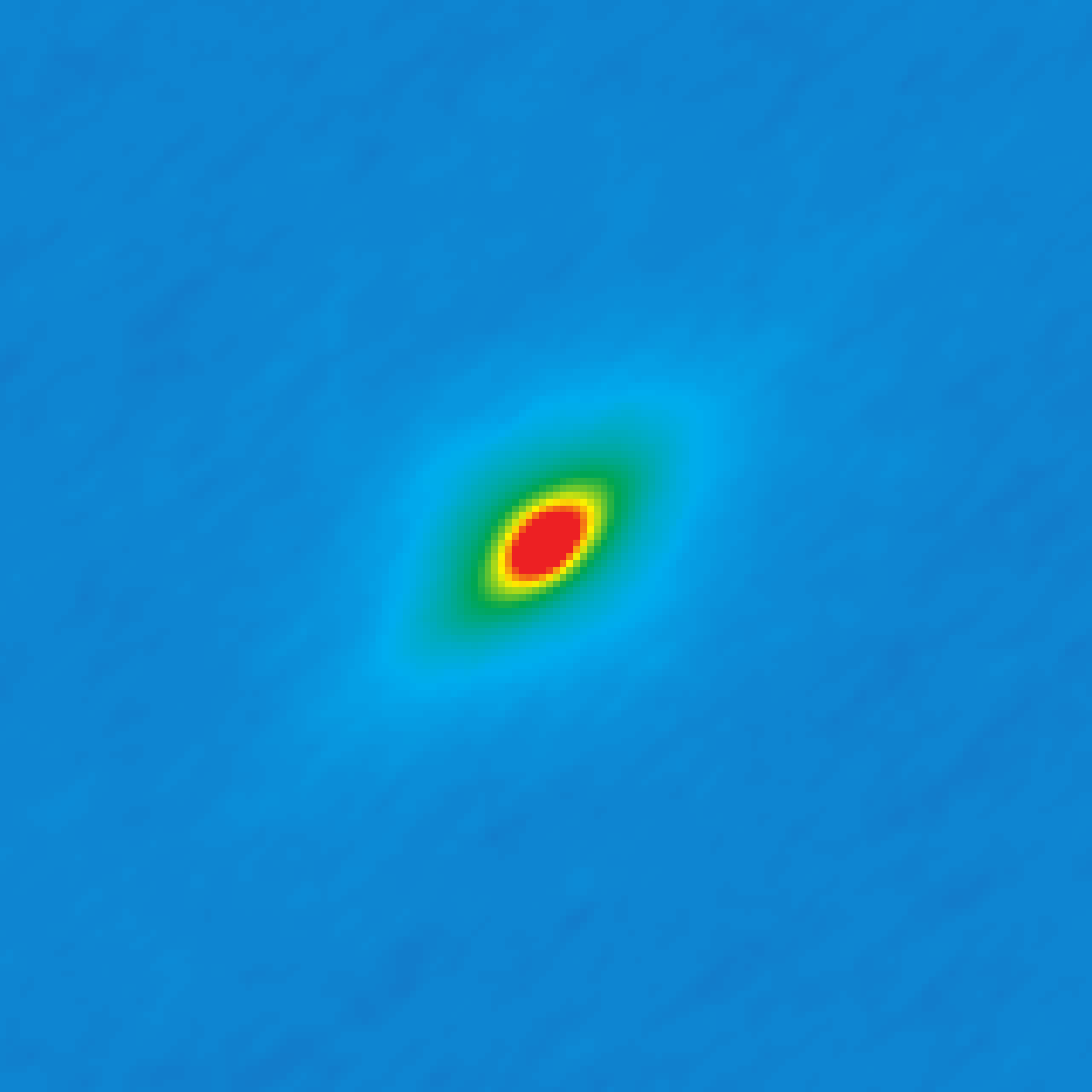}    
\includegraphics[width=4cm]{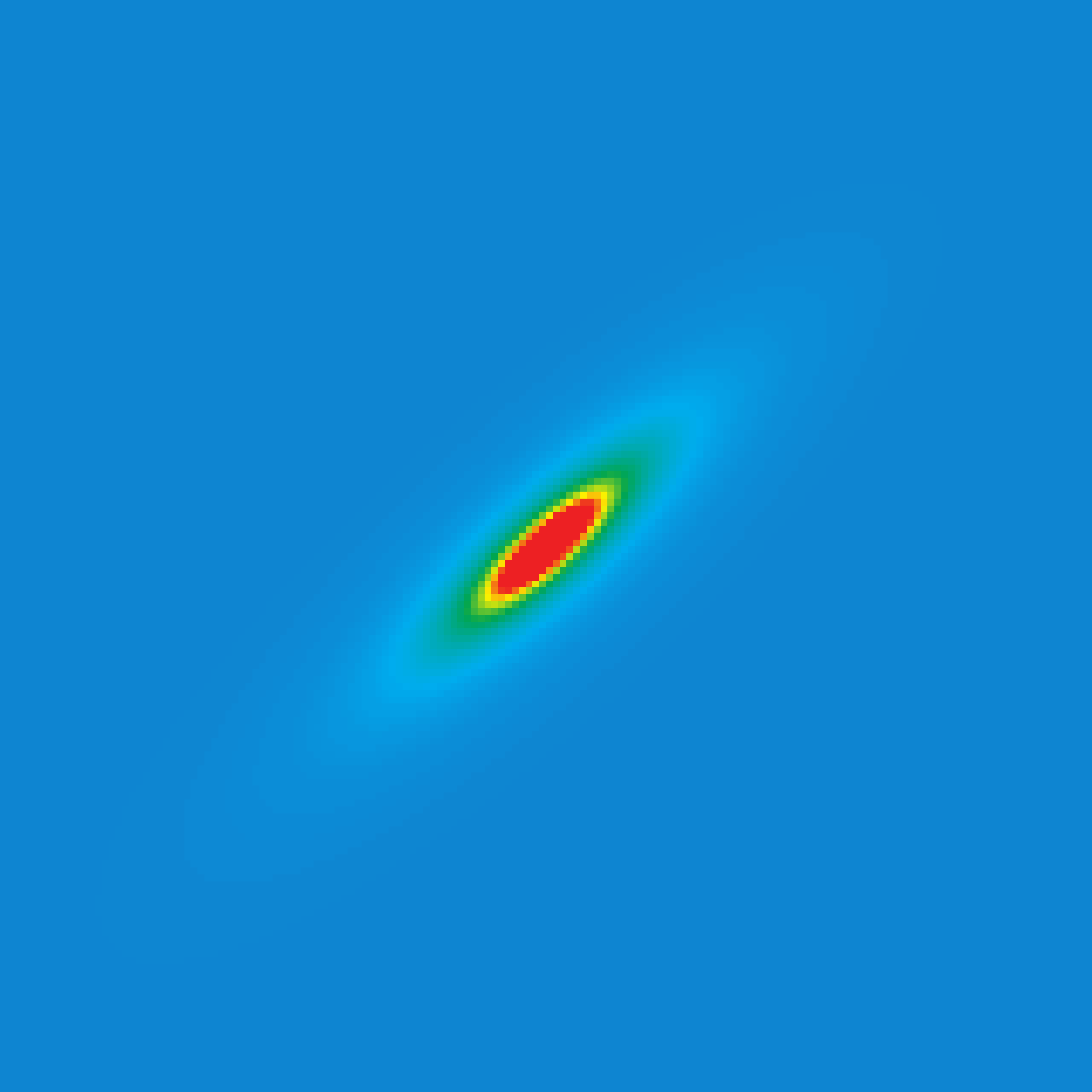}
\includegraphics[width=4cm]{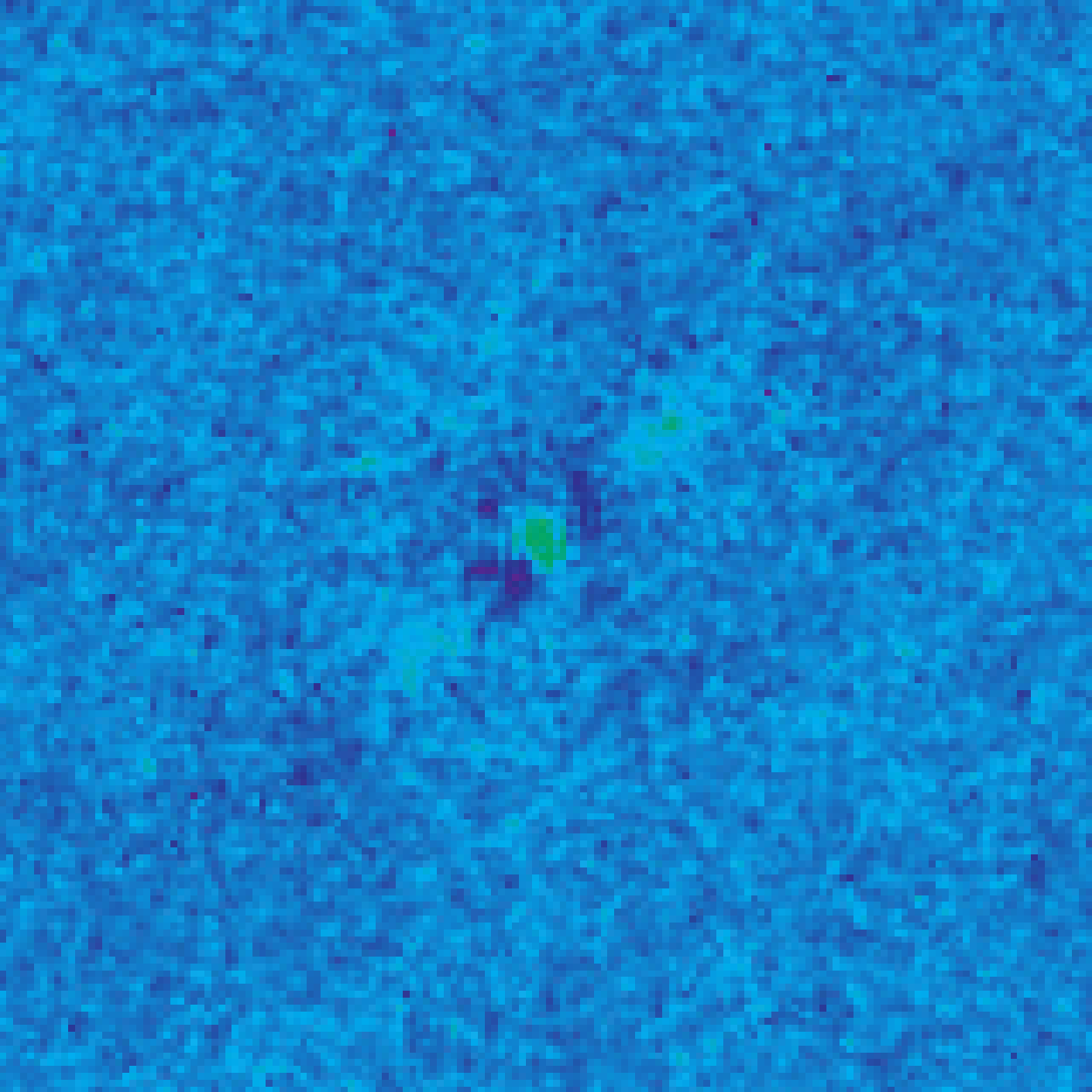}\\[1mm]
\hspace{40.7mm}\includegraphics[width=4cm]{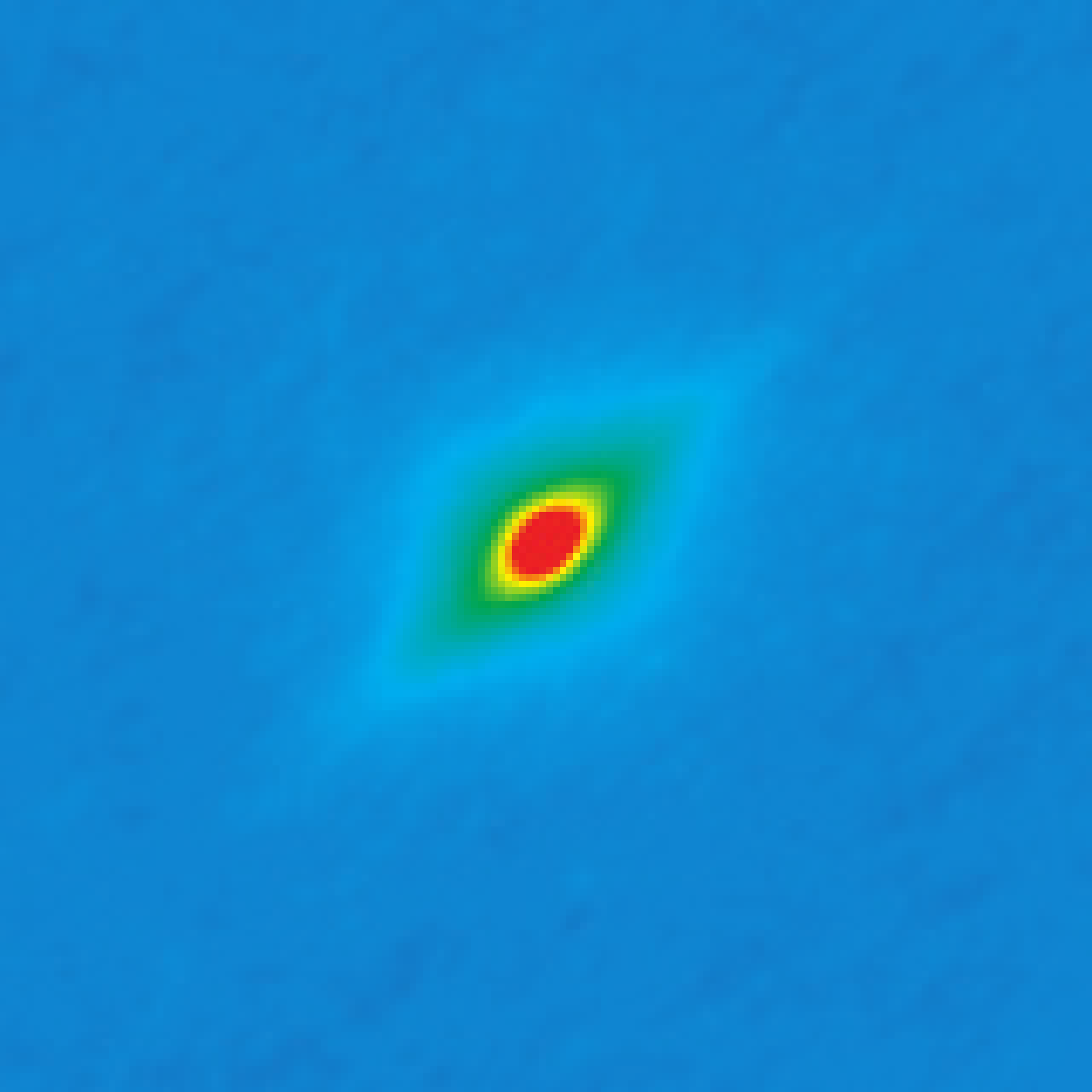}    
\includegraphics[width=4cm]{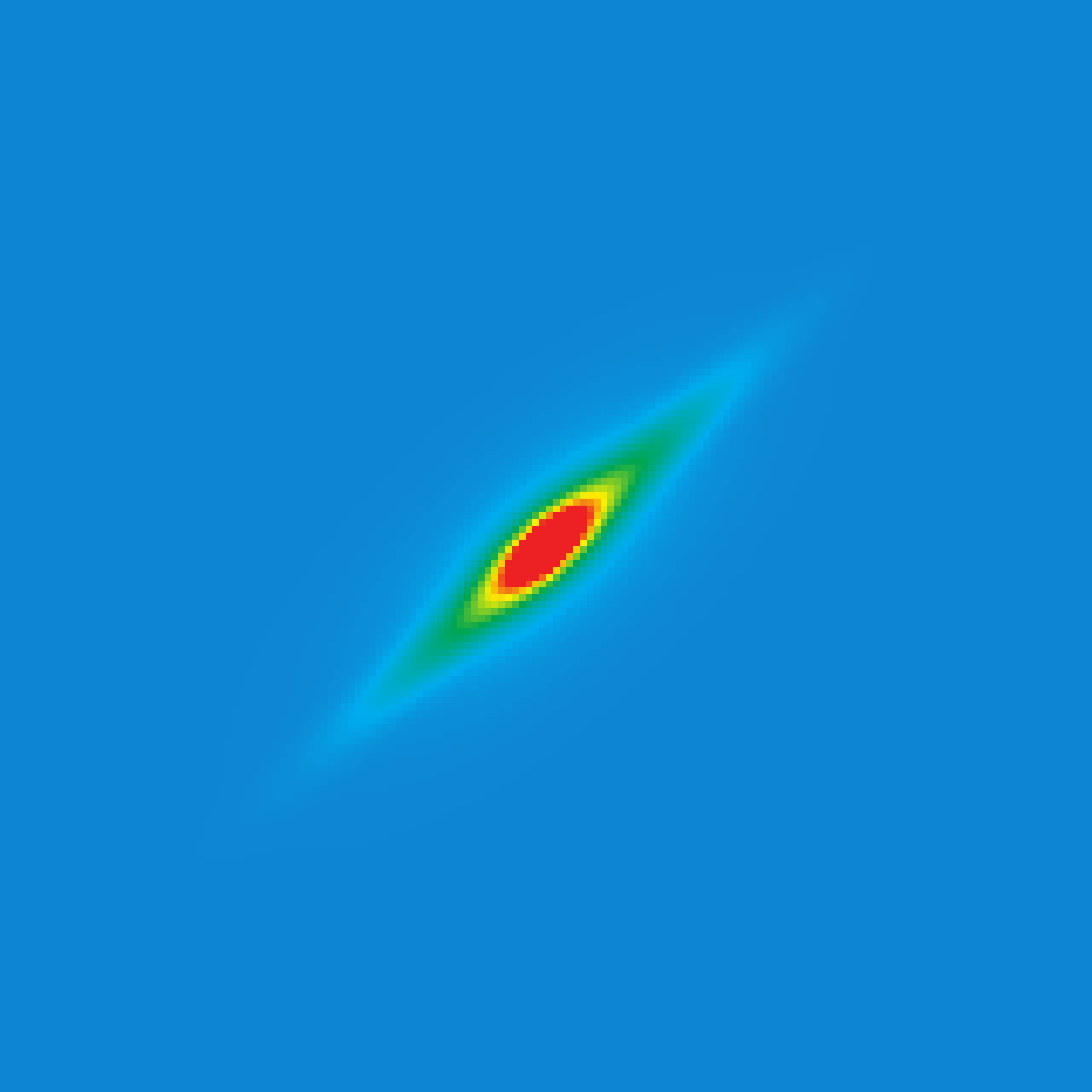}
\includegraphics[width=4cm]{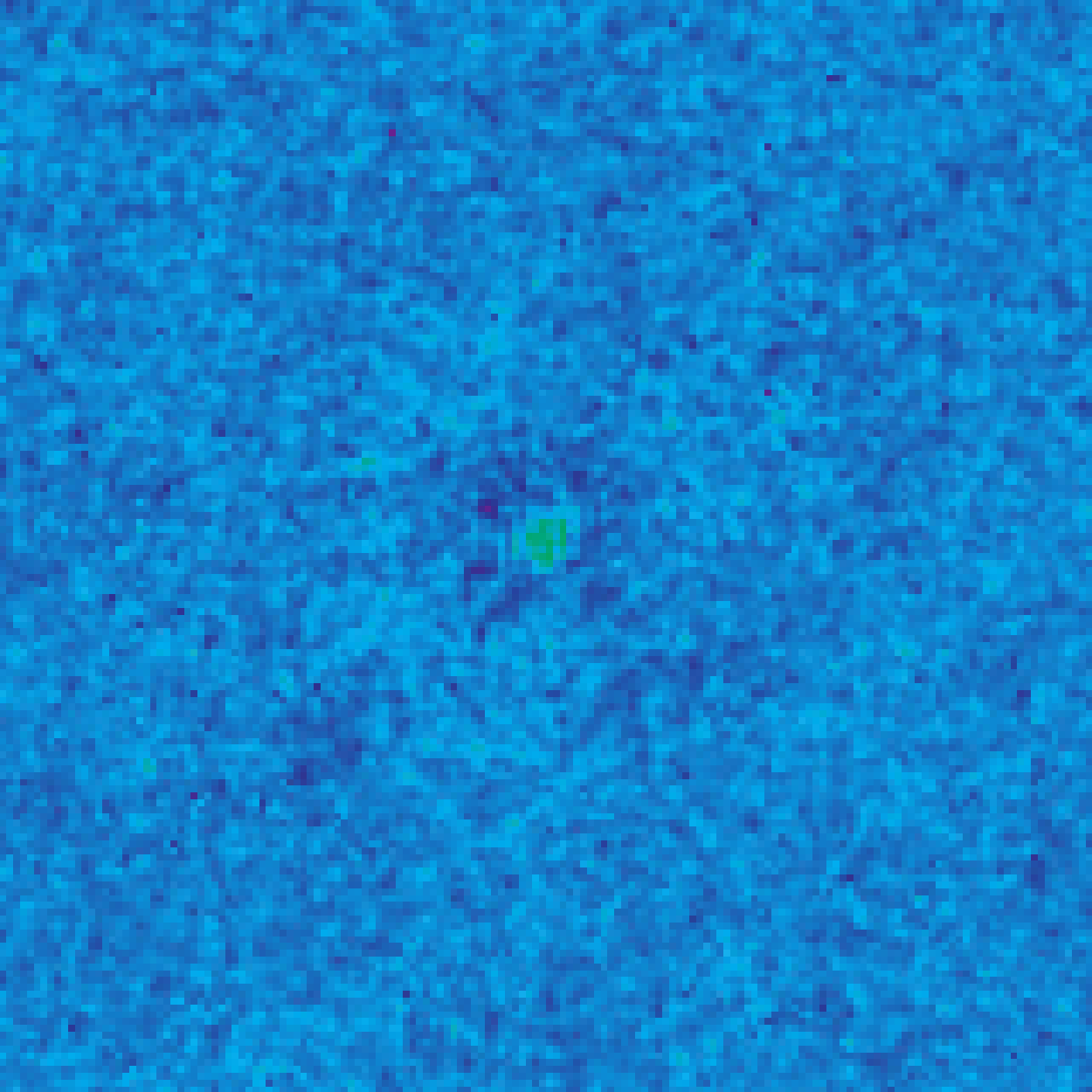}
\caption{Keck LGSAO false-color images and model fits, where the first row of a object represents a one-component fit, and the second row (if exists) represents a two-component fit. For each galaxy, the panels from left to right show the original image, the best-fit model, the model without PSF convolution, and the residual. The images and models are in power-law scale in order to show the faint outer profiles of the galaxies. The residuals are in linear scale with a different contrast in order to show the small variations across the fields. An one-arcsecond scale and the corresponding physical length at the galaxy redshift is shown in the lower-left corner of every galaxy image. North is up and east to the left 
for all images. For SDSS\,J011004.73+140933.0, we include another S${\rm \acute{e}}$rsic profile for the close companion.}
\label{keckimages}
\end{figure*}

\begin{figure*}
\centering
\includegraphics[width=4cm]{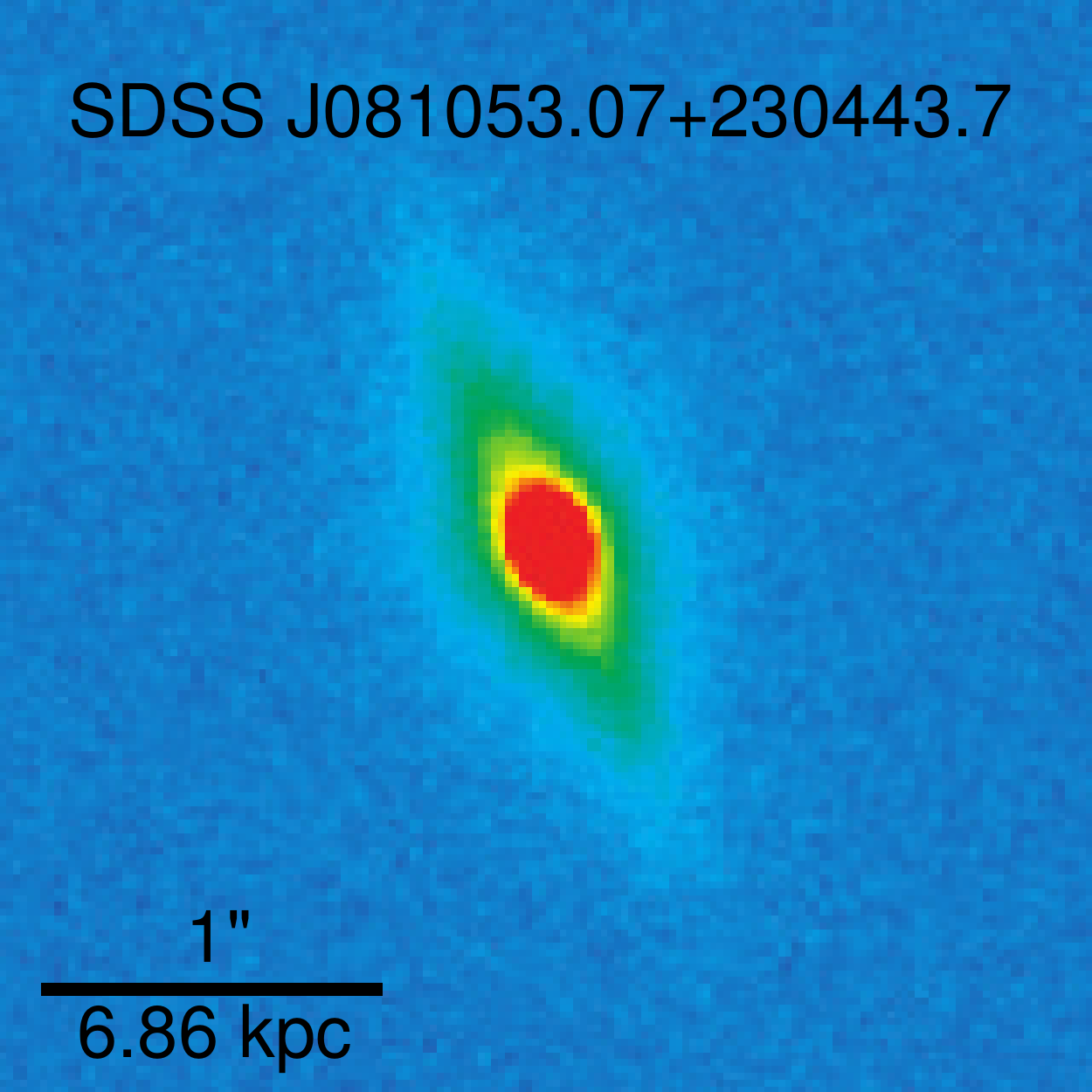}
\includegraphics[width=4cm]{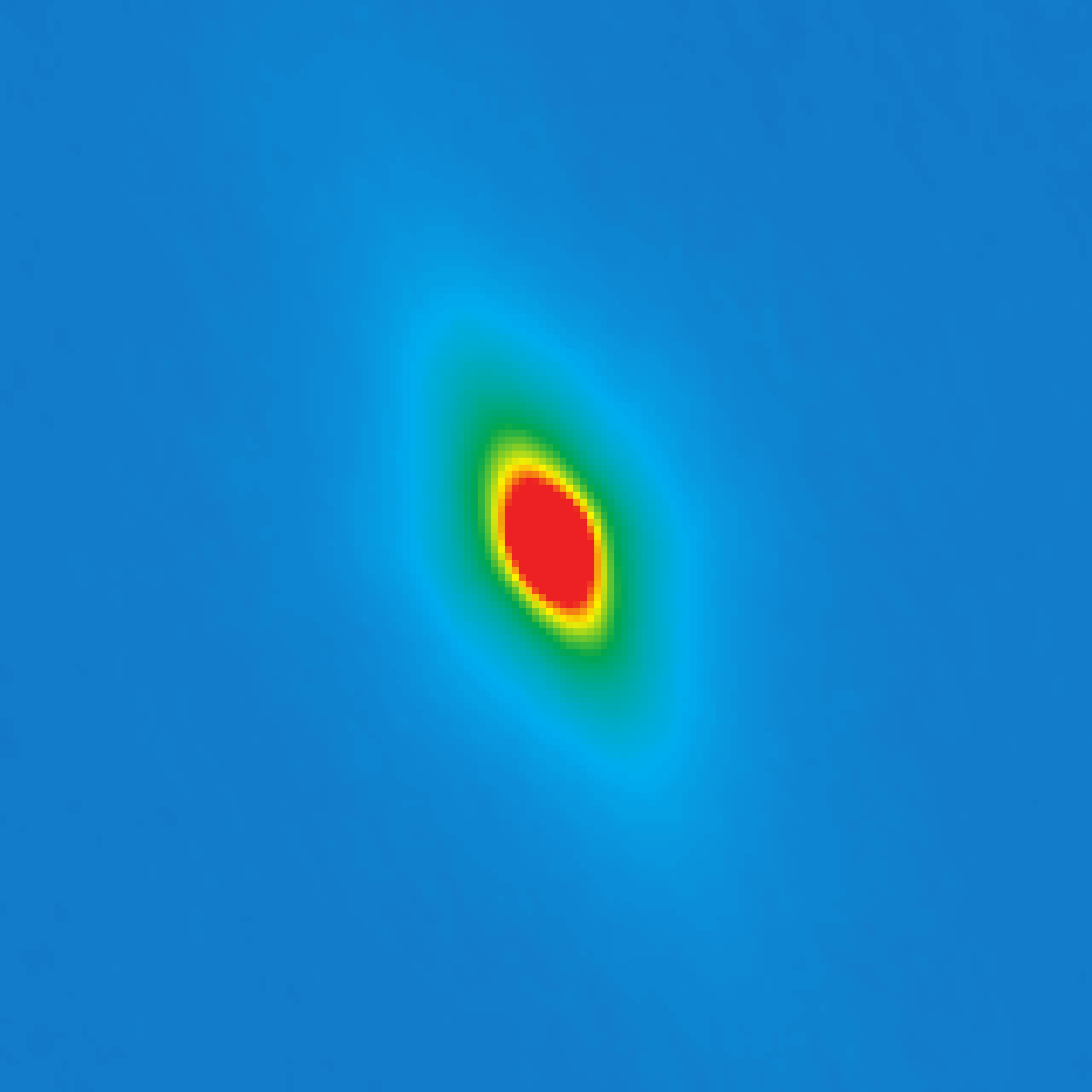}    
\includegraphics[width=4cm]{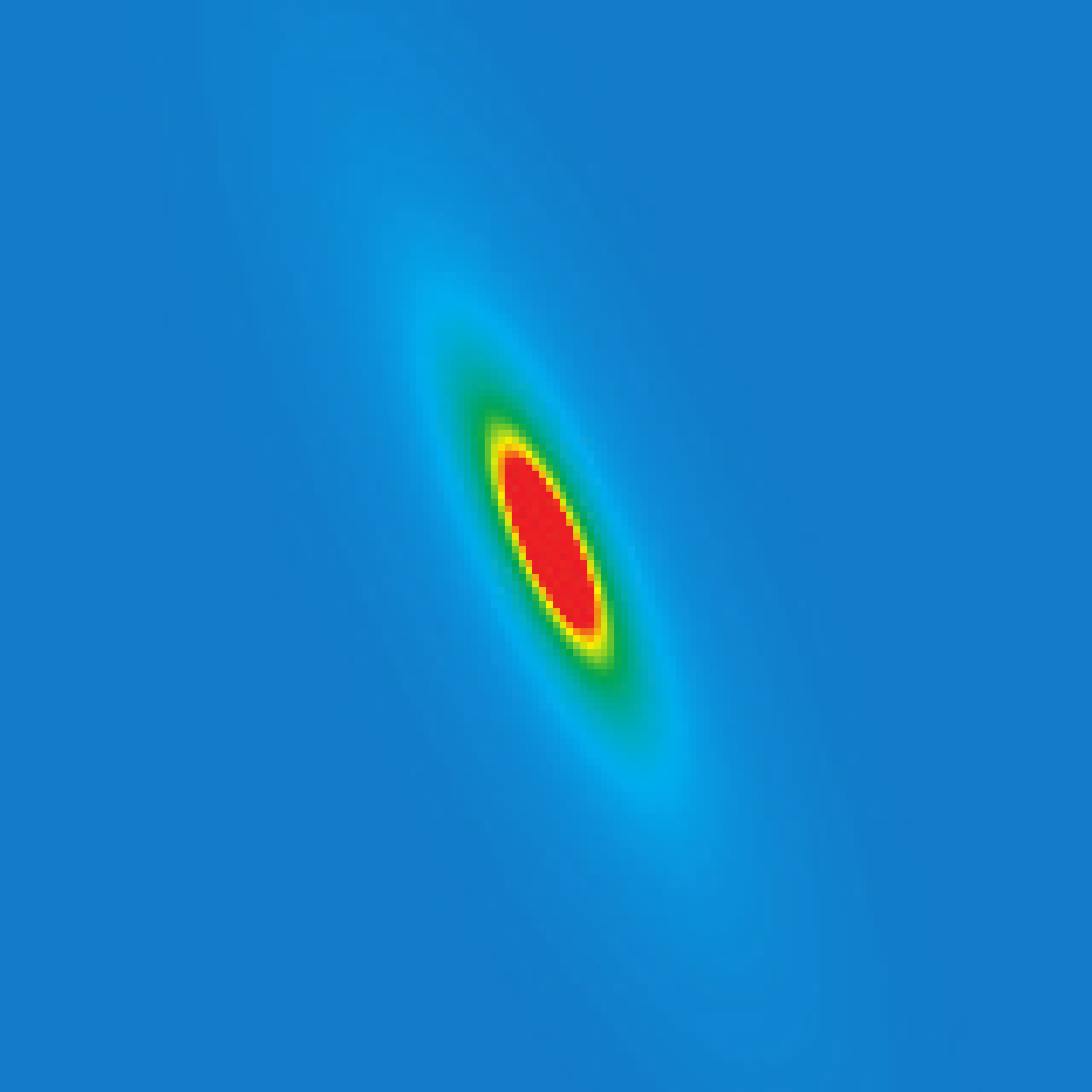}
\includegraphics[width=4cm]{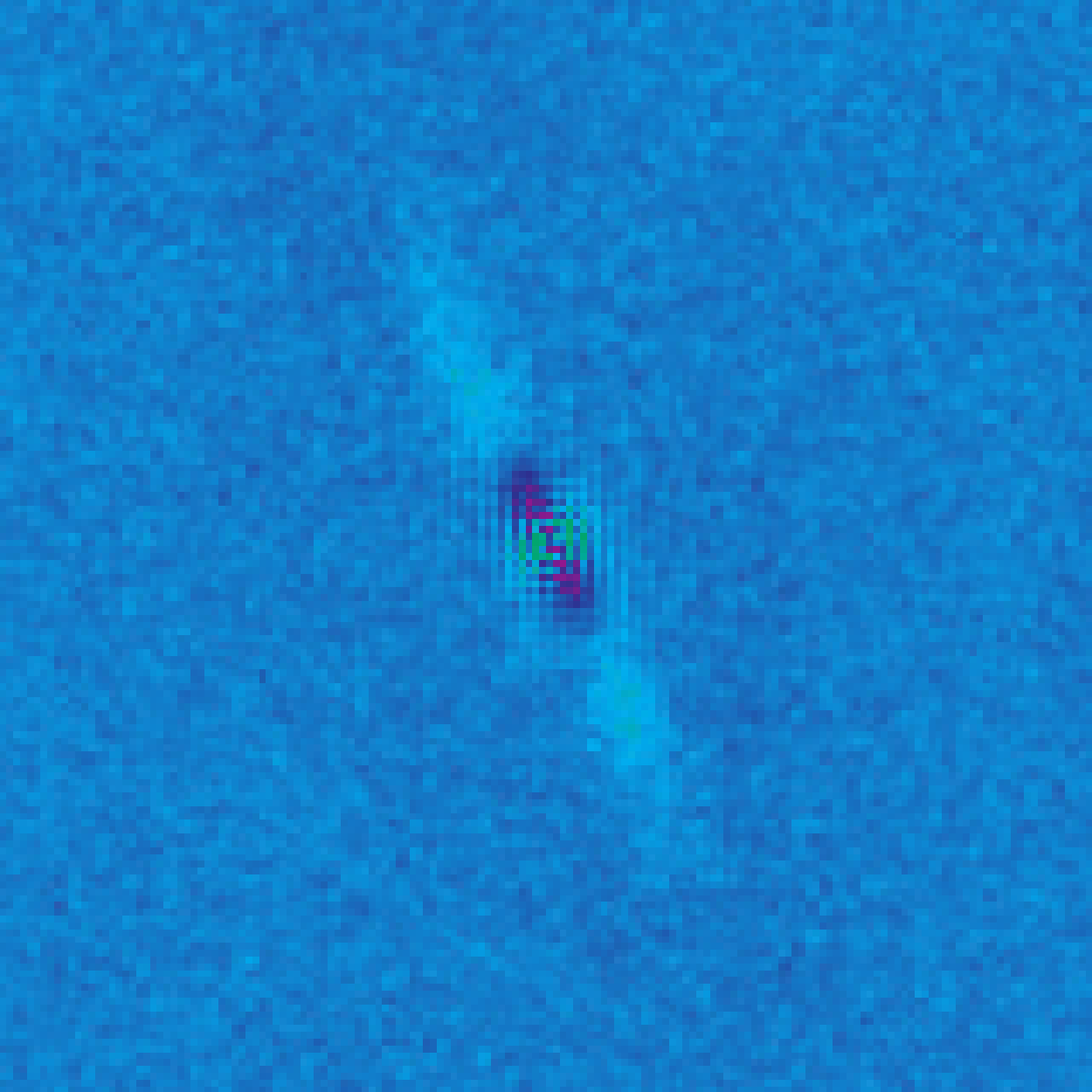}\\[1mm]
\hspace{40.7mm}\includegraphics[width=4cm]{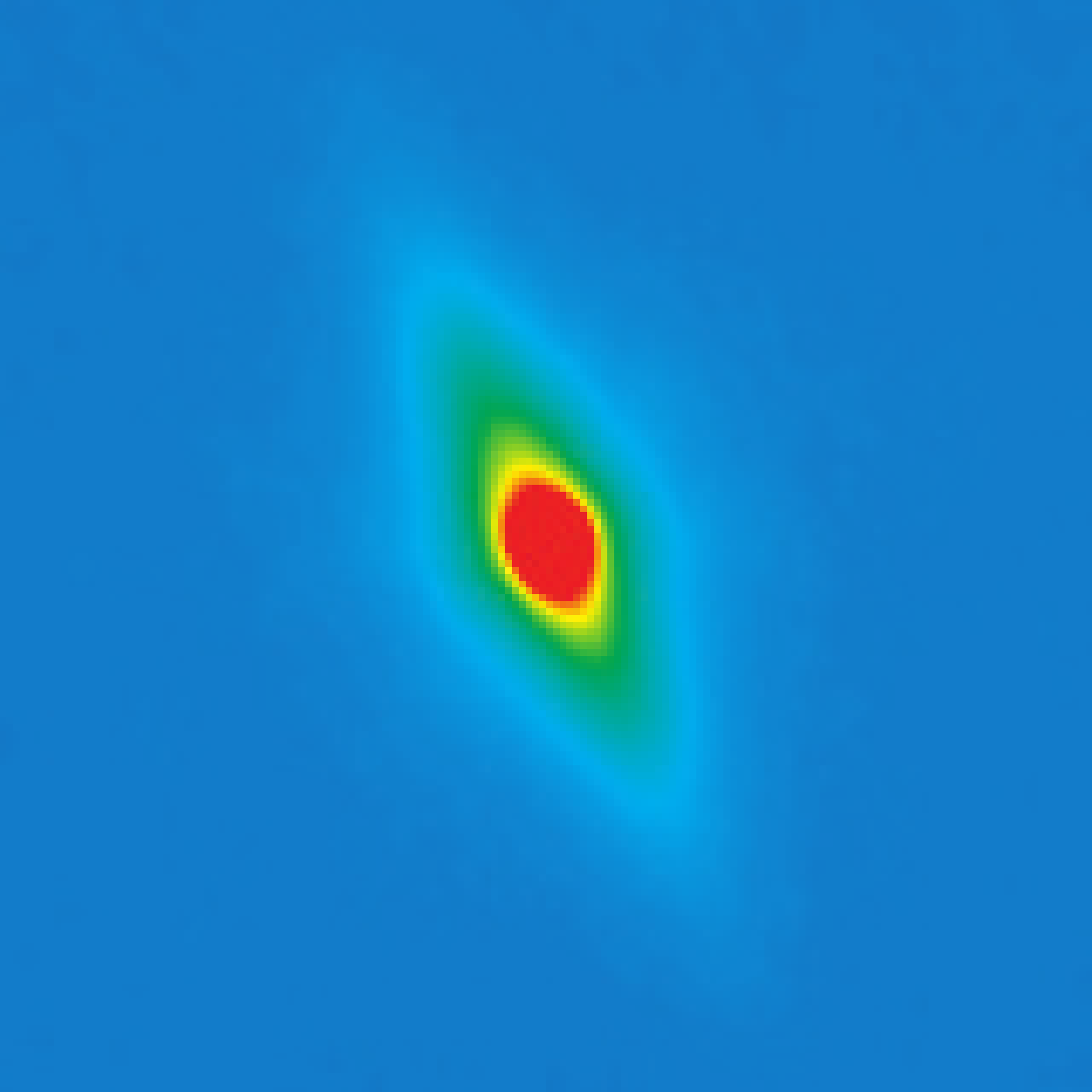}    
\includegraphics[width=4cm]{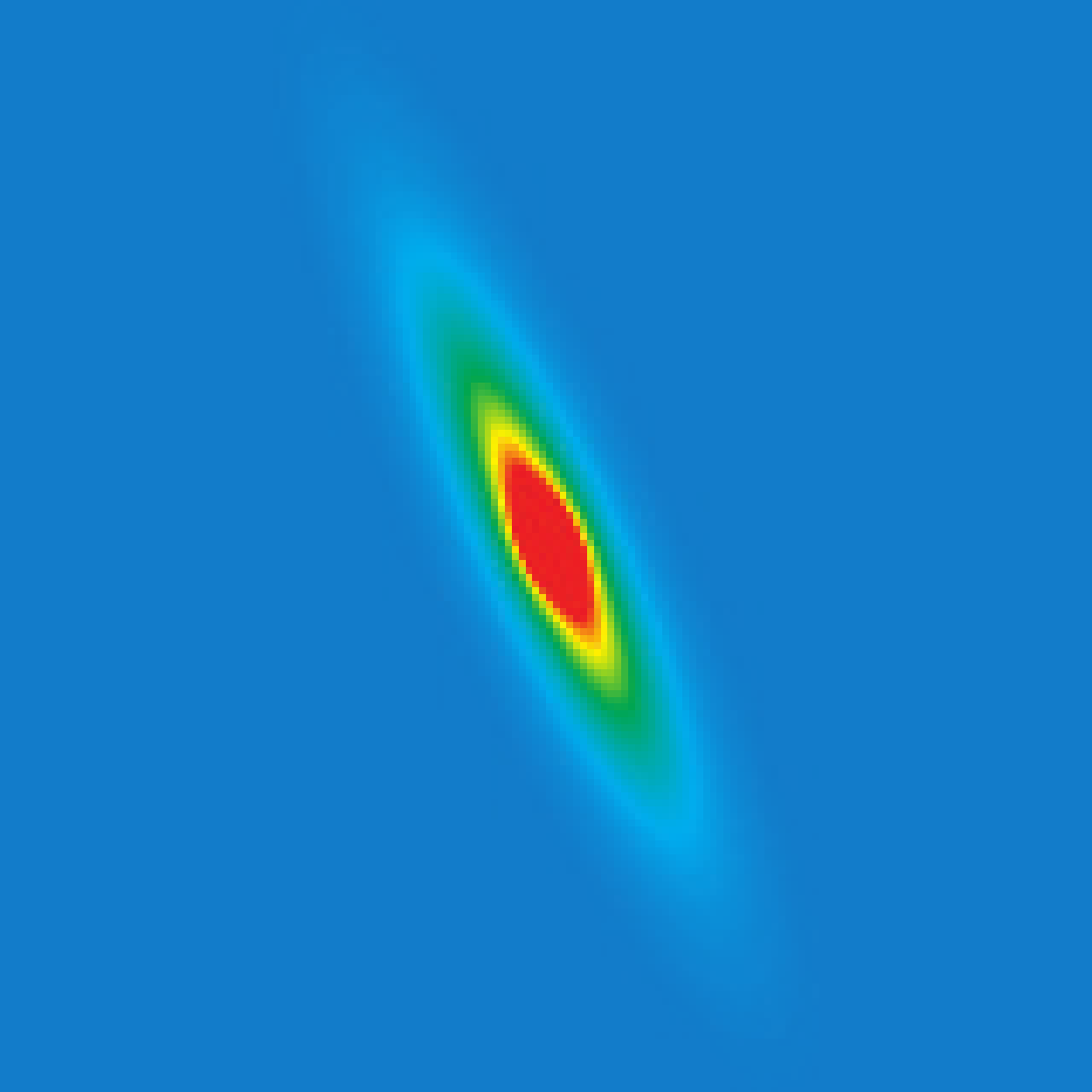}
\includegraphics[width=4cm]{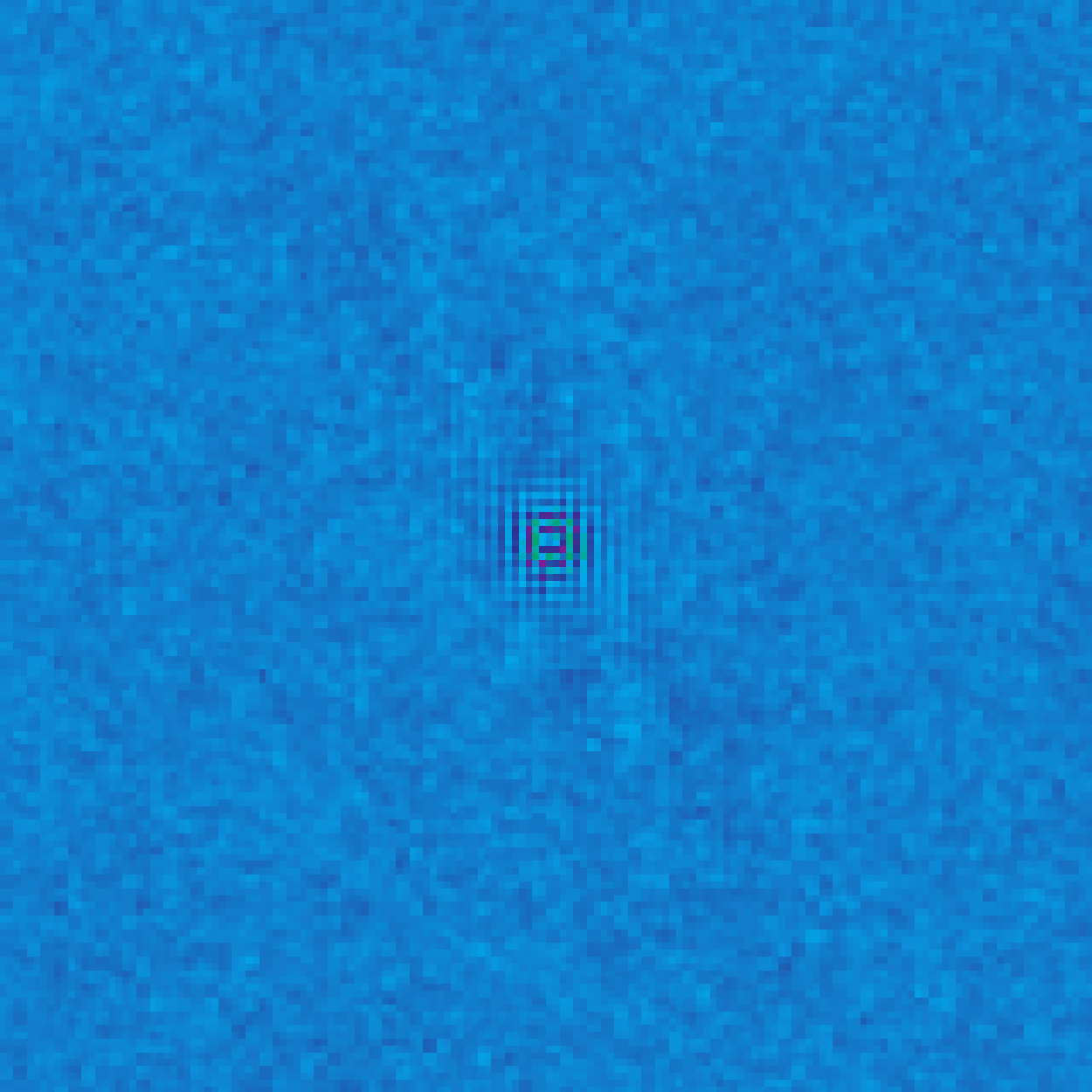}\\[2mm]
\includegraphics[width=4cm]{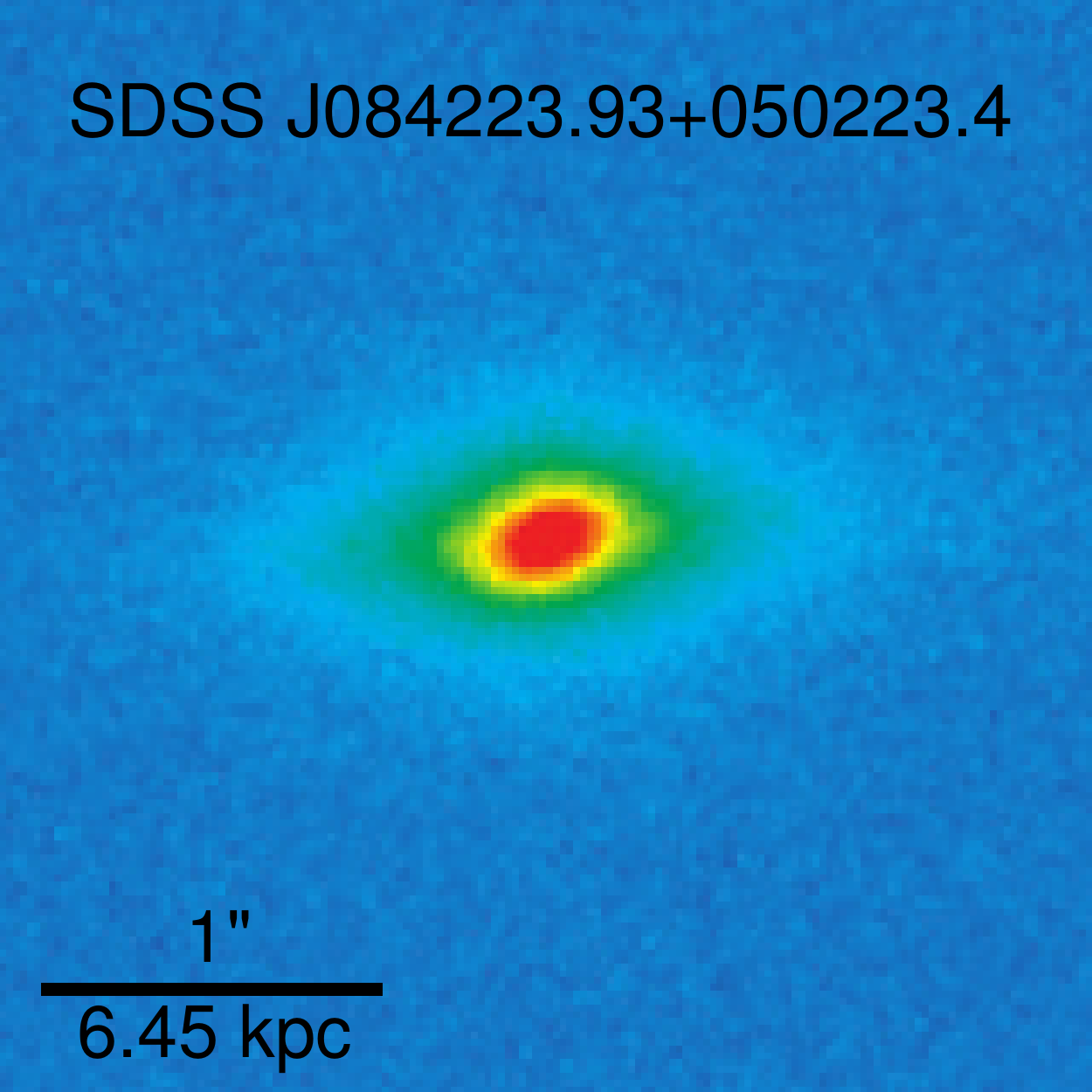}
\includegraphics[width=4cm]{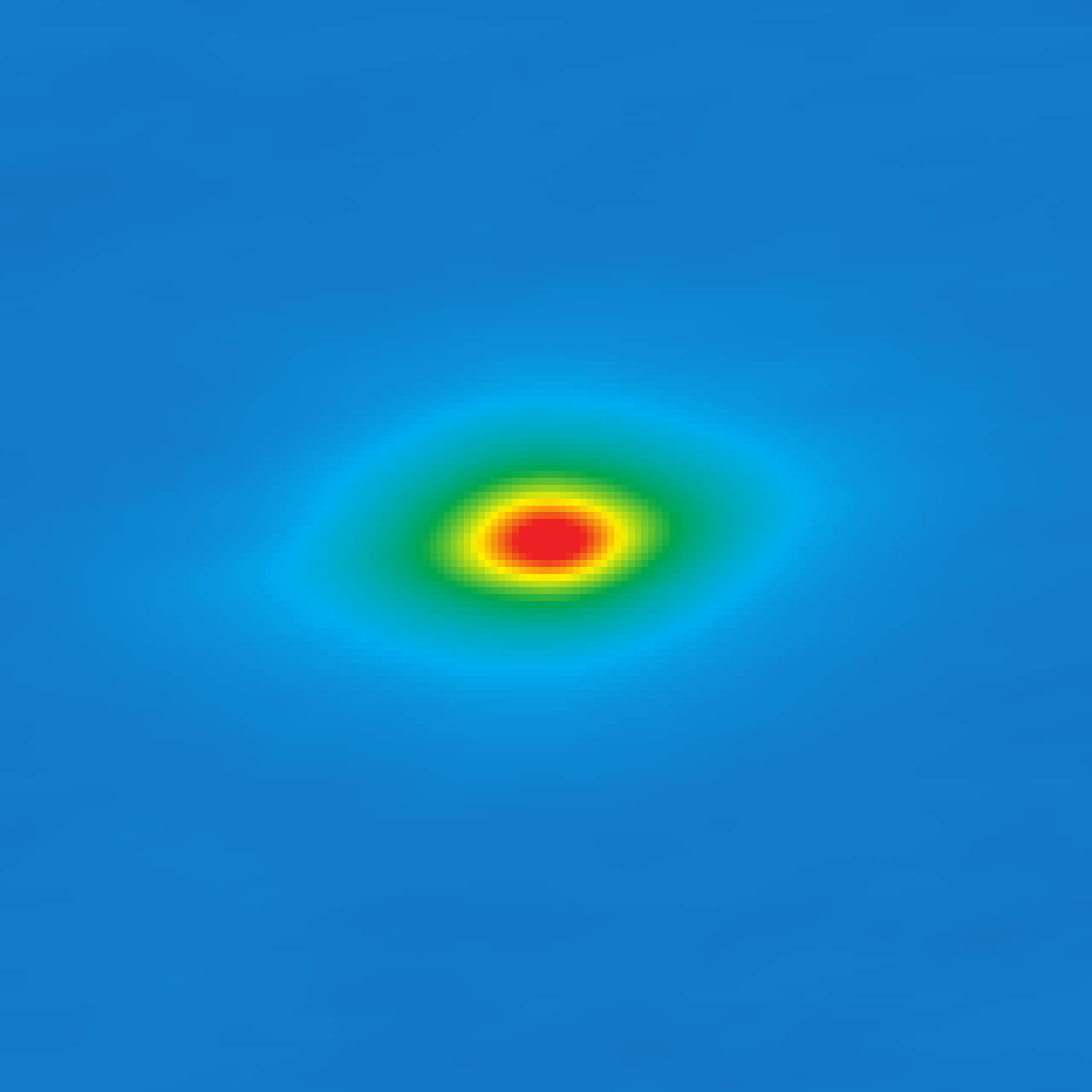}    
\includegraphics[width=4cm]{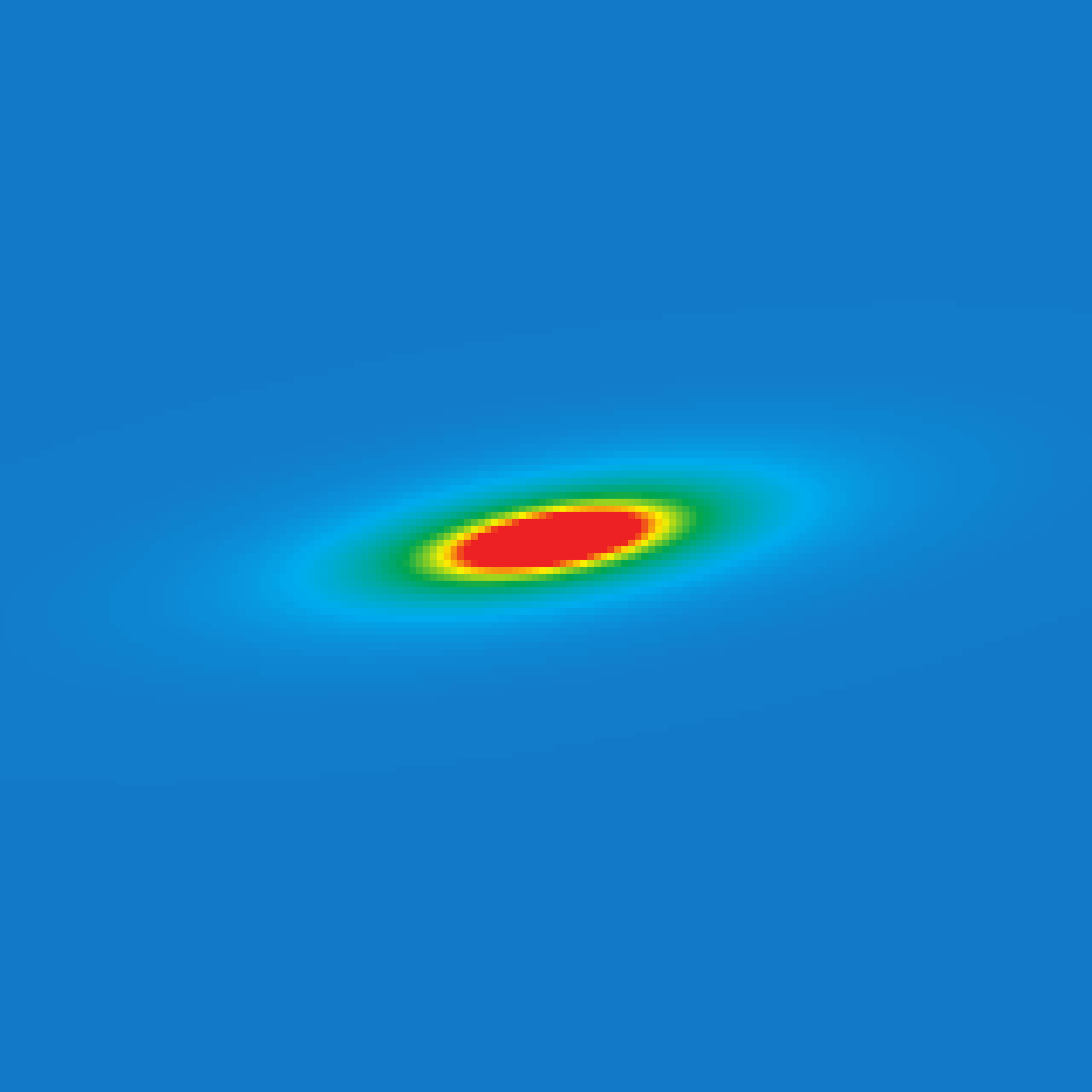}
\includegraphics[width=4cm]{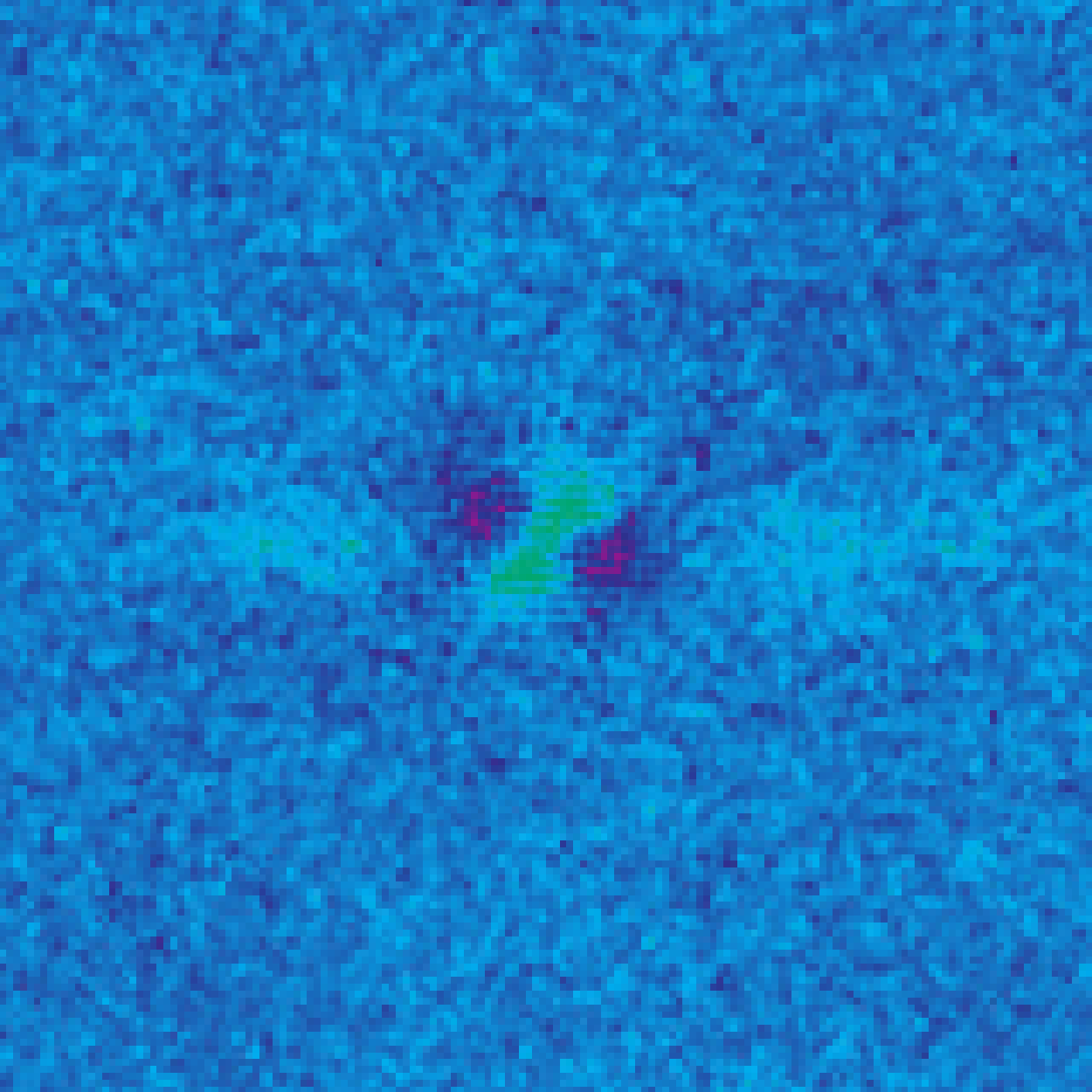}\\[1mm]
\hspace{40.7mm}\includegraphics[width=4cm]{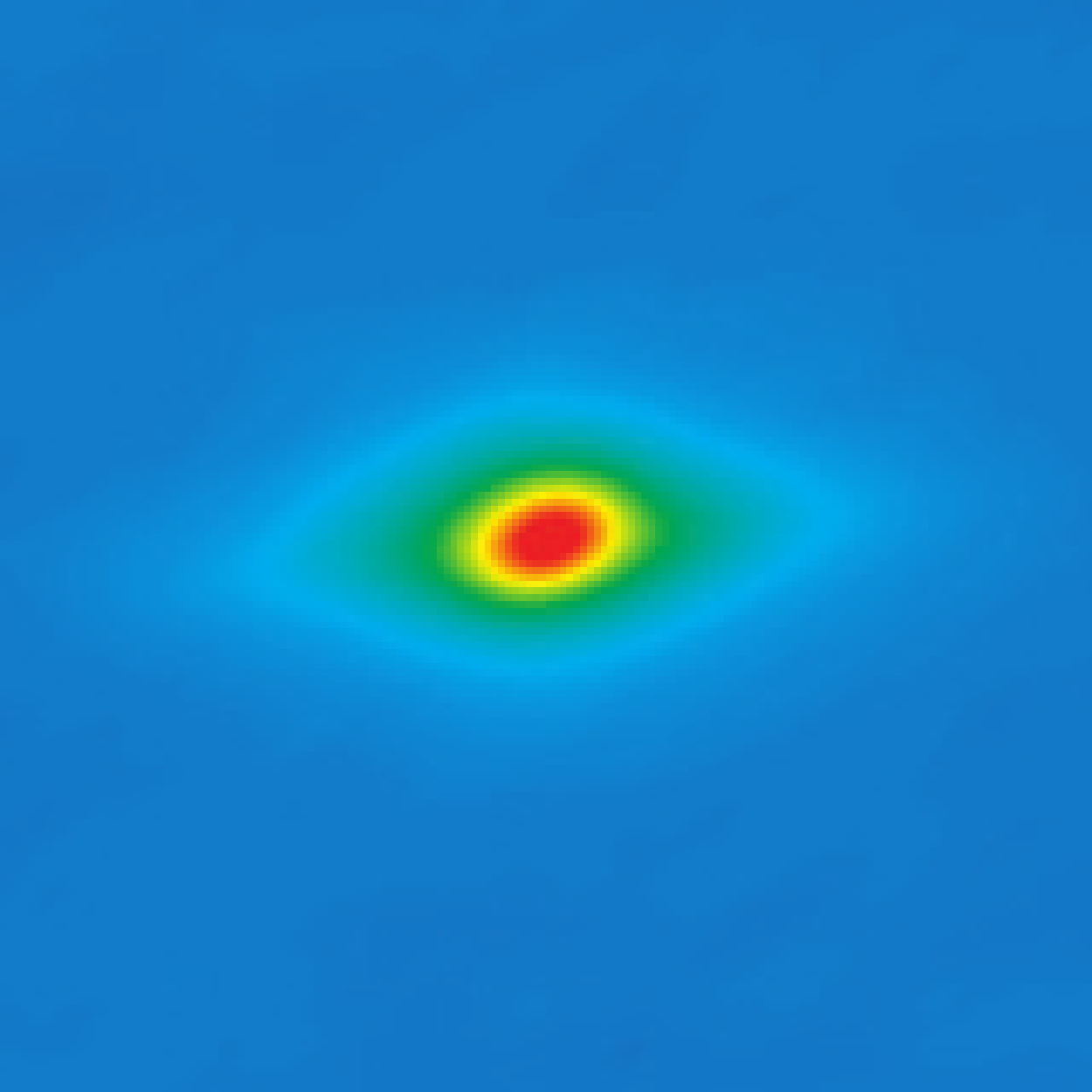}    
\includegraphics[width=4cm]{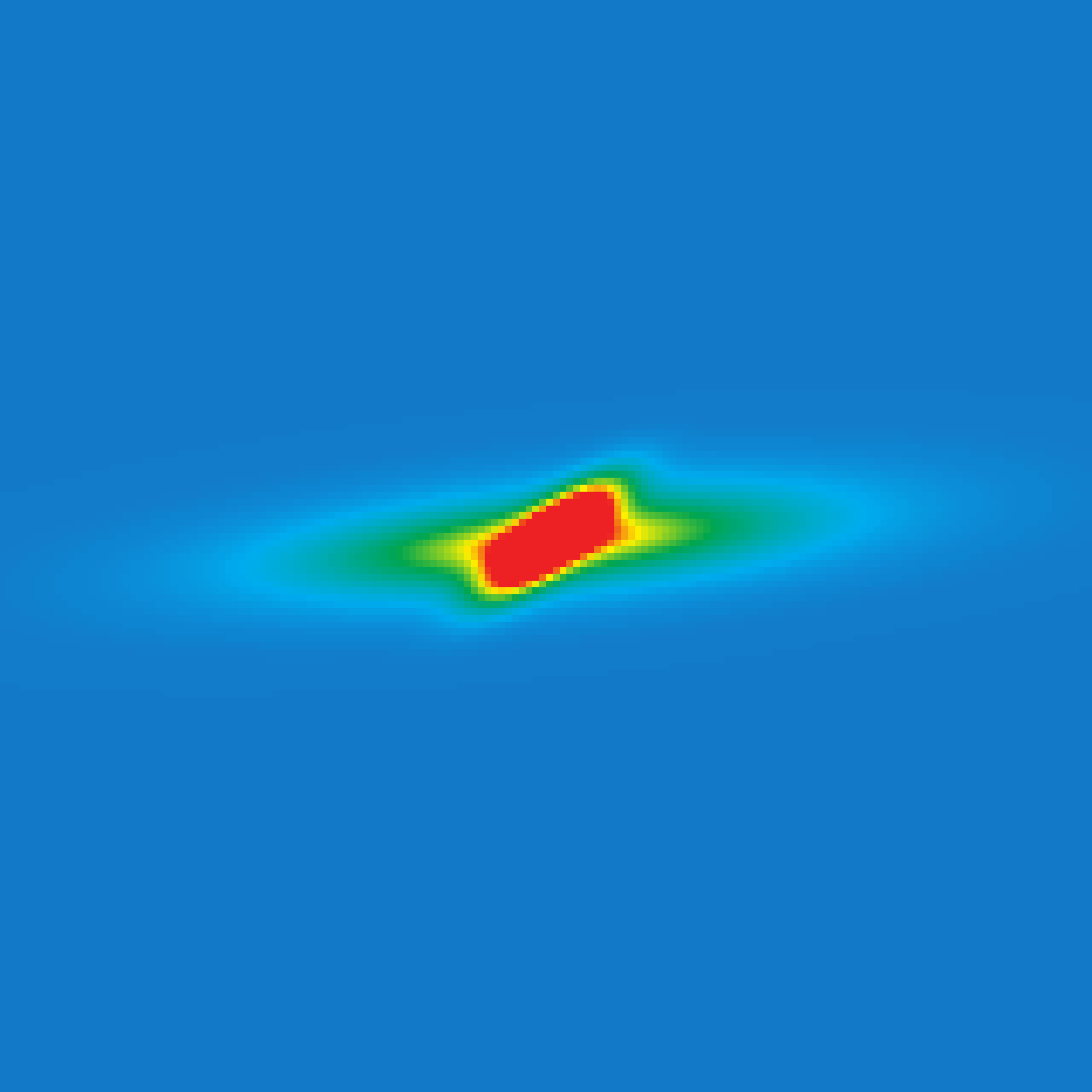}
\includegraphics[width=4cm]{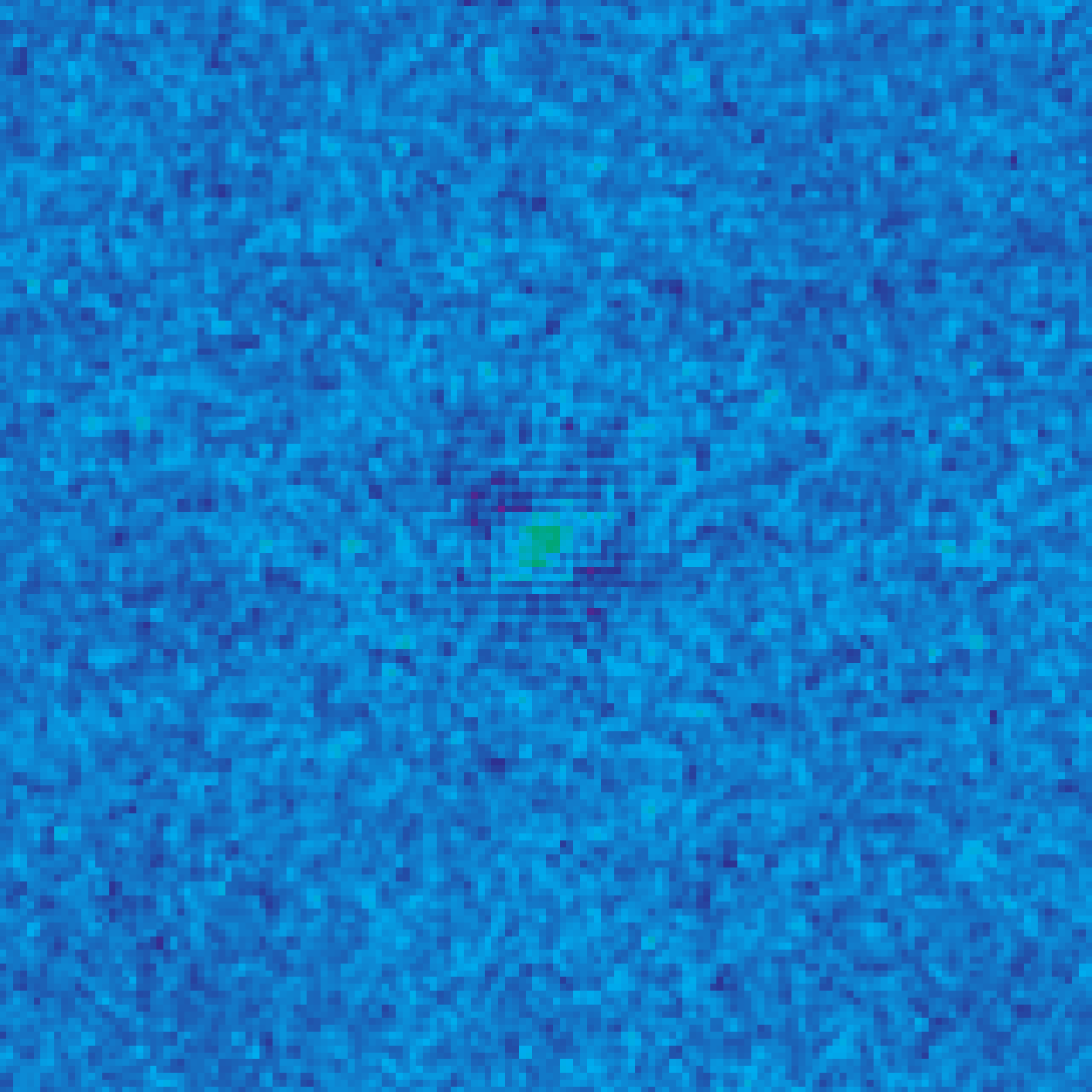}\\[2mm]
\includegraphics[width=4cm]{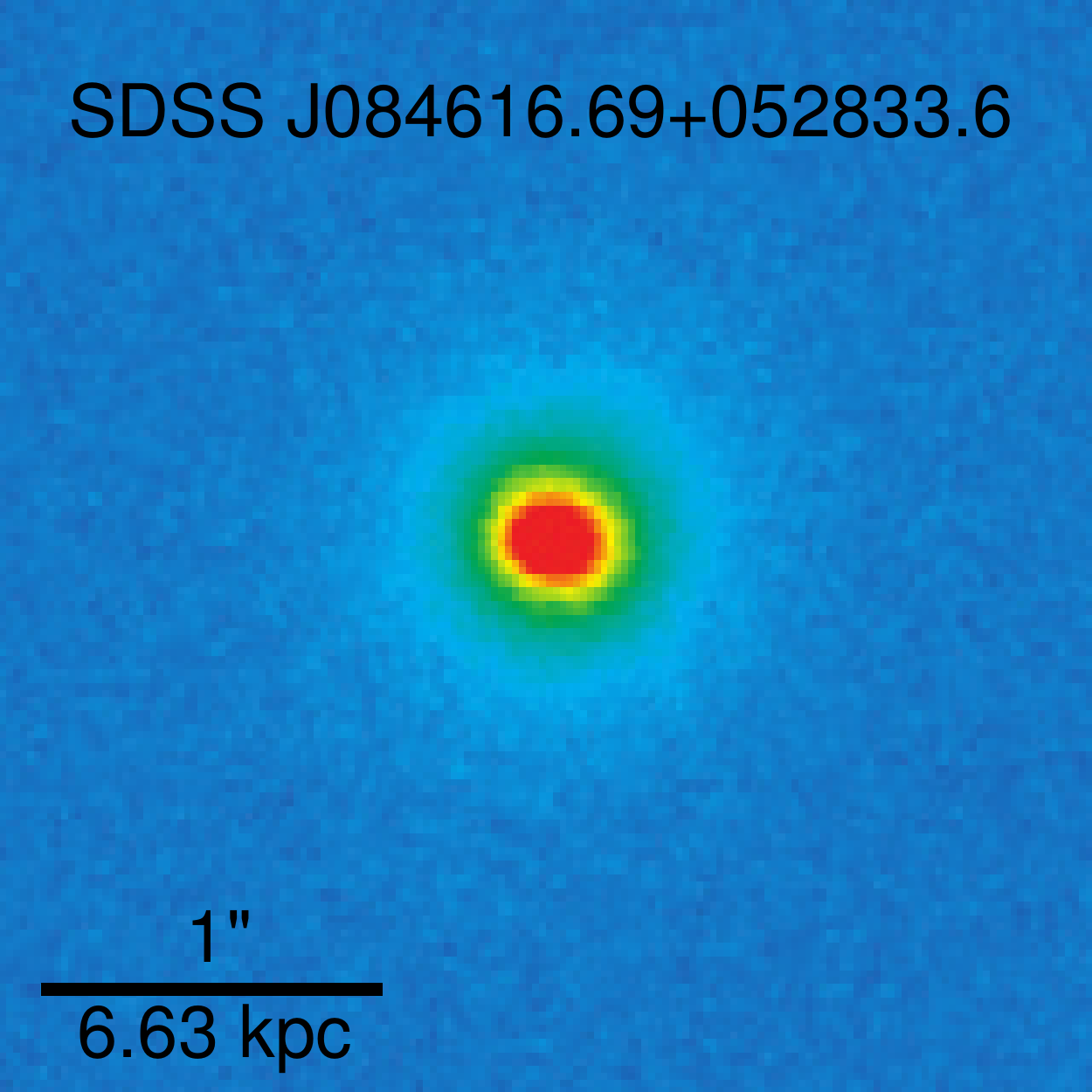}
\includegraphics[width=4cm]{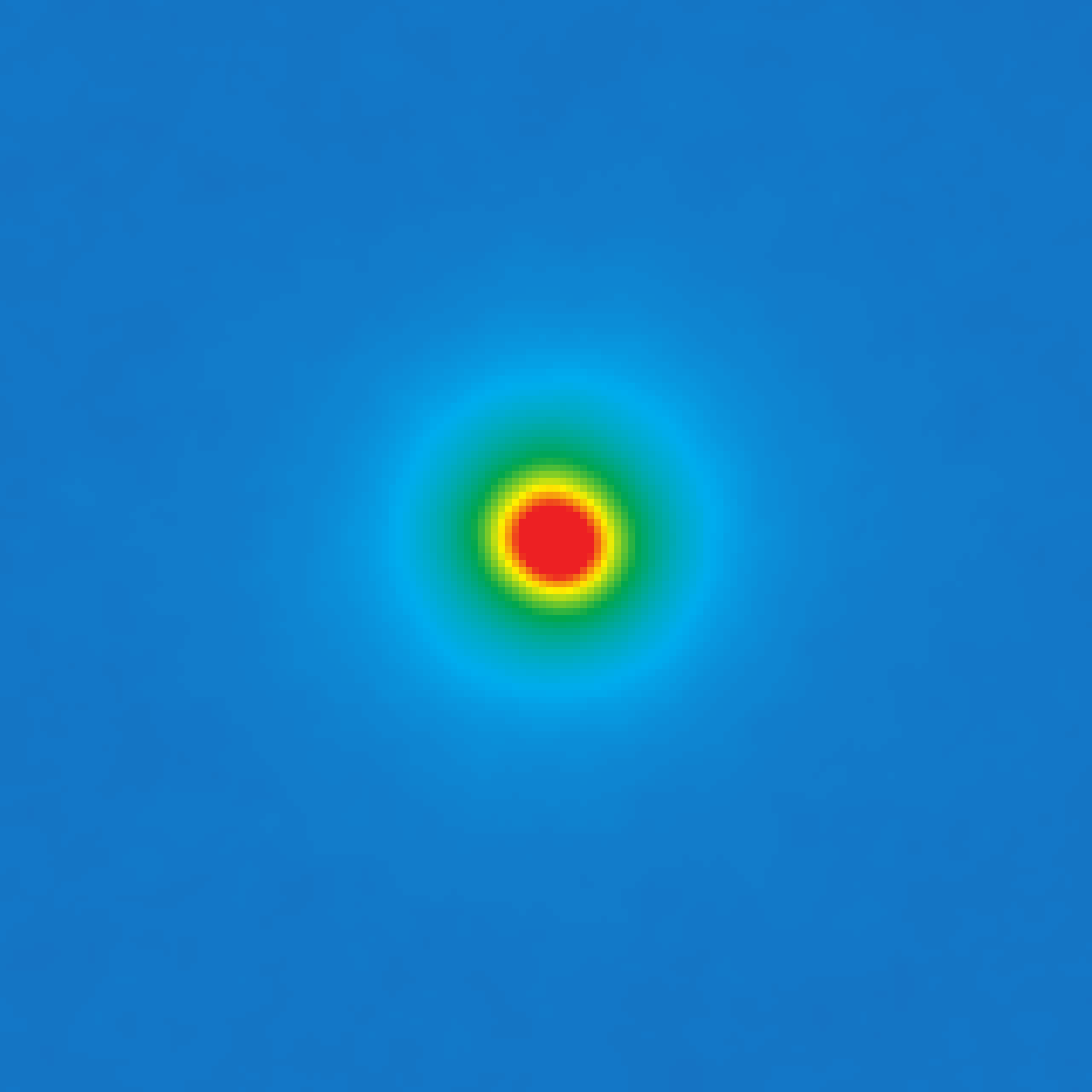}    
\includegraphics[width=4cm]{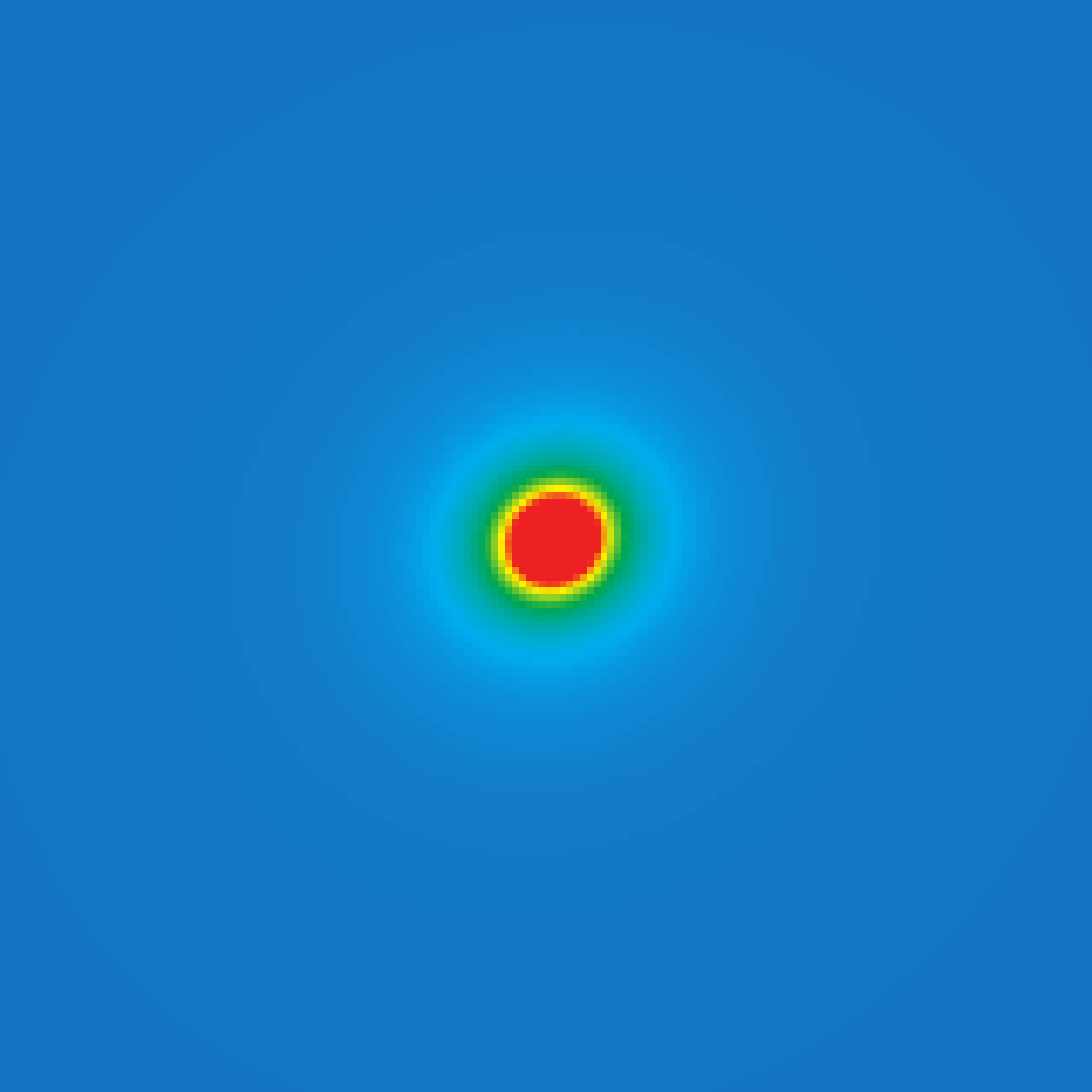}
\includegraphics[width=4cm]{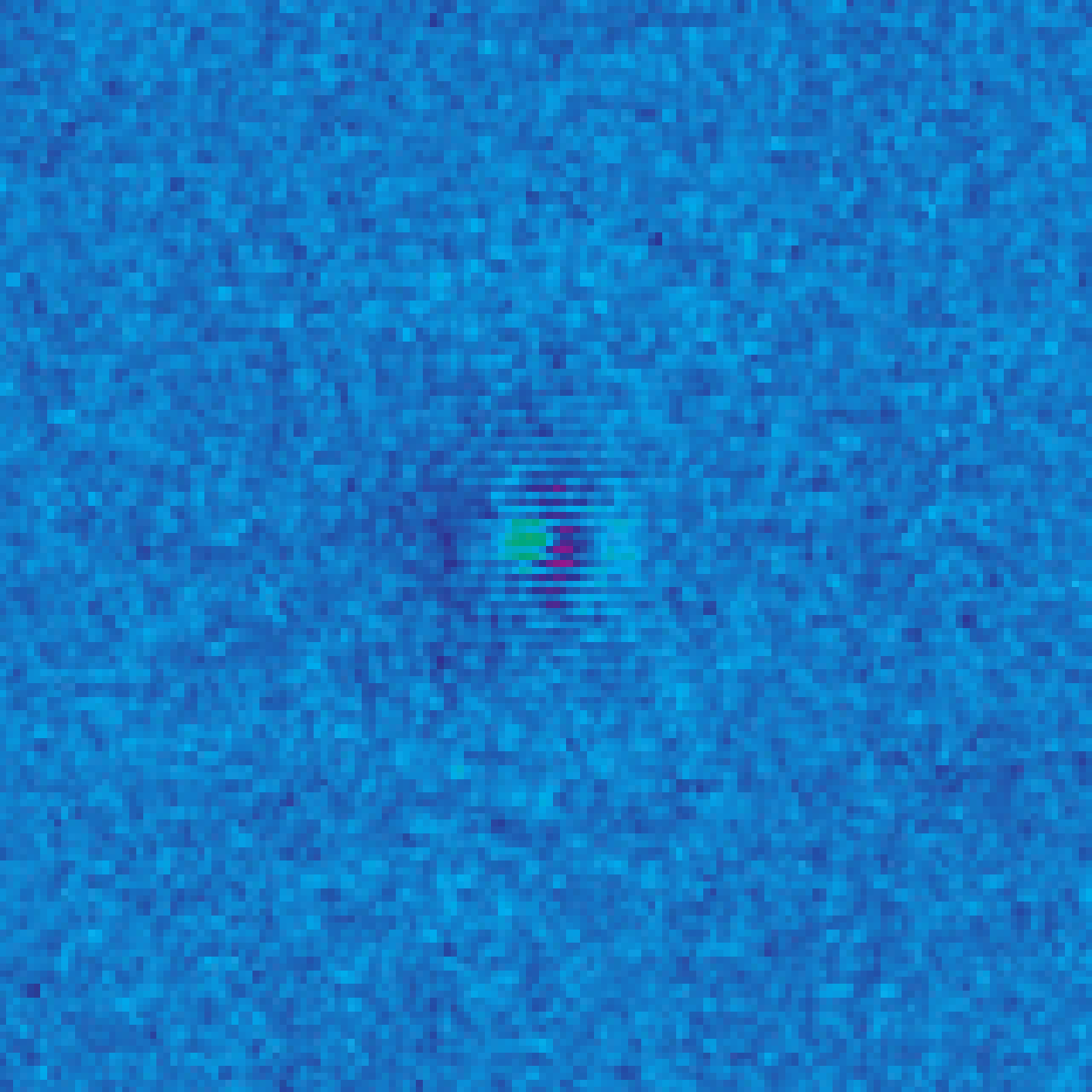}   
\caption{Continued. Notice that the PSF-deconvolved two-component model for SDSS\,J084223.93+050223.4 appears to be unphysical since the position angles of the two components do not align. However, we still keep this double-S${\rm \acute{e}}$rsic model because the galaxy image shows that the inner isophotes are indeed not aligned with the outer ones. Also, this model does much reduce the residual.}
\end{figure*}

\begin{figure*}
\centering
\includegraphics[width=4cm]{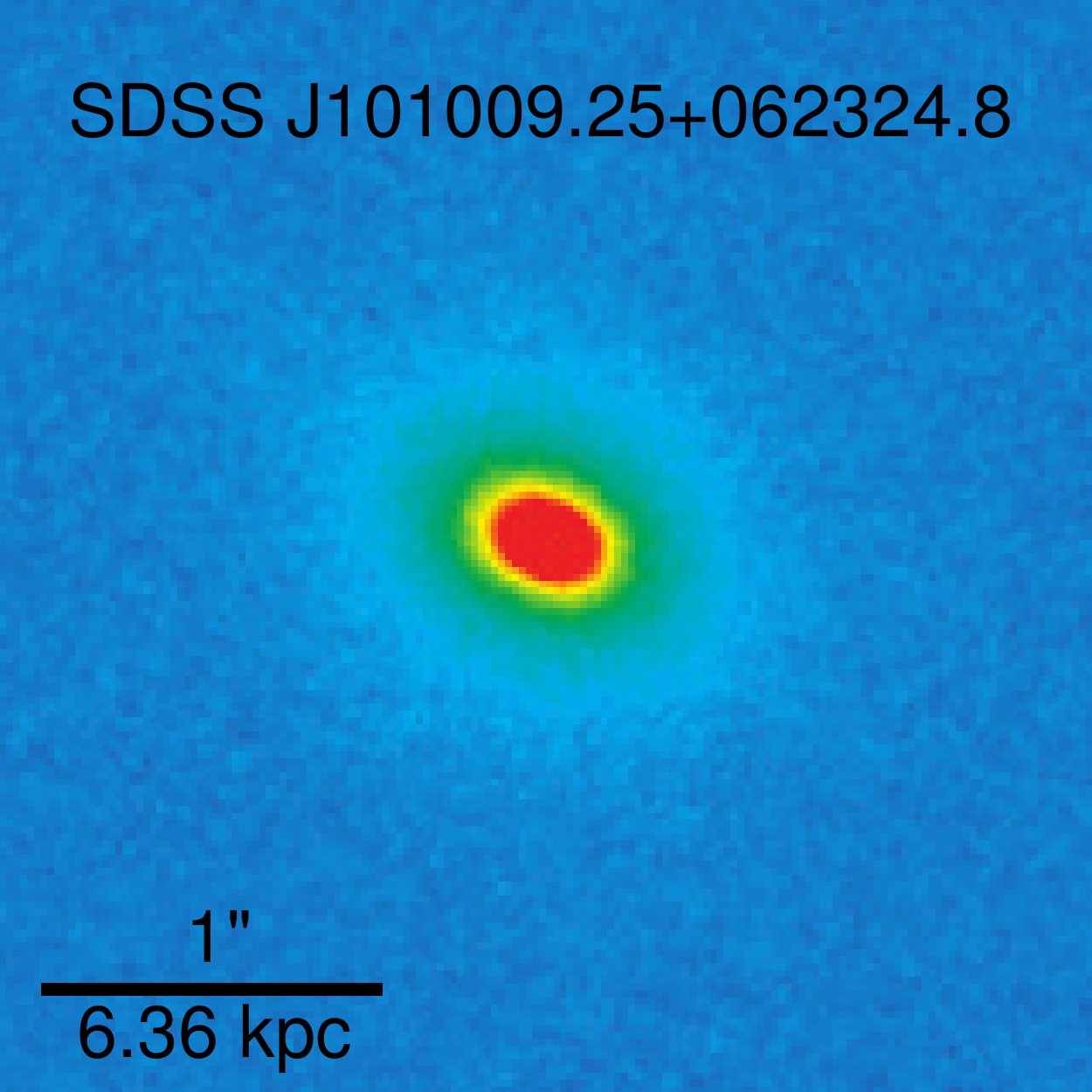}
\includegraphics[width=4cm]{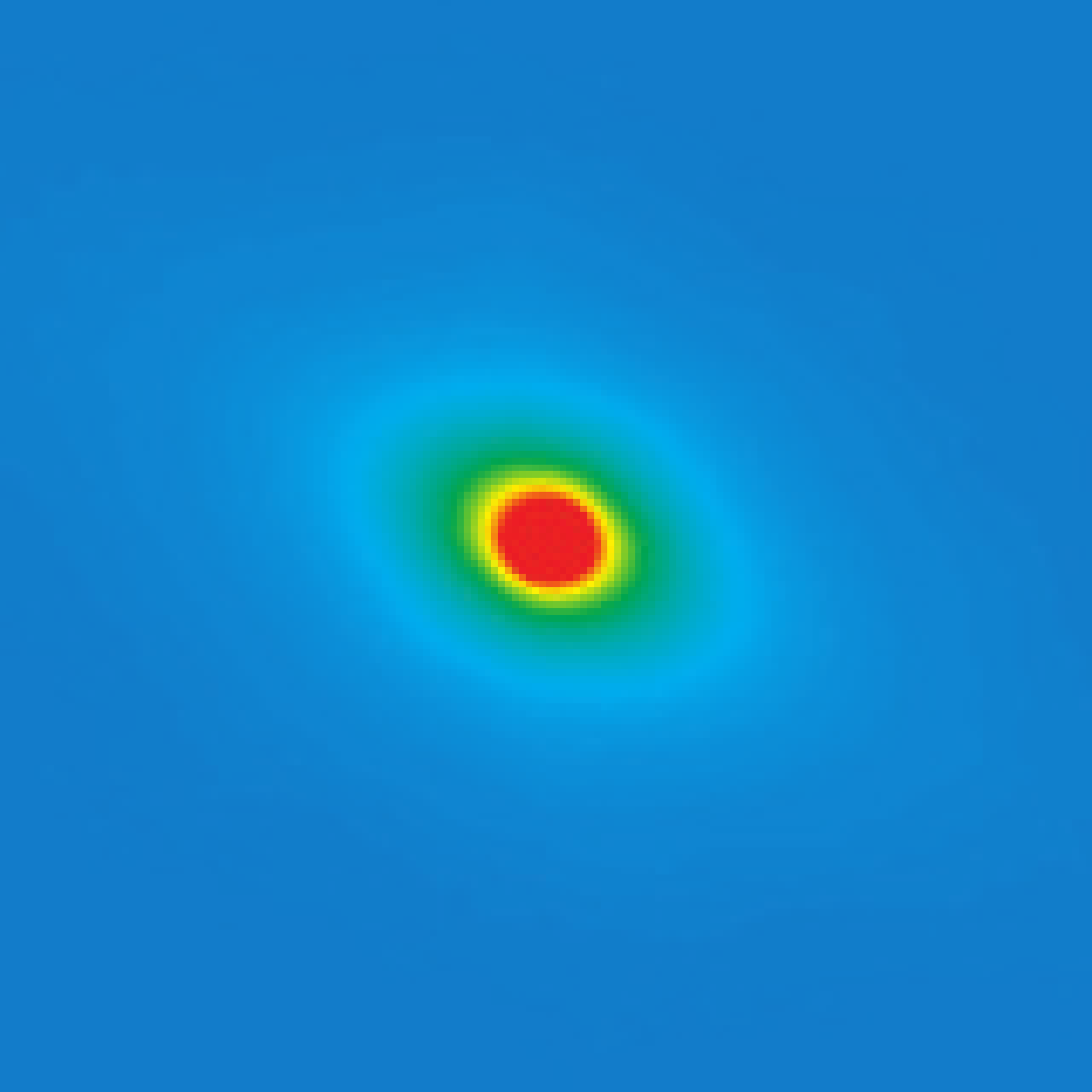}    
\includegraphics[width=4cm]{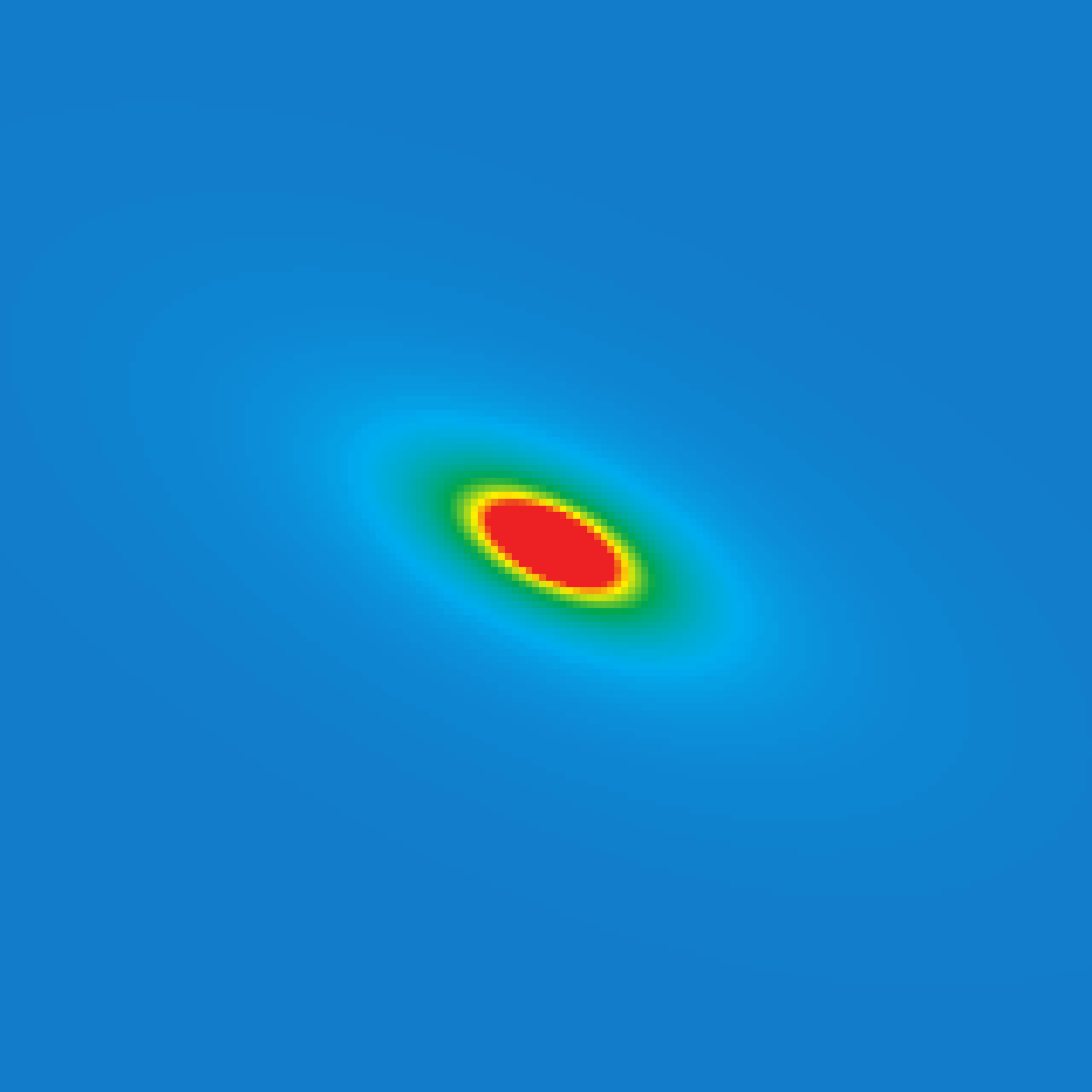}
\includegraphics[width=4cm]{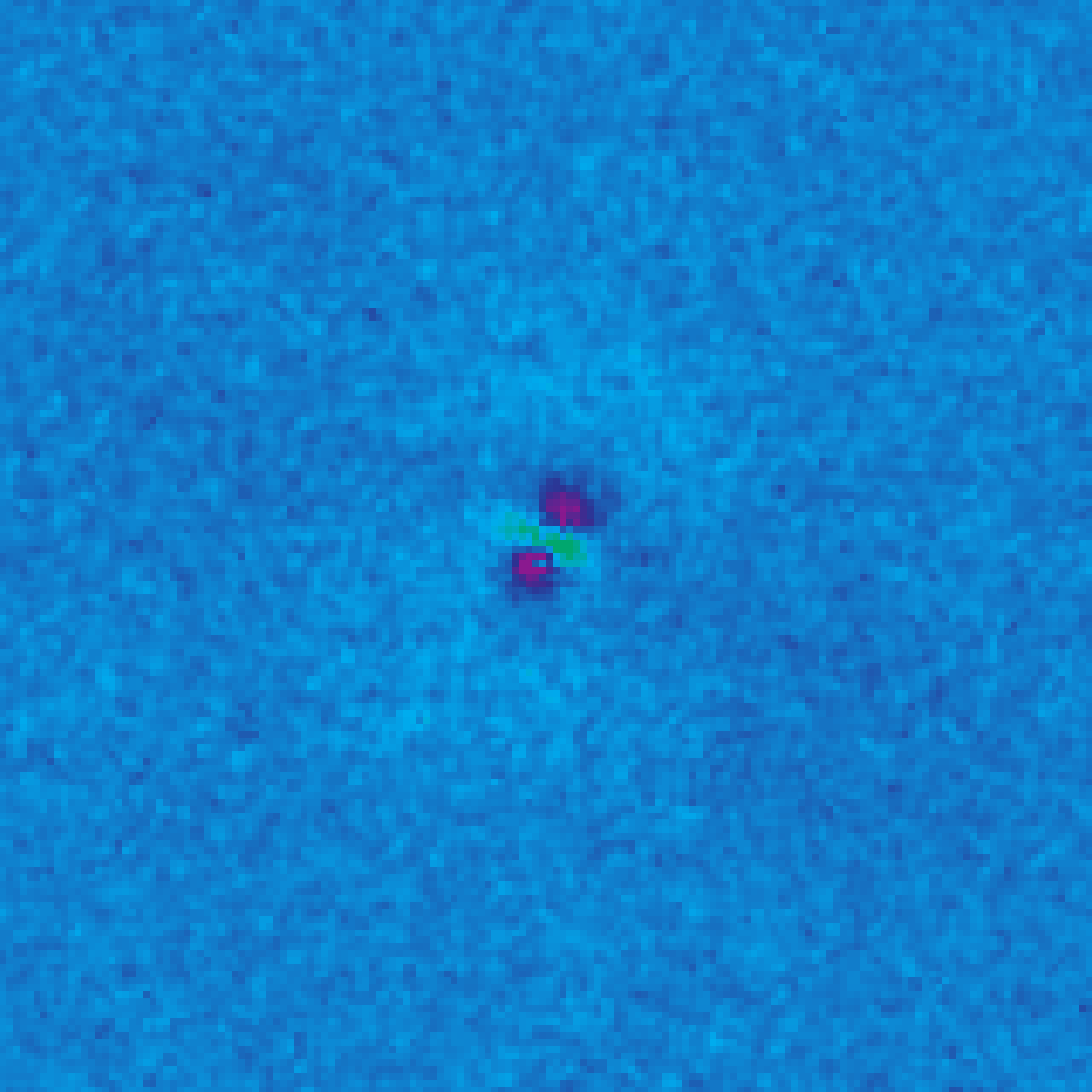}\\[2mm]
\includegraphics[width=4cm]{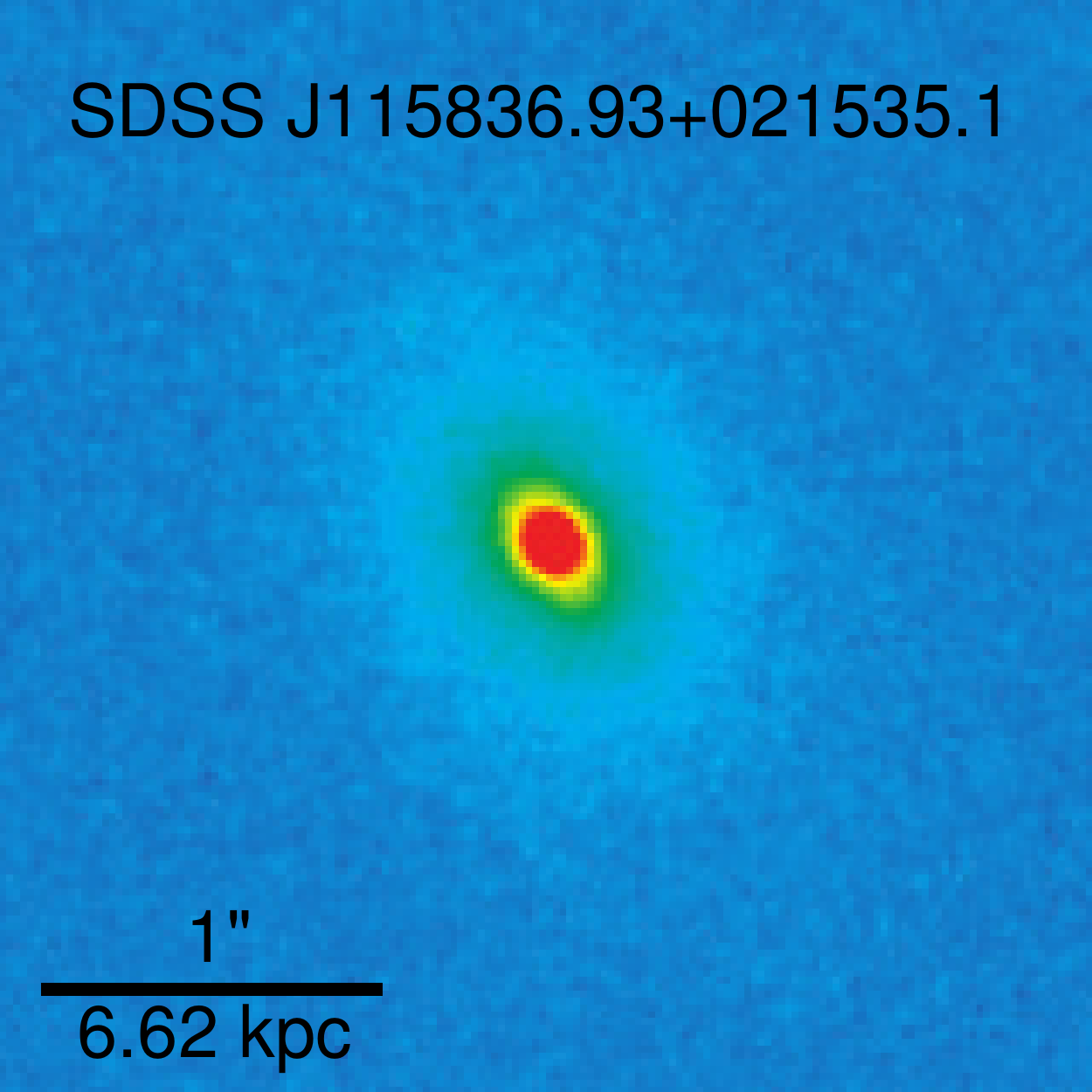}
\includegraphics[width=4cm]{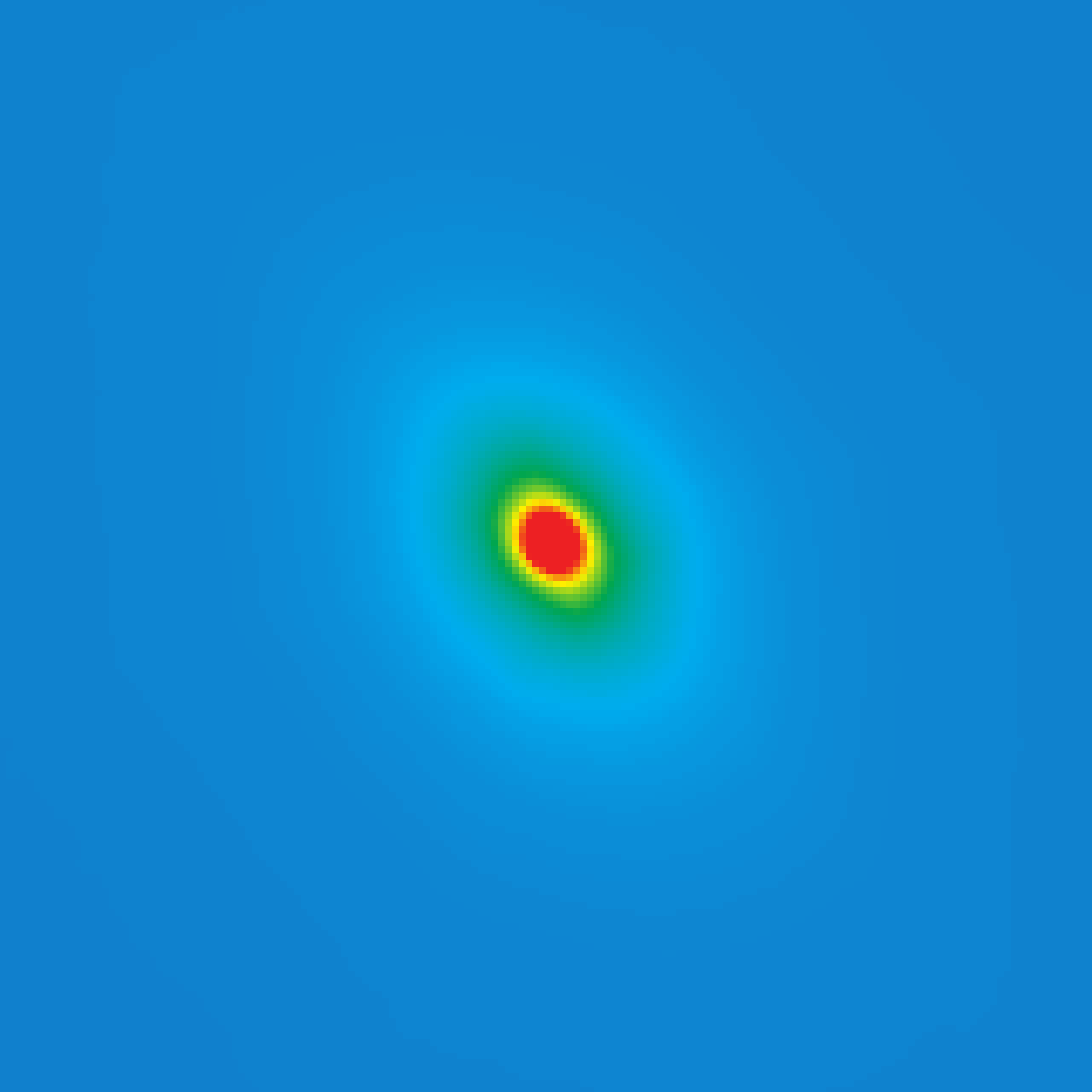}    
\includegraphics[width=4cm]{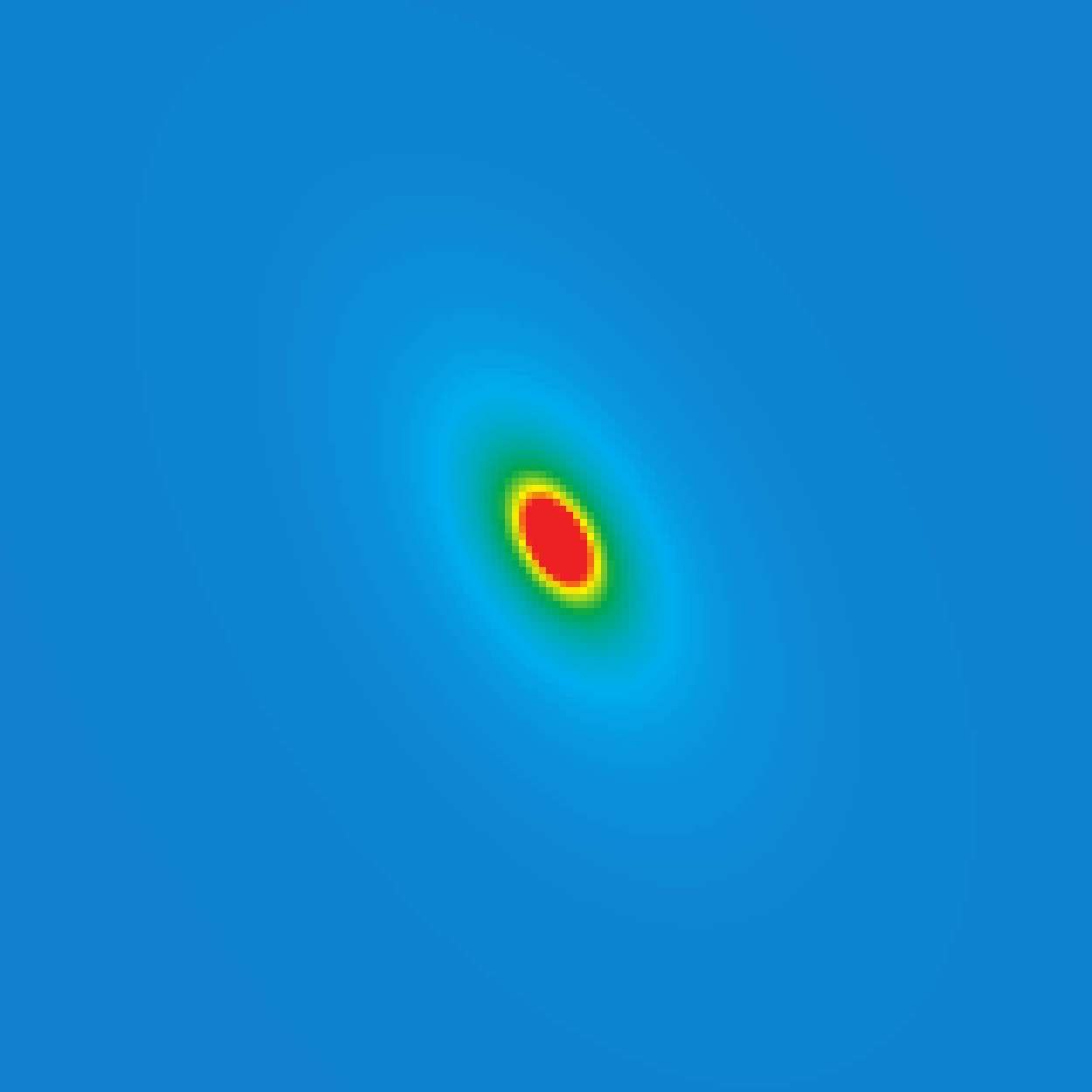}
\includegraphics[width=4cm]{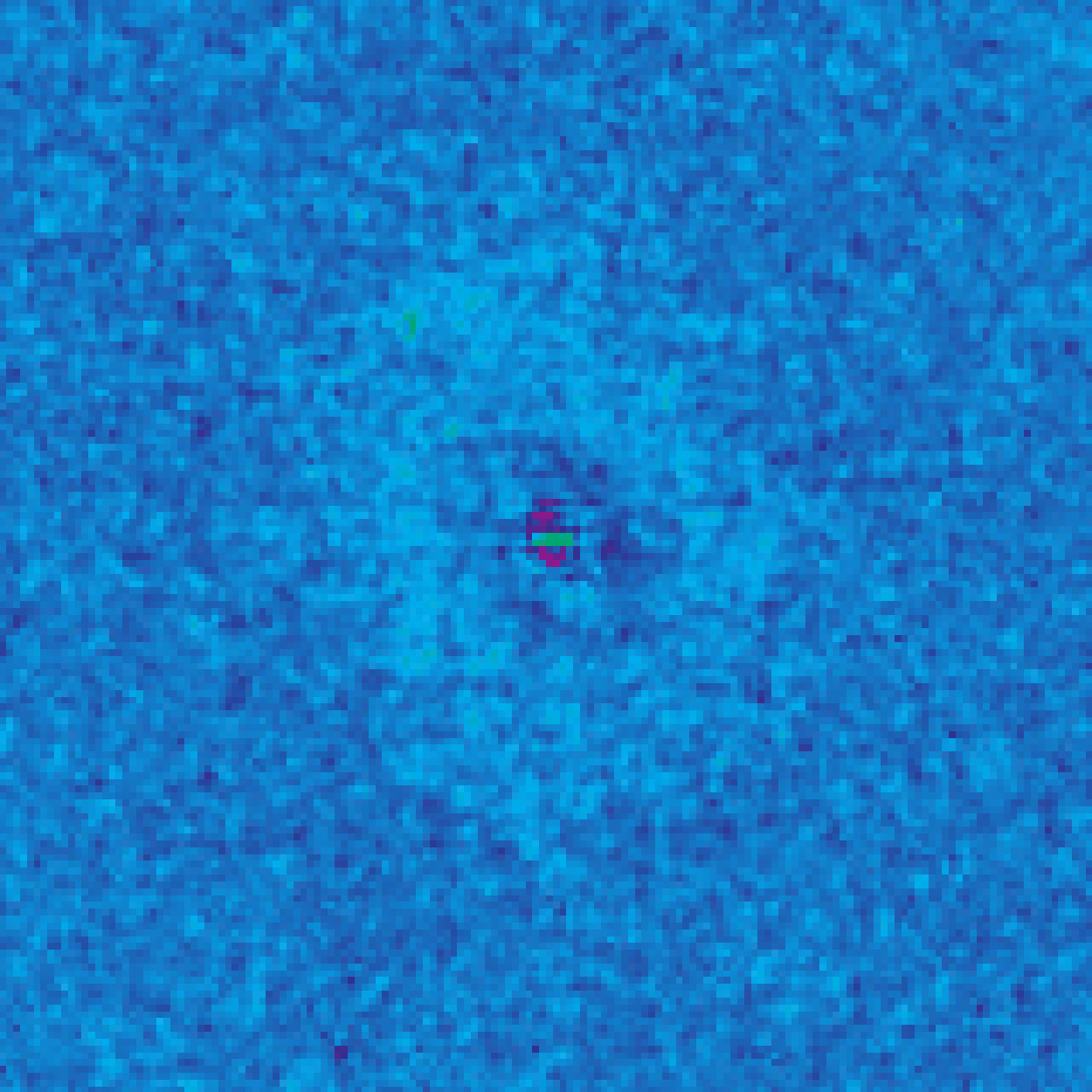}\\[1mm]
\hspace{40.7mm}\includegraphics[width=4cm]{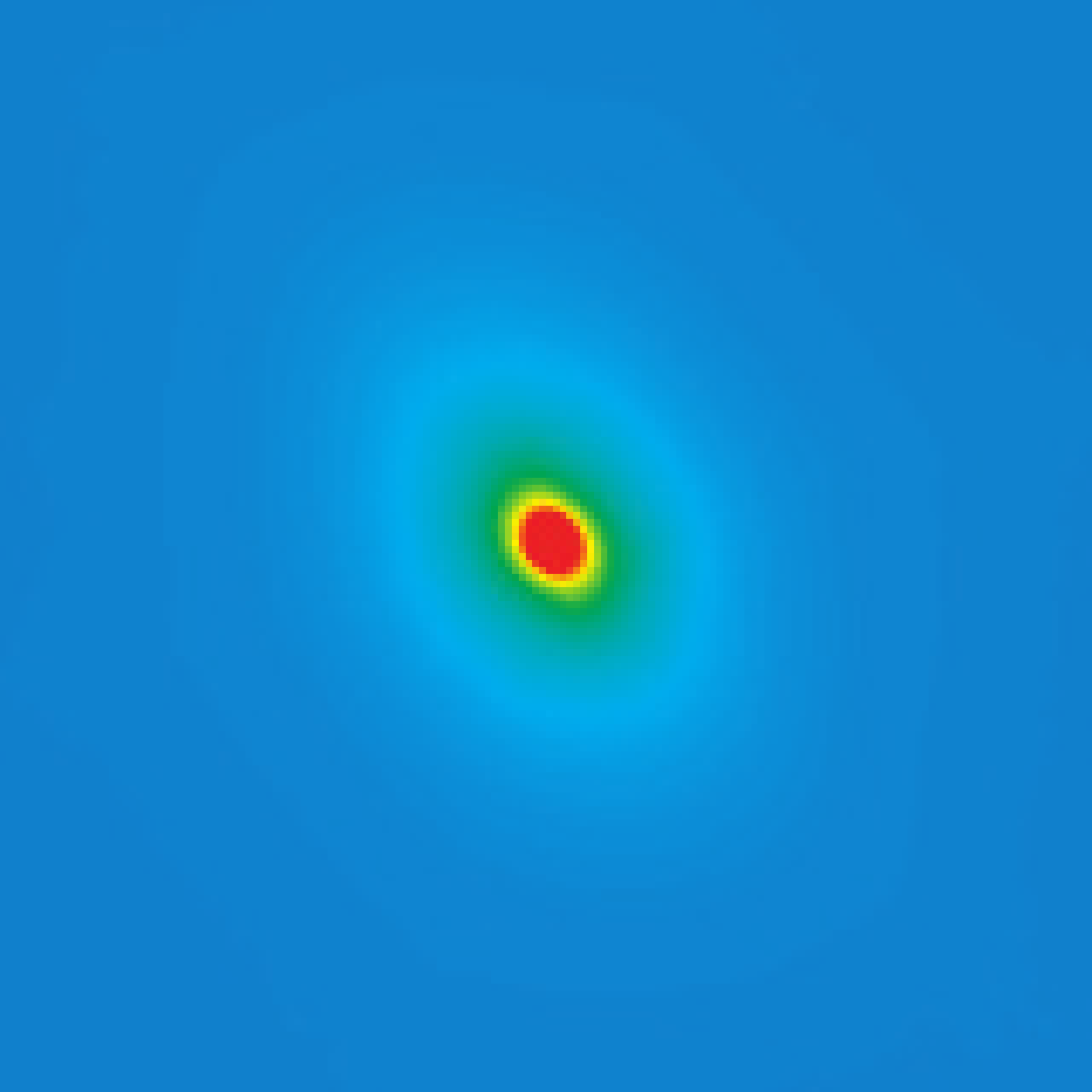}    
\includegraphics[width=4cm]{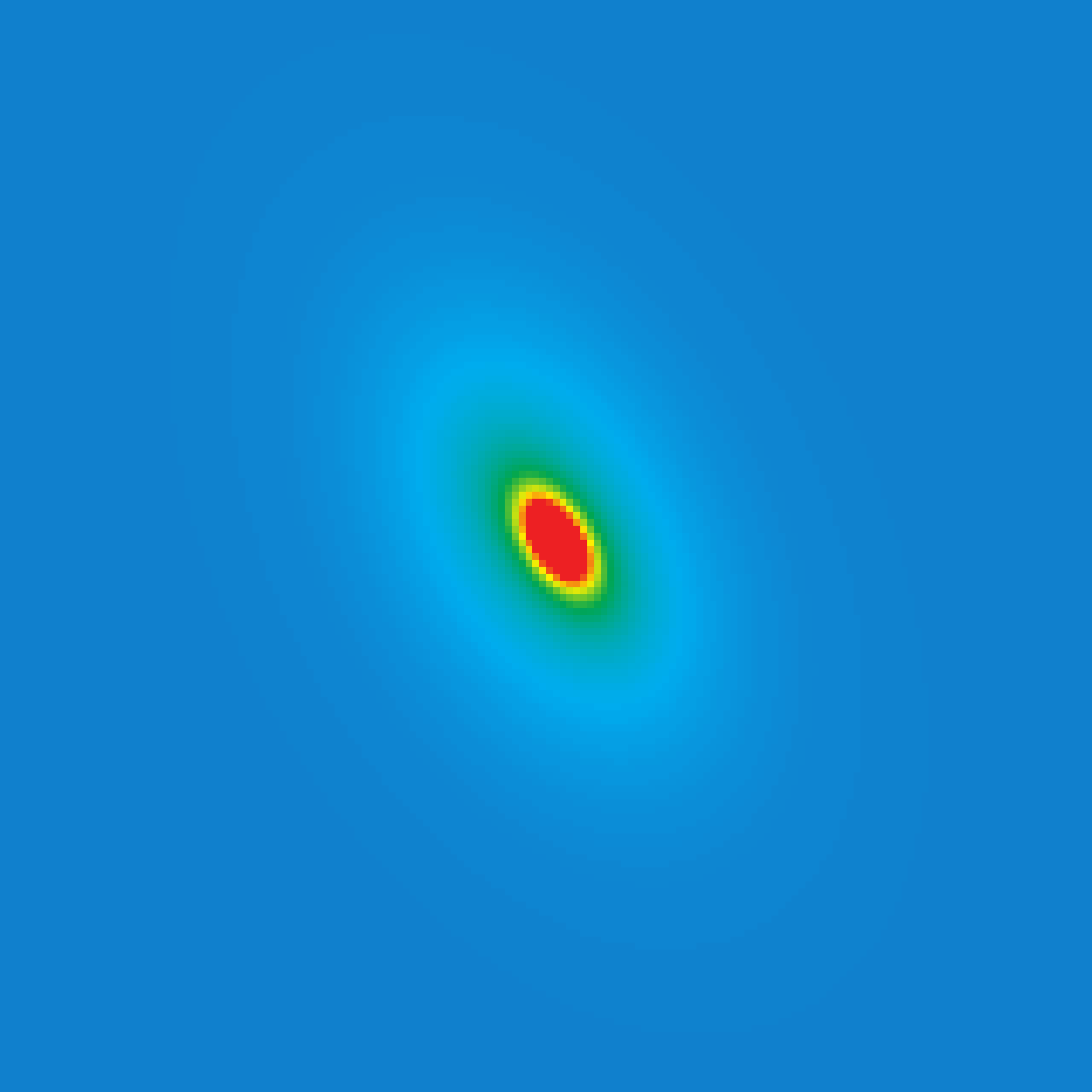}
\includegraphics[width=4cm]{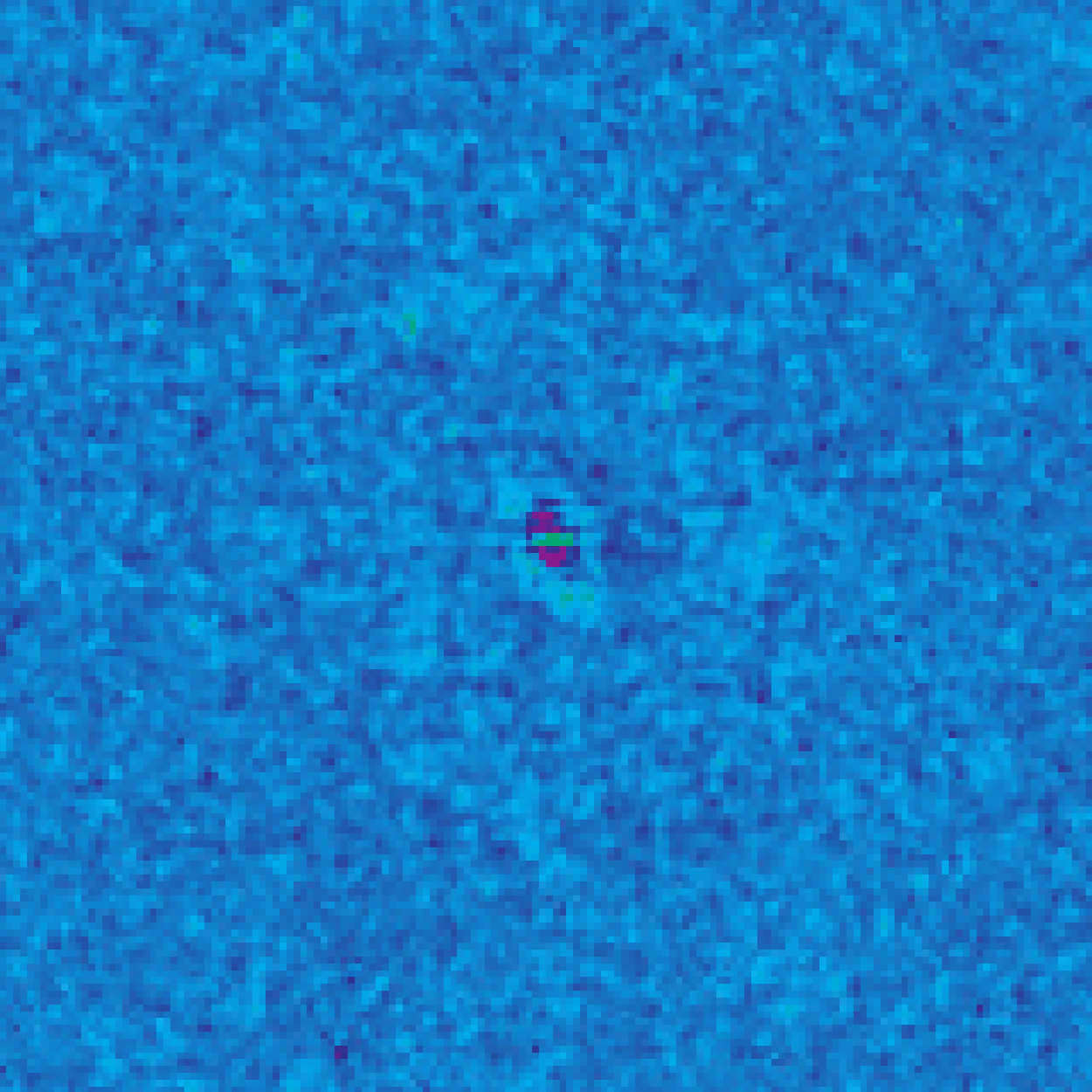}\\[2mm]
\includegraphics[width=4cm]{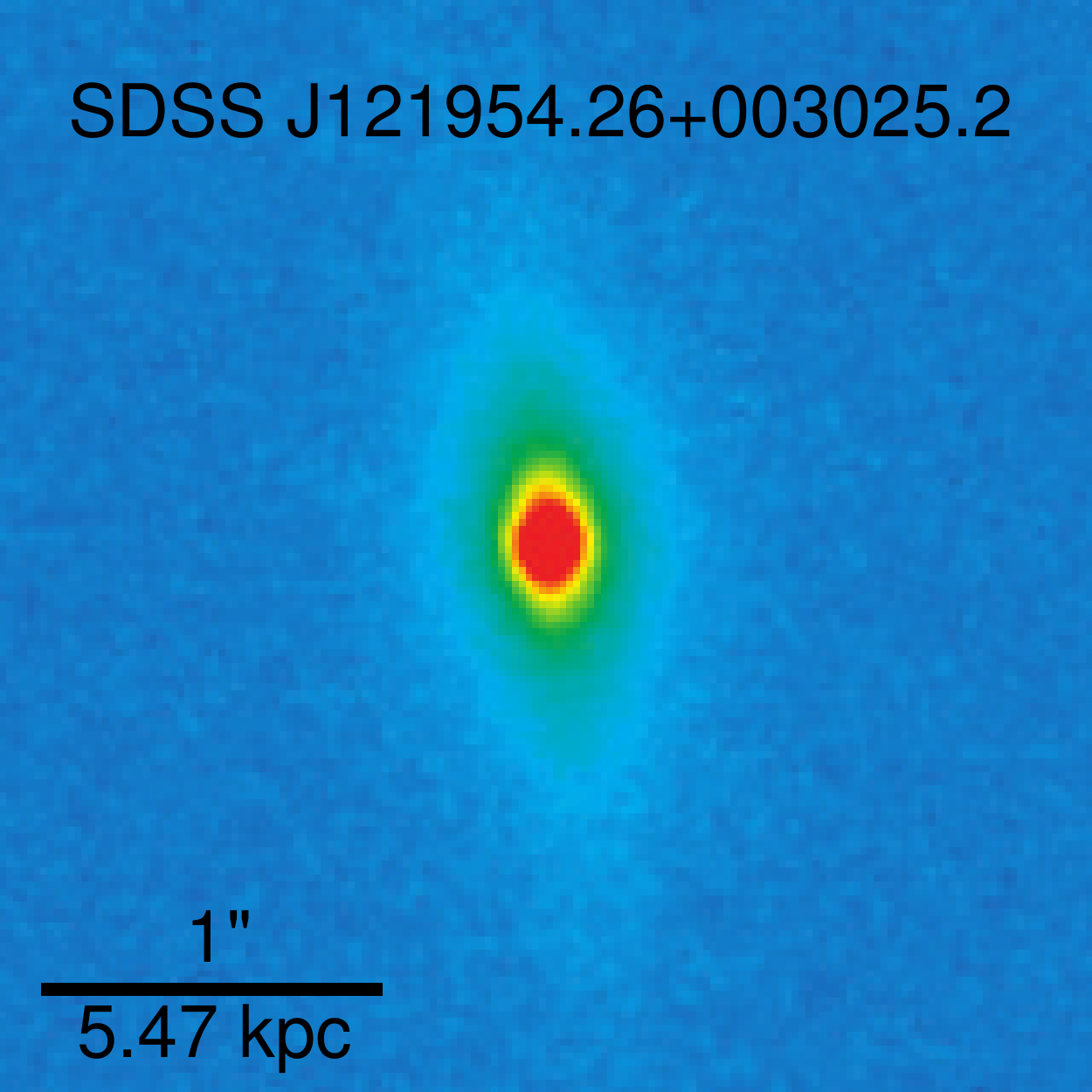}
\includegraphics[width=4cm]{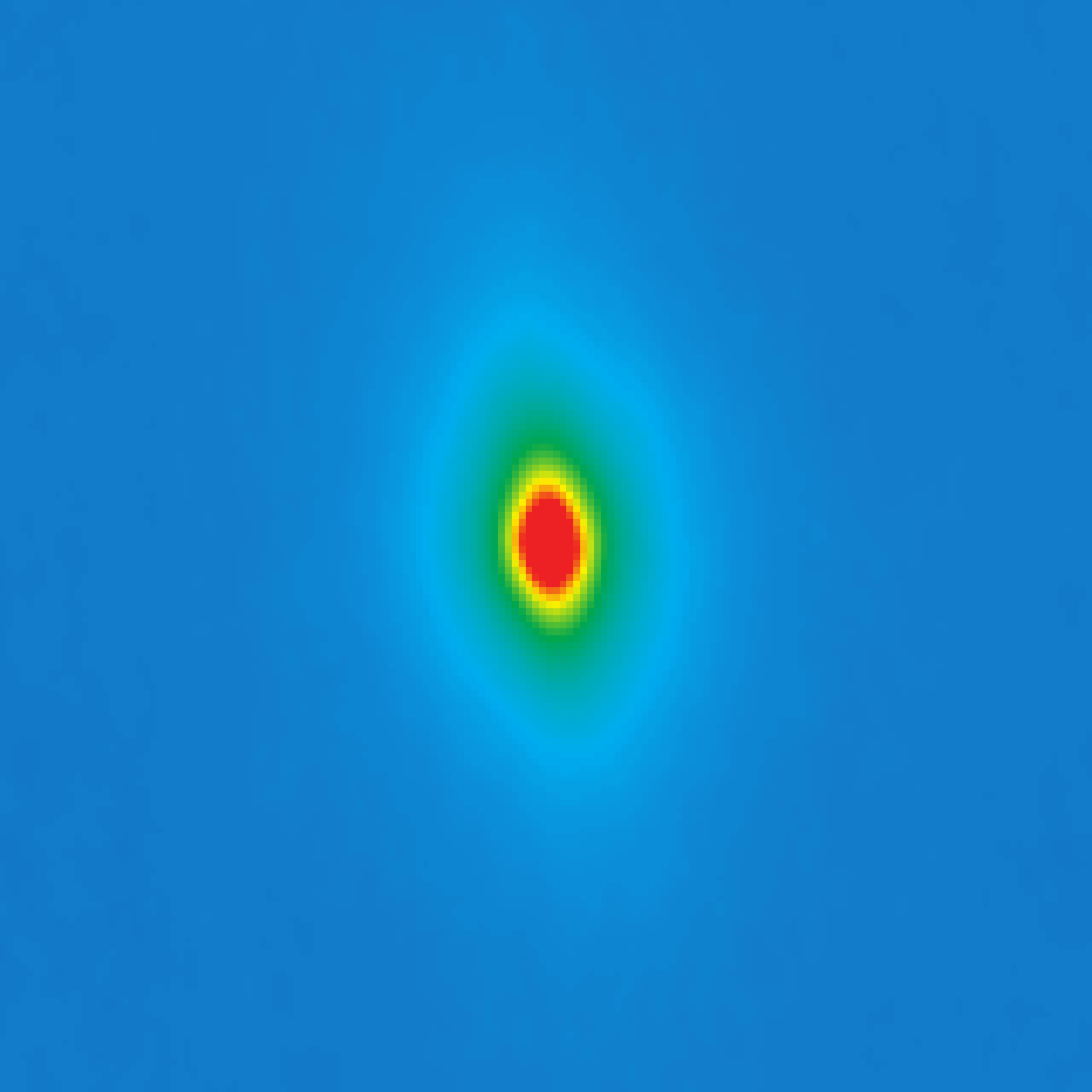}    
\includegraphics[width=4cm]{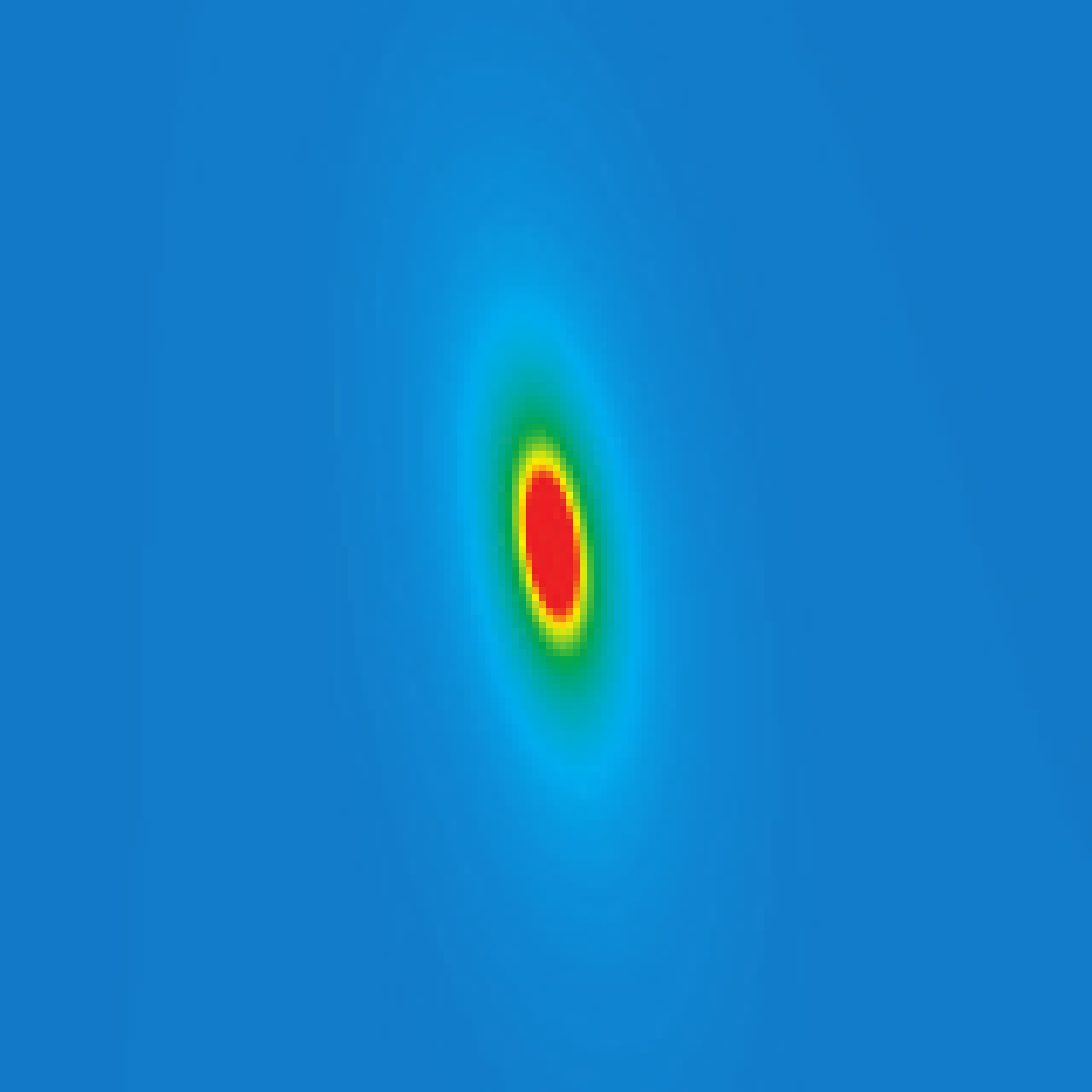}
\includegraphics[width=4cm]{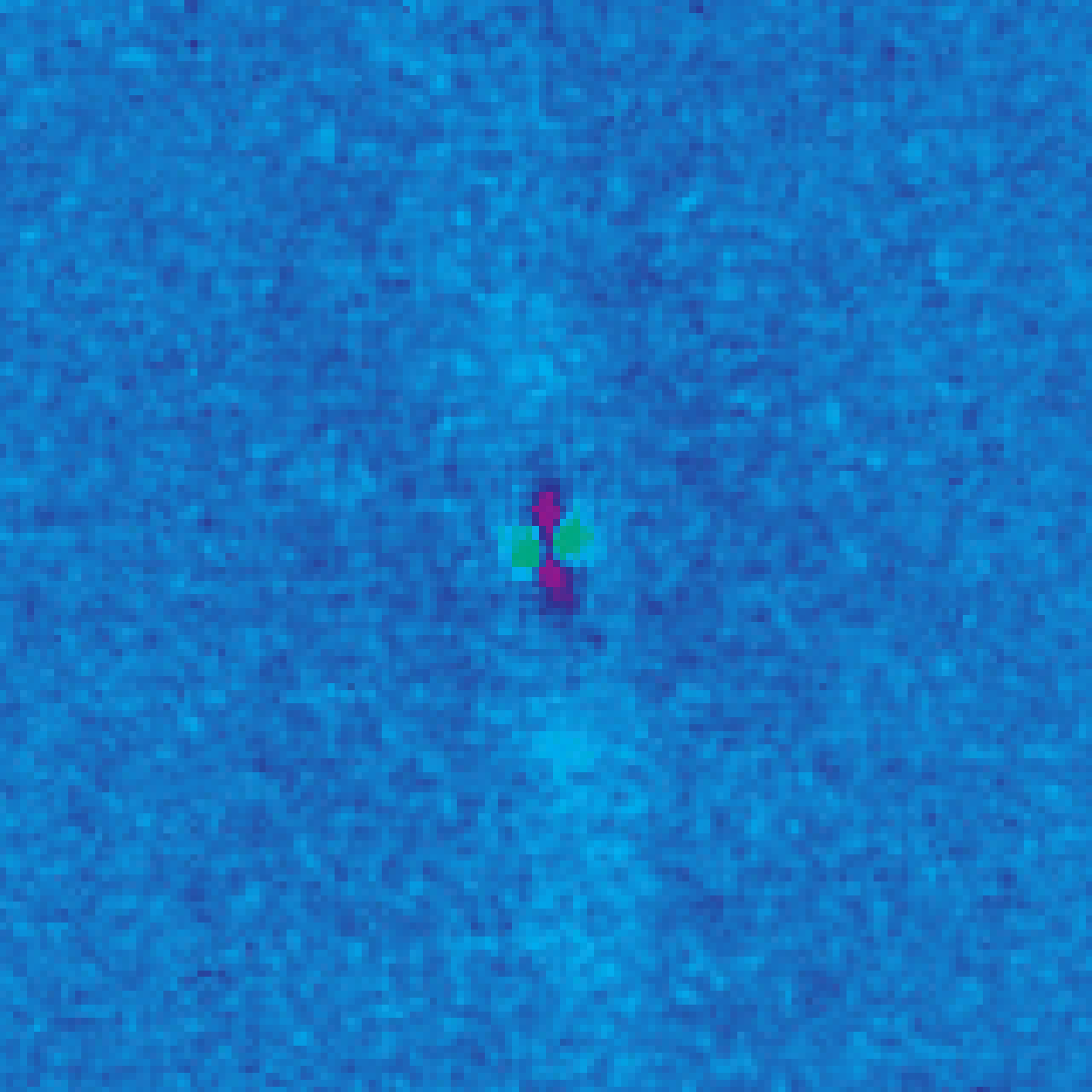}\\[1mm]
\hspace{40.7mm}\includegraphics[width=4cm]{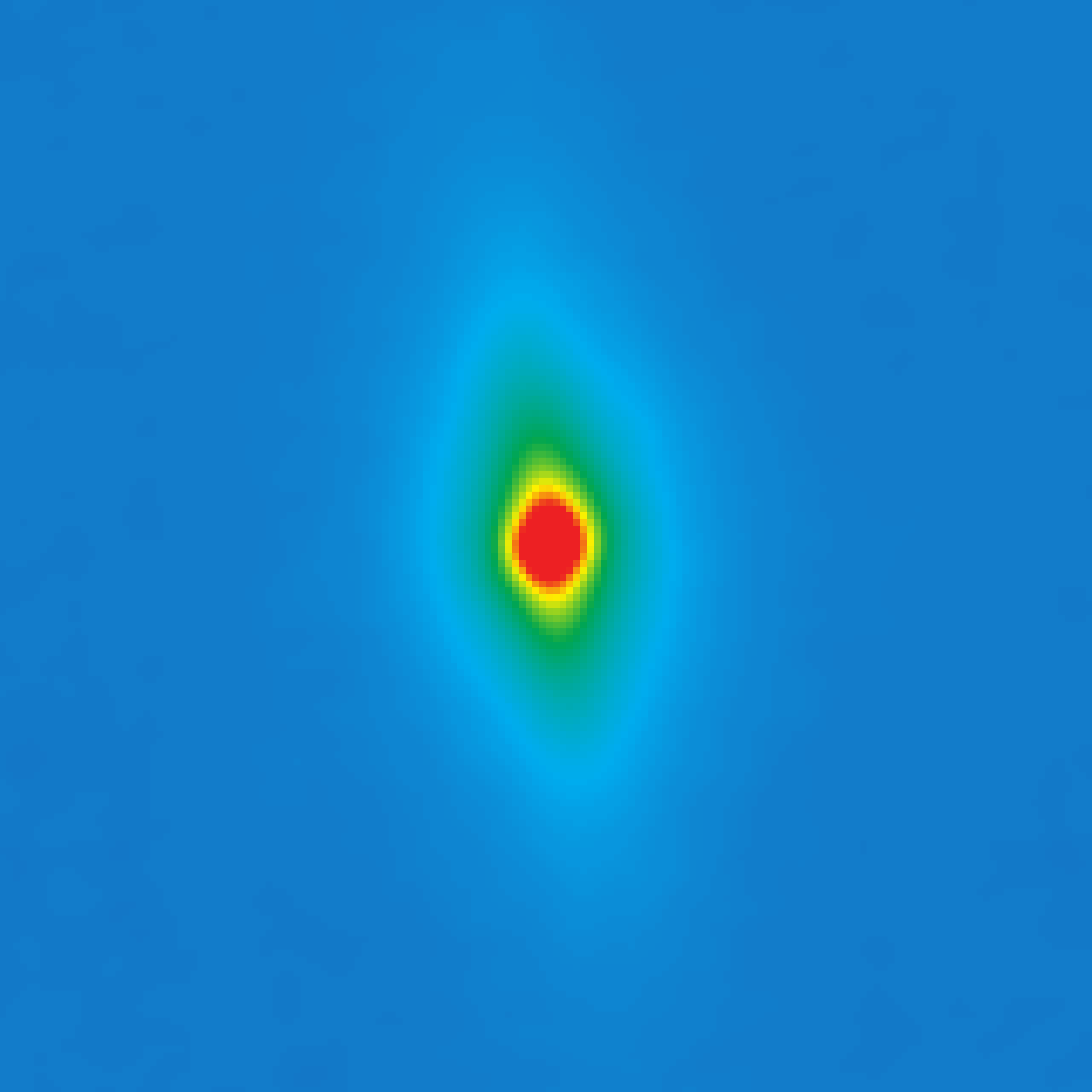}    
\includegraphics[width=4cm]{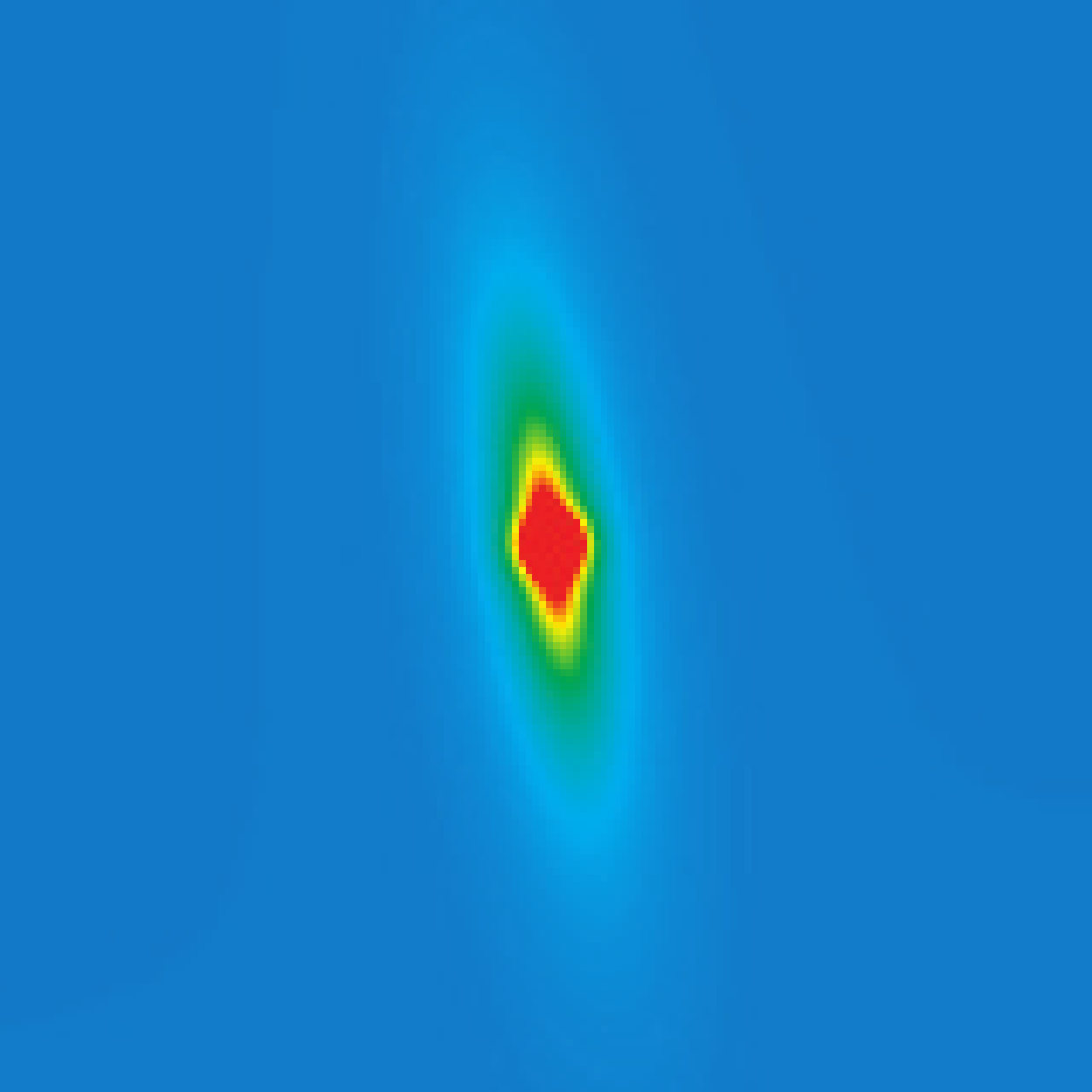}
\includegraphics[width=4cm]{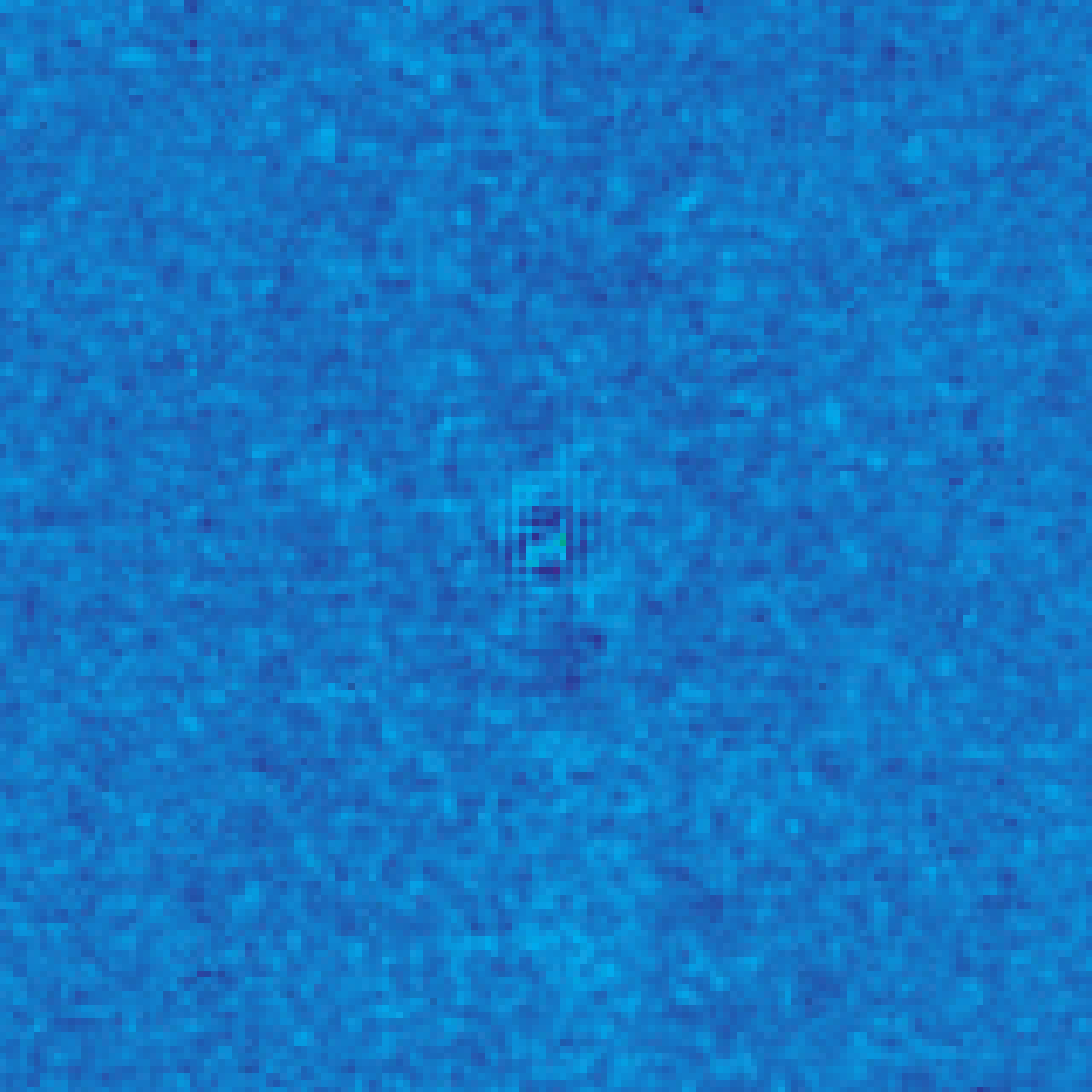}
%\caption{\centering Continued.}
\caption{Continued.}
\end{figure*}

\begin{figure*}
\centering
\includegraphics[width=4cm]{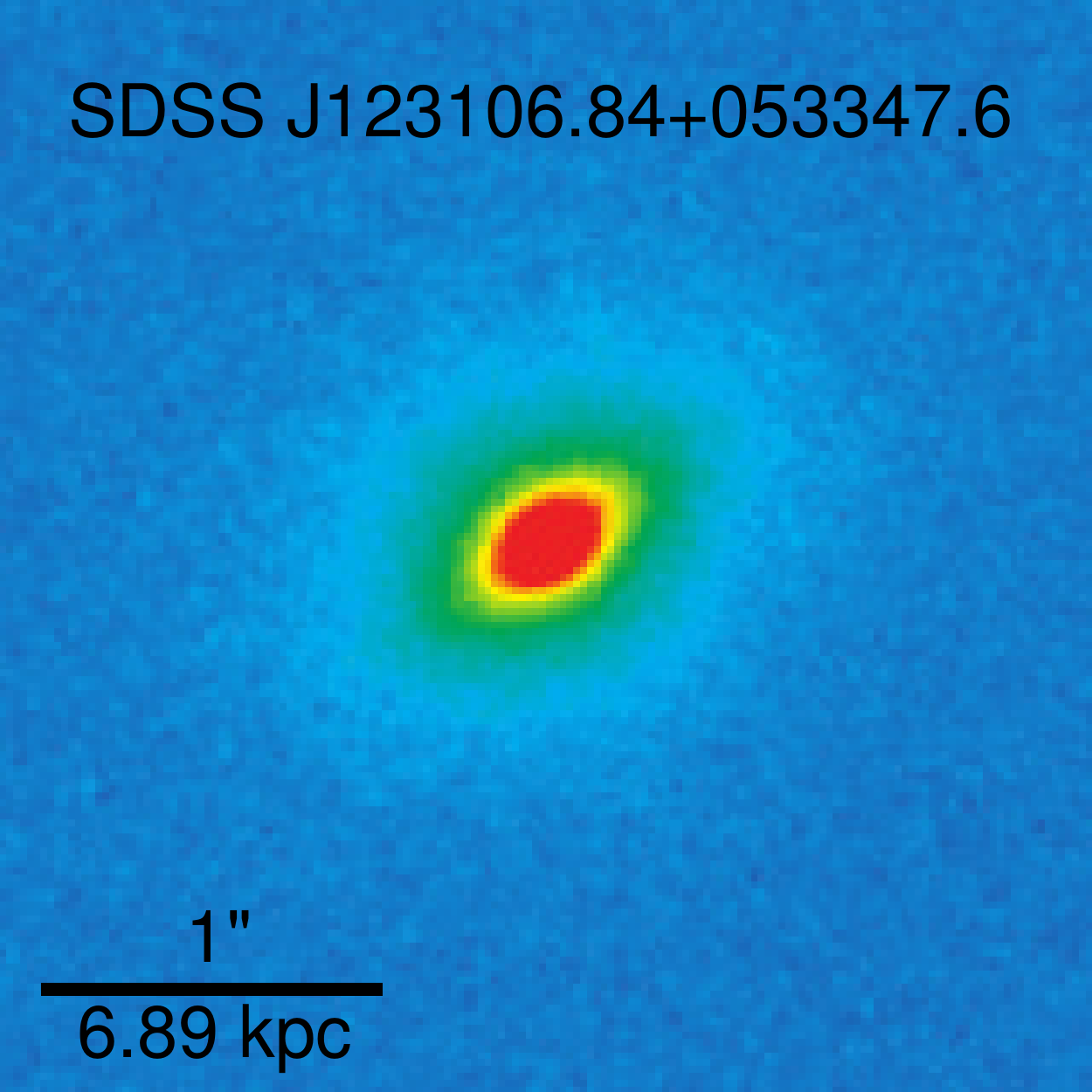}
\includegraphics[width=4cm]{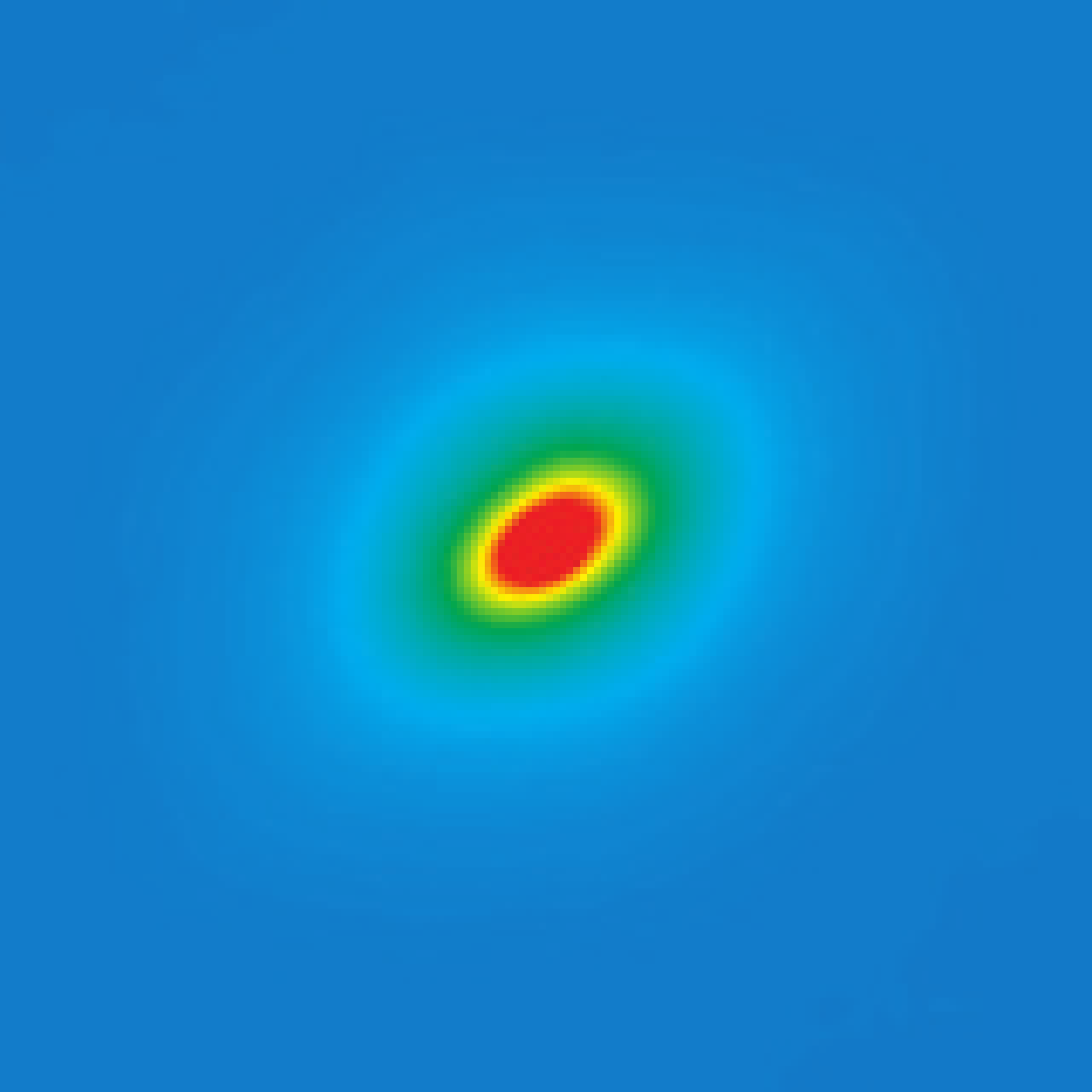}    
\includegraphics[width=4cm]{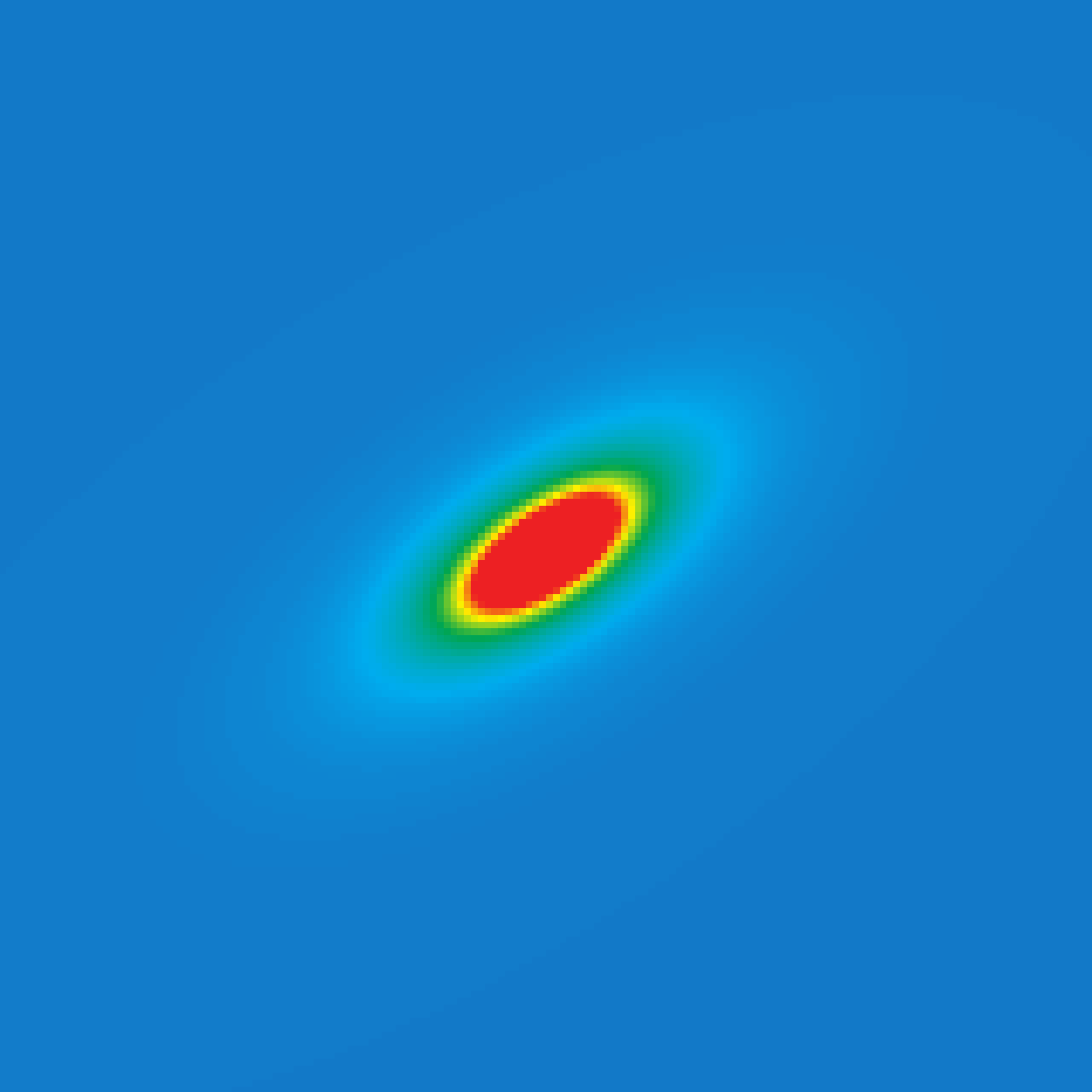}
\includegraphics[width=4cm]{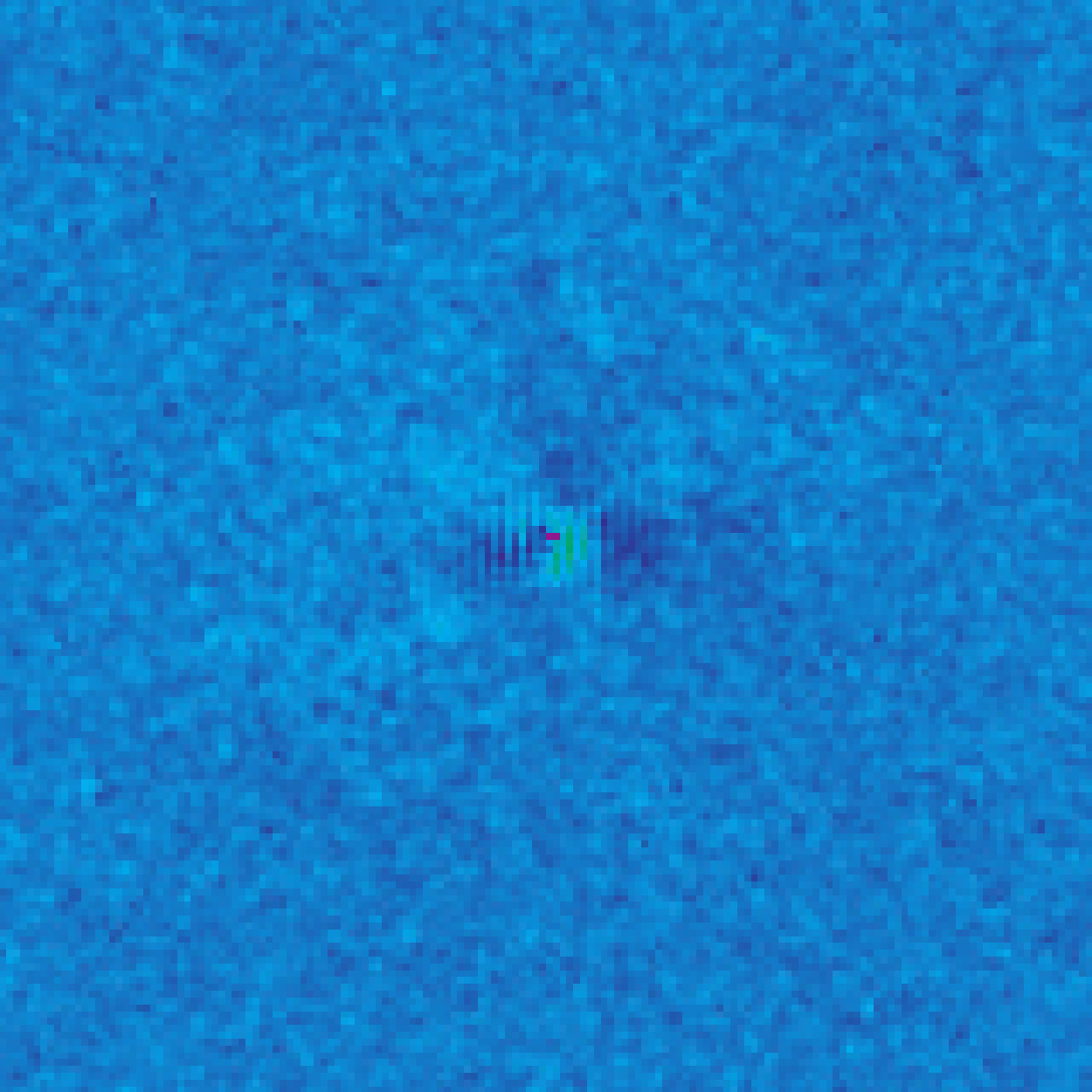}\\[2mm]
\includegraphics[width=4cm]{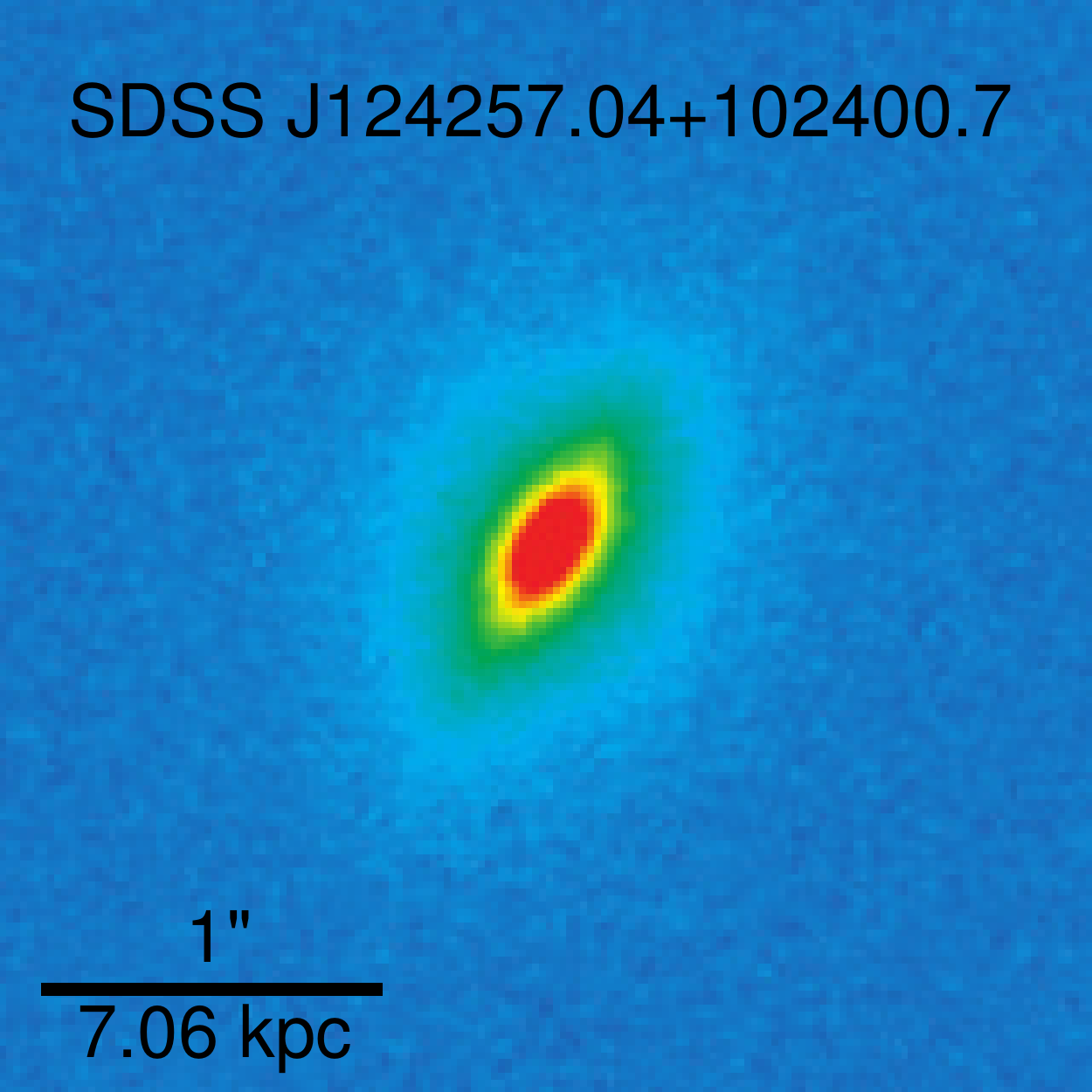}
\includegraphics[width=4cm]{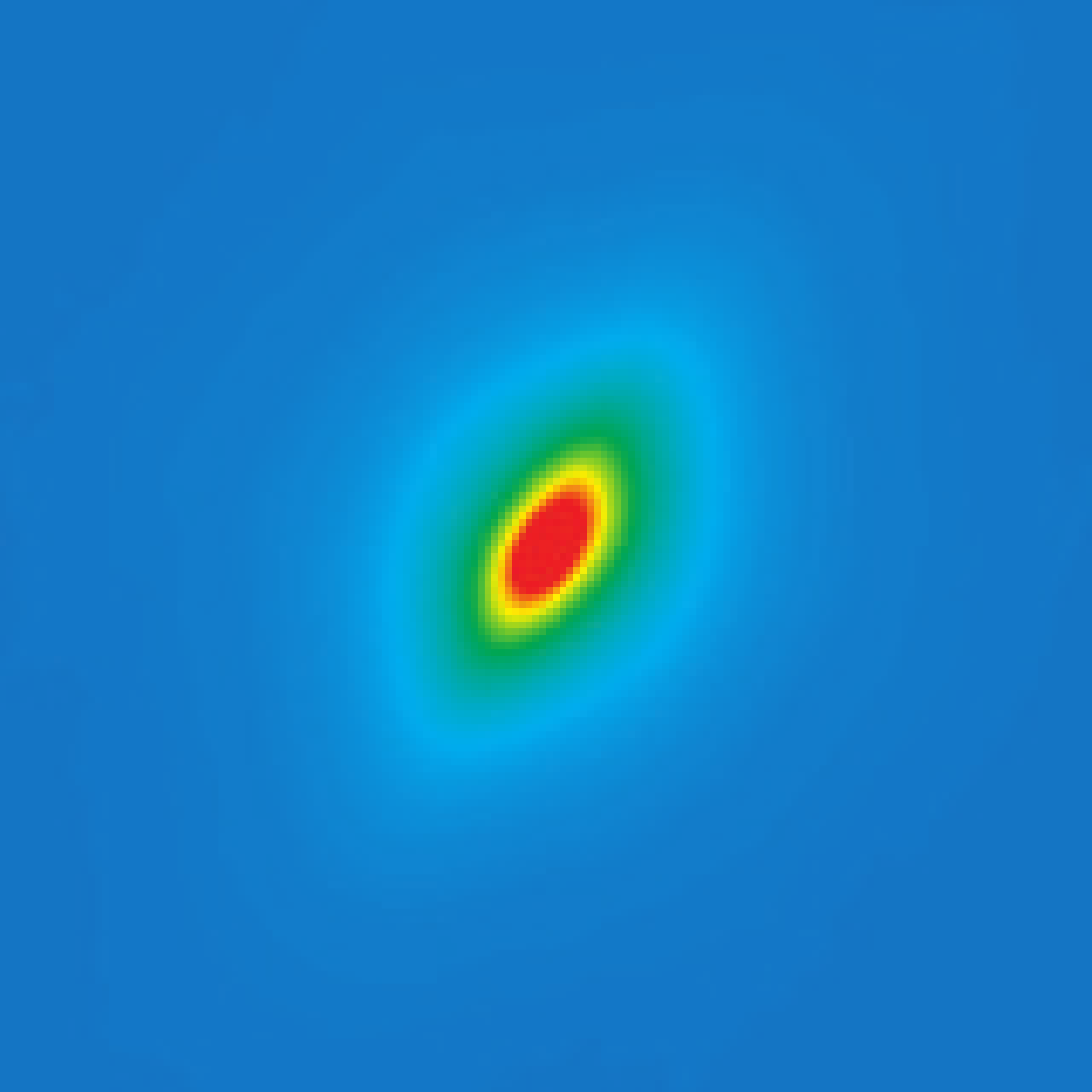}    
\includegraphics[width=4cm]{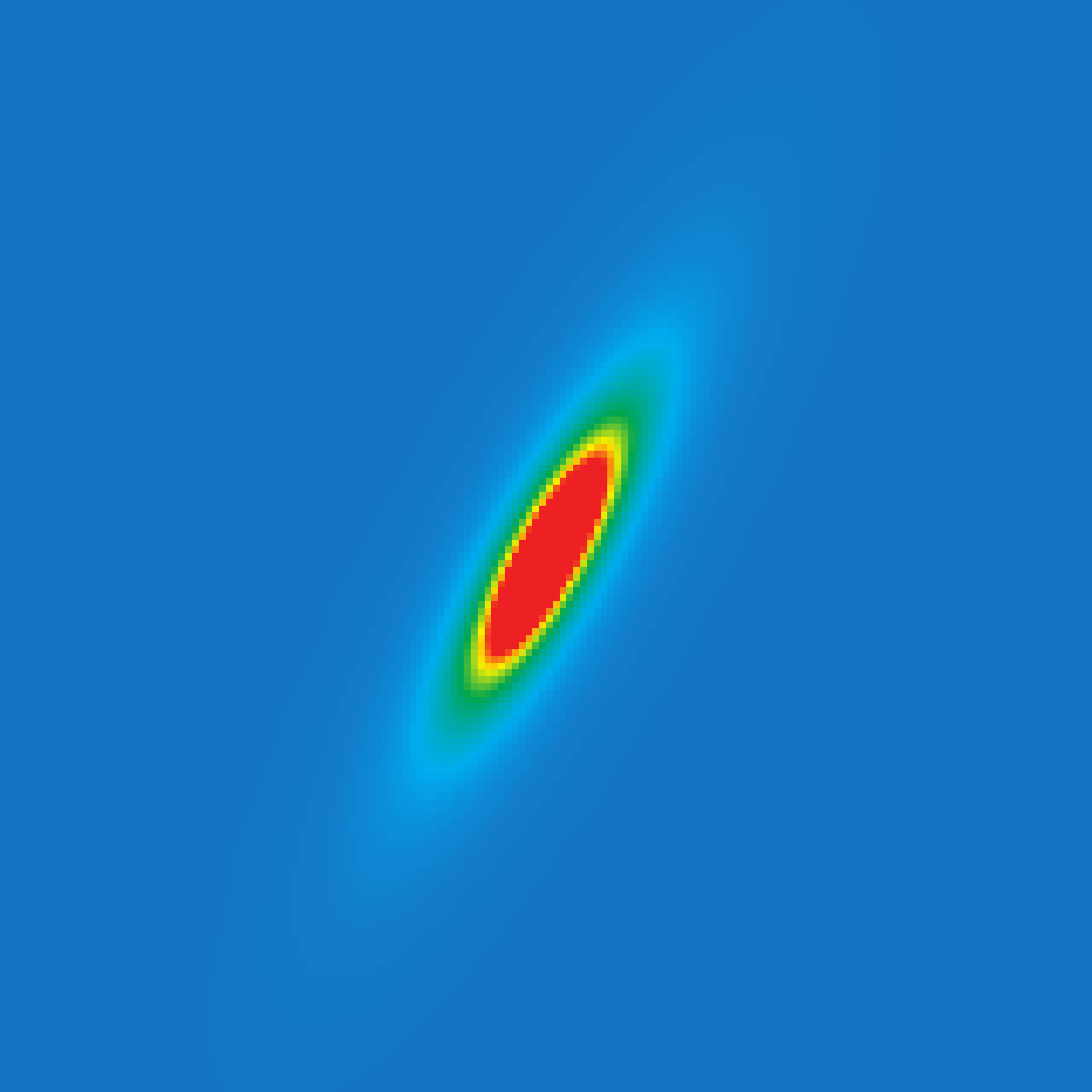}
\includegraphics[width=4cm]{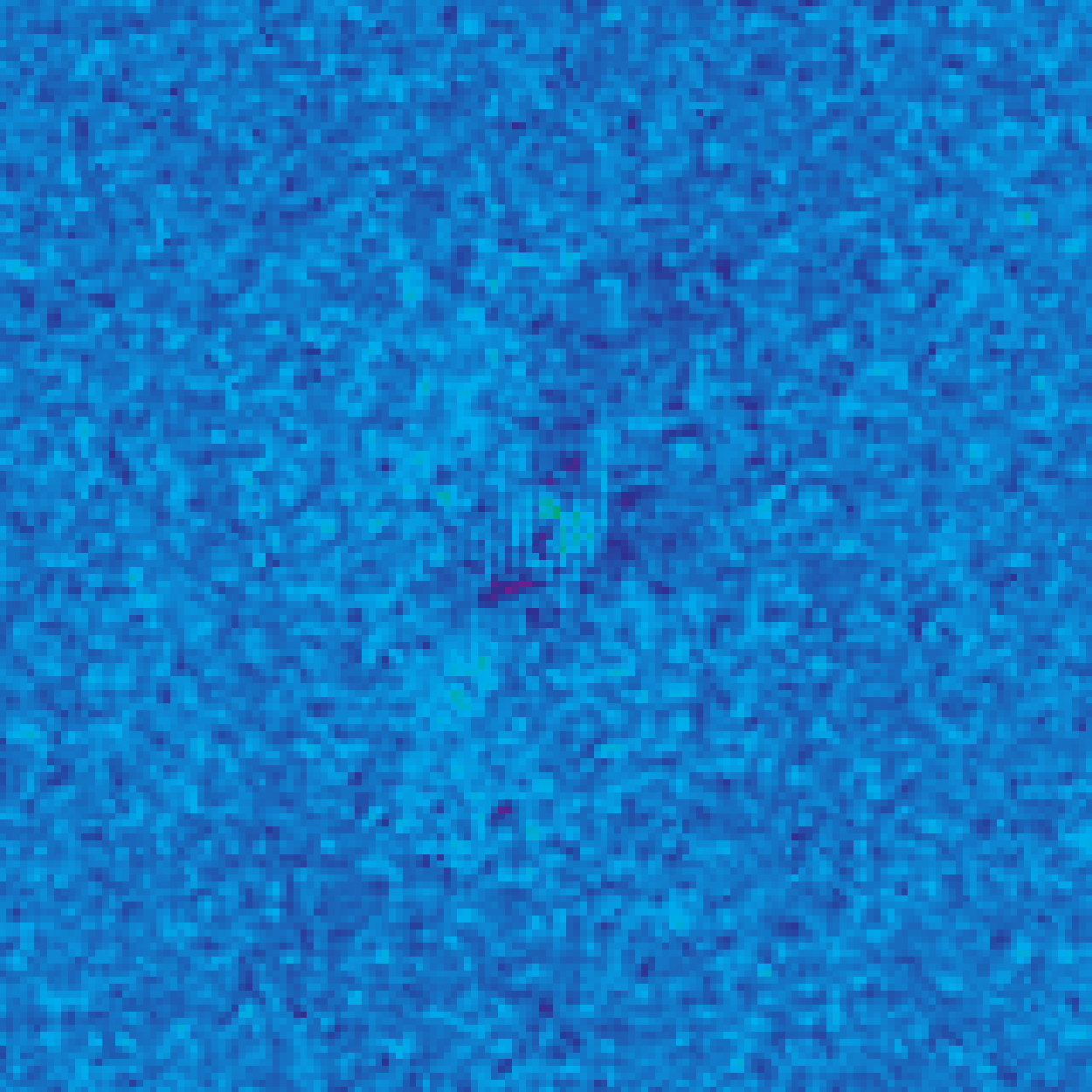}\\[2mm]
\includegraphics[width=4cm]{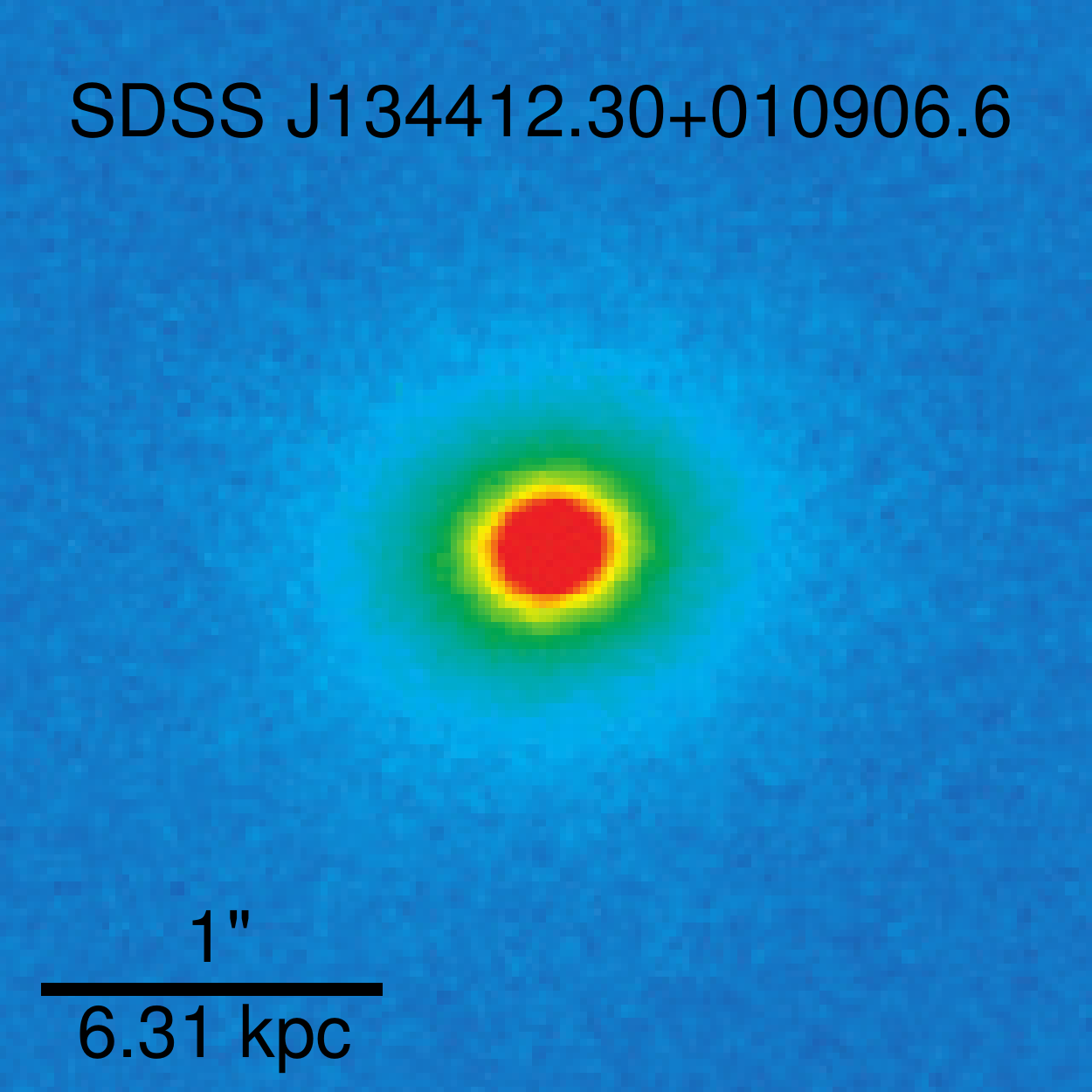}
\includegraphics[width=4cm]{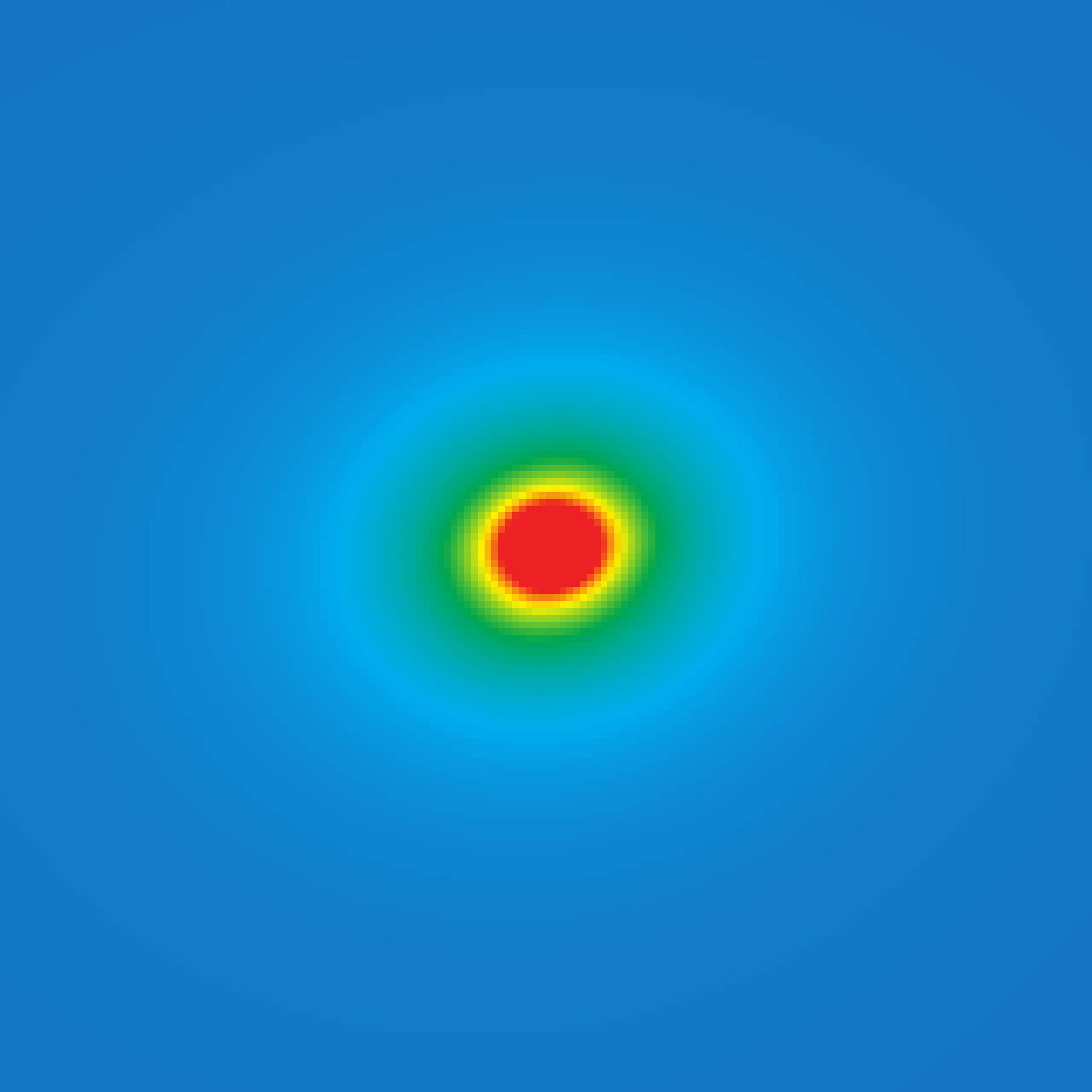}    
\includegraphics[width=4cm]{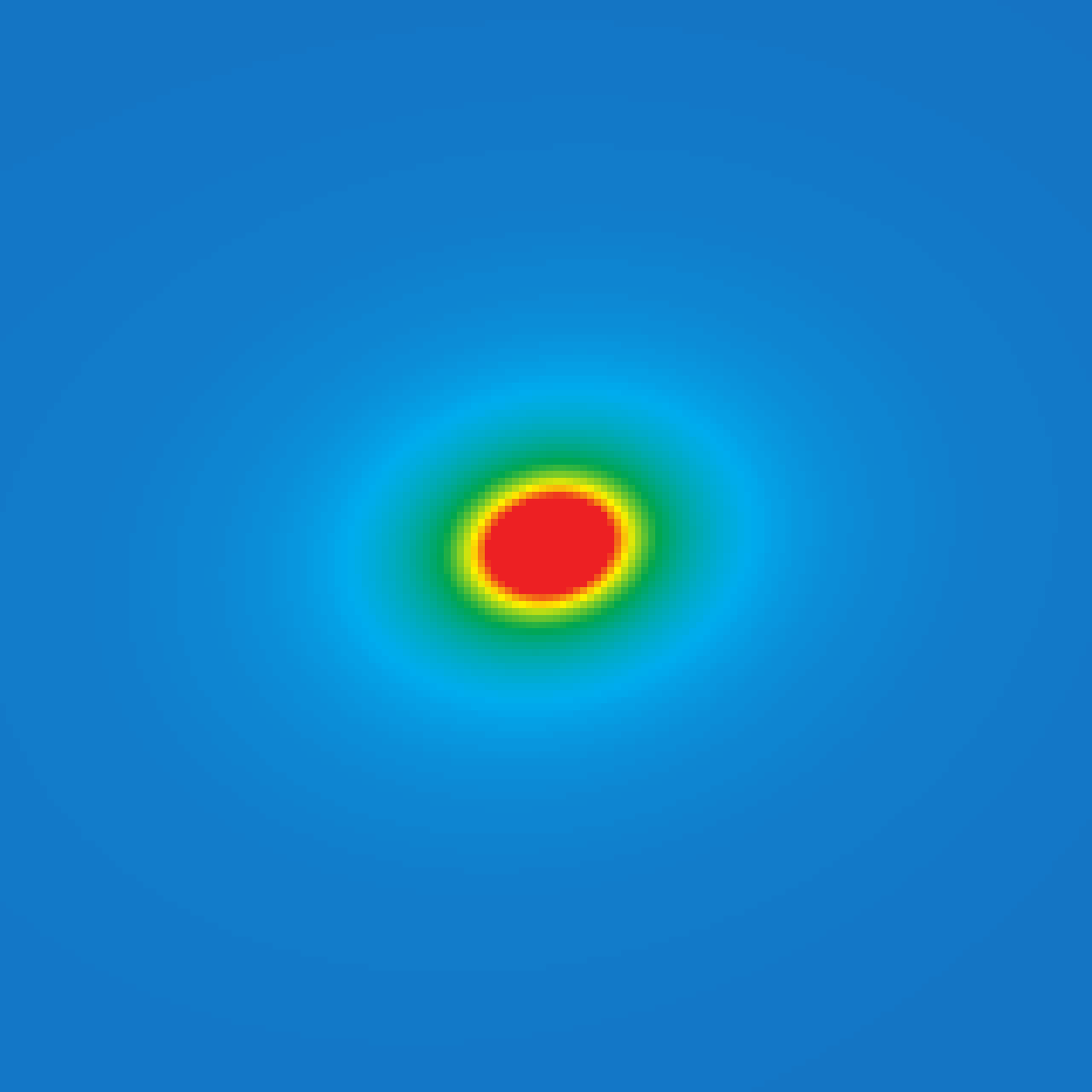}
\includegraphics[width=4cm]{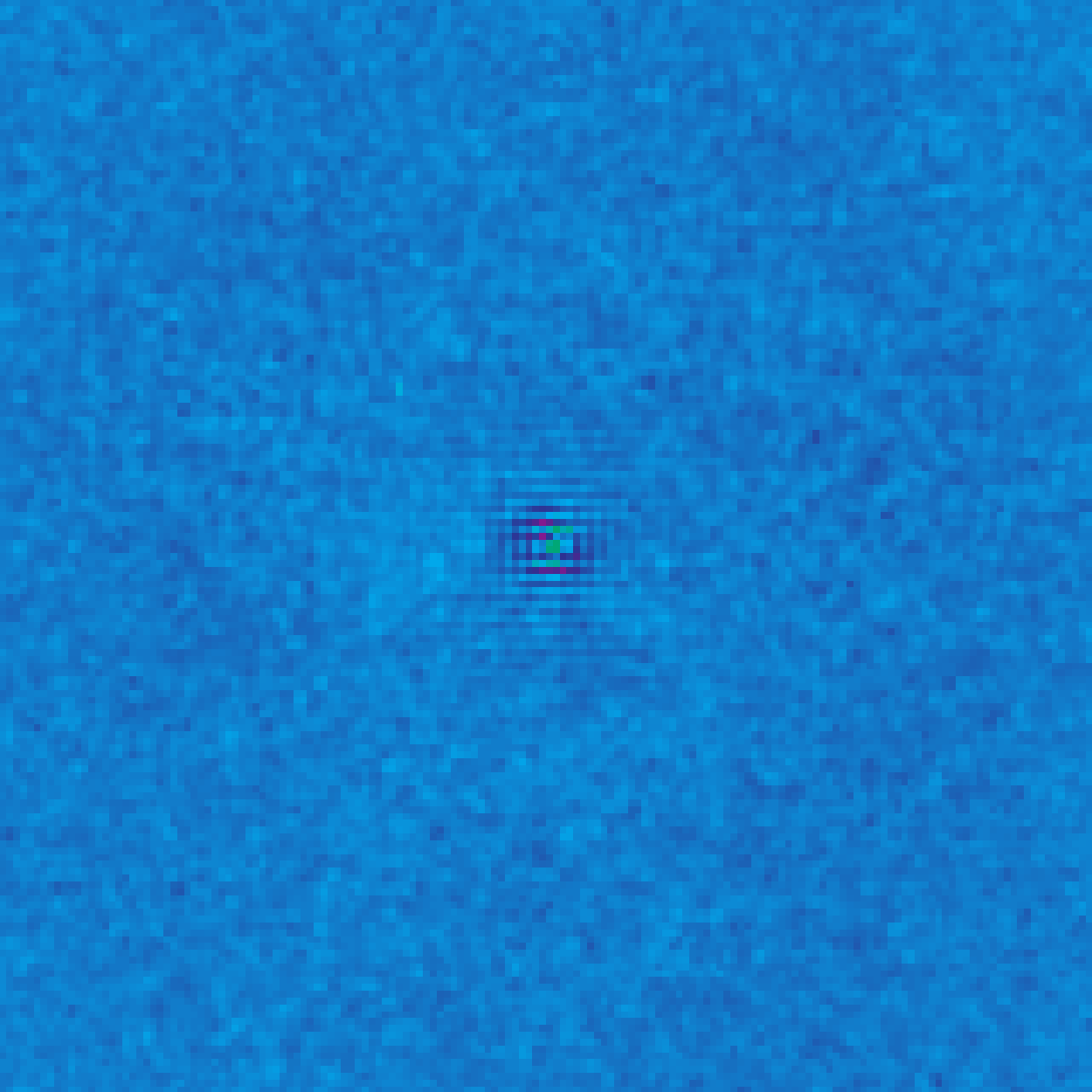}\\[2mm]
\includegraphics[width=4cm]{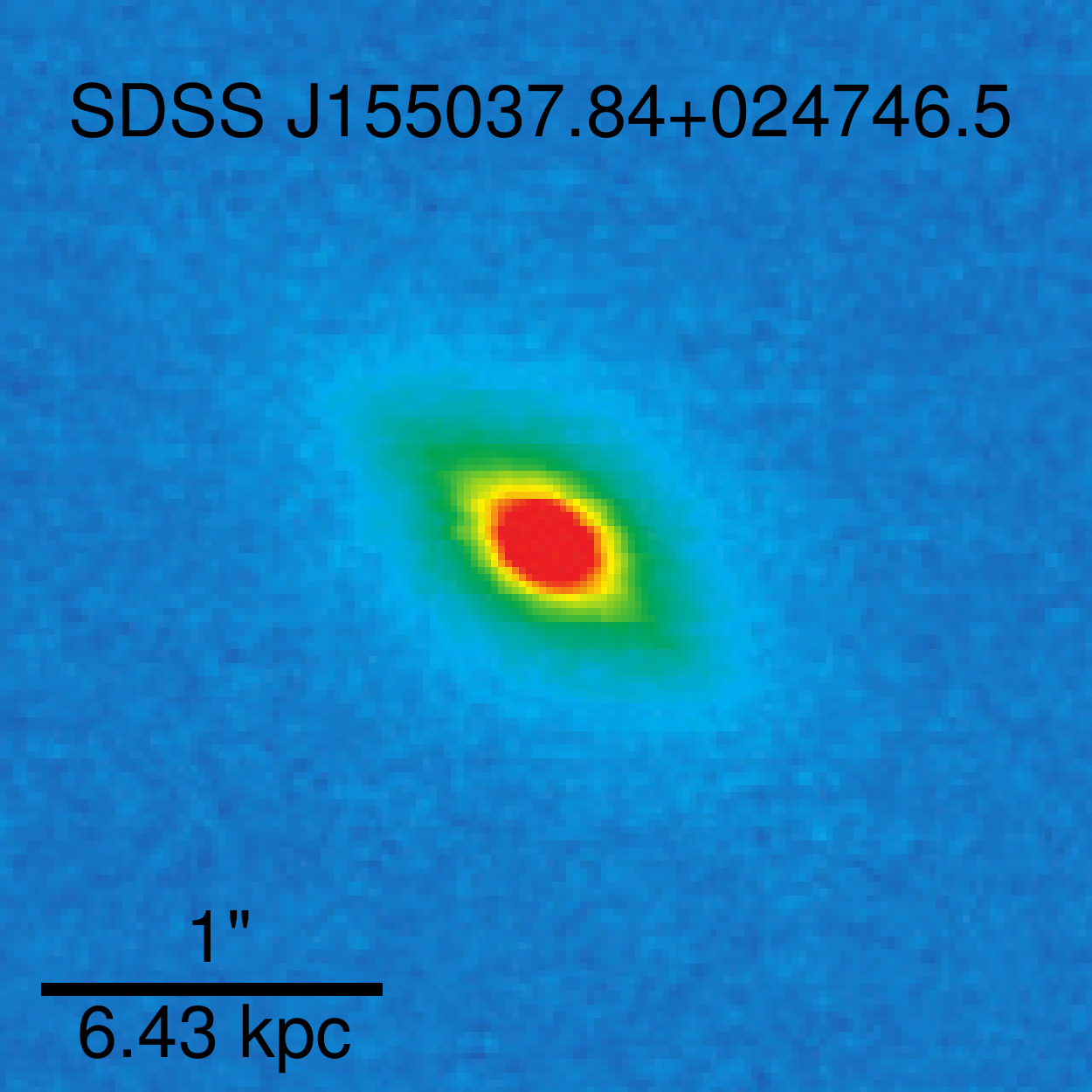}
\includegraphics[width=4cm]{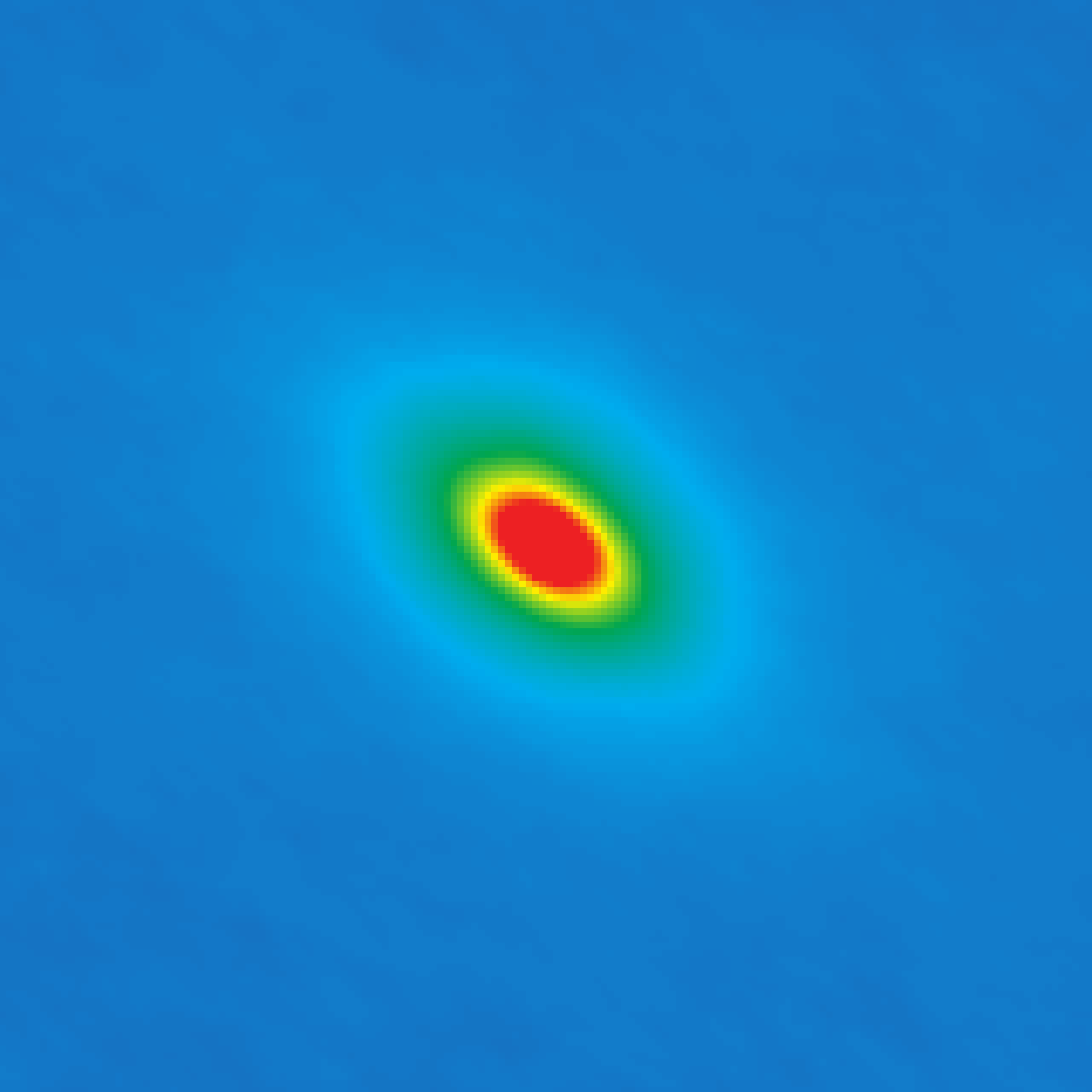}    
\includegraphics[width=4cm]{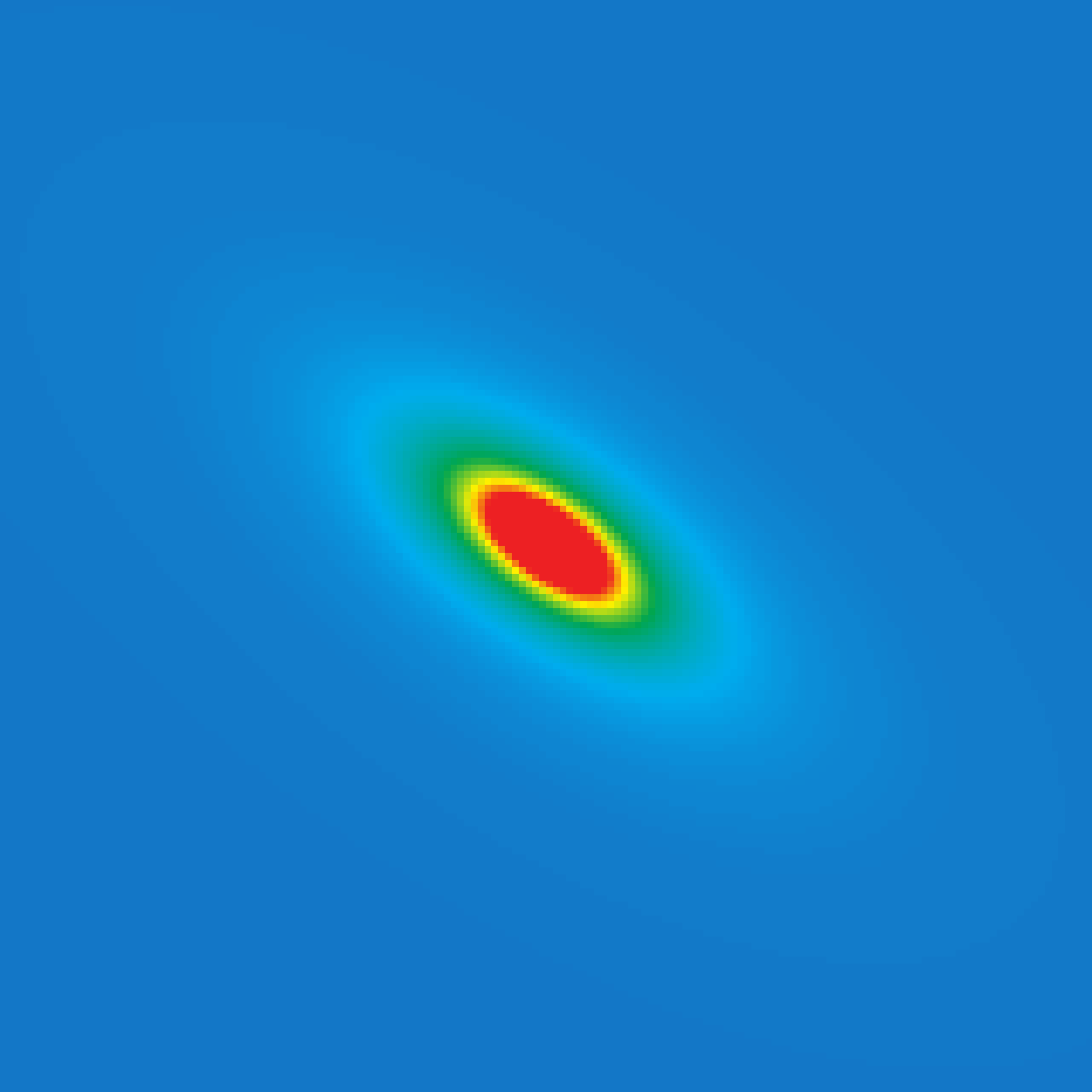}
\includegraphics[width=4cm]{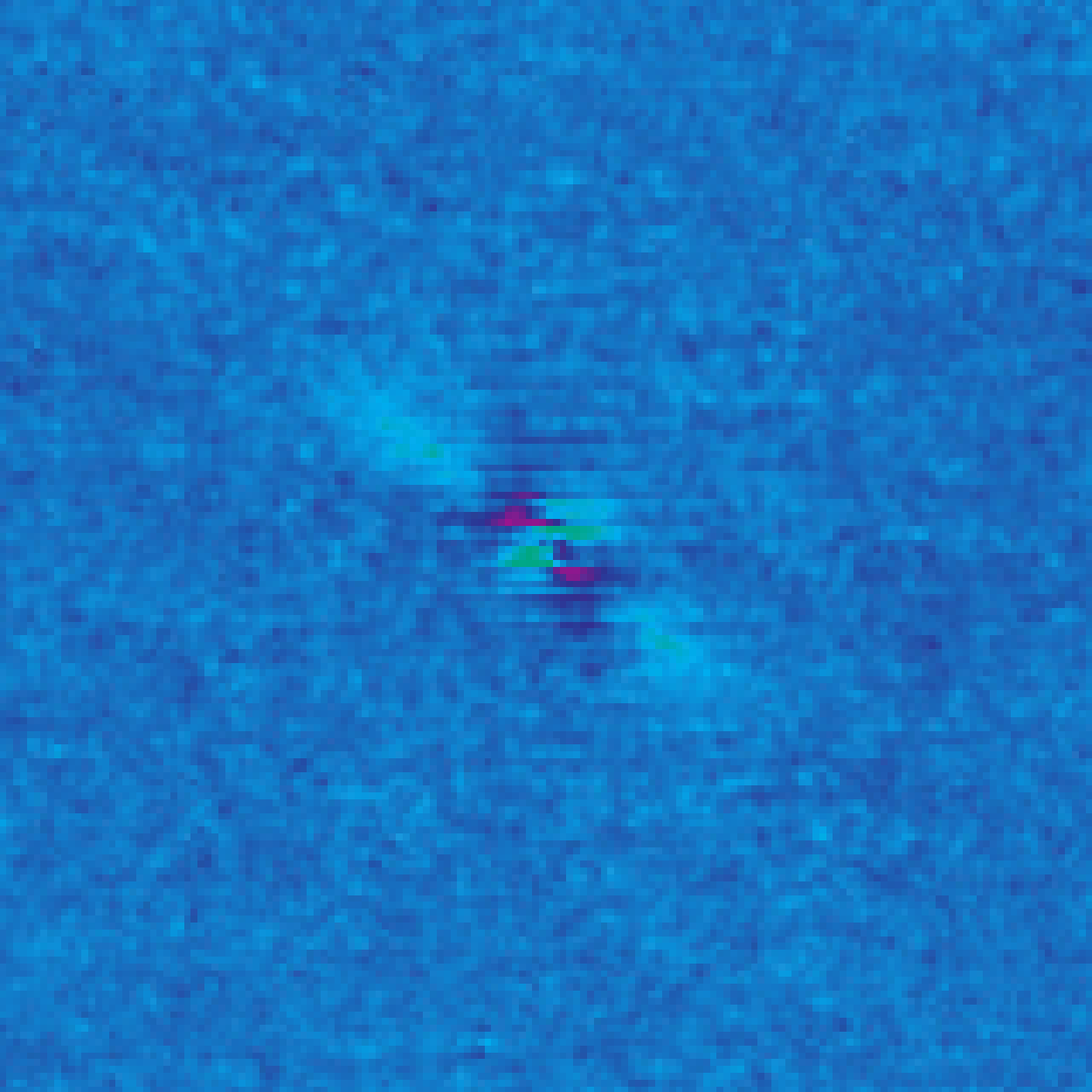}\\[1mm]
\hspace{40.7mm}\includegraphics[width=4cm]{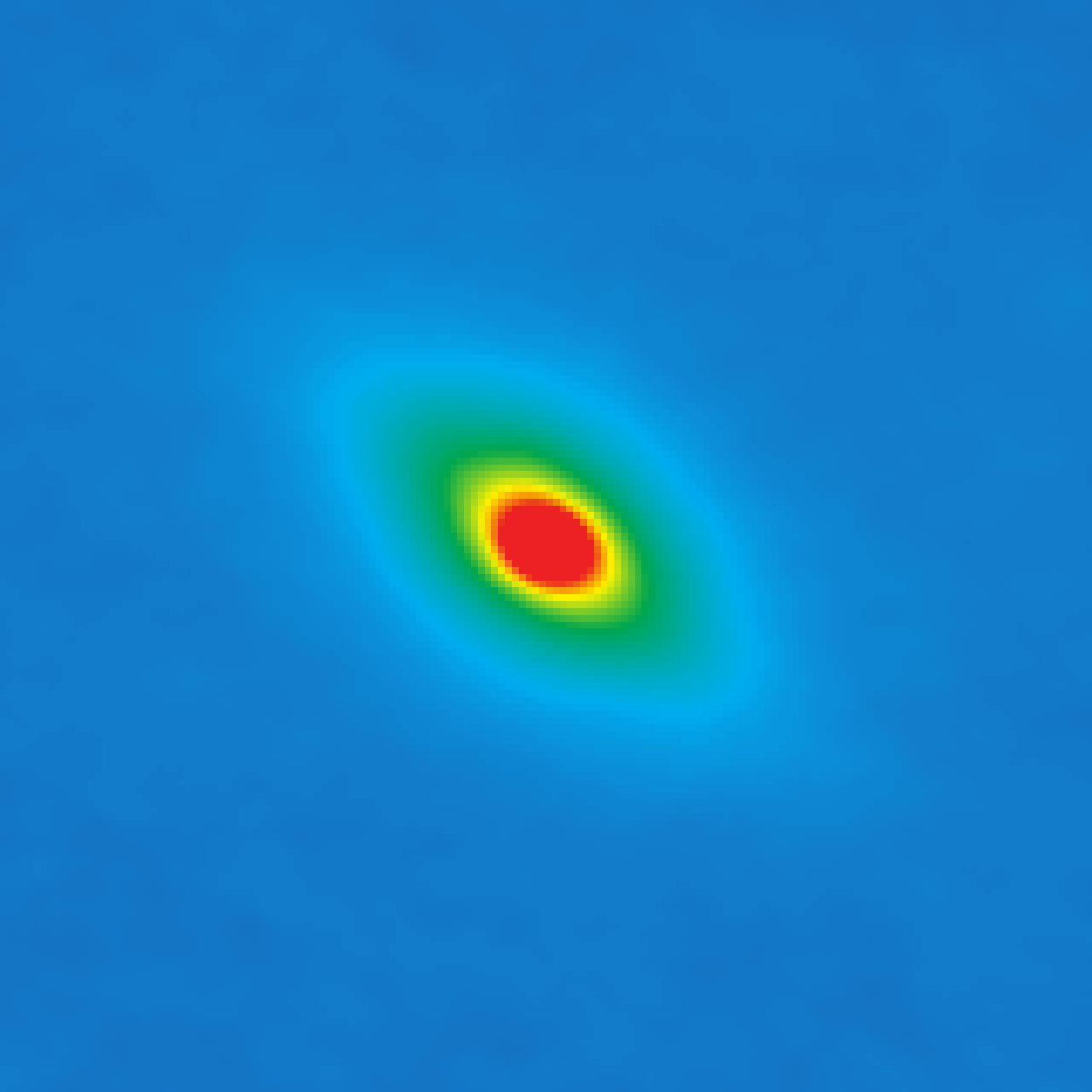}    
\includegraphics[width=4cm]{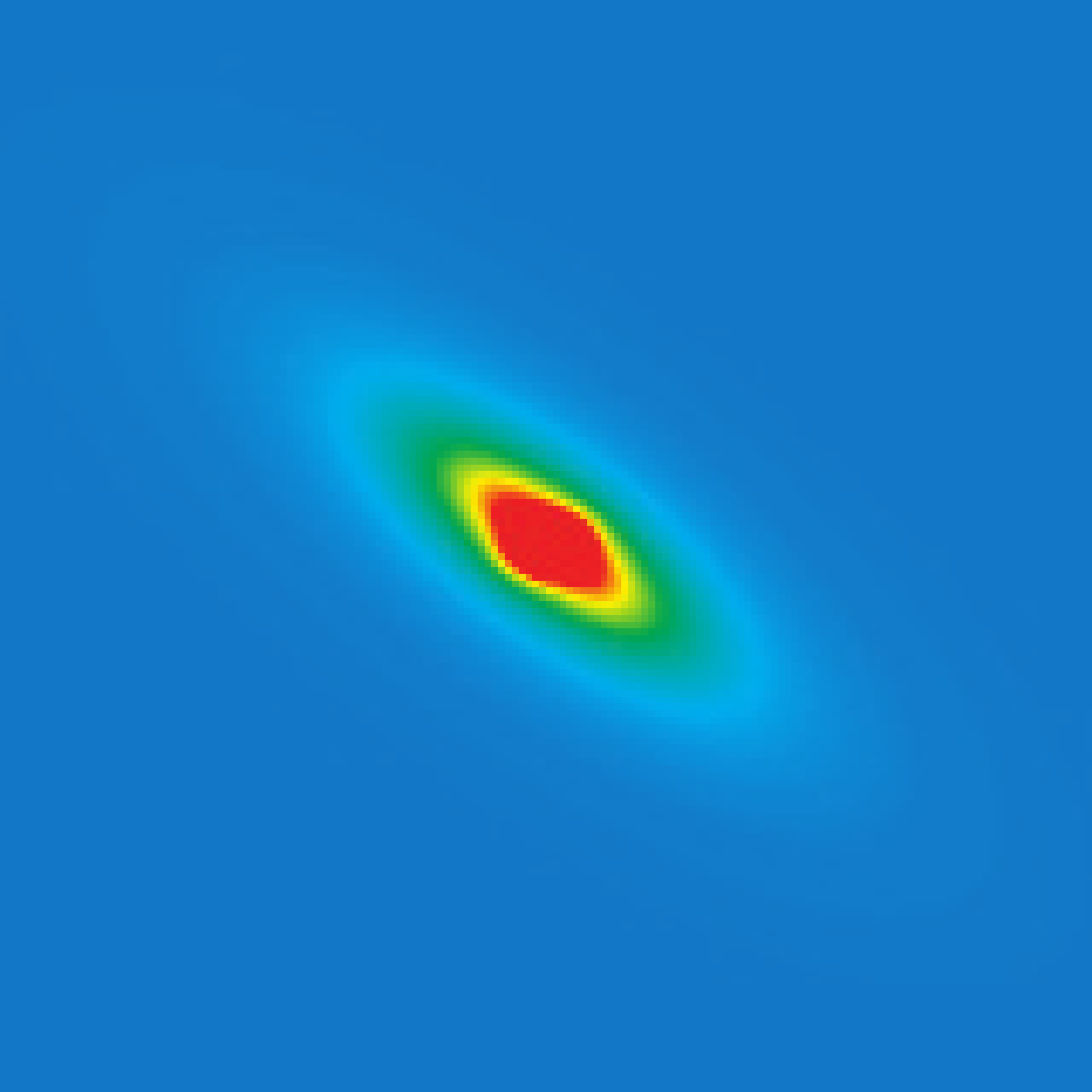}
\includegraphics[width=4cm]{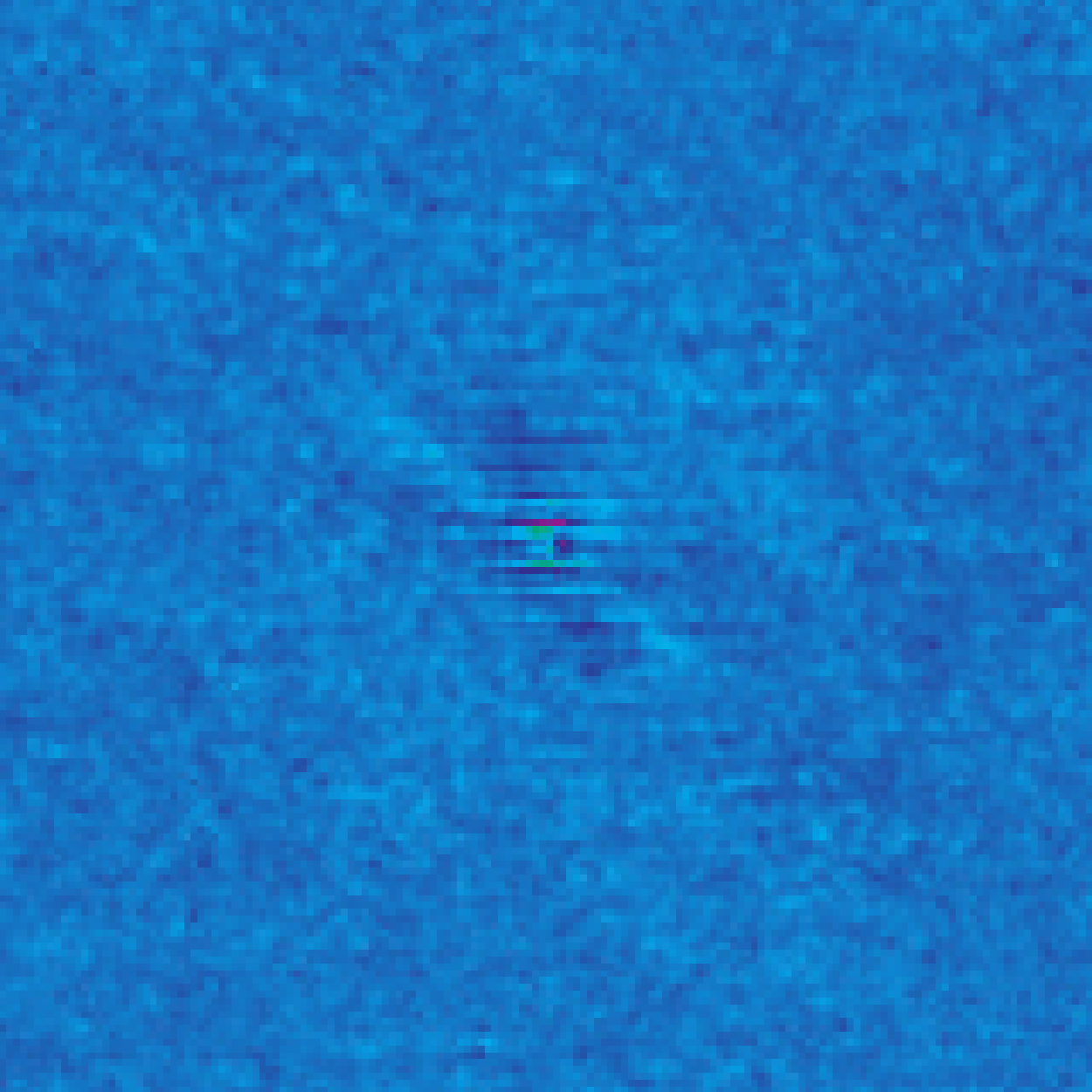}
%\caption{\centering Continued.}
\caption{Continued.}
\end{figure*}

\begin{figure*}
\centering
\includegraphics[width=4cm]{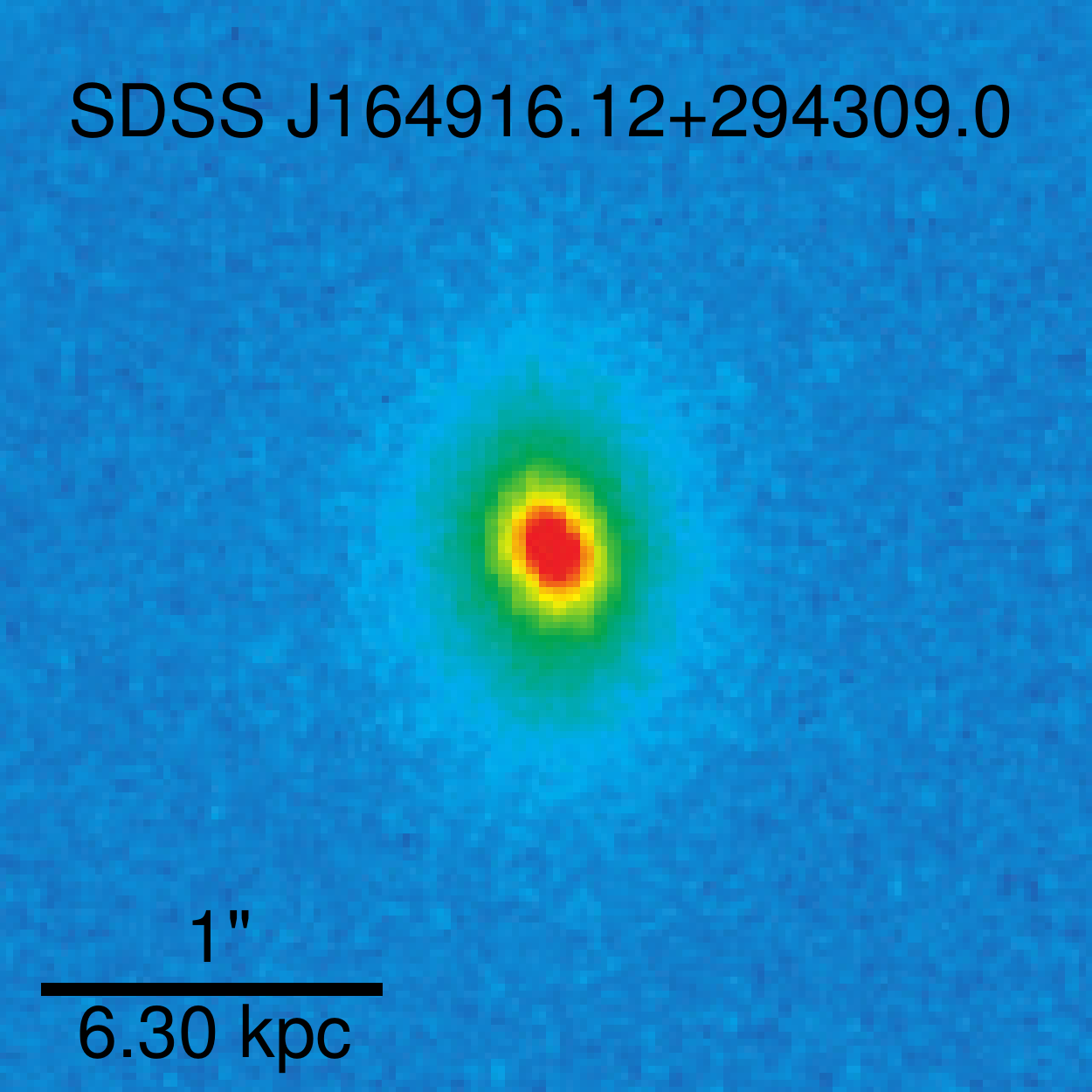}
\includegraphics[width=4cm]{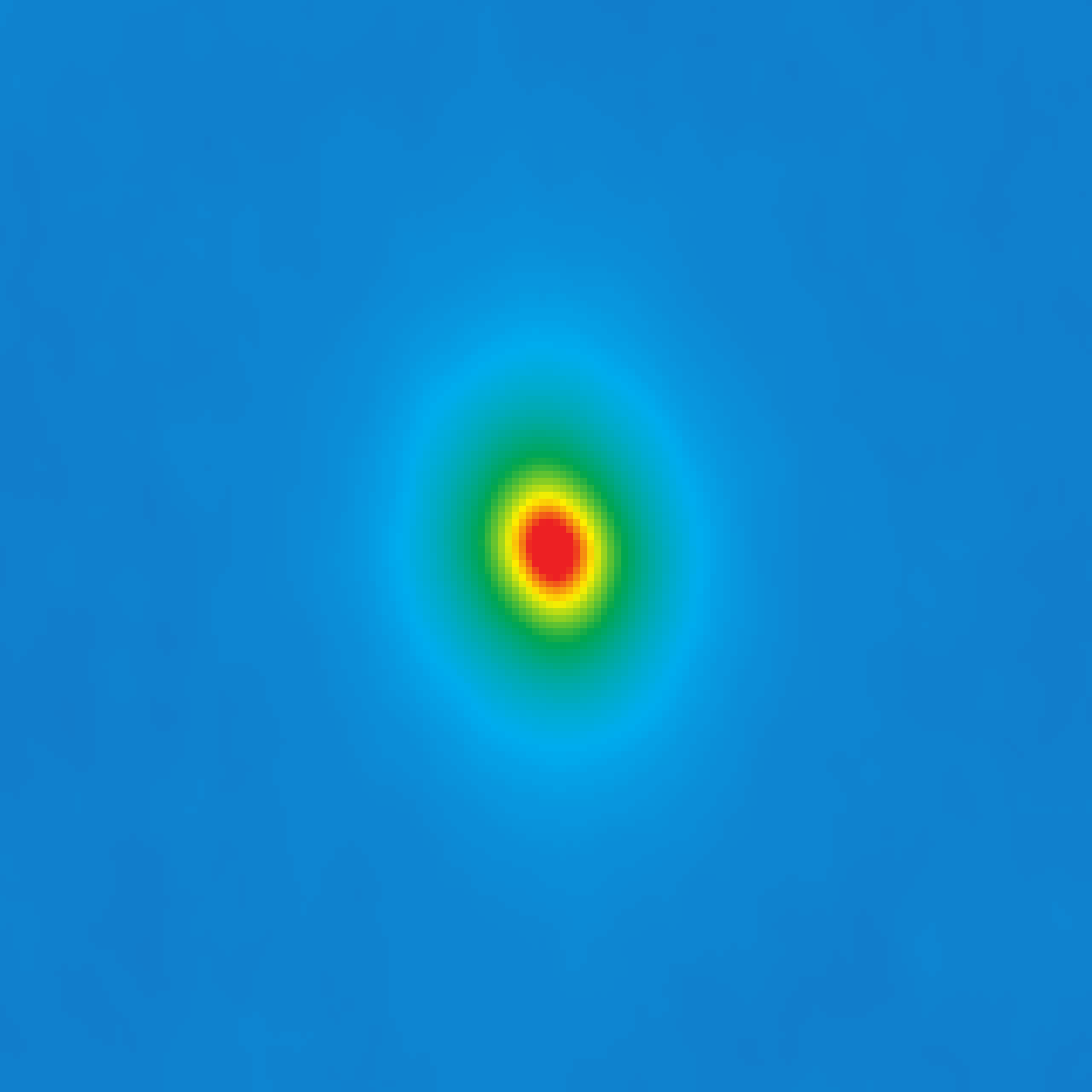}    
\includegraphics[width=4cm]{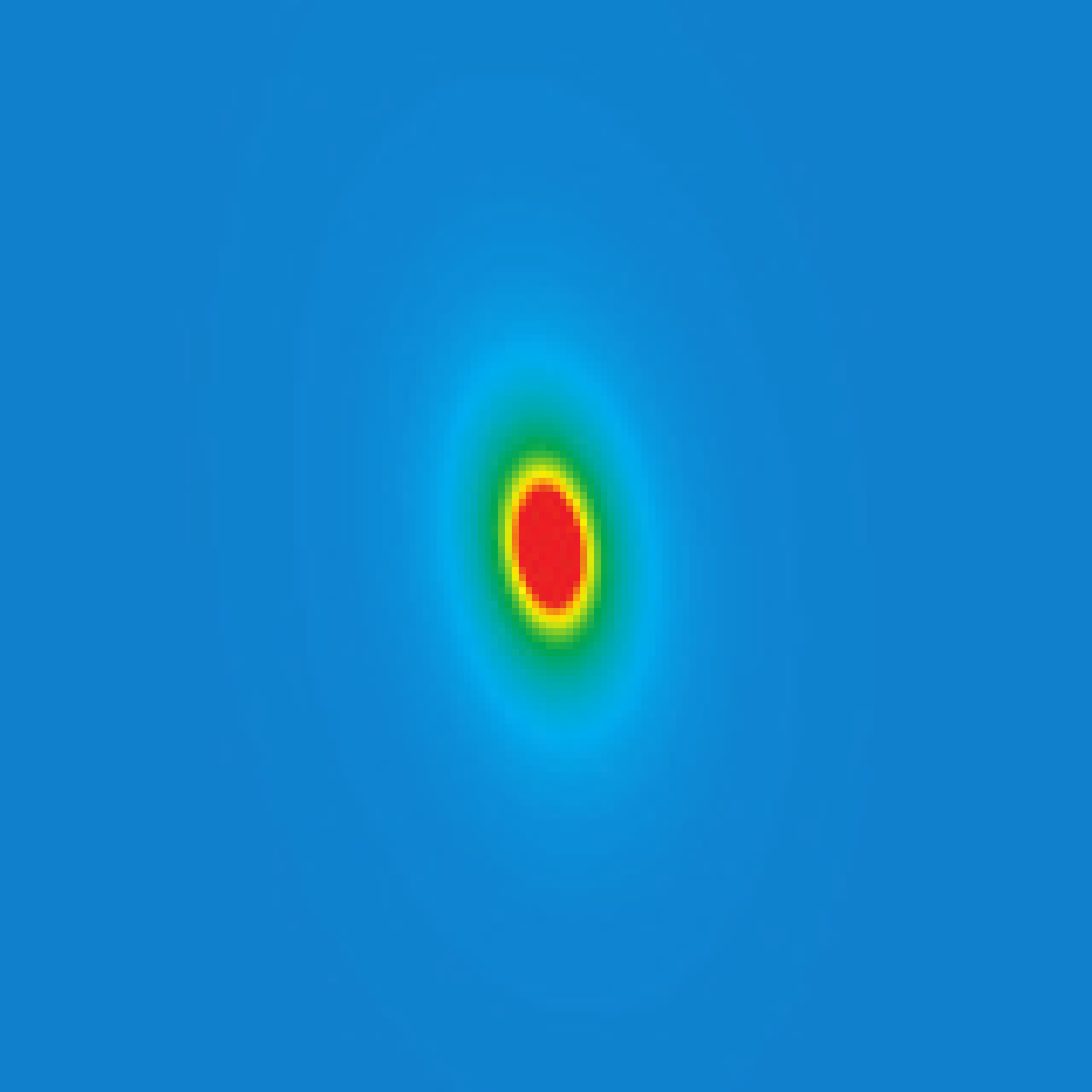}
\includegraphics[width=4cm]{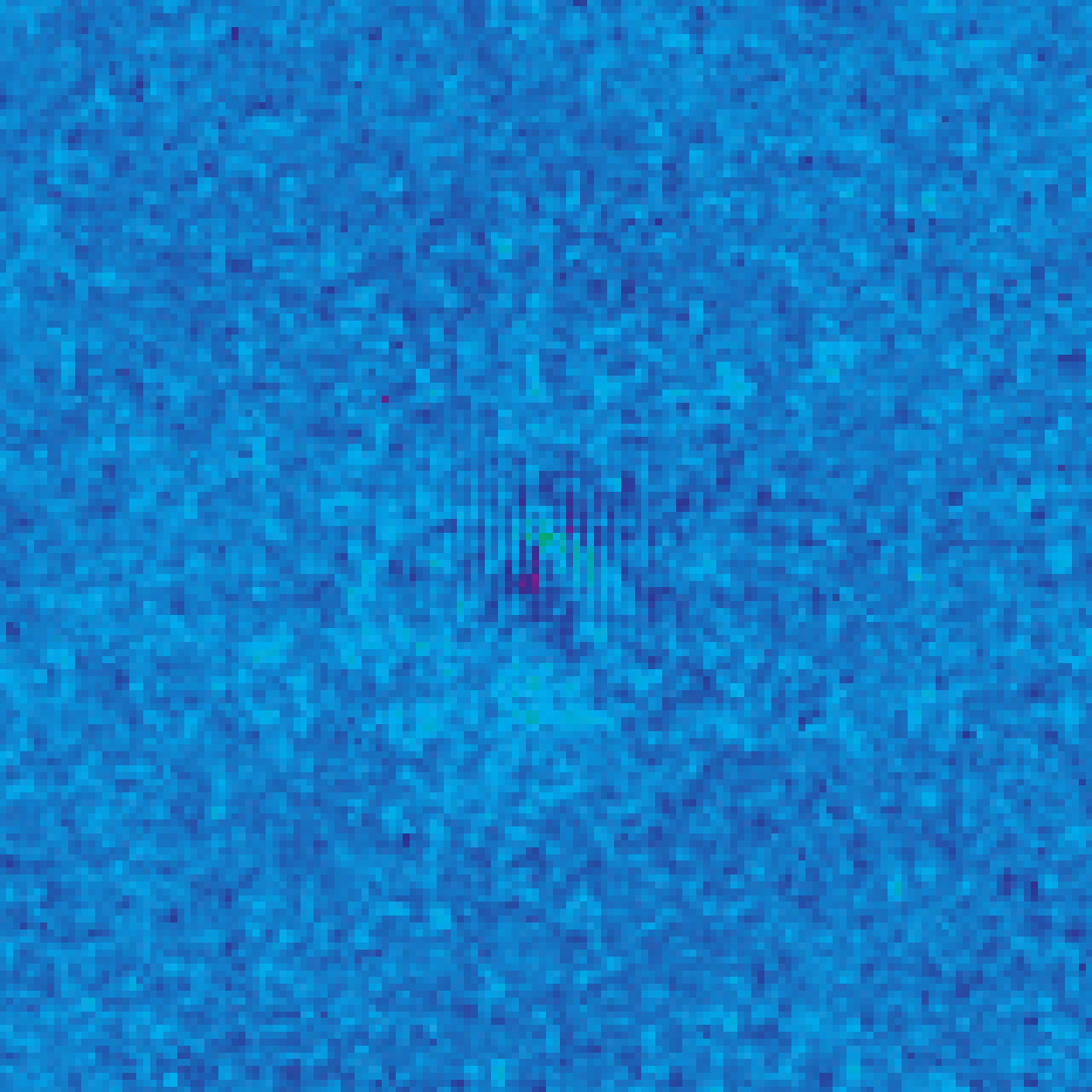}\\[2mm]
\includegraphics[width=4cm]{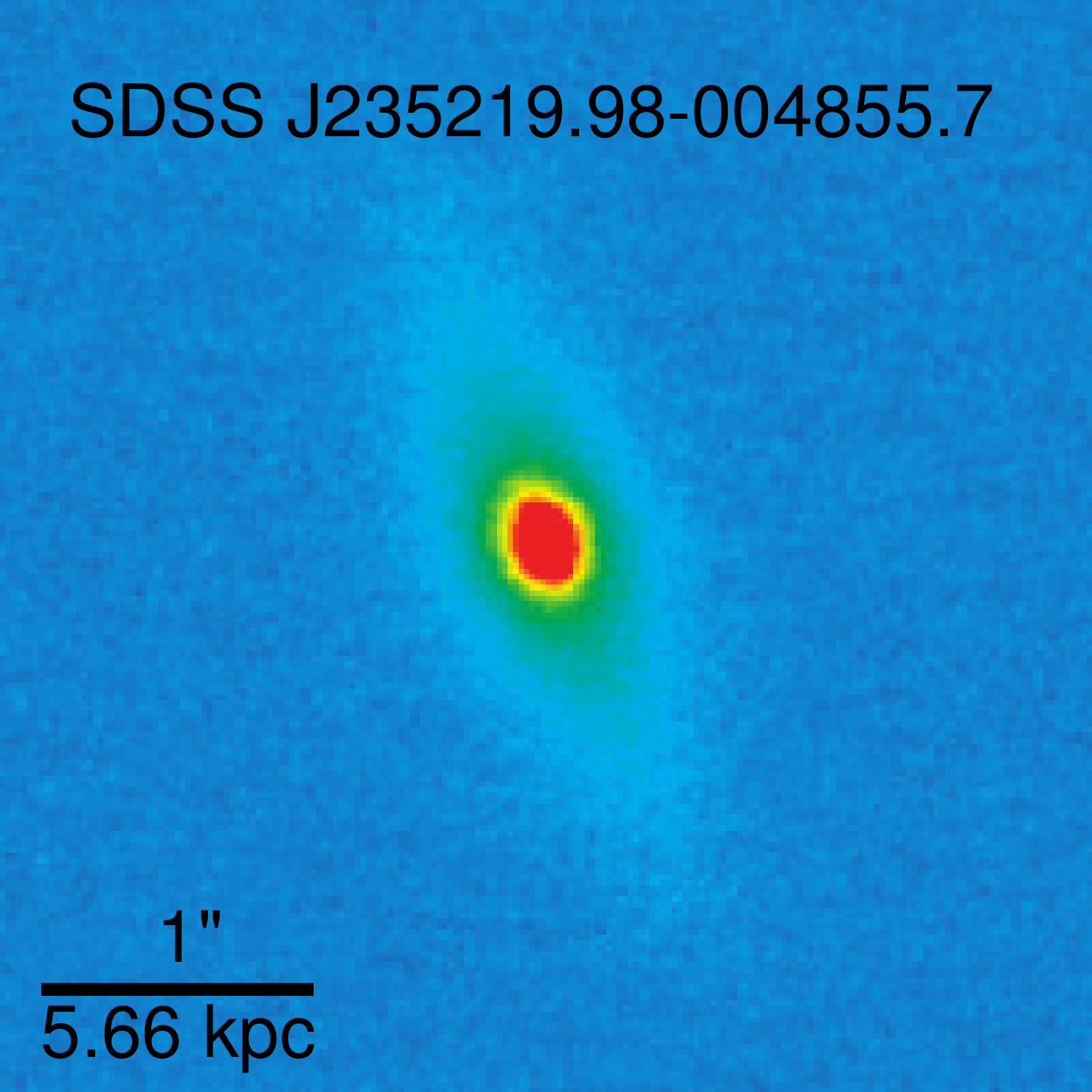}
\includegraphics[width=4cm]{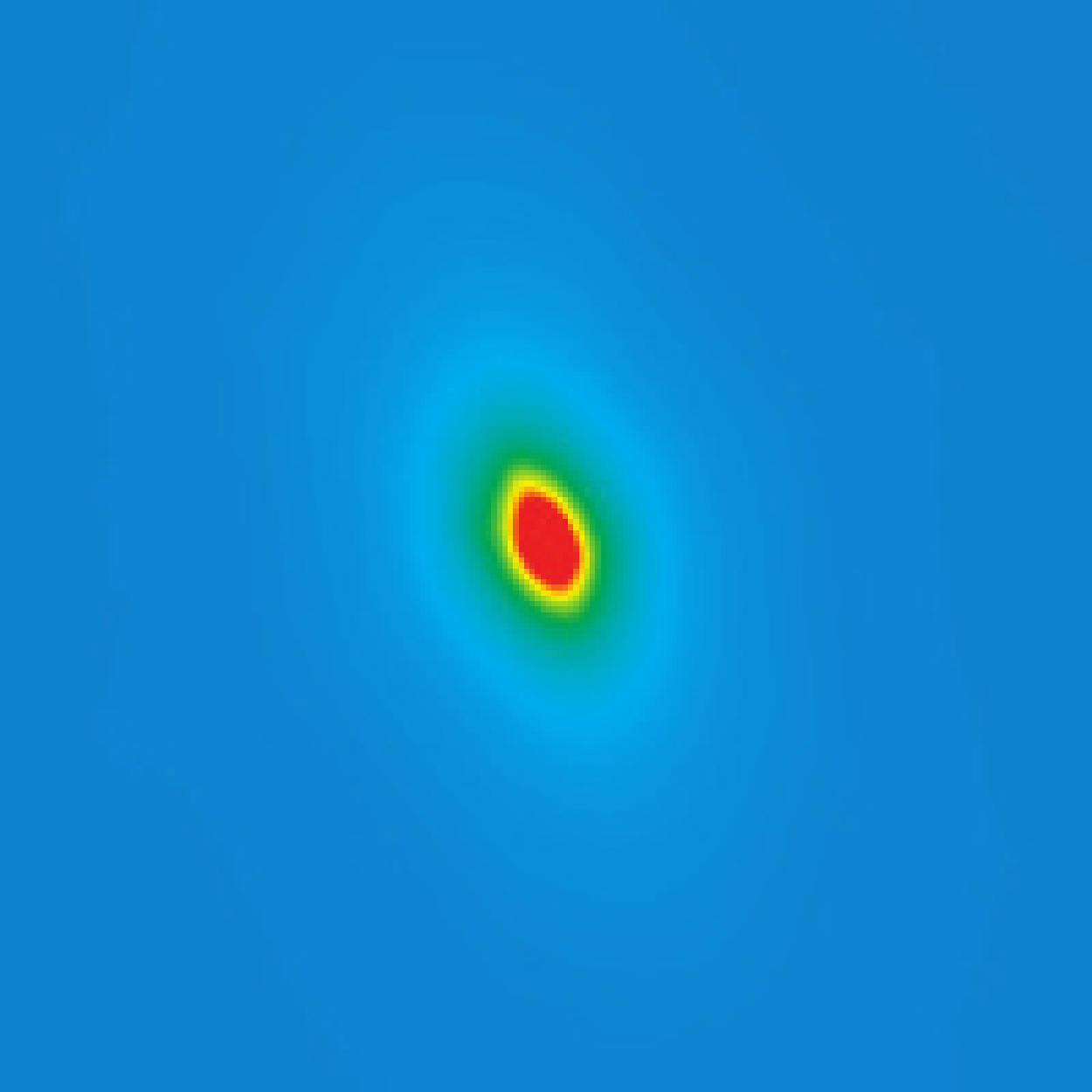}    
\includegraphics[width=4cm]{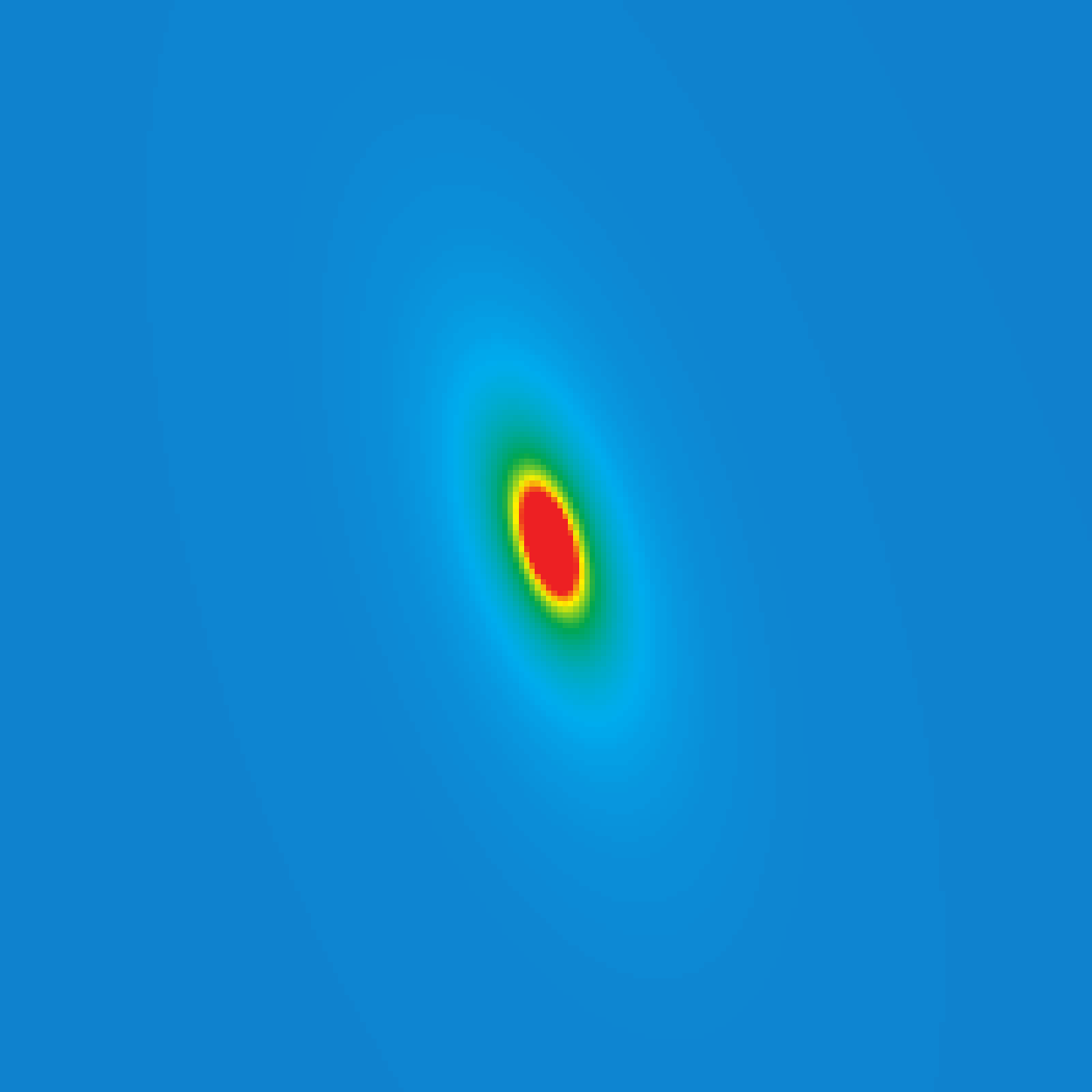}
\includegraphics[width=4cm]{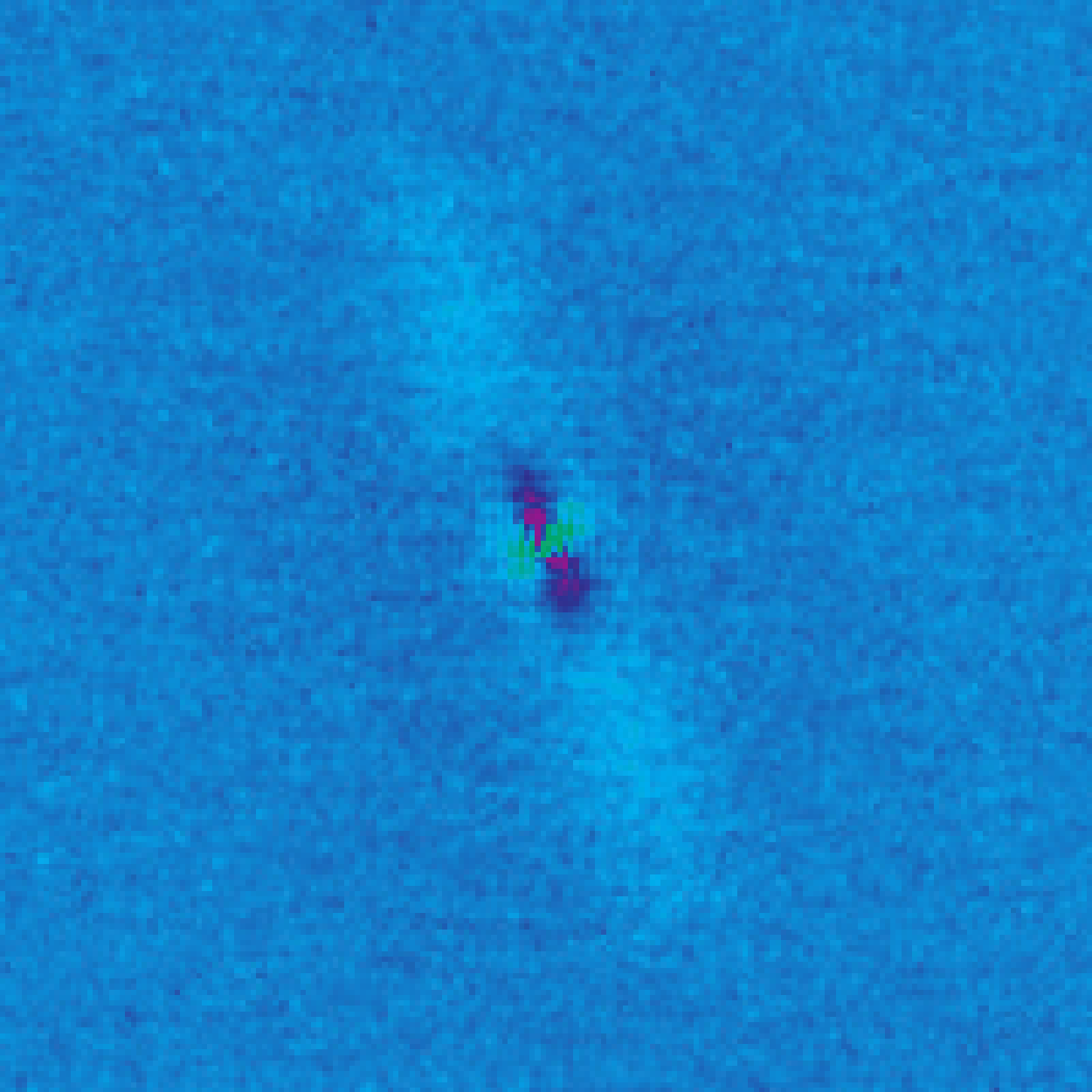}\\[1mm]
\hspace{40.7mm}\includegraphics[width=4cm]{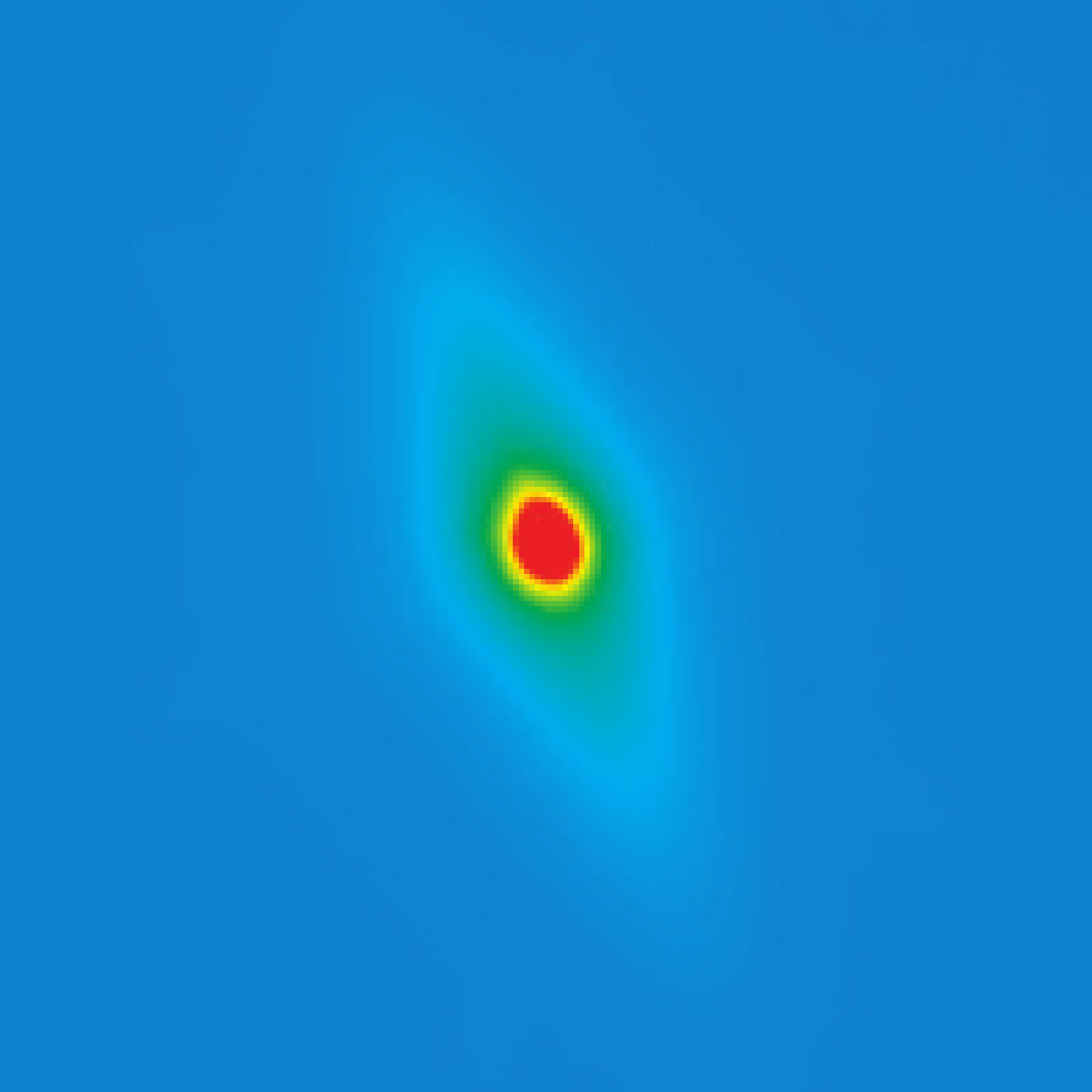}    
\includegraphics[width=4cm]{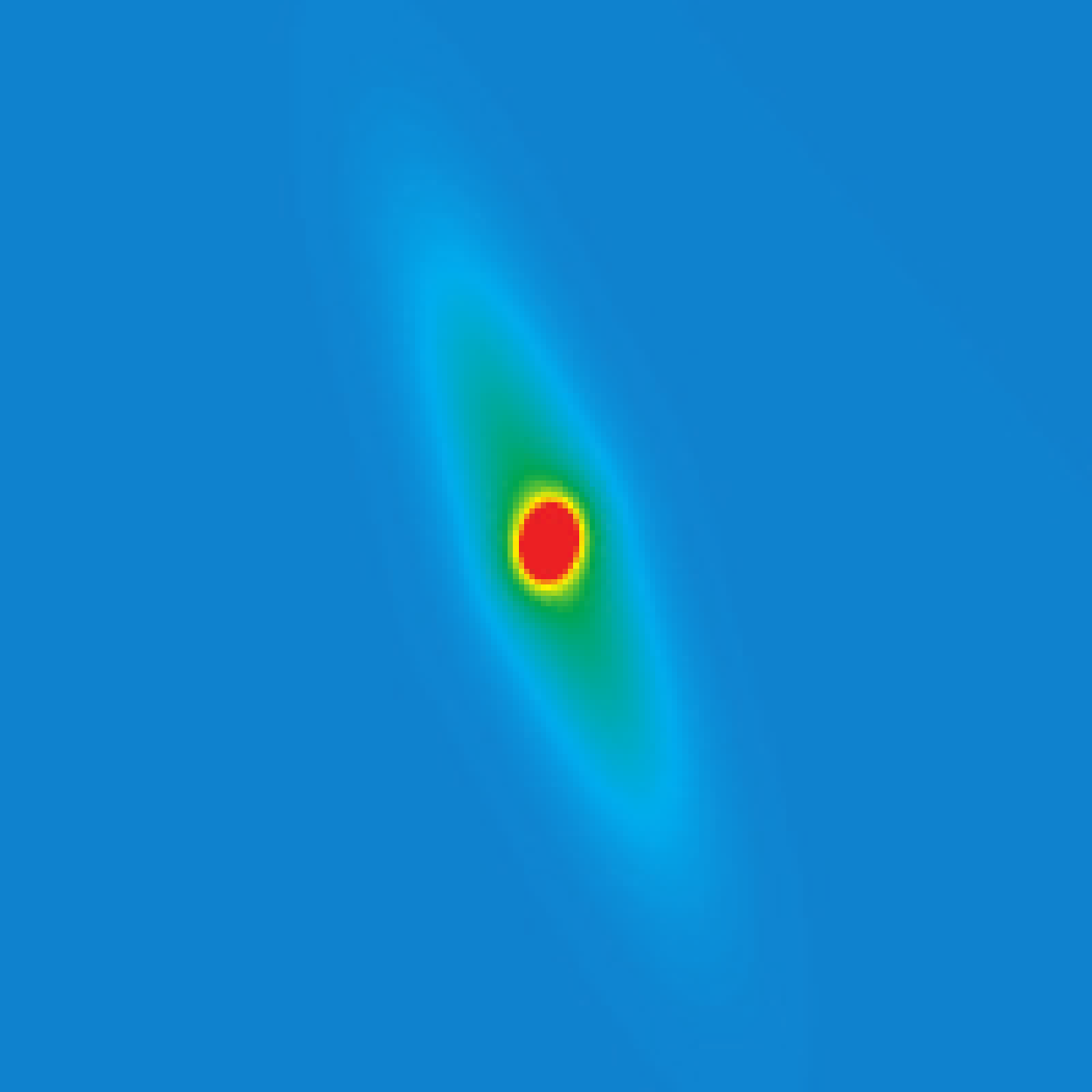}
\includegraphics[width=4cm]{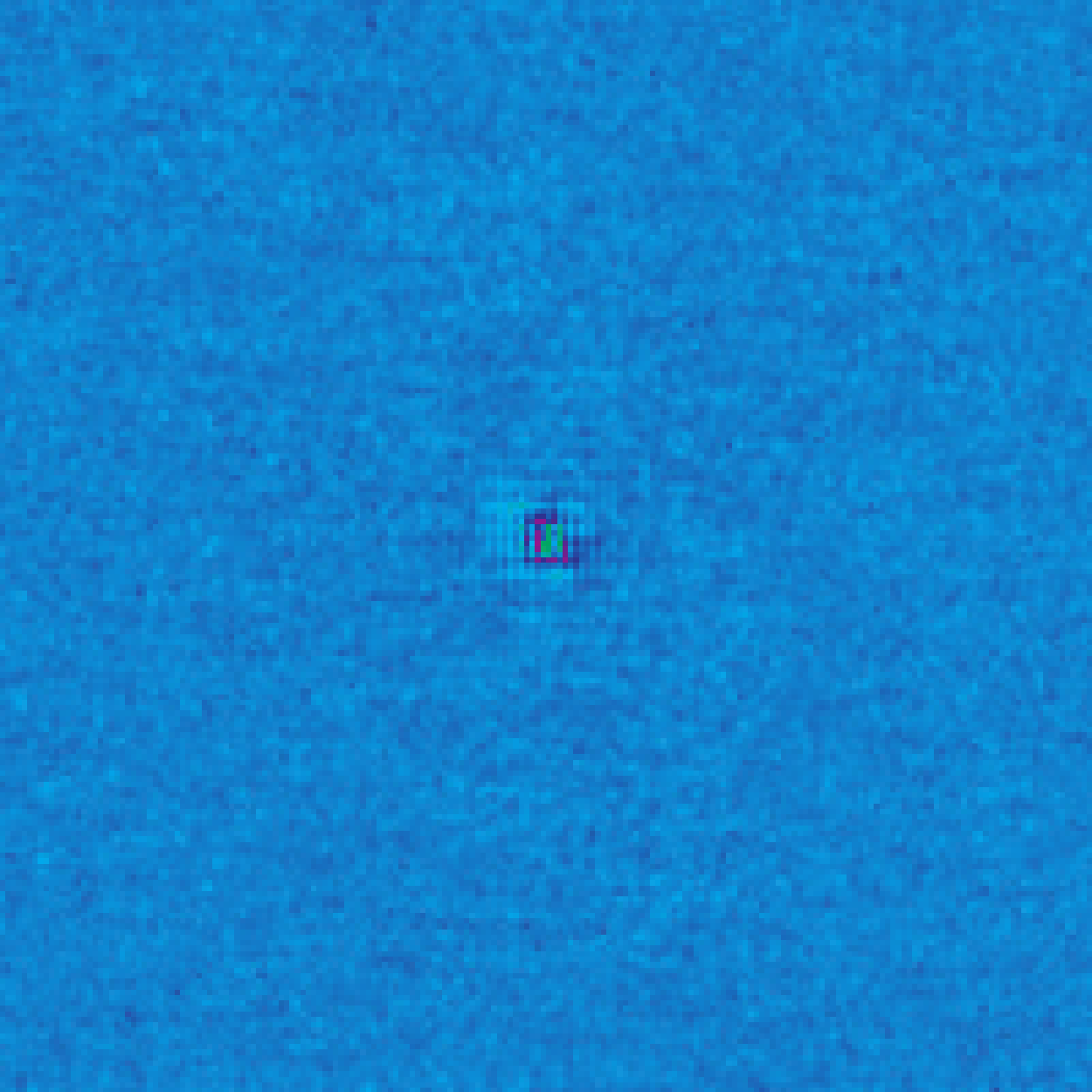}
%\caption{\centering Continued.}
\caption{Continued.}
\end{figure*}

\renewcommand{\thefigure}{A2}

\begin{figure*}
\centering
\includegraphics[width=4cm]{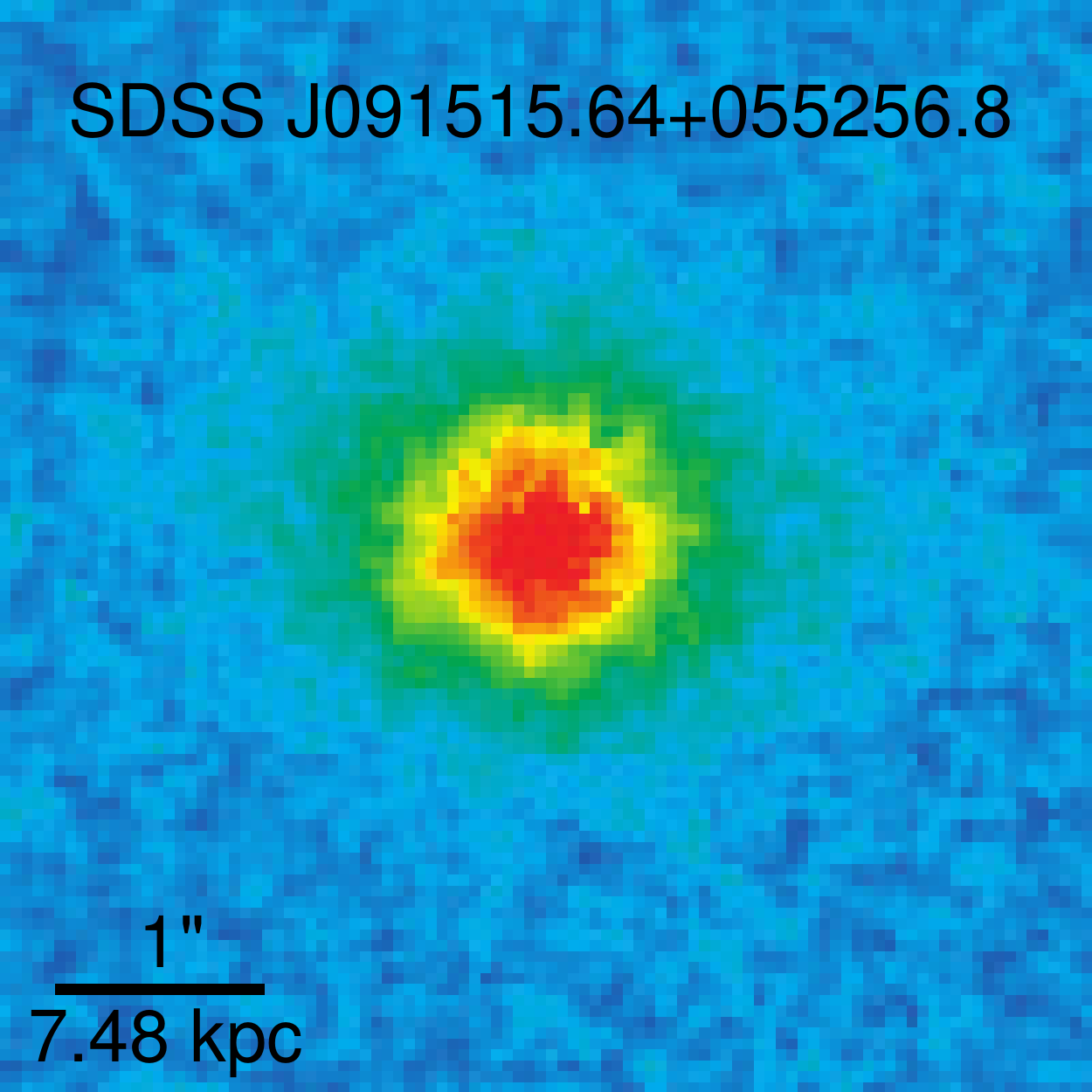}
\includegraphics[width=4cm]{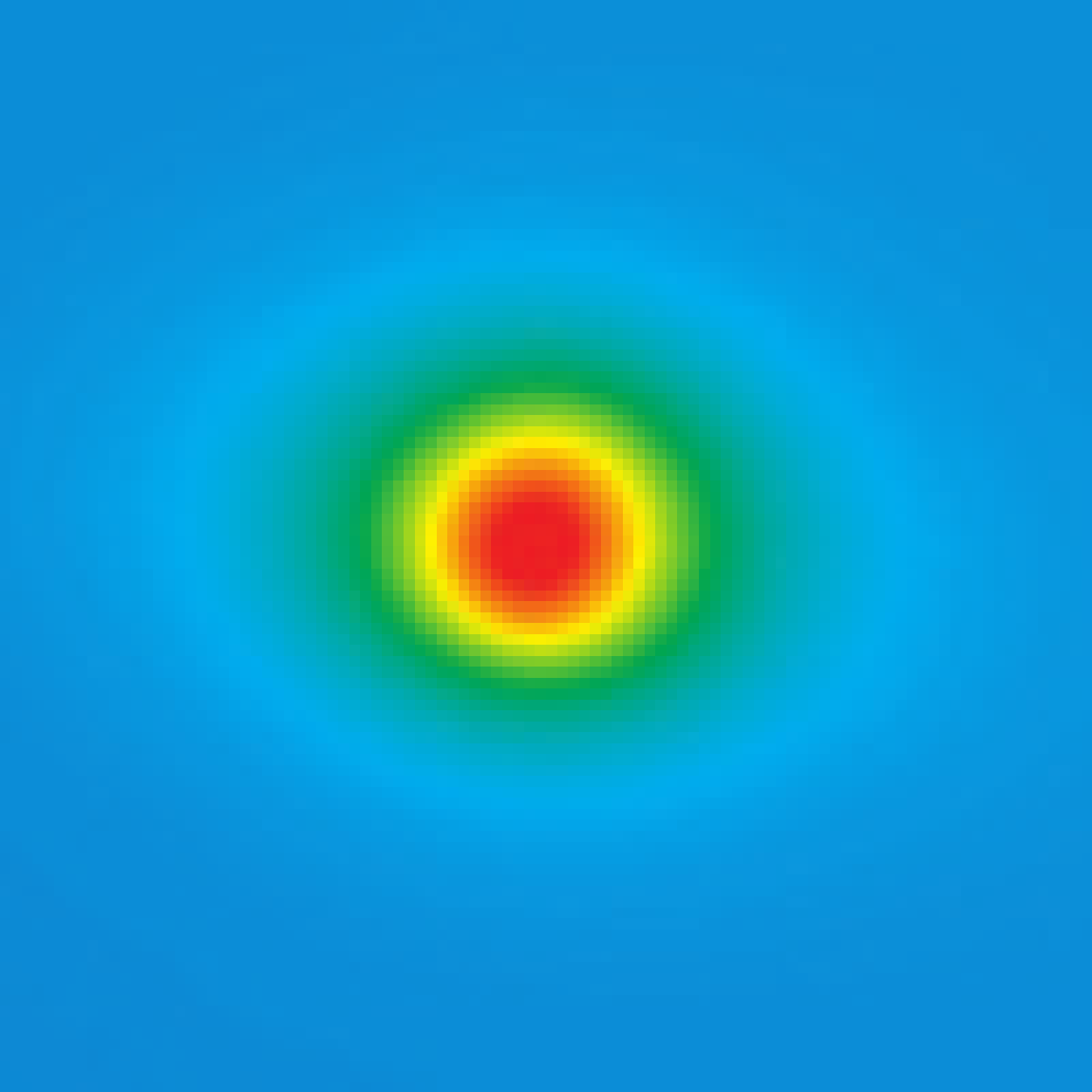}   
\includegraphics[width=4cm]{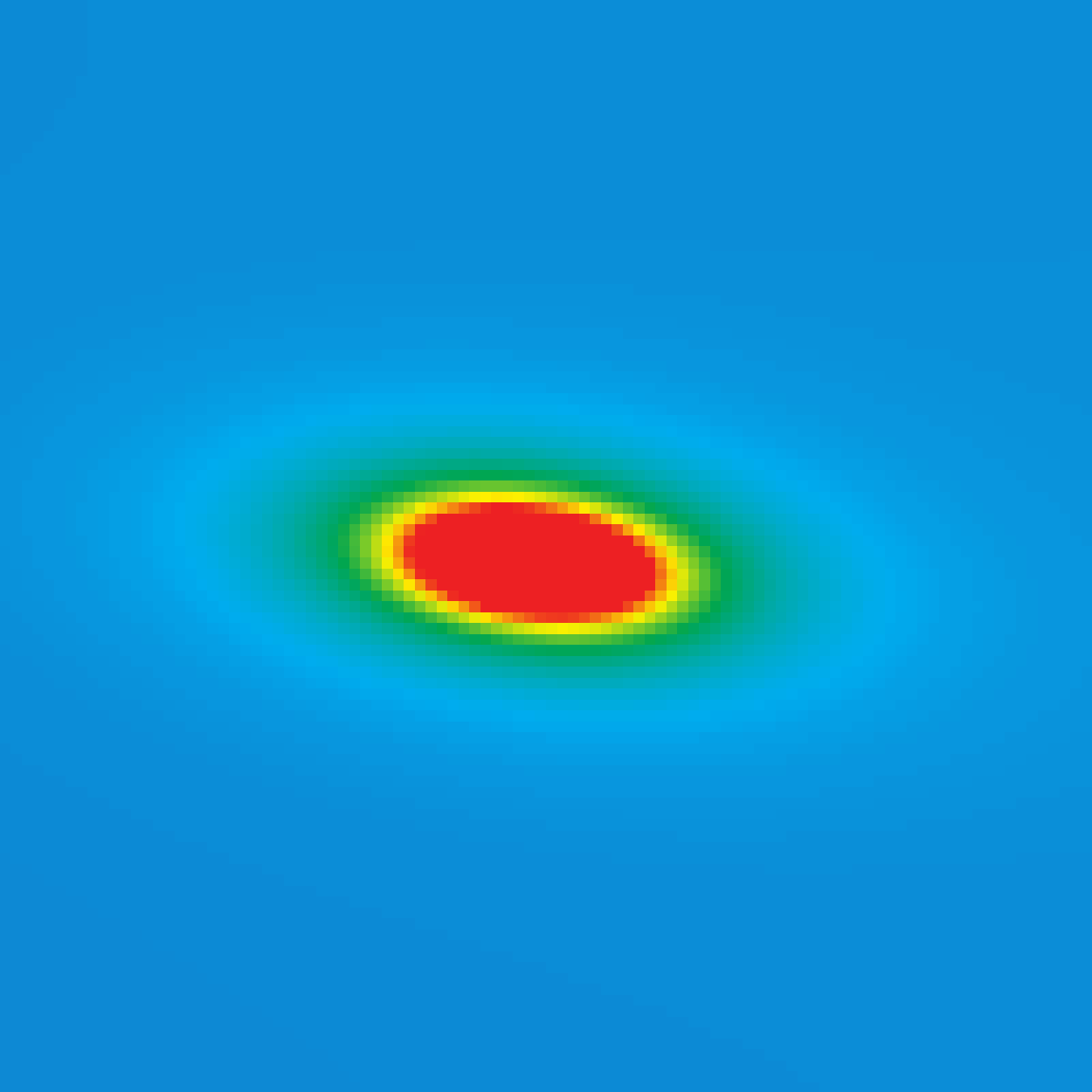}
\includegraphics[width=4cm]{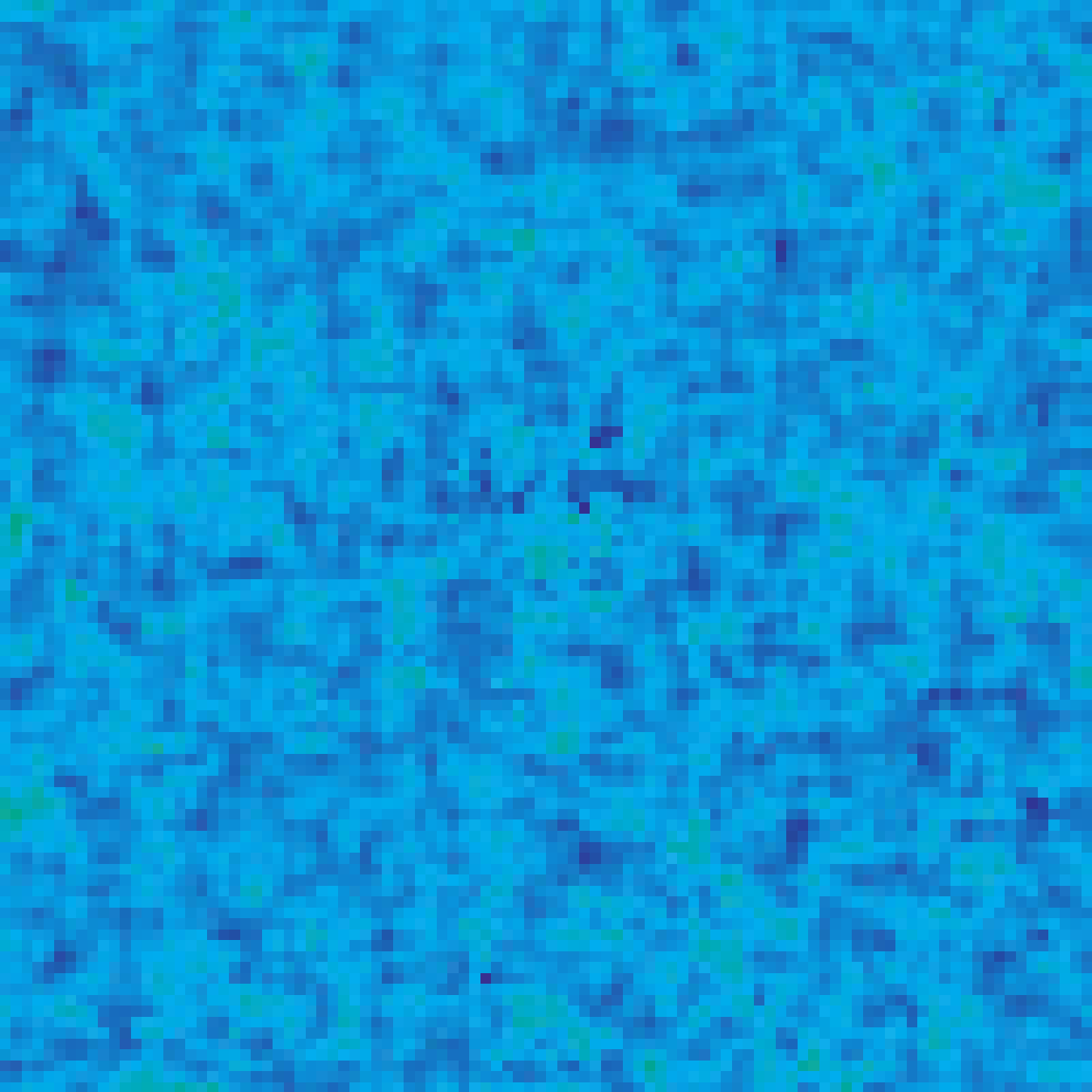}\\[2mm]
\includegraphics[width=4cm]{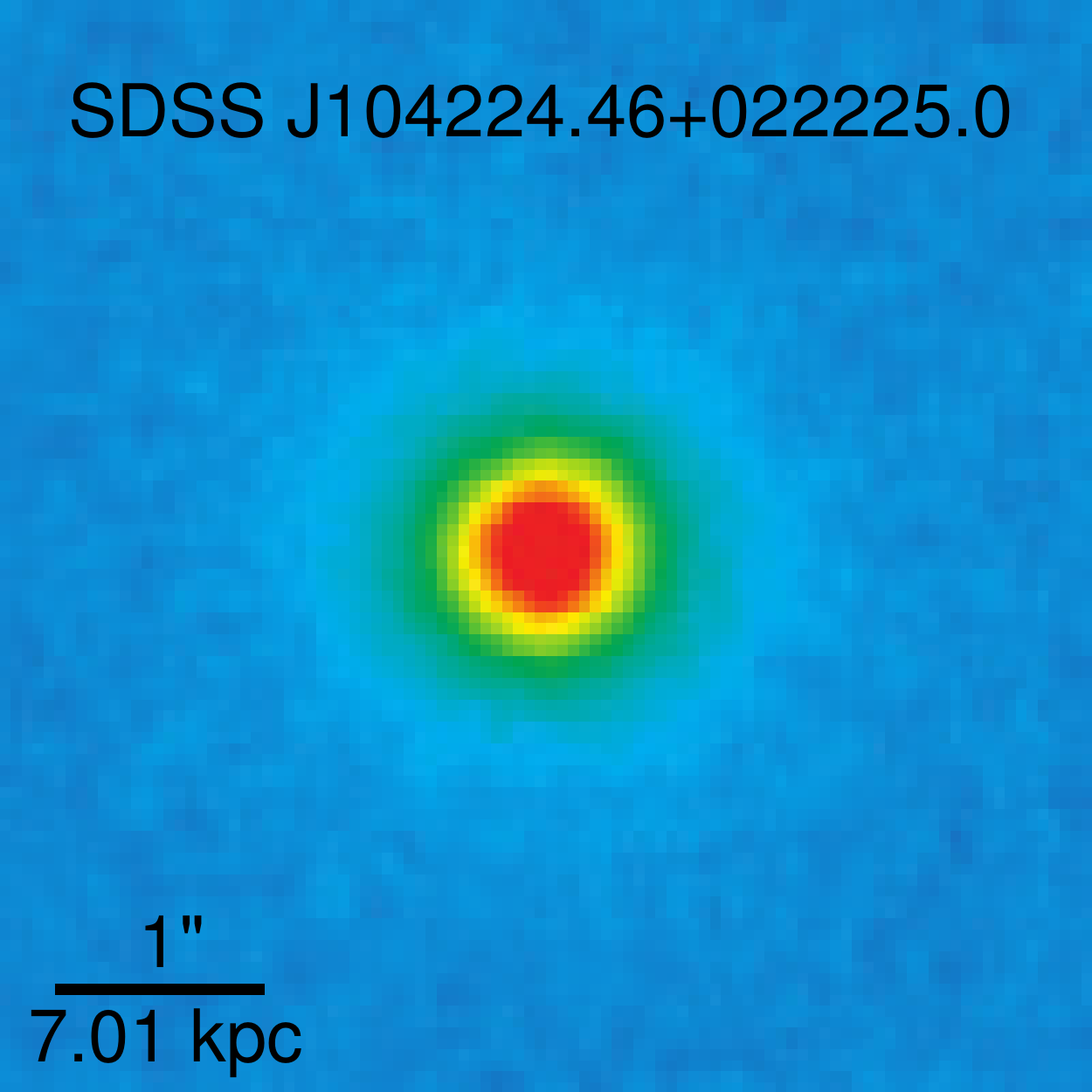}
\includegraphics[width=4cm]{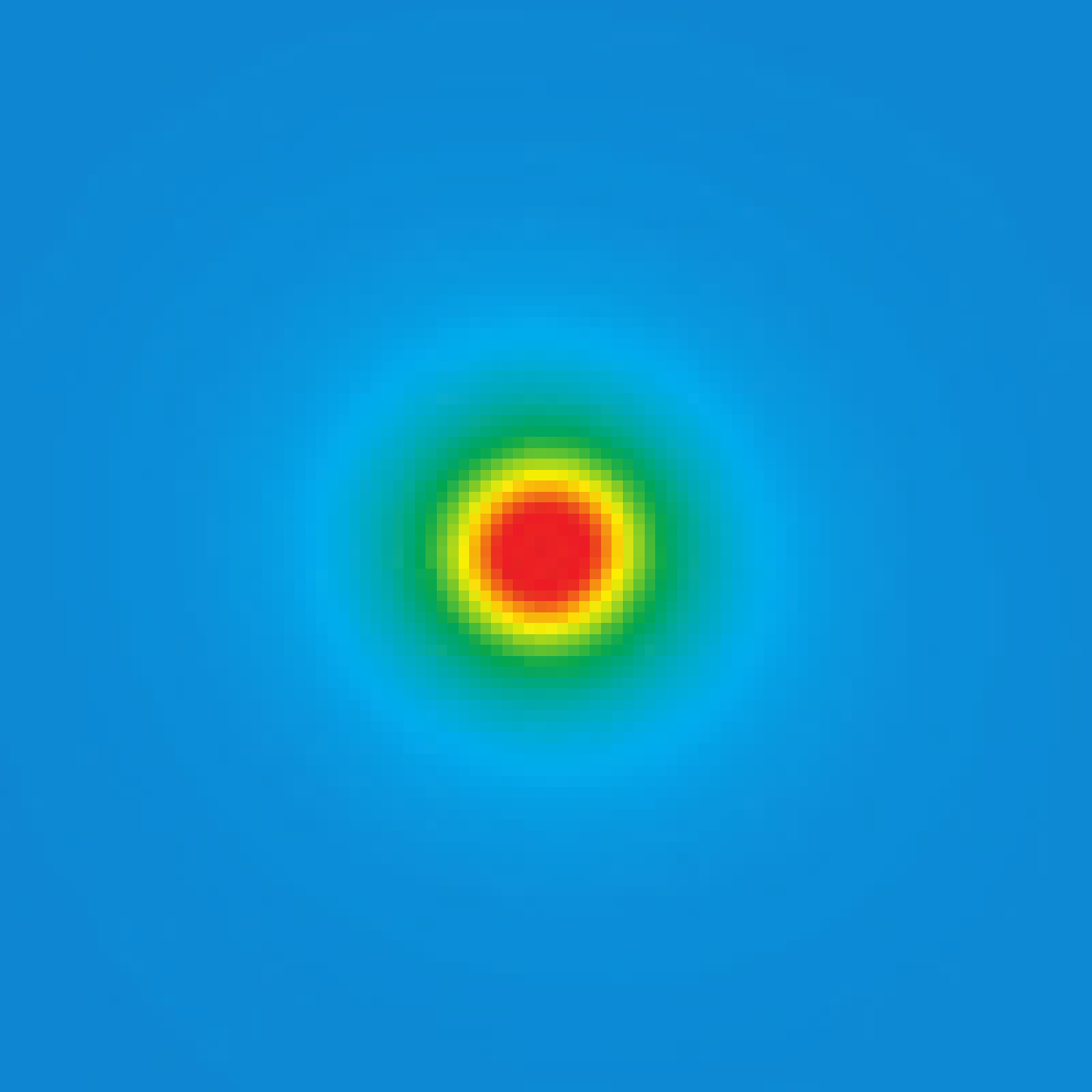}    
\includegraphics[width=4cm]{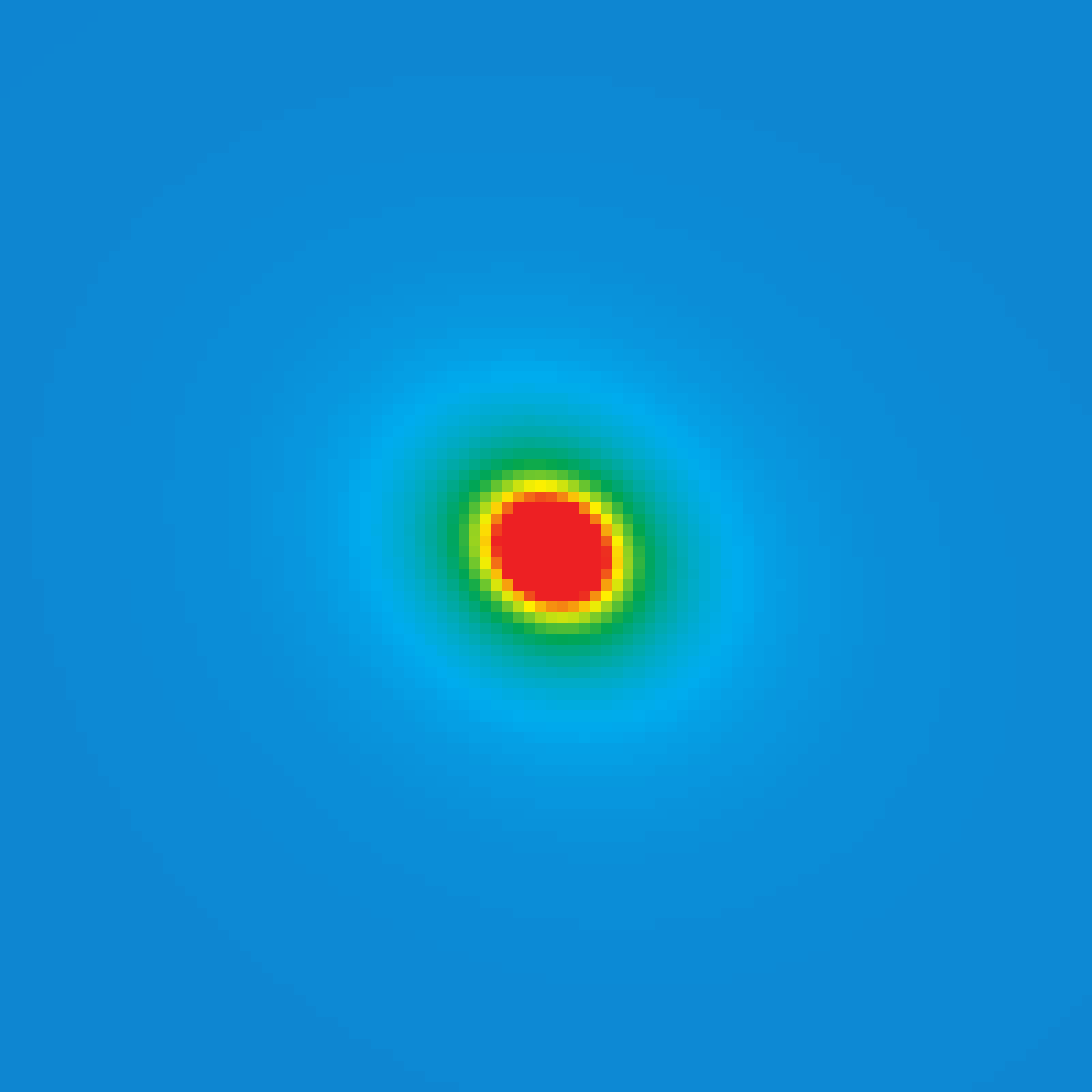}
\includegraphics[width=4cm]{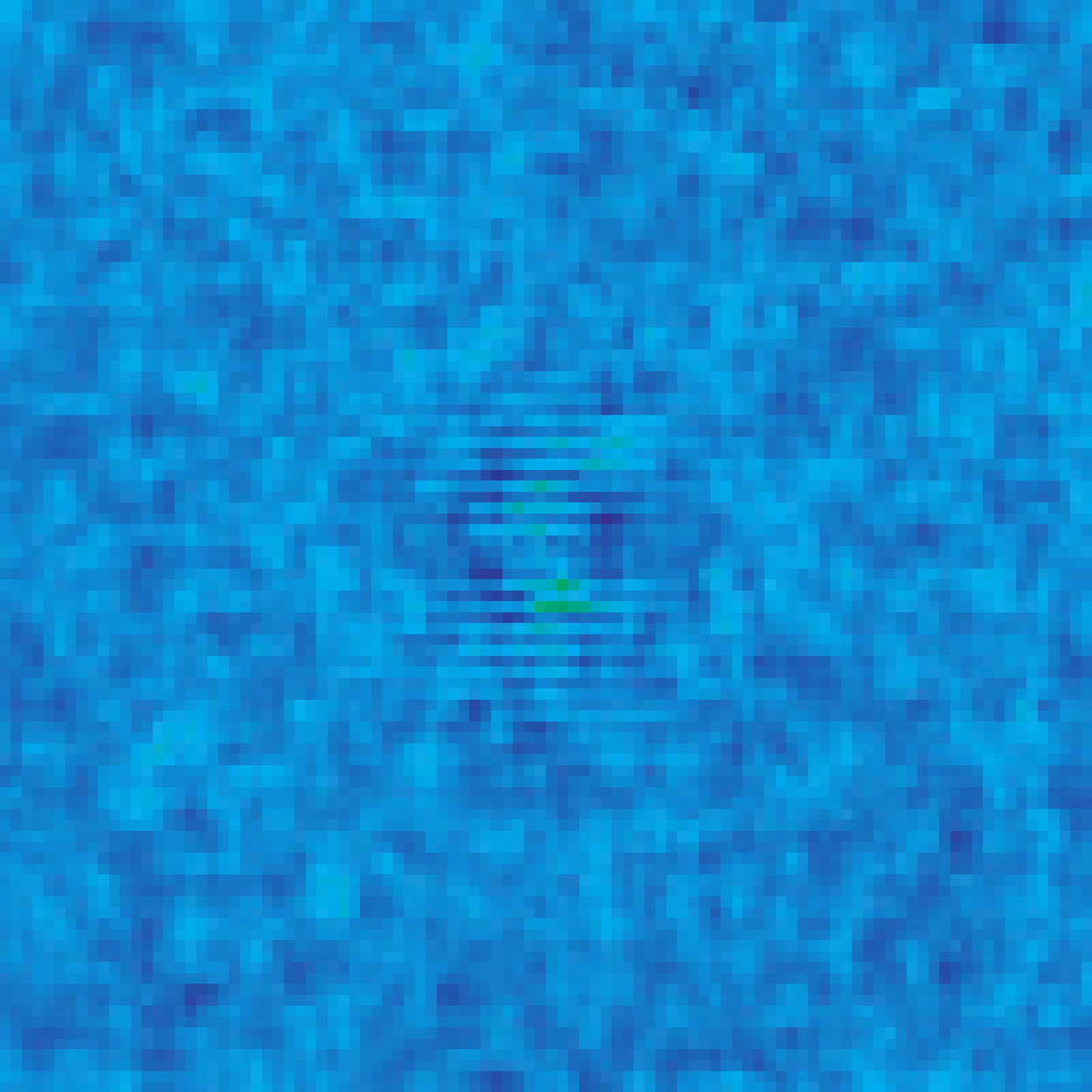}\\[2mm]   
\includegraphics[width=4cm]{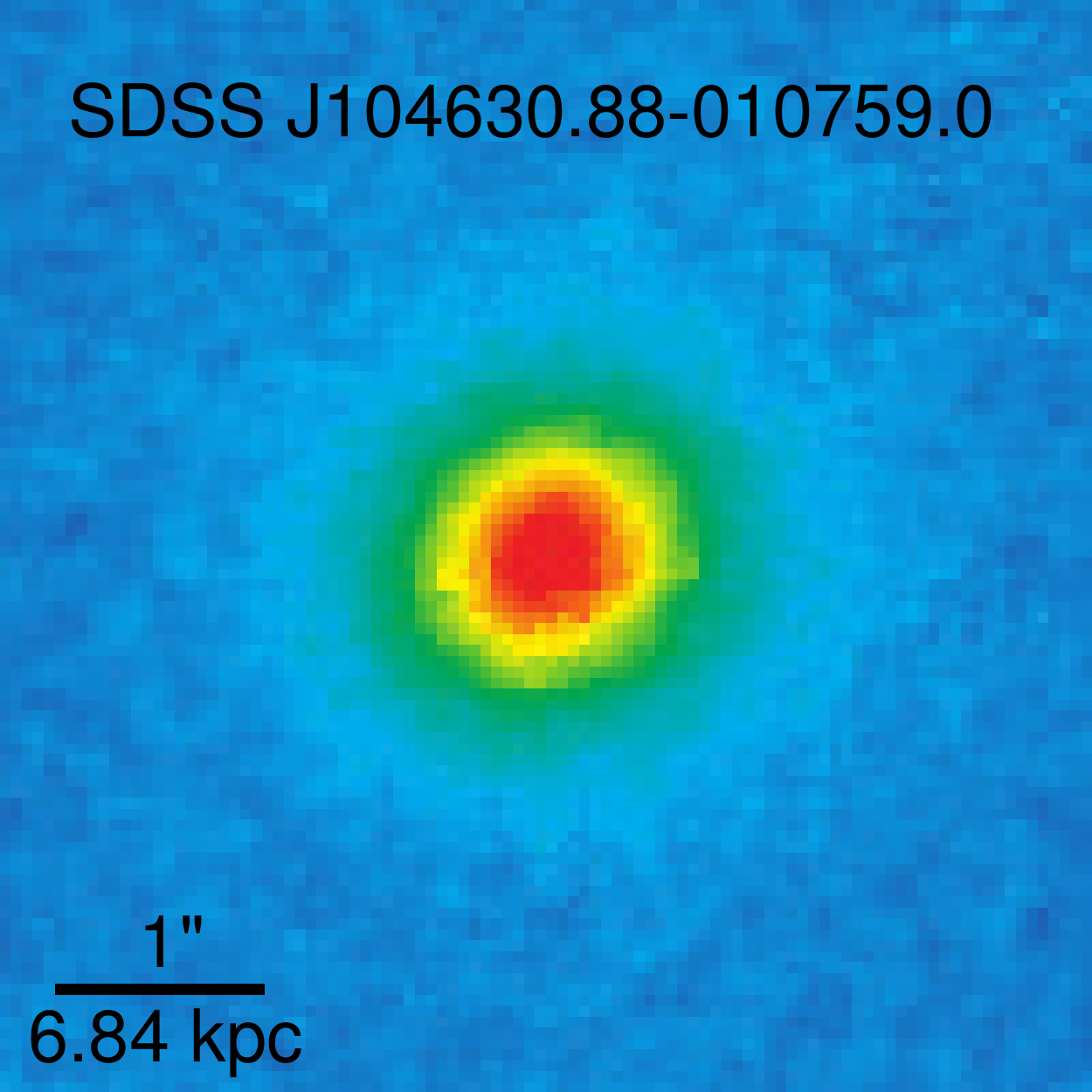}
\includegraphics[width=4cm]{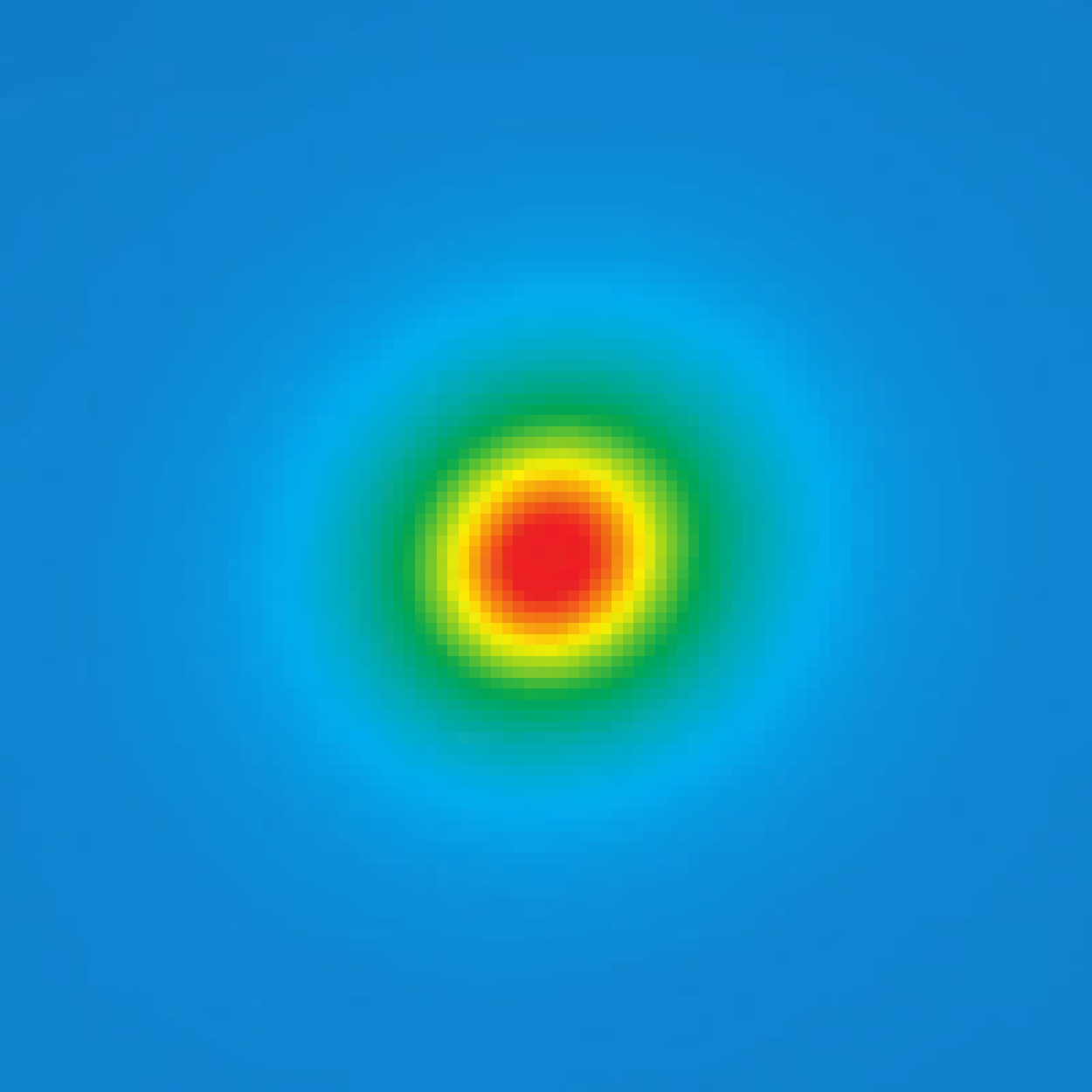}    
\includegraphics[width=4cm]{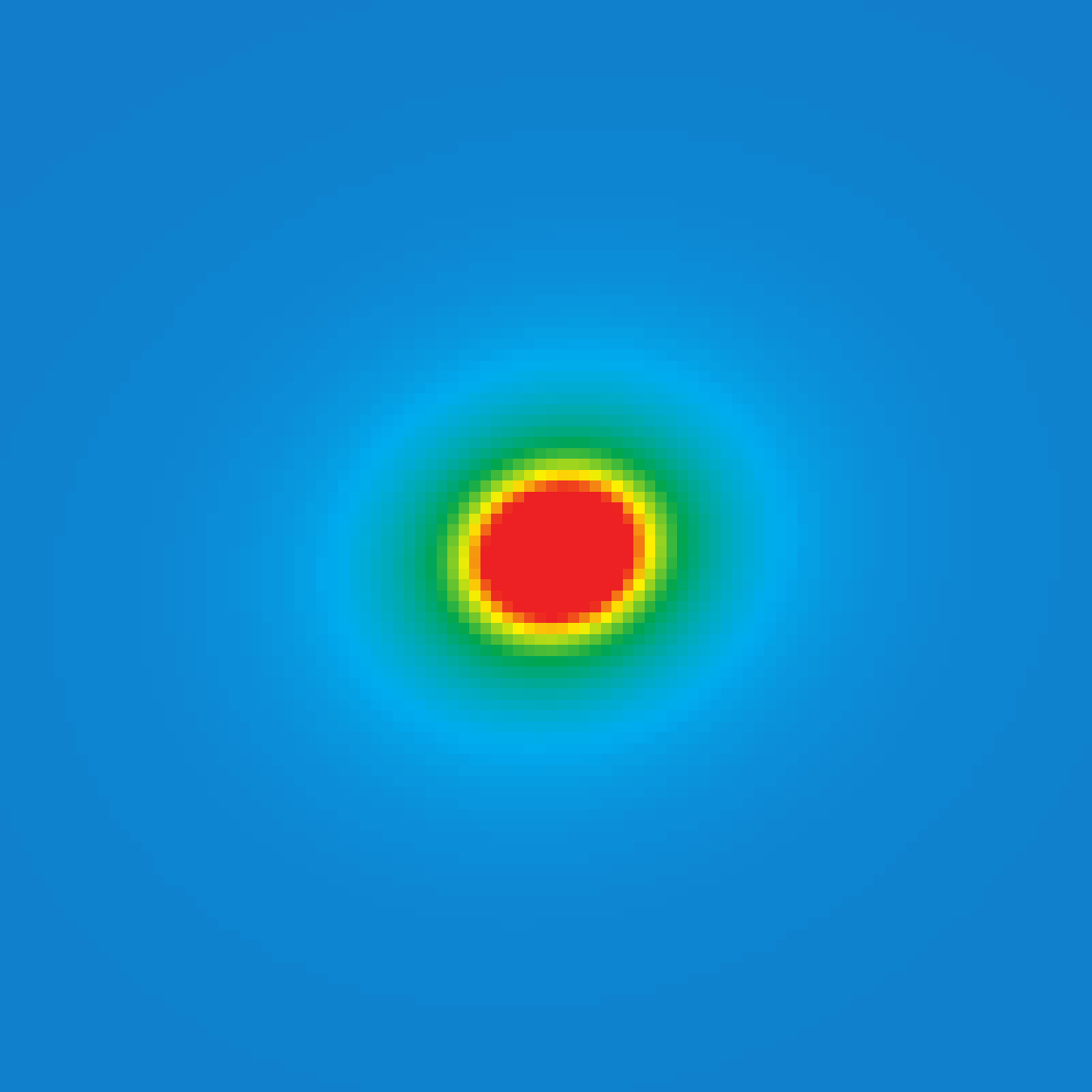}
\includegraphics[width=4cm]{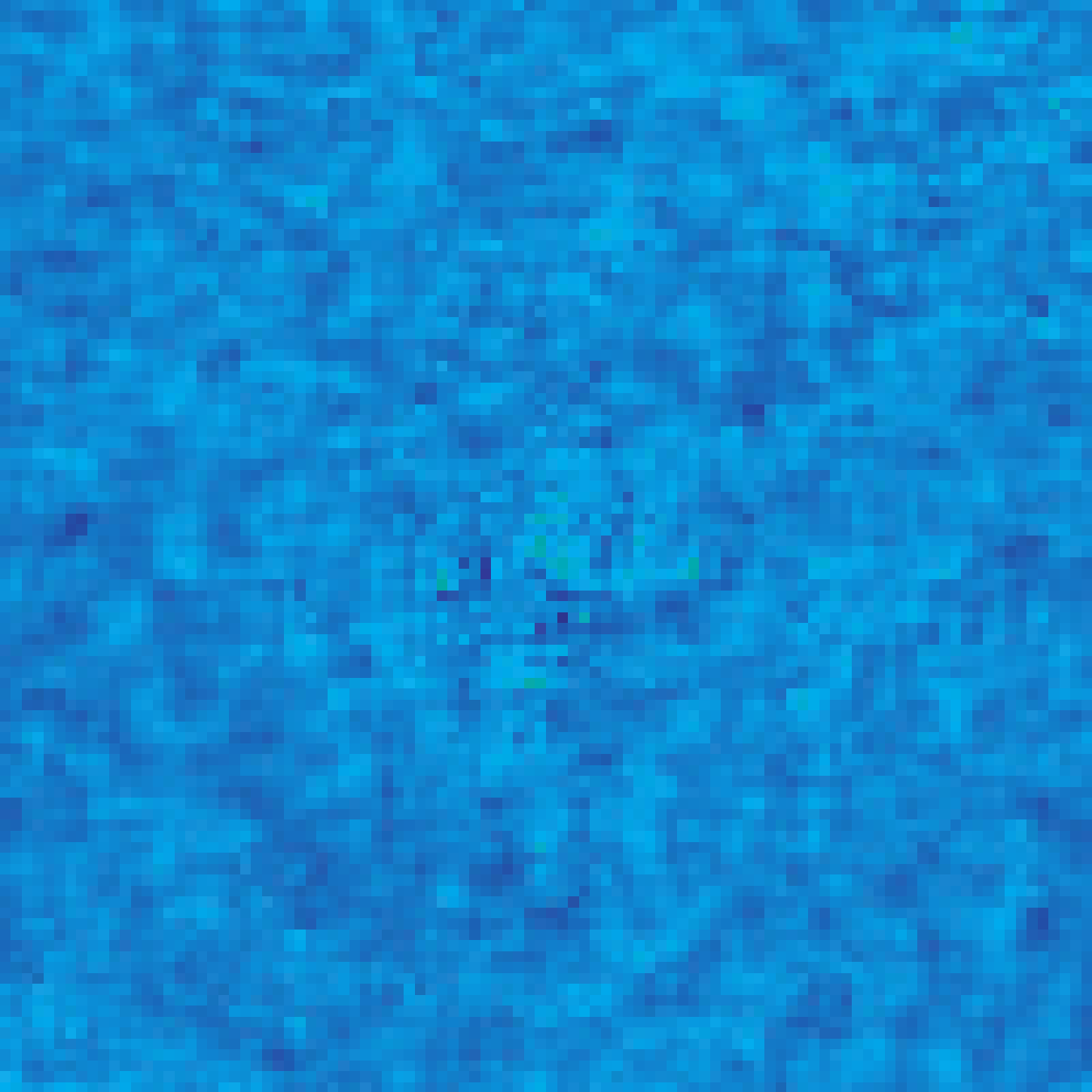}\\[2mm]    
\includegraphics[width=4cm]{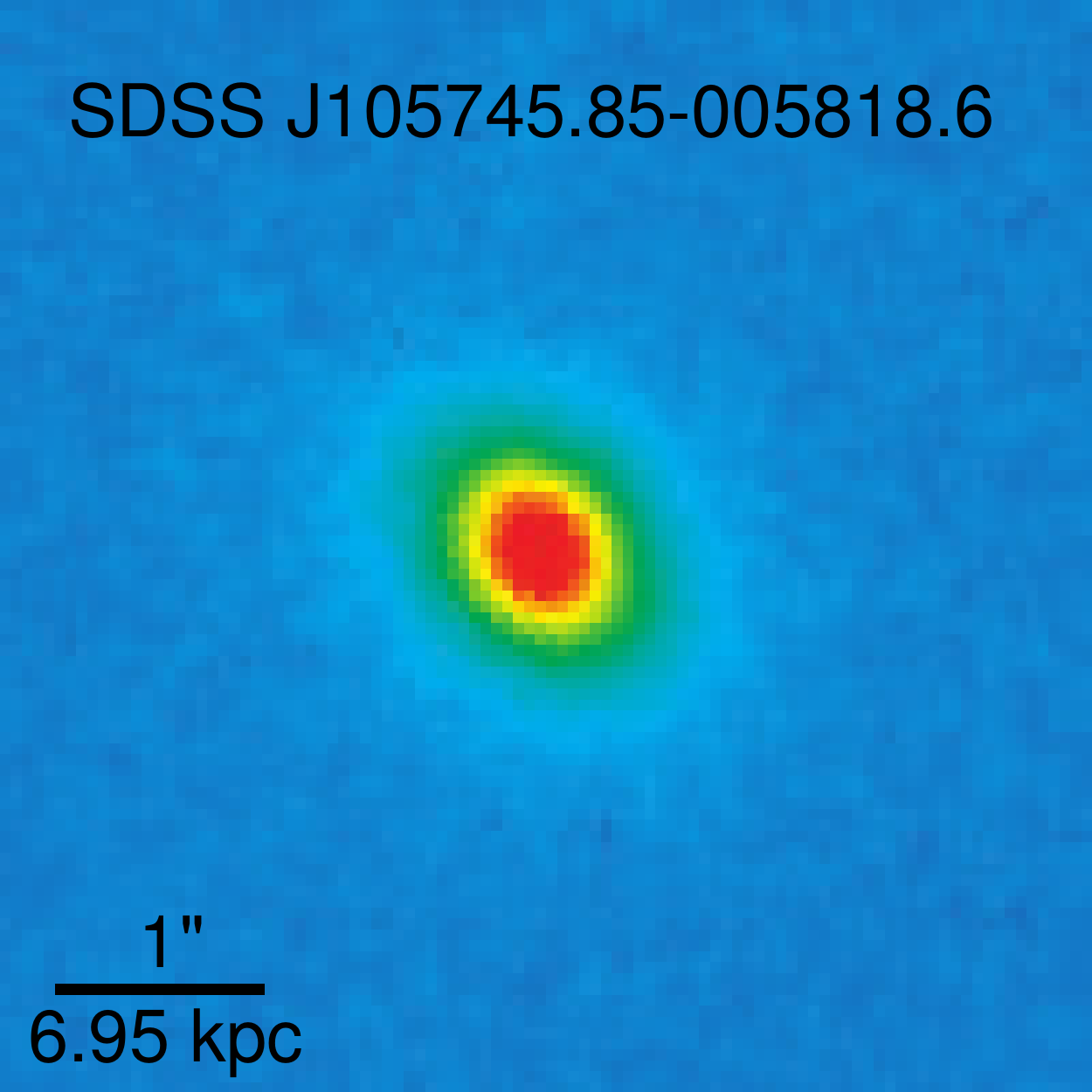}
\includegraphics[width=4cm]{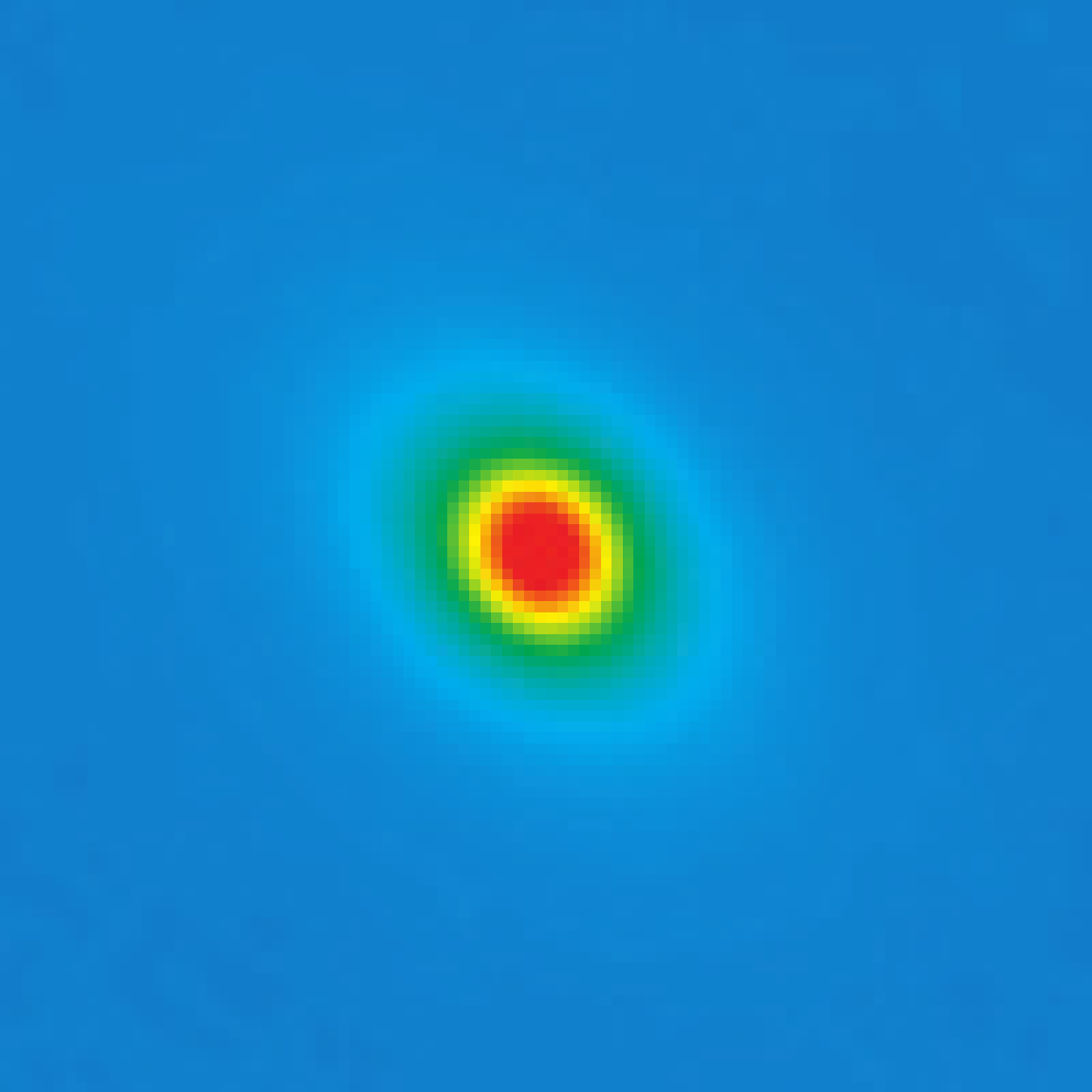}    
\includegraphics[width=4cm]{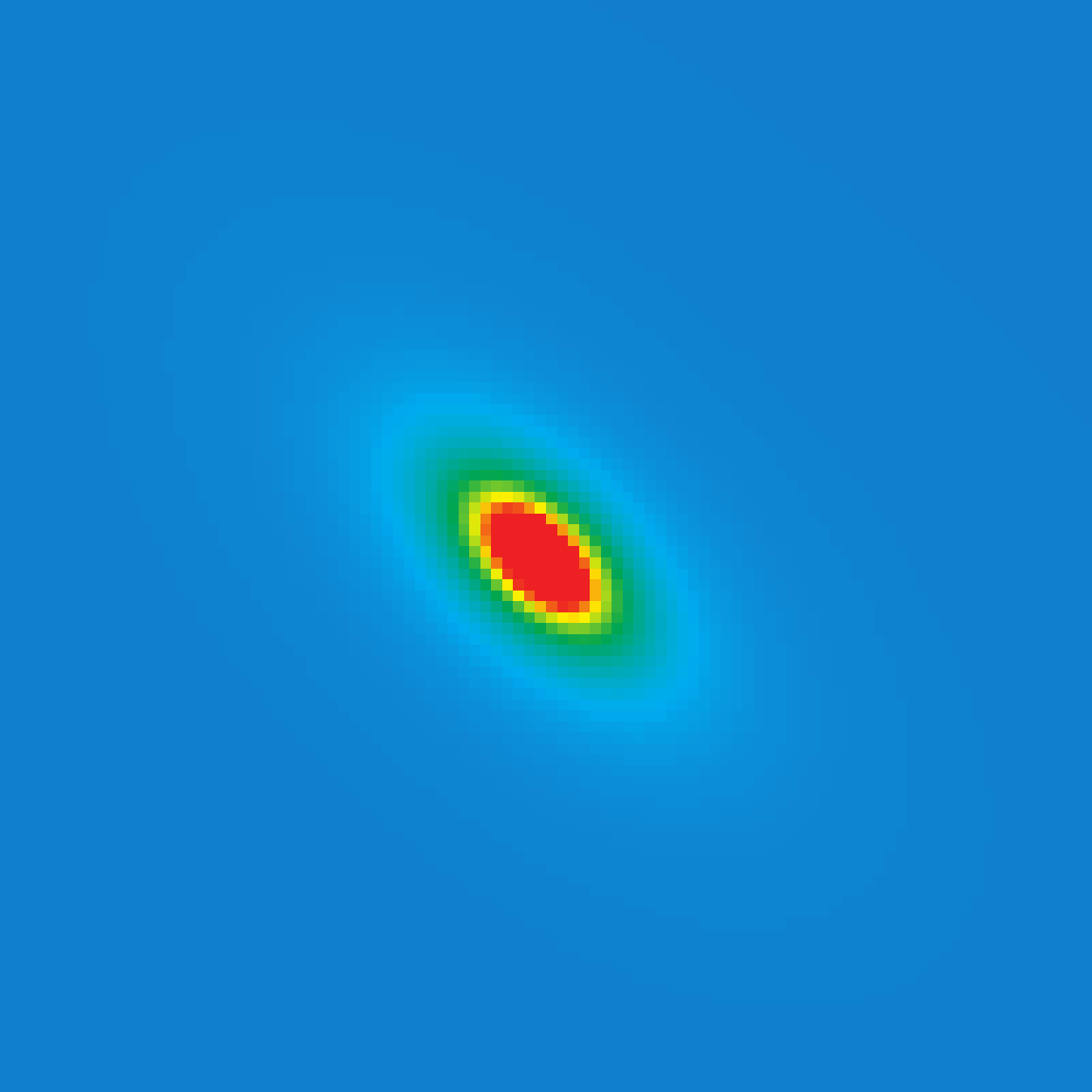}
\includegraphics[width=4cm]{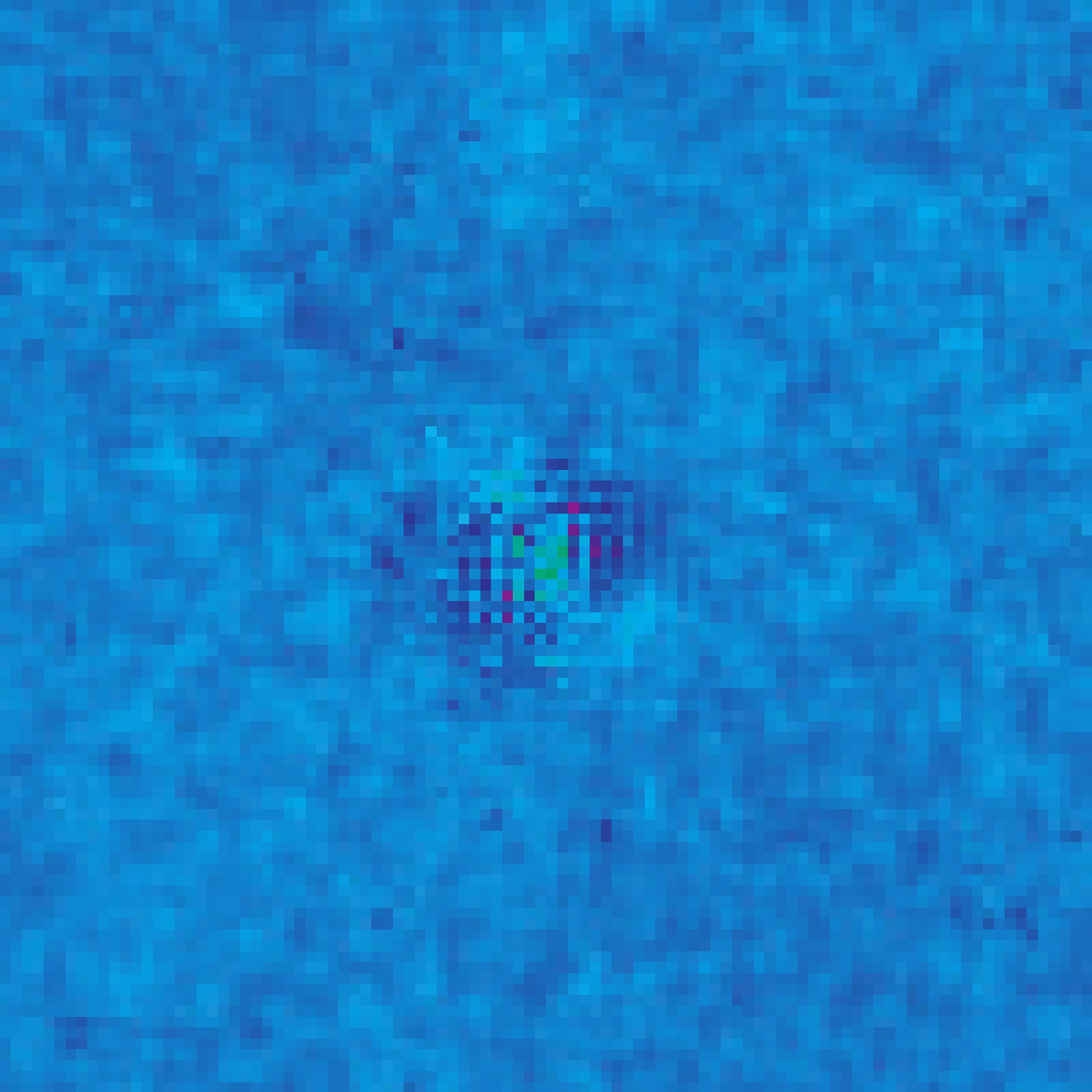} \\[2mm]   
\includegraphics[width=4cm]{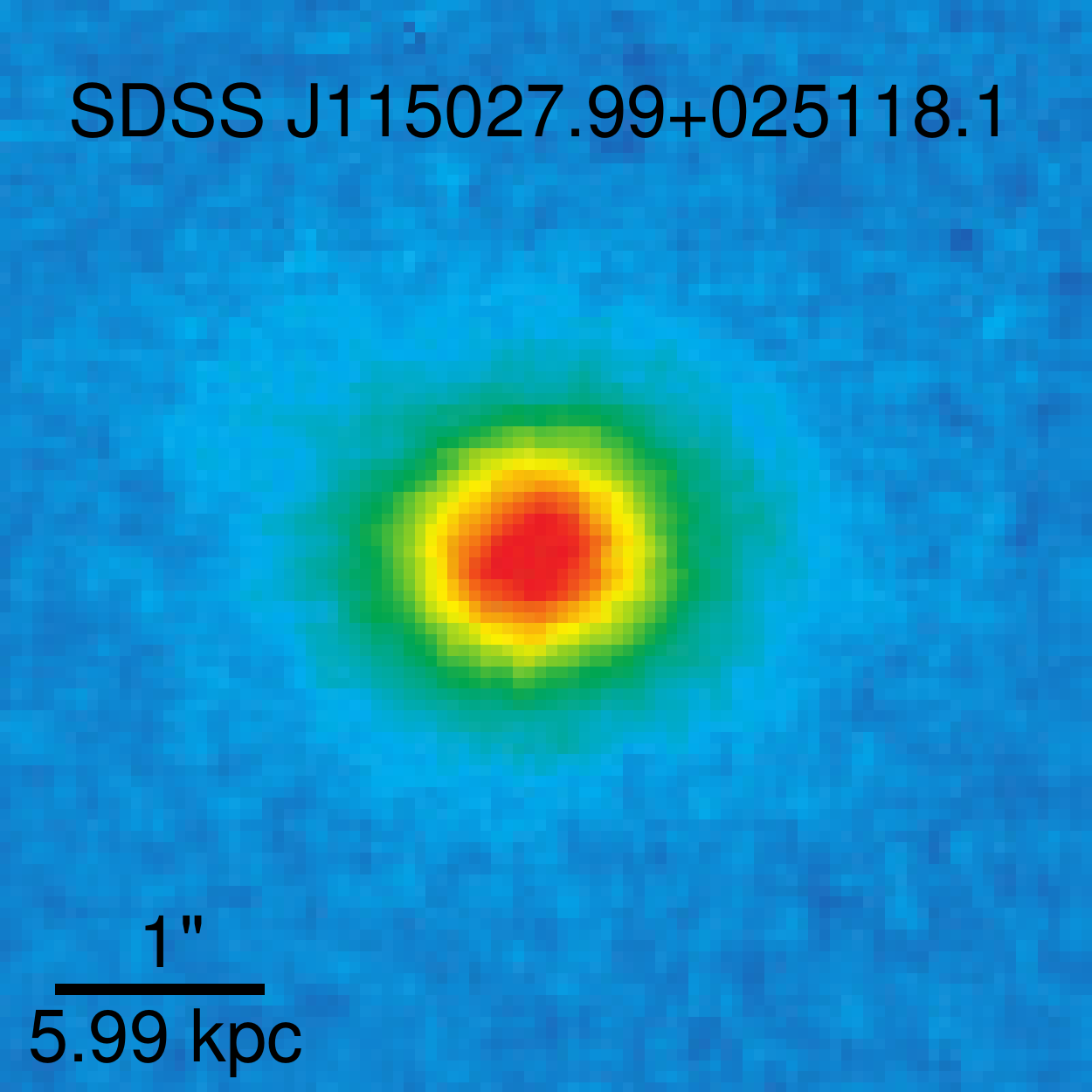}
\includegraphics[width=4cm]{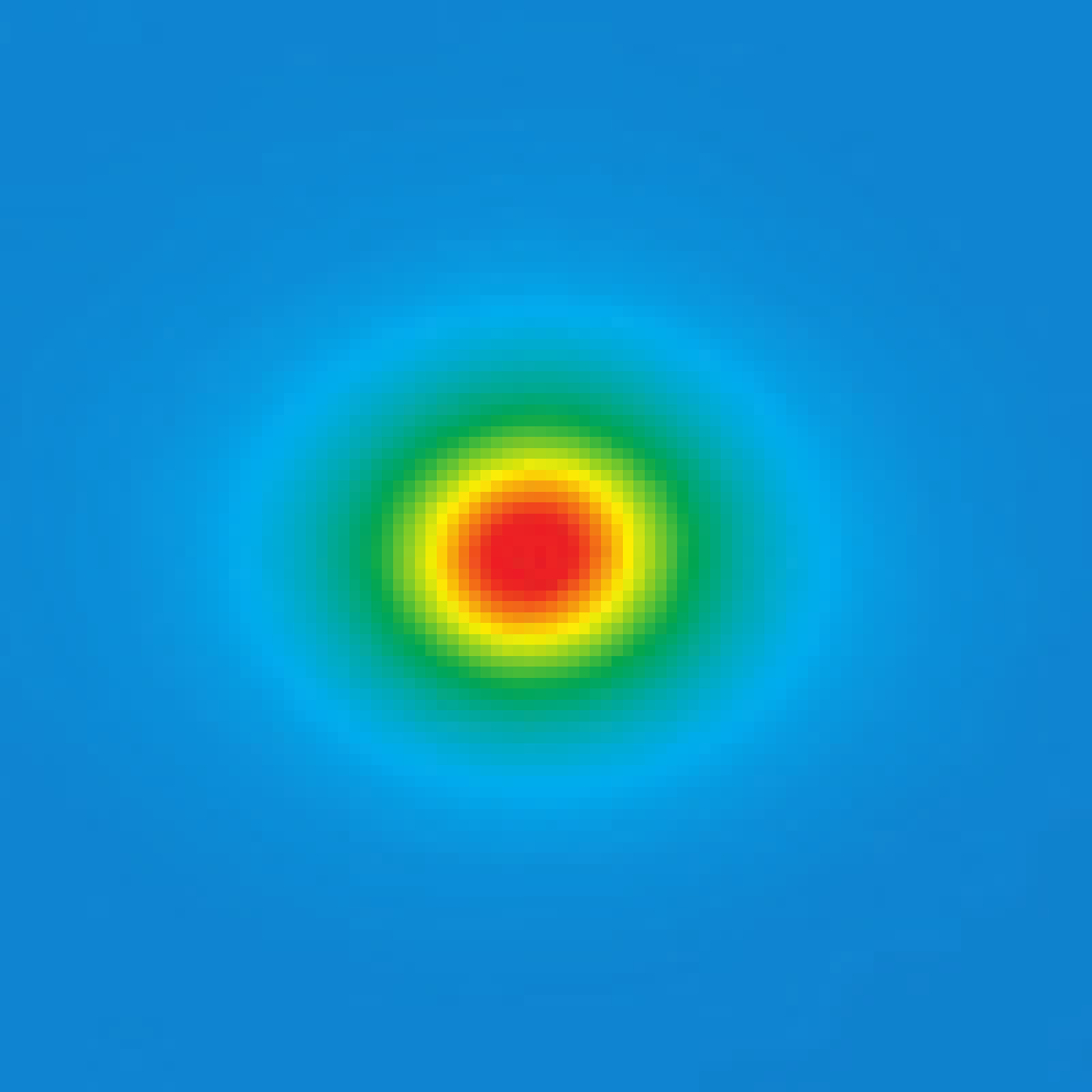}    
\includegraphics[width=4cm]{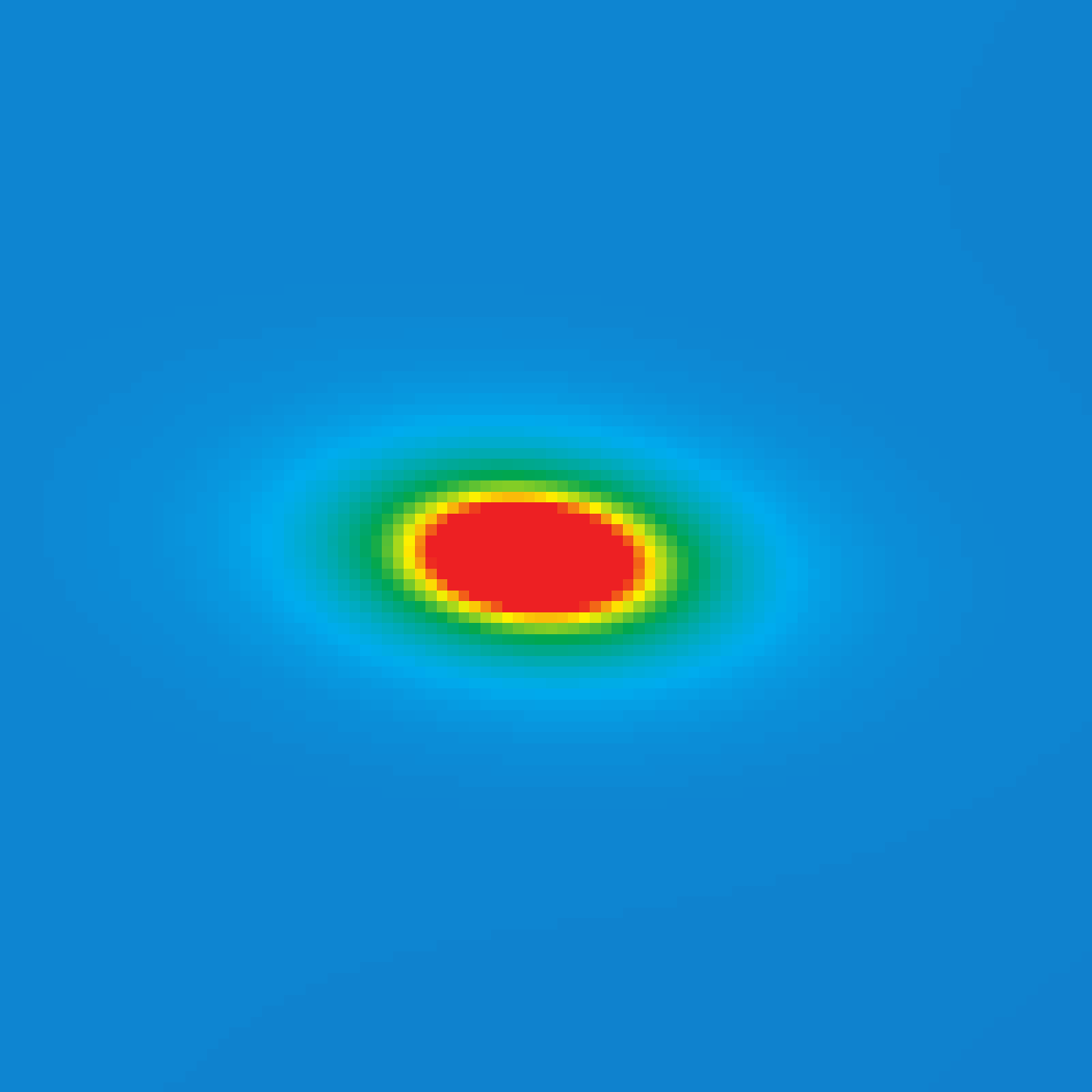}
\includegraphics[width=4cm]{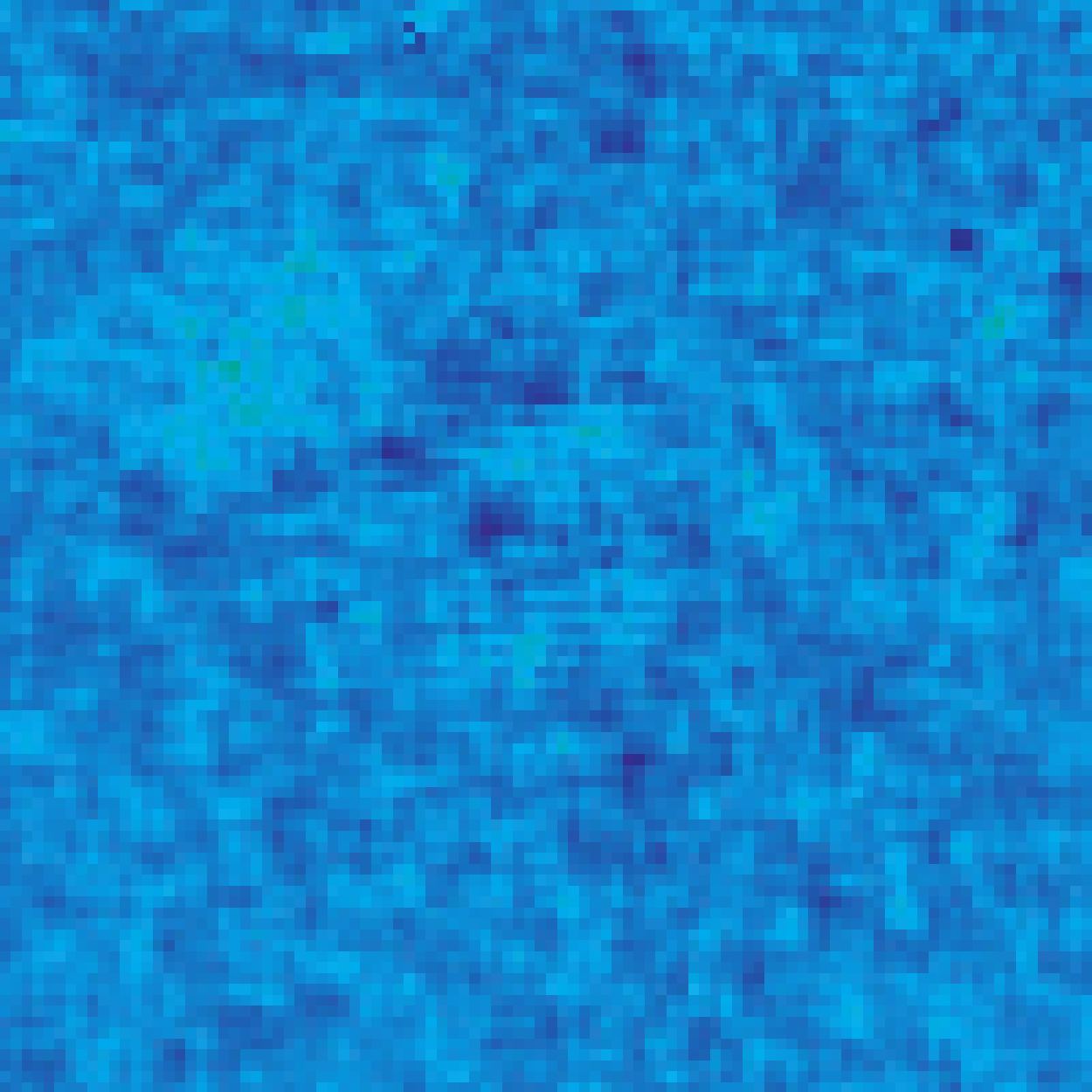}
\caption{Subaru/FOCAS false-color images and model fits. For each galaxy, the panels from left to right show the original image, the best-fit model, the model without PSF convolution, and the residual. The images and models are in power-law scale in order to show the faint outer profiles of the galaxies. The residuals are in linear scale with a different contrast in order to show the small variations across the fields. An one-arcsecond scale and the corresponding physical length at the galaxy redshift is shown in the lower-left corner of every galaxy image. North is up and east to the left for all images.}
\label{focasimages}
\end{figure*}

\begin{figure*}
\centering
\includegraphics[width=4cm]{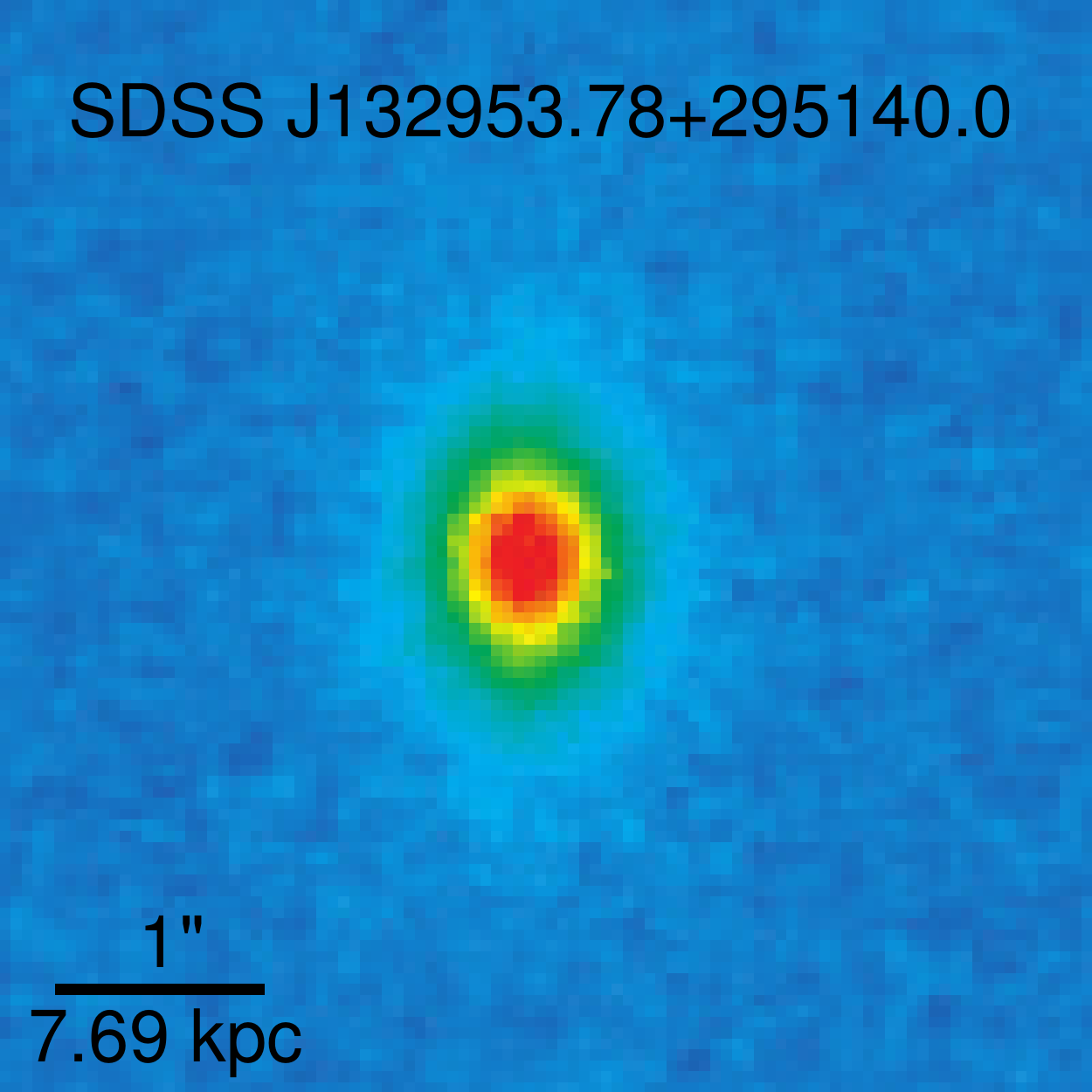}
\includegraphics[width=4cm]{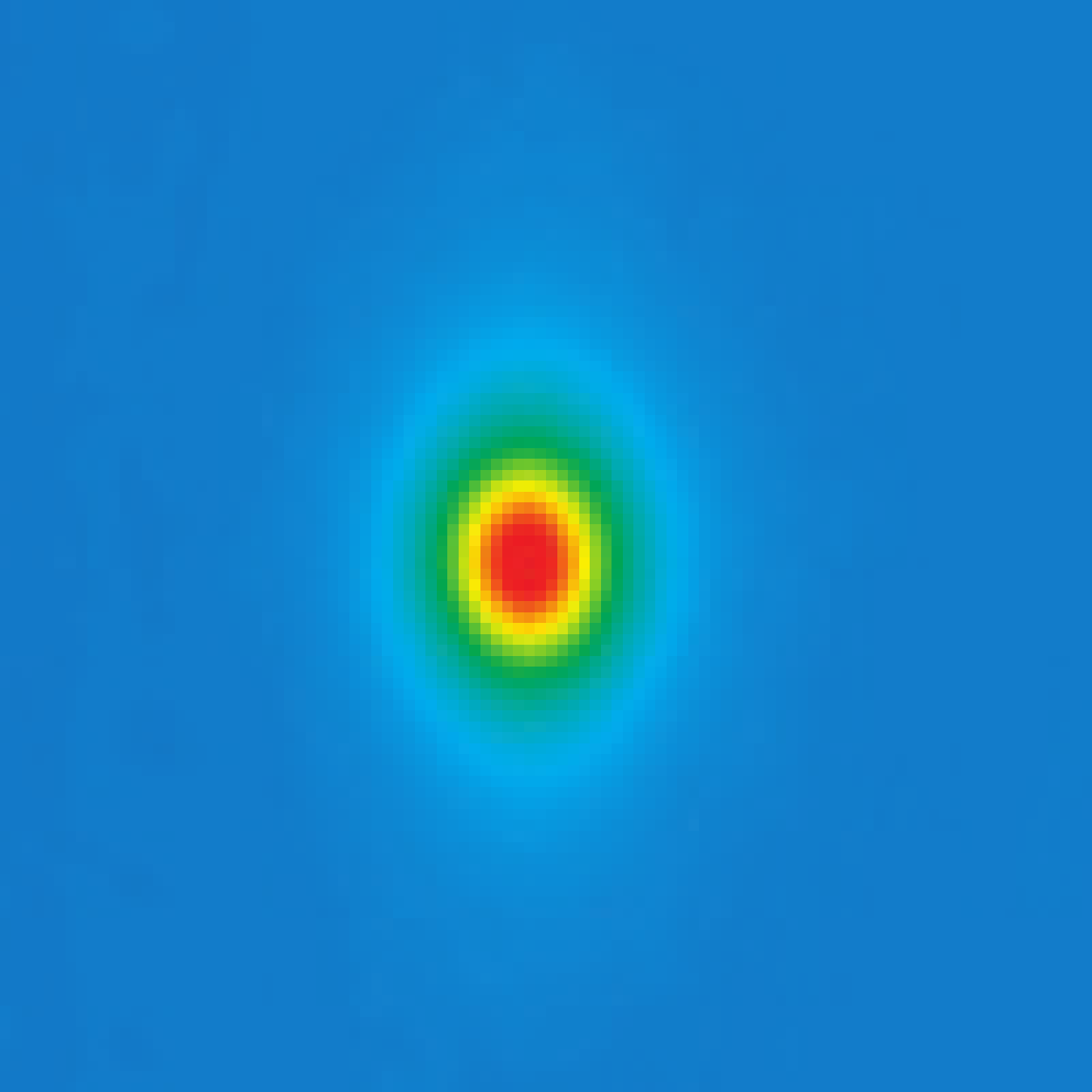}    
\includegraphics[width=4cm]{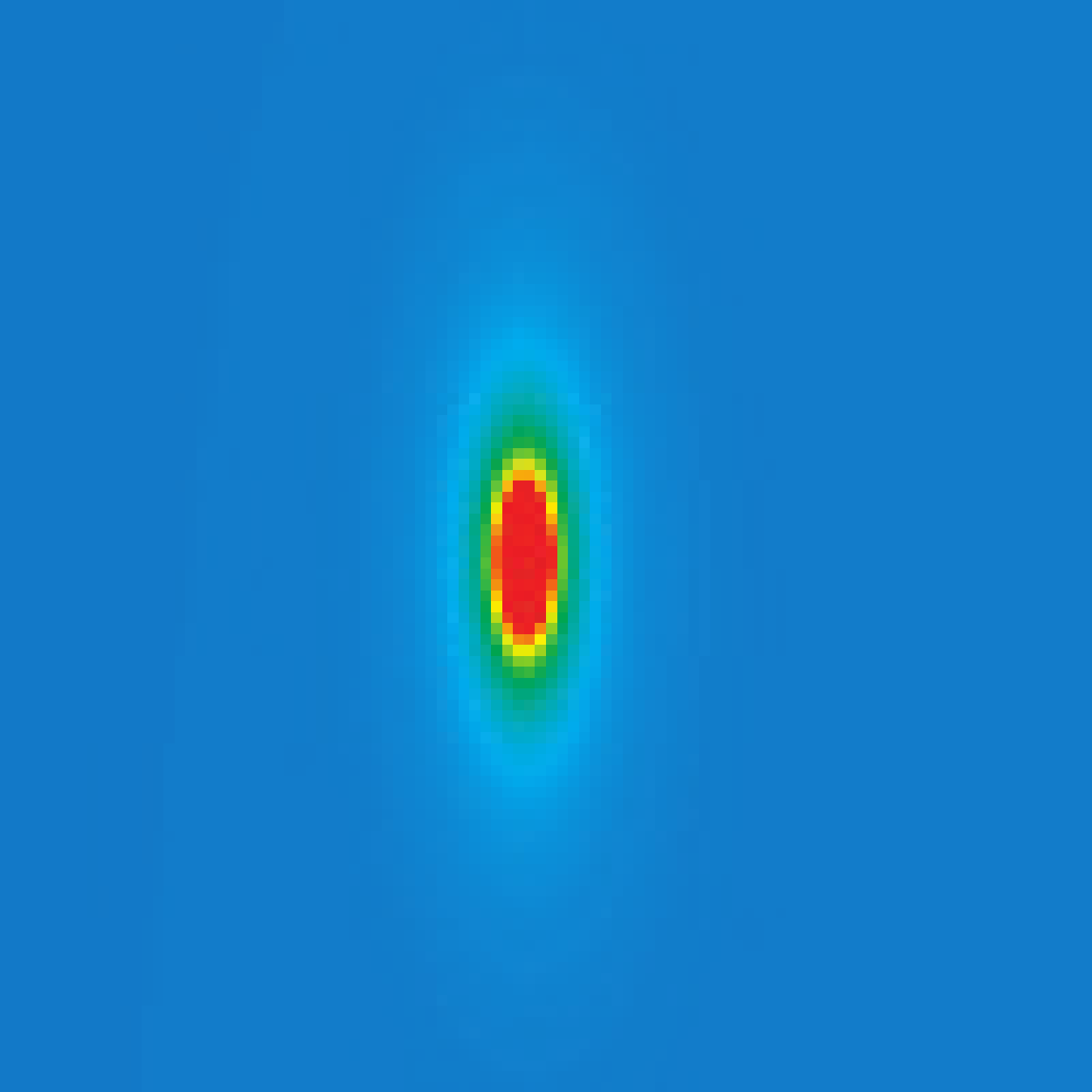}
\includegraphics[width=4cm]{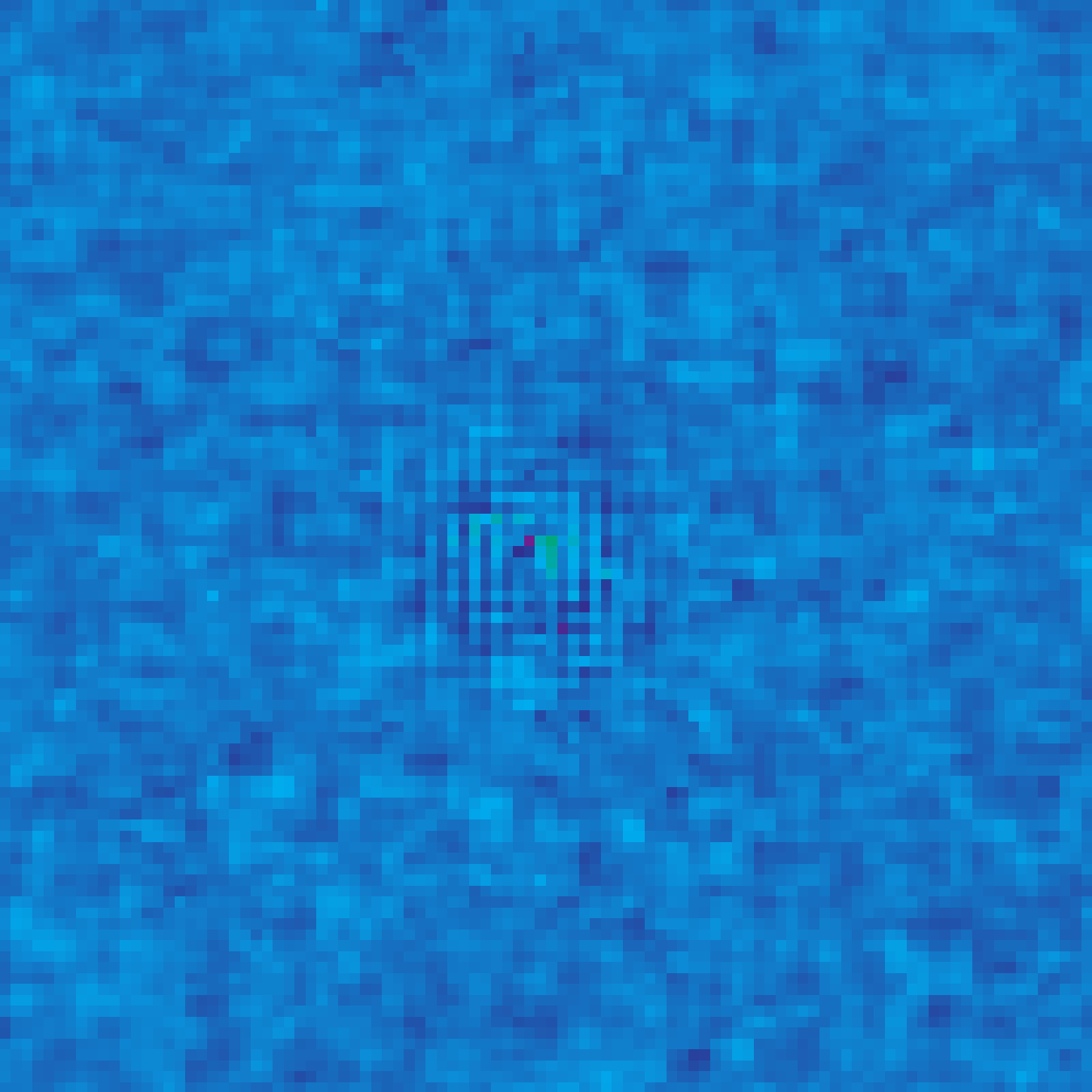}\\[2mm]   
\includegraphics[width=4cm]{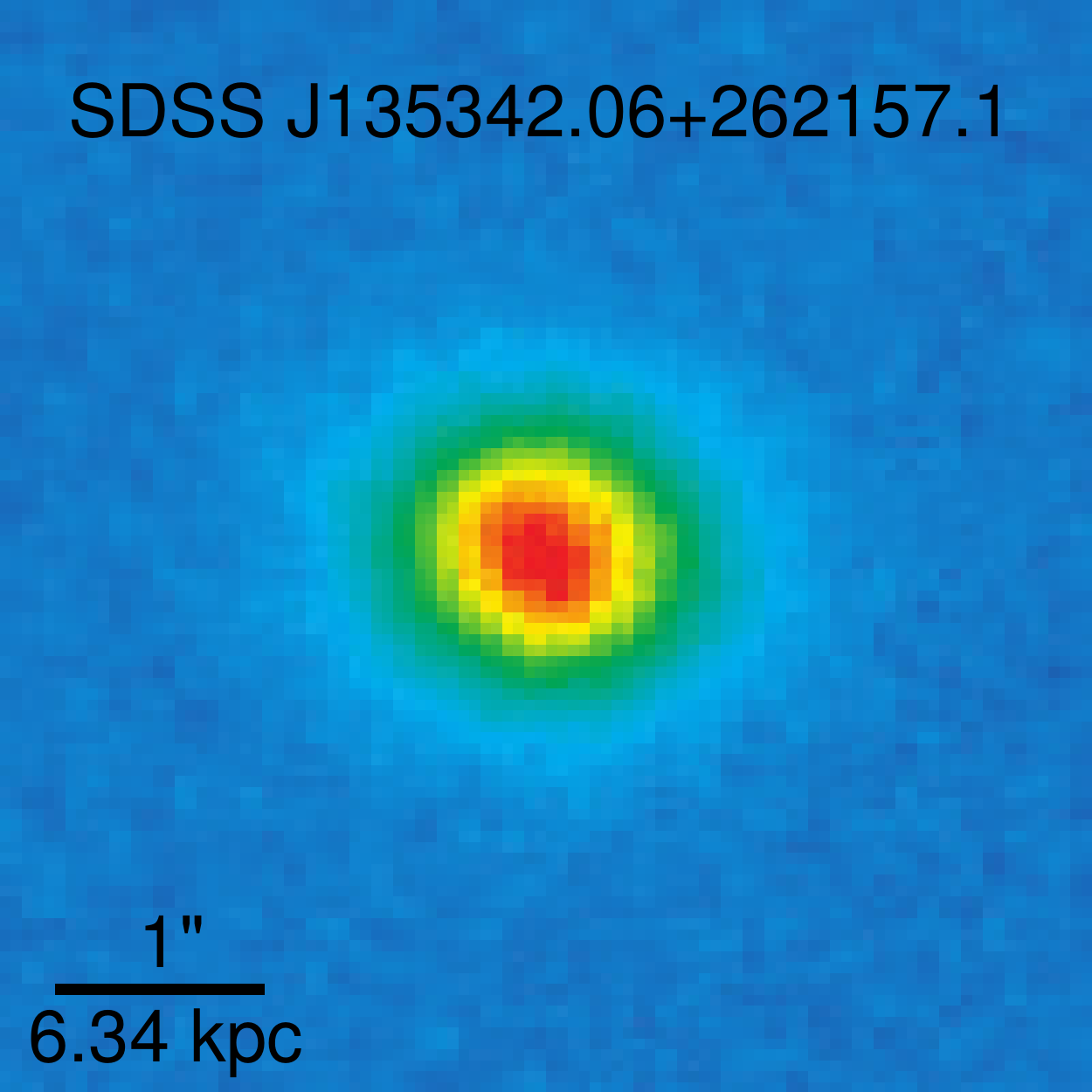}
\includegraphics[width=4cm]{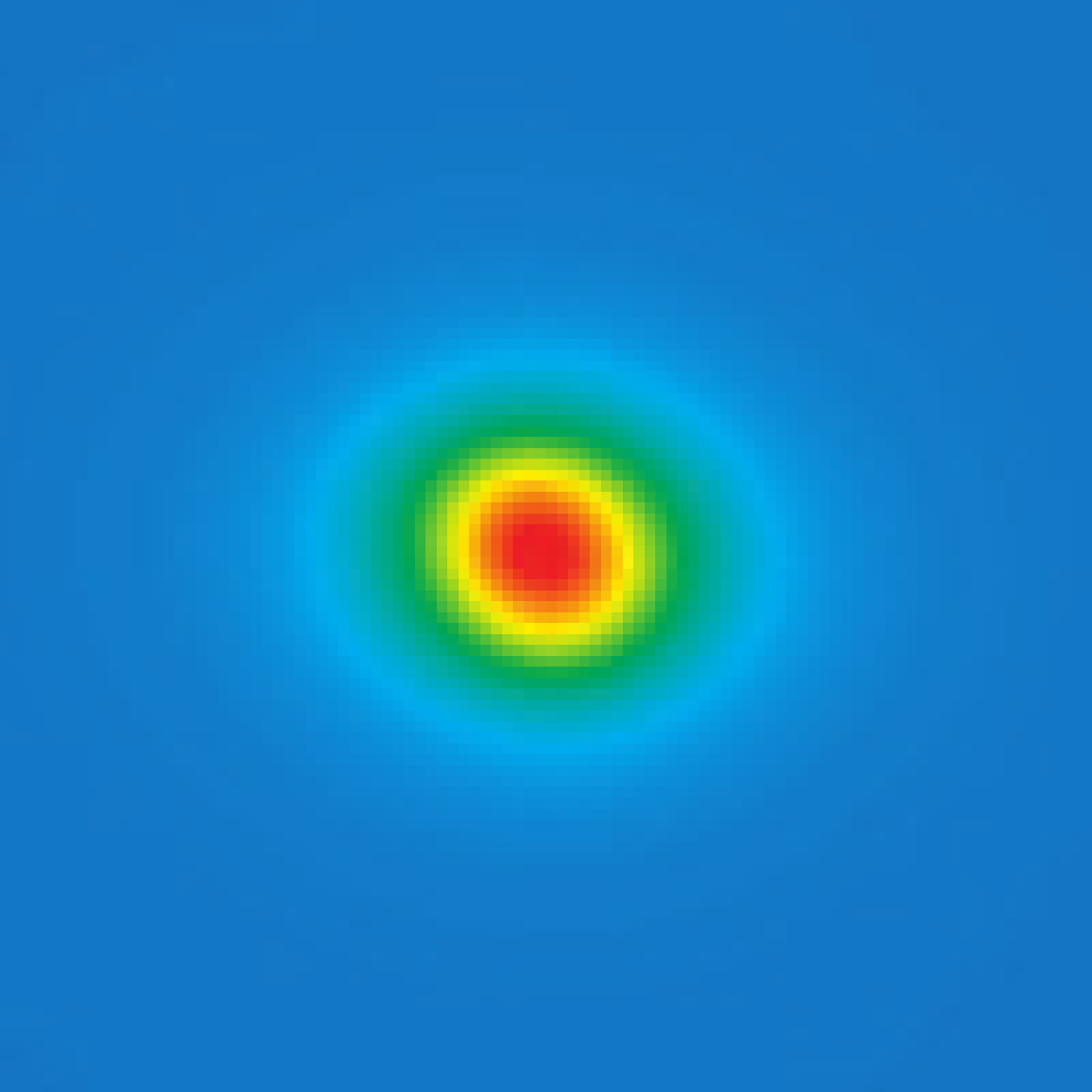}    
\includegraphics[width=4cm]{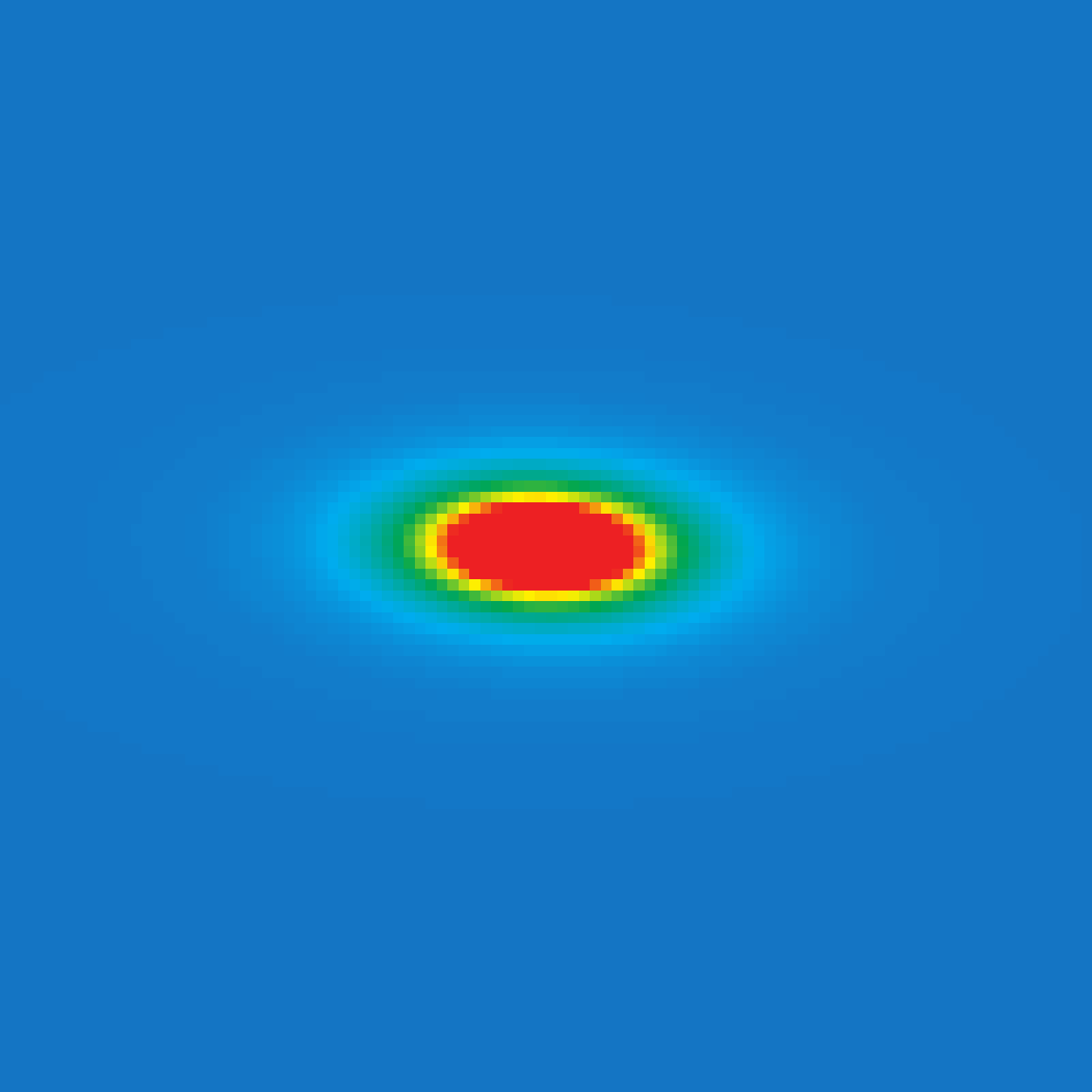}
\includegraphics[width=4cm]{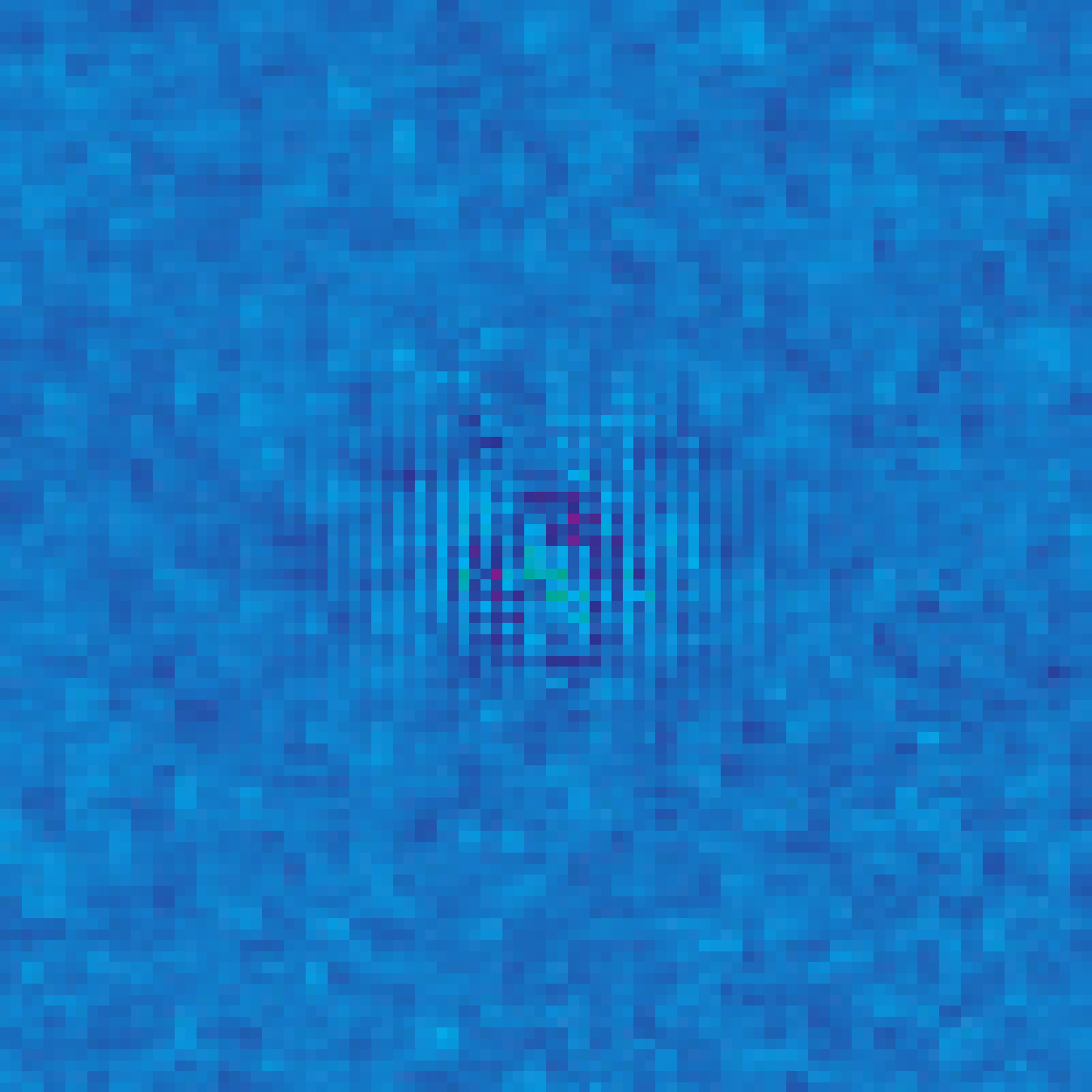}
%\caption{\centering Continued.}
\caption{Continued.}
\end{figure*}

\renewcommand{\thefigure}{A3}

\begin{figure*}
\centering
\includegraphics[width=15.5cm]{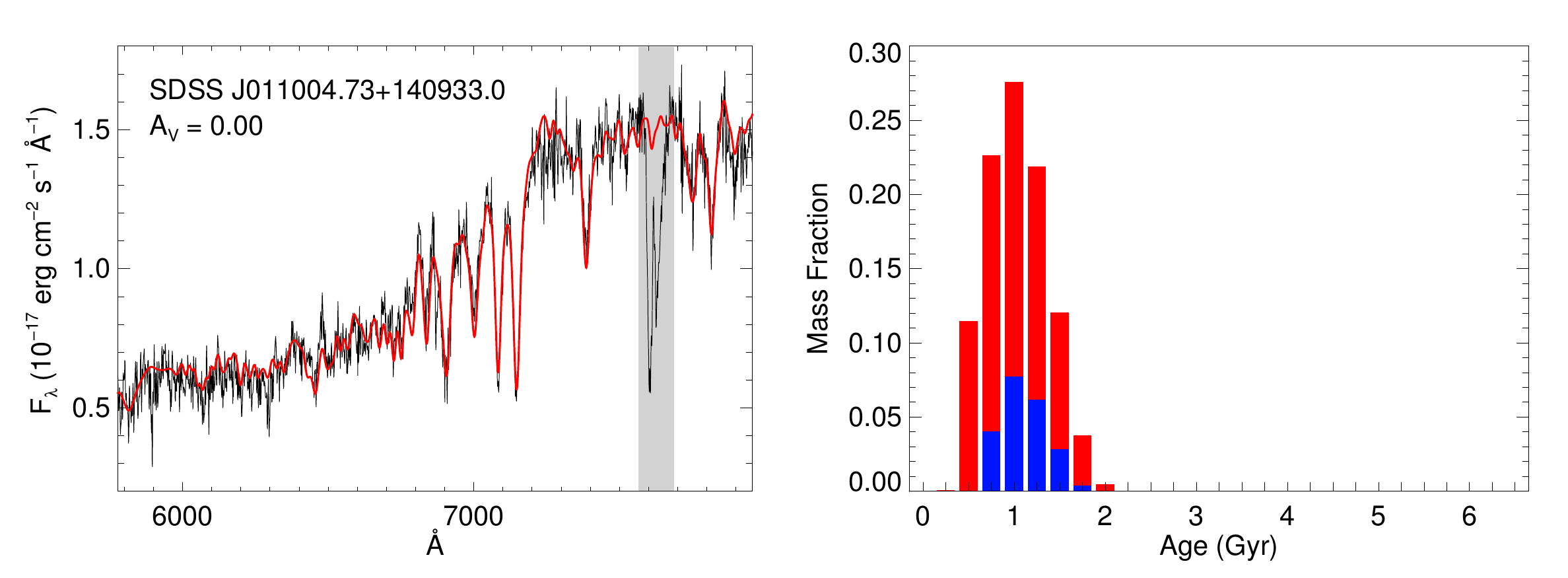}   
\includegraphics[width=15.5cm]{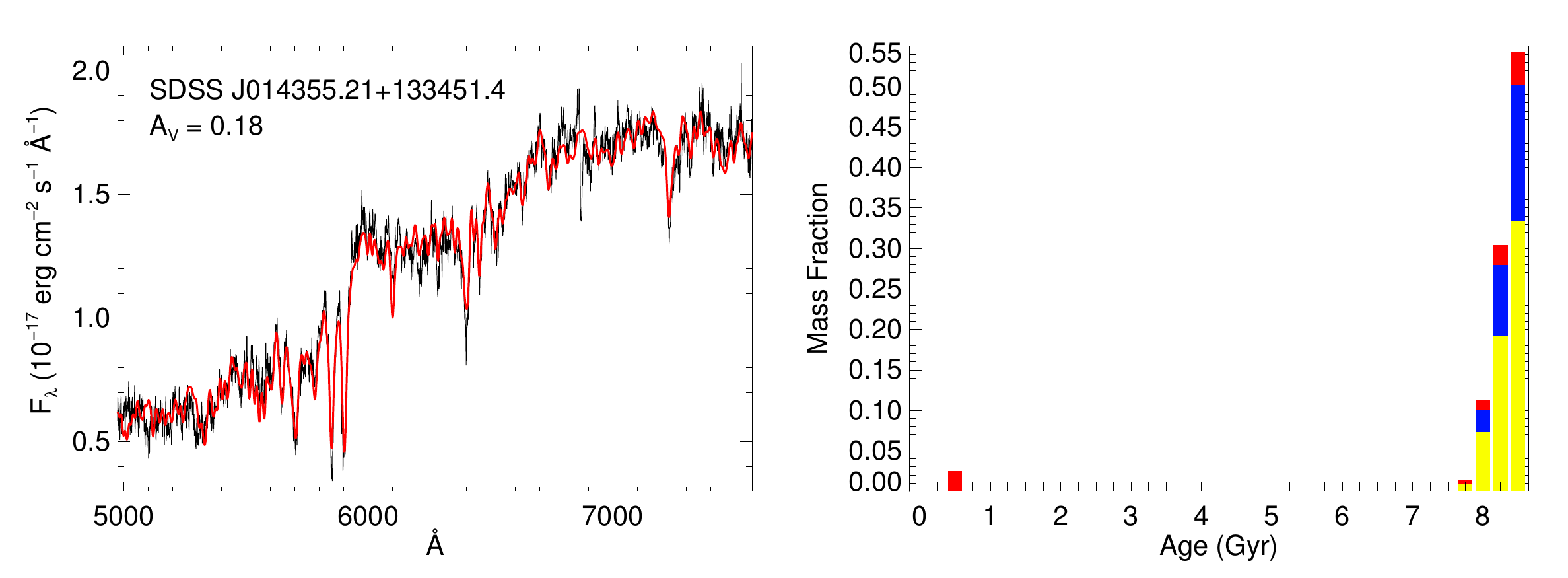}   
\includegraphics[width=15.5cm]{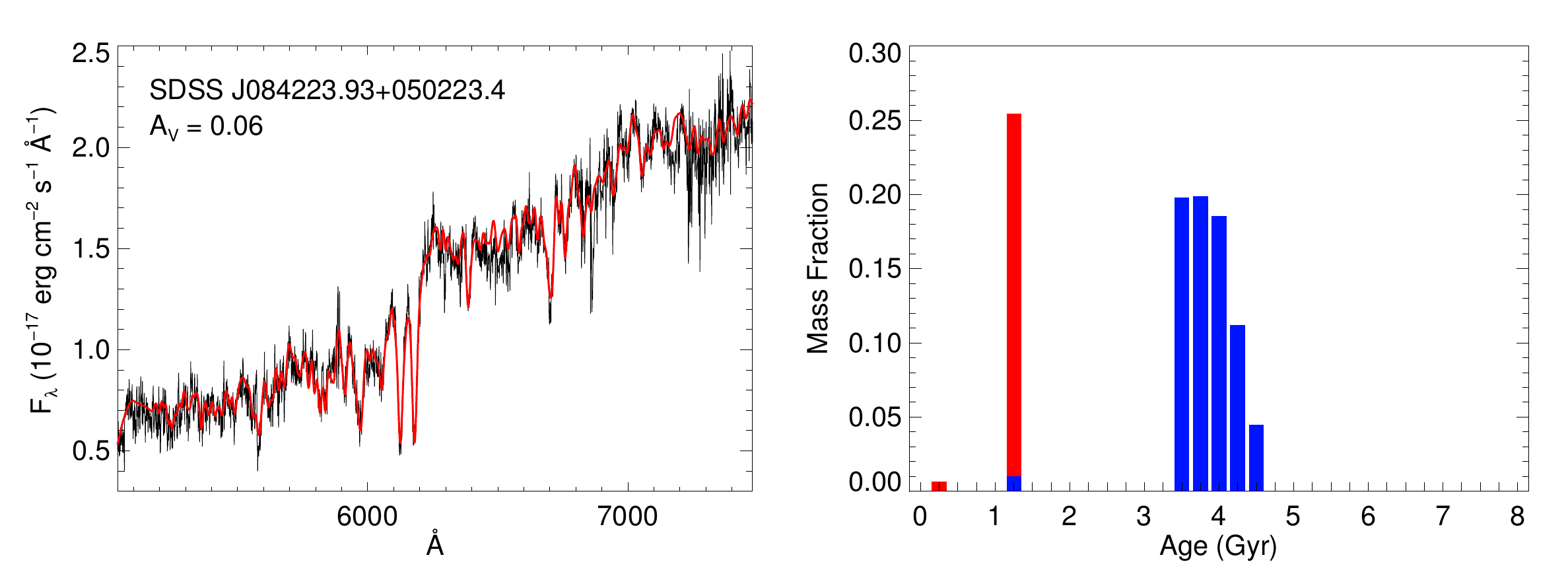}   
\includegraphics[width=15.5cm]{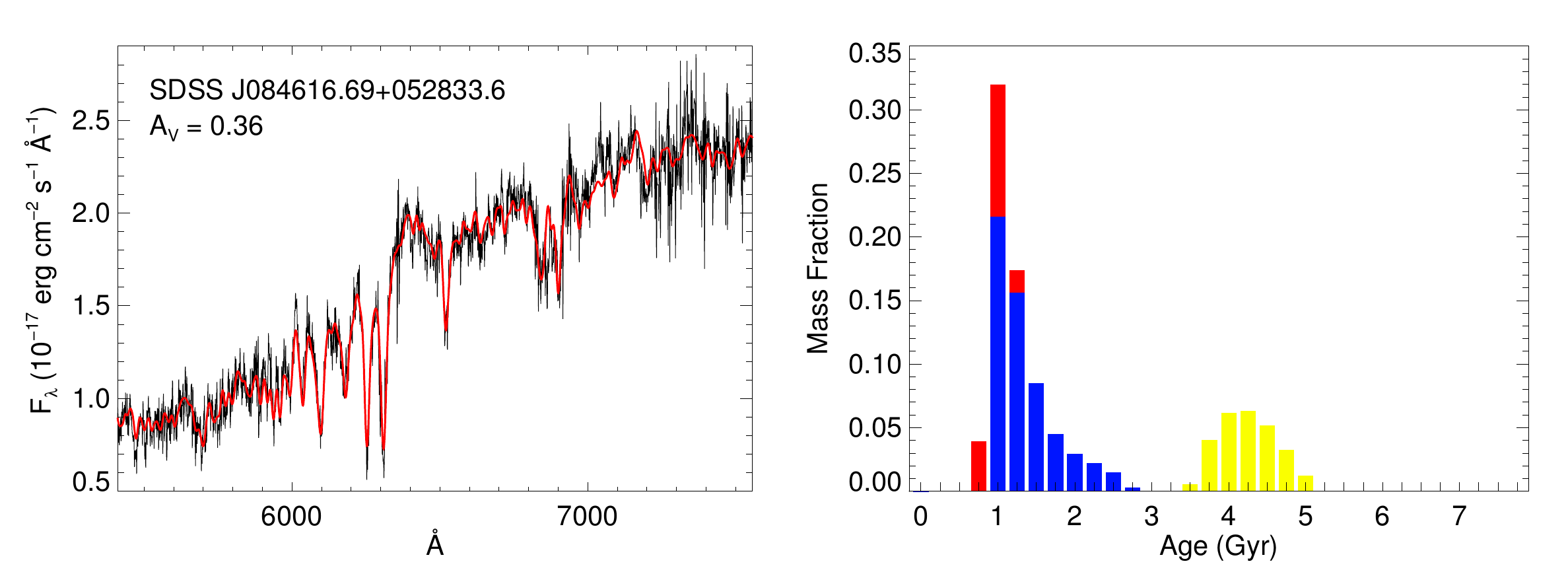}     
\caption{Left panels: full-spectrum fits using pPXF, where black lines are the observed spectra and red lines are the best-fit models. Gray areas represent the masks of bad pixels. Right panels: The best-fit stellar populations, illustrated by the fraction of star formation at a given look-back time. Red, blue and yellow portions of the bars represent the additive mass fractions with metallicity [Z/H] of 0.4, 0.0 and -0.4, respectively (i.e., the mass fraction at a given time is represented by the height of the corresponding column, which is the sum of red, blue and yellow portions of the bar). The sum of all the columns equals one. The last age grid is the oldest age younger than the age of universe at the galaxy redshift.}
\label{spectra}
\end{figure*}

\begin{figure*}
\centering
\includegraphics[width=15.5cm]{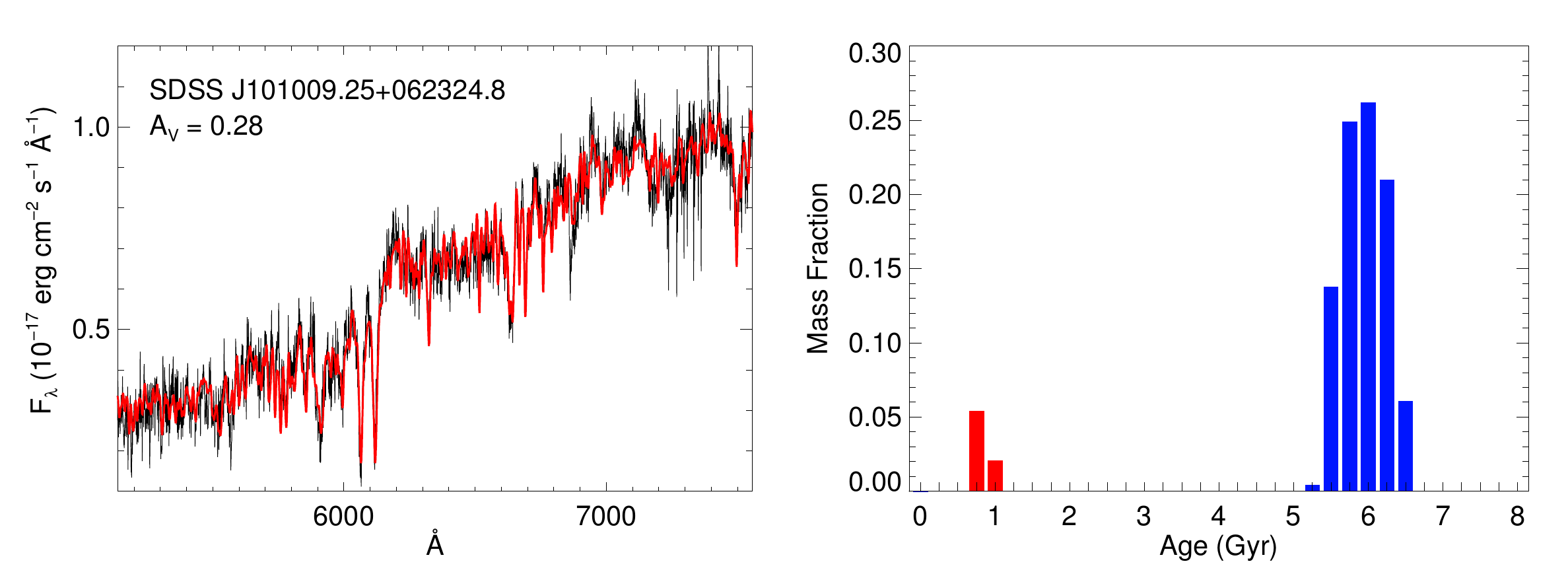} 
\includegraphics[width=15.5cm]{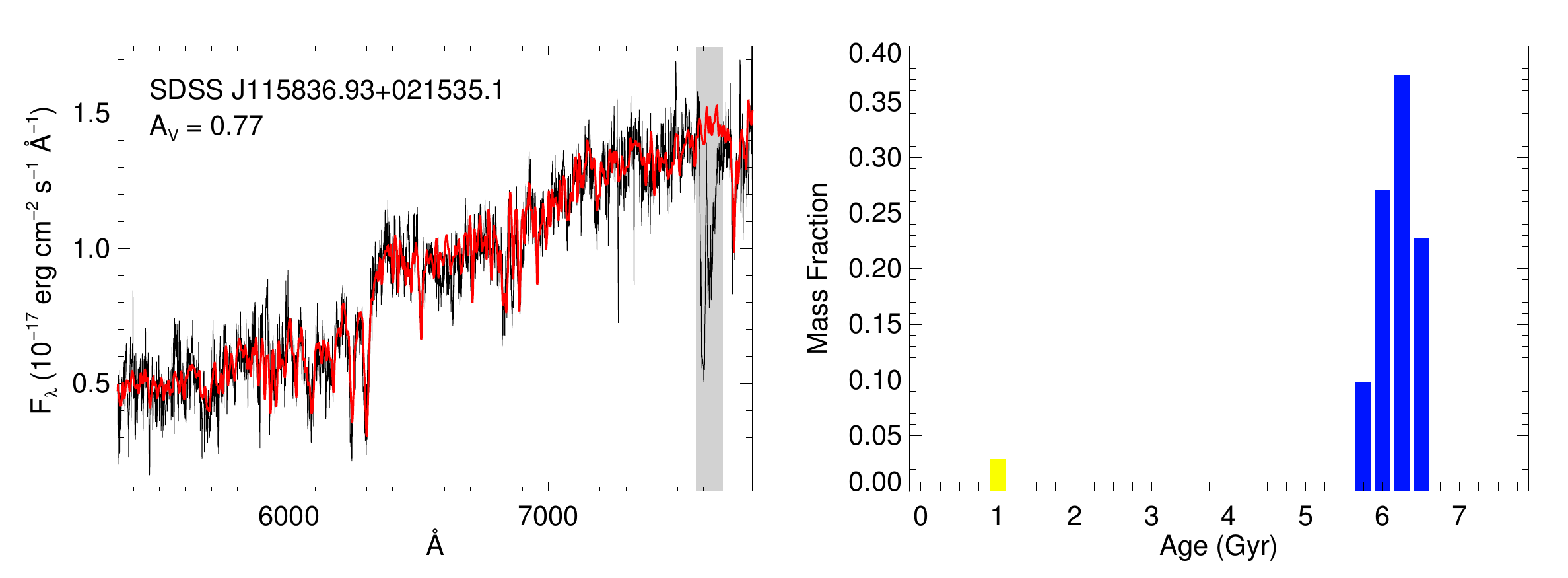}   
\includegraphics[width=15.5cm]{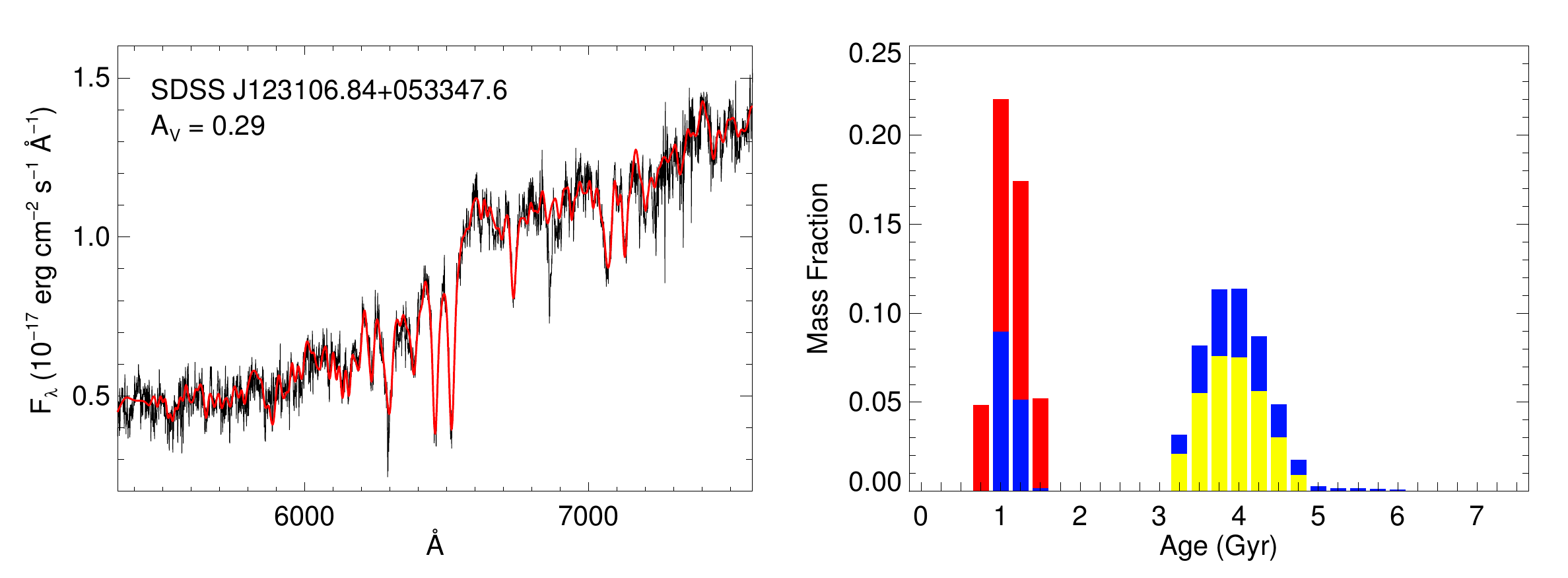}       
\includegraphics[width=15.5cm]{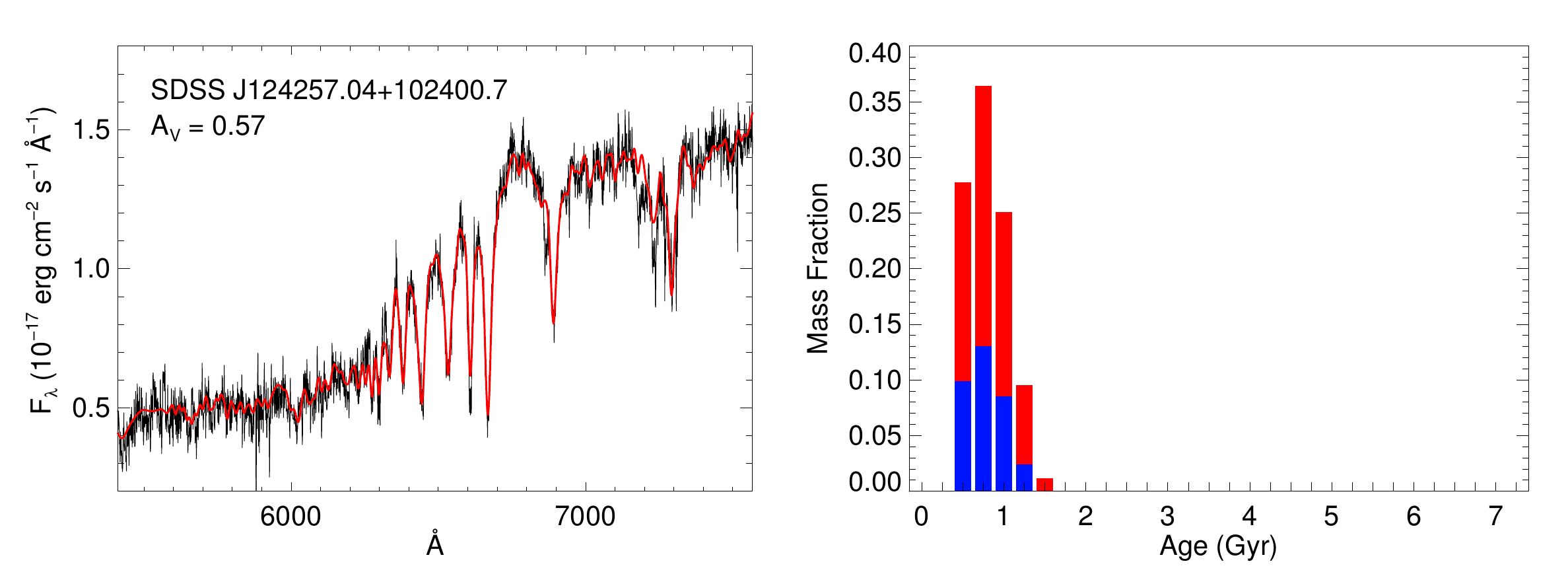}  
%\caption{\centering Continued.}
\caption{Continued.}
\end{figure*}

\begin{figure*}
\centering
\includegraphics[width=15.5cm]{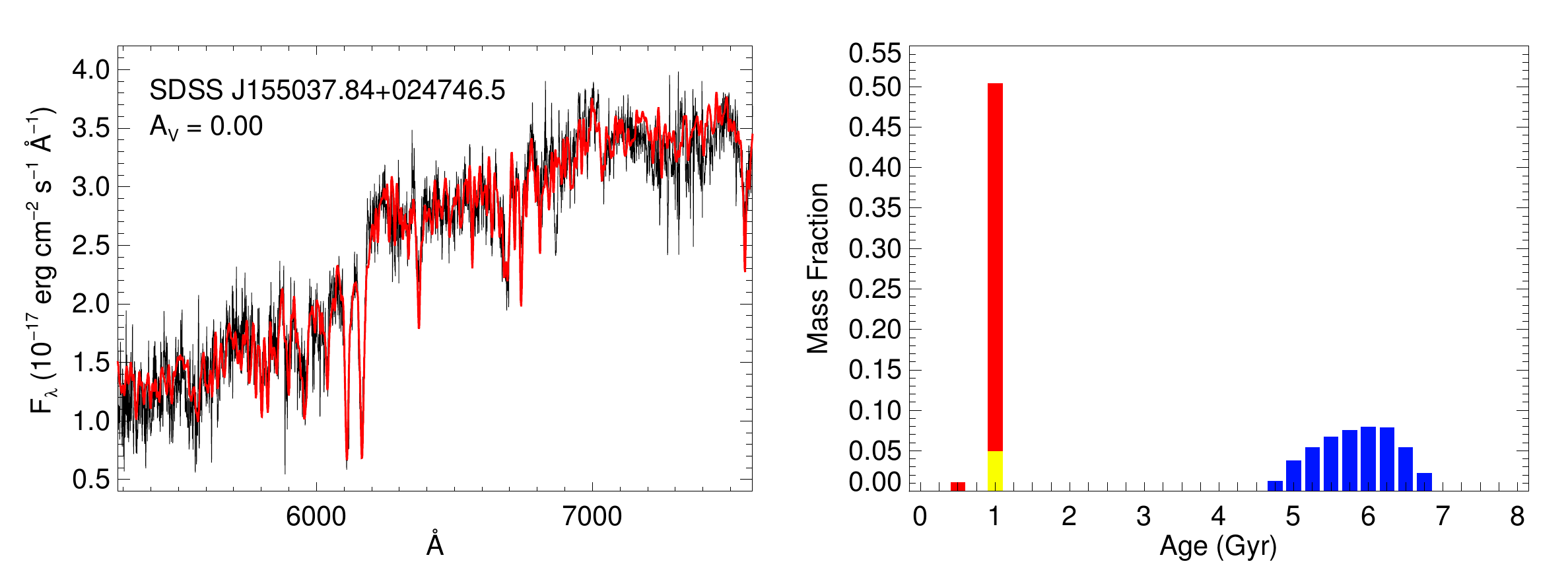} 
\includegraphics[width=15.5cm]{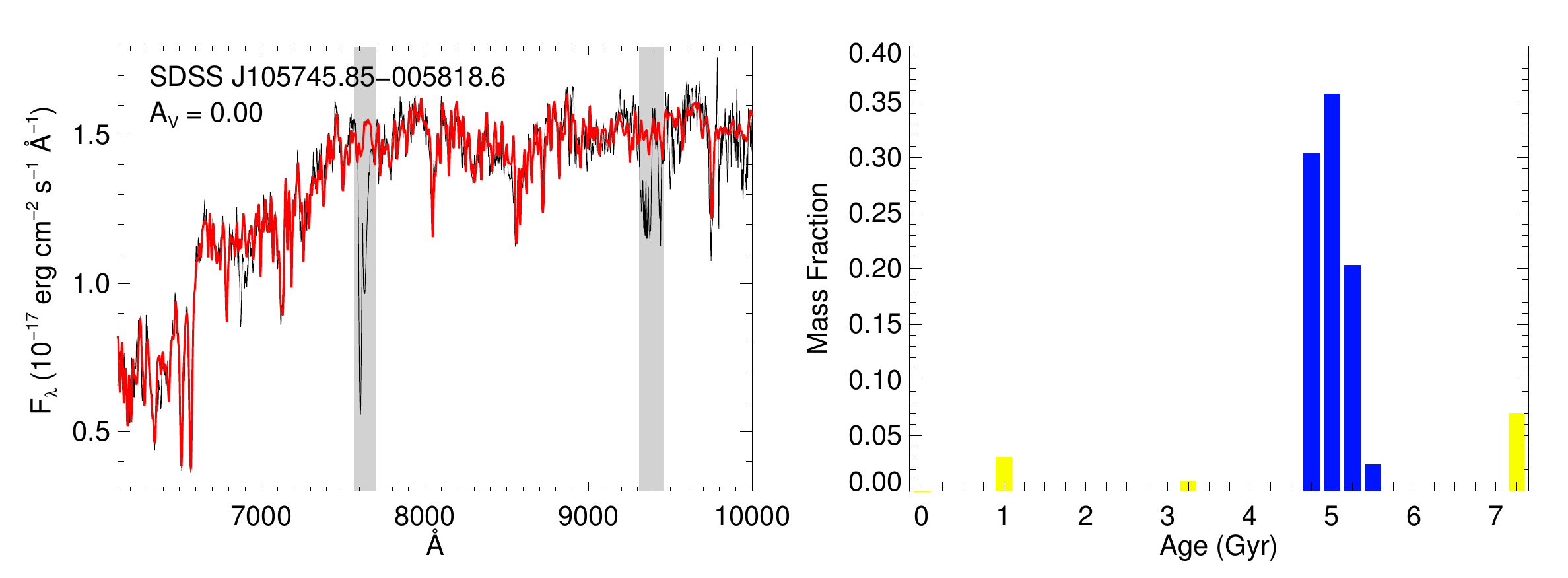} 
%\caption{\centering Continued.}
\caption{Continued.}
\end{figure*}

\end{document}